# Efficient simulation techniques for biochemical reaction networks

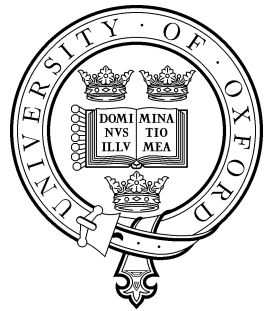

Christopher Lester

Wadham College

University of Oxford

A thesis submitted for the degree of

*Doctor of Philosophy in Mathematics*

Trinity Term, 2017

# Abstract

Discrete-state, continuous-time Markov models are becoming commonplace in the modelling of biochemical processes. The mathematical formulations that such models lead to are opaque, and, due to their complexity, are often considered analytically intractable. As such, a variety of Monte Carlo simulation algorithms have been developed to explore model dynamics empirically. Whilst well-known methods, such as the Gillespie Algorithm, can be implemented to investigate a given model, the computational demands of traditional simulation techniques remain a significant barrier to modern research.

In order to further develop and explore biologically relevant stochastic models, new and efficient computational methods are required. In this thesis, high-performance simulation algorithms are developed to estimate summary statistics that characterise a chosen reaction network. The algorithms make use of variance reduction techniques, which exploit statistical properties of the model dynamics, so that the statistics can be computed efficiently.

The multi-level method is an example of a variance reduction technique. The method estimates summary statistics of well-mixed, spatially homogeneous models by using estimates from multiple ensembles of sample paths of different accuracies. In this thesis, the multi-level method is developed in three directions: firstly, a nuanced implementation framework is described; secondly, a reformulated method is applied to stiff reaction systems; and, finally, different approaches to variance reduction are implemented and compared.

The variance reduction methods that underpin the multi-level method are then re-purposed to understand how the dynamics of a spatially-extended Markov model are affected by changes in its input parameters. By exploiting the inherent dynamics of spatially-extended models, an efficient finite difference scheme is used to estimate parametric sensitivities robustly.

The new simulation methods are tested for functionality and efficiency with a range of illustrative examples. The thesis concludes with a discussion of our findings, and a number of future research directions are proposed.

# Acknowledgements

I would like to acknowledge the unwavering help, encouragement and support provided by Professor Ruth E. Baker and Dr Christian A. Yates, who jointly supervised the research leading to this thesis.

In addition, support from the Mathematical Institute, Oxford, as well as Wadham, St Hilda's, St Peter's and Brasenose Colleges, is gratefully acknowledged.

Additional travel funds were provided by the Society for Mathematical Biology, United States; the Centre National de la Recherche Scientifique, France; the Isaac Newton Institute, Cambridge; and the International Centre for Mathematical Sciences, Edinburgh.

The friends and family that supported me throughout the past four years, have been thanked privately.

# Contents

# Chapter 1

# Introduction

By drawing on computational and mathematical techniques, the field of Systems Biology seeks to understand the dynamics of complicated, but co-ordinated, systems comprising many biological components. Some researchers take the view that such efforts will ultimately be futile [1], whilst others argue that a symbiosis between experimental and theoretical techniques leads to new insights and developments [2, 3, 4, 5, 6]. The mathematical formulations that Systems Biology approaches lead to are often complicated, and can involve large numbers of variables and parameters represented in an opaque, and perhaps intractable, formula. In order to understand, further develop, and explore such mathematical formulations, new and effective computational methods are required.

By relying on probability distributions to explain the temporal evolution of a biological system of interest, a stochastic model can effectively describe a wide range of biological phenomena [7, 8, 9]. Experimental researchers have, for example, demonstrated the stochastic nature of gene regulatory networks [10, 11, 12]. Stochastic models have also been used to replicate and explain inter- and intra-cellular kinet-ics [13, 14, 15], and to study population dynamics in ecology [16]. In particular, stochastic effects often affect systems characterised by low molecular populations [17], but systems with large molecular populations can also be affected under certain circumstances [18]. Whilst traditional, deterministic models are able to describe a wide range of biological processes satisfactorily [19], such models may provide misleading predictions as they are unable to account for phenomena such as dynamic system bistability [17], stochastic focussing [20] and stochastic resonance [21].

This thesis will focus on the design and development of simulation algorithms that can be used to explore the dynamics of stochastic models.

A multitude of stochastic modelling frameworks have been developed [8]. In particular, advances in computational power have made it possible to develop individual-based models that enable the study of the behaviour of individuals 'particles' (often described as 'agents') that comprise a biological system of interest [9, 22]. In this thesis, the models that we will study will be described by the 'chemical master equation' (the 'CME') and the 'reaction-diffusion master equation' (the 'RDME'). The CME is a population-level model that records the total number of particles of a particular type over a time-period of interest [23, 24]. The RDME extends the CME framework, as the RDME additionally records the physical locations of the system particles, to a suitable level of accuracy [25, 26]. The RDME therefore describes a spatially-inhomogeneous model, whilst the CME describes a spatially-homogeneous or 'well-stirred' model.

The CME and RDME frameworks are both examples of discrete-state, continuous-time Markov chains [27]. For very simple systems, closed-form, analytic solutions to the CME and RDME can be obtained [28]; however, any complication (for example, a bimolecular reaction) is likely to frustrate an analytic approach [9, 23, 24]. Under particular circumstances, specialised numerical approaches may be feasible [28, 29, 30, 31, 32], but, usually, the high dimensionality of the problem remains a challenge. Therefore, the dynamics described by the CME and RDME are widely considered to be both analytically and numerically intractable for the vast majority of physically-relevant models.

In this thesis, Monte Carlo simulation will be used to understand the dynamics of CME and RDME models. Through the use of a stochastic simulation algorithm (an 'SSA'), an *ensemble* of sample paths is generated. Each sample path provides a possible realisation or example of how the model might evolve over time. The resultant ensemble of sample paths is then used to characterise the model dynamics. Whilst it is possible to describe the data qualitatively, we will focus on a quantitative description of CME or RDME model behaviour by considering the values of summary statistics of interest. By way of example, we might be interested in the mean number of particles of a specific type at a given time. Other summary statistics, such as the mean extinction time of a particular particle type, might also be of interest.

The mostly widely used SSA is the Gillespie Direct Method [33] (the 'DM'). The DM can be implemented in a straightforward way, but requires a relatively high level of computer resources to generate the required sample paths. Therefore, a wide miscellany of highly efficient and improved SSA methods have been developed [23, 24].

Many of the improved SSAs are formally equivalent to the DM because the statistics of any sample paths that are generated share the same distribution as the statistics of sample paths generated with the DM. For systems with multiple, concurrent timescales, in the interests of computational efficiency, a range of bespoke SSAs have been developed [34, 35, 36, 37], though these methods are not formally equivalent to the DM.

Numerous SSAs have now been included in user-friendly toolkits [38, 39, 40], including software packages that can be accessed on the internet [41]. Nonetheless, in order to arrive at a Monte Carlo estimate quickly, it might still be necessary to rely on central processing unit (CPU) or graphics processing unit (GPU) parallelisation techniques [42, 43], and, absent substantial computational resources, we remain at a stage where characterising the full range of behaviours of many stochastic biochemical models lies beyond our reach.

The aim of this thesis is to reduce the computational cost of Monte Carlo simulation through the use of variance reduction techniques. Variance reduction techniques exploit statistical properties of the dynamics of a model of interest so that the required summary statistics can be estimated by using relatively small-sized ensembles of sample paths [44].

By reducing the computational cost of Monte Carlo simulation, we make the following contributions to the field. Firstly, where the computational cost of simulating a stochastic model had previously been prohibitive, a deterministic, ordinary differential equation model had to be used instead. It might now possible to use a stochastic, individual-based model, which provides a more accurate representation of the biological process of interest. Secondly, given a constant level of computing resources, tighter confidence intervals on summary statistics can be obtained through the use of our new simulation methods. Thirdly, the interactions between components of our model can be better understood. For example, through the use of our new algorithms, parameter sweeps can be carried out in greater detail, and model robustness can be more clearly assessed and understood.

This thesis is structured as follows:

**Chapter 2: Introduction.** In this chapter, we introduce the CME and RDME

modelling frameworks. Established SSA methods, such as the Gillespie Direct [33] and tau-leap [45] methods, are then discussed.

**Chapter 3: The multi-level method.** The multi-level method is introduced and described [46, 47]. We explain that the multi-level method uses estimates from multiple ensembles of sample paths of different accuracies to estimate a summary statistic of interest. We demonstrate that the multi-level method can reduce the CPU time required to estimate a given summary statistic by orders of magnitude. The time-savings brought about by using the method can be attributed to the repeated application of a variance reduction technique; therefore, a key aim of this chapter is to explain and implement the variance reduction principle. A variety of implementation issues are carefully considered, and numerical case studies are presented.

**Chapter 4: The *adaptive* multi-level method.** The multi-level method presented in Chapter 3 can result in poor performance when the reaction activity of a system changes substantially over the time-scale of interest. We therefore set out a new, generalised, *adaptive* multi-level method. Our new method is particularly suited to studying the summary statistics of stiff systems, and can be run with only a minimal amount of user input. Further case studies are presented.

**Chapter 5: The *robust* multi-level method.** In this chapter, we develop the multi-level method in two directions: (1) to increase the robustness, reliability and performance of the multi-level method, we suggest an improved variance reduction method for generating the sample paths of each ensemble; and (2) to improve computational performance, we provide a different mechanism for choosing which ensembles should be included in the multi-level algorithm.

**Chapter 6: Parameter sensitivity analysis for spatially-extended models.** In order to understand how the dynamics of a reaction-diffusion model are affected by changes in its input parameters, efficient methods for computing parametric sensitivities are required. Whilst highly efficient parameter sensitivity methods have been developed for well-mixed, CME models, the theory is less well-developed for RDME models. In this chapter, we exploit the characteristic dynamics of spatially-extended reaction networks to efficiently and robustly estimate parametric sensitivities in such models. Different variance reduction approaches are discussed, and we then describe a hybrid technique that dynamically chooses the most appropriate simulation method.

**Chapter 7: Discussion.** In this final chapter, the findings (and limitations) of this thesis are discussed in greater detail. A number of directions for future work are highlighted, and the thesis is then drawn to a close.

The following three themes characterise the approach taken throughout this thesis:

**Usability.** Throughout, we seek to develop methods that can be widely applicable. As far as possible, our approach will be modular, so that individual elements of our work can be re-purposed and re-used as required. As part of our modular approach, our focus is on developing simulation algorithms that can applied to any appropriately formulated biological model.

**Efficiency.** The simulation algorithms that we describe aim to be efficient. Therefore, with a constant level of computing resources, more detailed and in-depth information regarding a biological model can be obtained.

**Understanding.** As part of our development of efficient simulation algorithms, we seek to understand the role that stochastic noise plays in influencing the behaviour of a CME or RDME model. This deeper comprehension of our modelling framework is essential in understanding how different variance techniques might be used to accelerate our algorithms.

The material contained within this thesis has been adapted from the following four manuscripts:

1. Lester, C., Baker, R. E., Giles, M. B., and Yates, C. A. Extending the multi-level method for the simulation of stochastic biological systems. *Bulletin of Mathematical Biology*, **78**(8):1640–1677, 2016;

2. Lester, C., Yates, C. A., Giles, M. B., and Baker, R. E. An adaptive multi-level simulation algorithm for stochastic biological systems. *Journal of Chemical Physics*, **142**(2):024113, 2015;

3. Lester, C., Yates, C. A., and Baker, R. E. Efficient parameter sensitivity computation for spatially-extended reaction networks. *Journal of Chemical Physics*, **146**(4):044106, 2017;

4. Lester, C., Yates, C. A., and Baker, R. E. Robustly simulating chemical reaction kinetics with multi-level Monte Carlo. *arXiv preprint arXiv:1707.09284*, 2017.

By developing new and efficient computational methods, this thesis provides significant increases in our means to understand and explore complicated biological models, thereby providing the potential to investigate models that would otherwise be seen as intractable.

In order to establish our new approaches to Monte Carlo simulation, we now start by introducing the CME and RDME modelling frameworks, and describing existing simulation tools.

# Chapter 2

# Stochastic modelling and simulation

In this chapter, we discuss and extend the theory that underpins the research chapters of this thesis. We introduce the chemical master equation and the random time change representation (the 'RTCR'), and we explain that the CME and RTCR provide equivalent mathematical descriptions of a biochemical reaction network. We then motivate the need for Monte Carlo simulation, and proceed to discuss a range of simulation algorithms that can be implemented to estimate suitable summary statistics. We discuss both 'exact' and 'approximate' algorithms, and undertake numerical investigations. Finally, we describe an extension of the CME that provides a framework for the stochastic modelling of spatially-inhomogeneous biochemical networks.

## 2.1 Well-stirred biochemical reaction networks

We initially concentrate on spatially-homogeneous, population-level models, which record only the numbers of each molecule type within the system. The temporal evolution of the molecular populations is fully described by a CME, which comprises a system of ordinary differential equations (ODEs). For each possible system state, the CME provides an ODE describing how the probability of the system being in this particular state changes over time [8, 9, 23, 24, 52].

We consider a biochemical network comprising $N$ species, $S_1,\ldots,S_N$, that may each be involved in $M$ possible interactions, $R_1,\ldots,R_M$, which are referred to as reaction channels. Initially, we assume that the system is 'well-stirred', which means that we do not take into account the evolution of molecule positions [8]. Strictly speaking, the system is well-stirred if the particles are randomly (and uniformly) distributed throughout the volume of interest. Assuming that the system is well-stirred would be reasonable if, for example, the diffusion of molecules occurs on a time-scale far faster than that of the reactions between molecules [33, 53]. The population size of $S_i$ (for $i = 1,\ldots,N$) is known as its copy number and is denoted by $X_i(t)$ at time $t$, where $t \geq 0$. The state vector is then defined as [33]

$$\boldsymbol{X}(t) := \begin{bmatrix} X_1(t) \\ \vdots \\ X_N(t) \end{bmatrix}. \qquad (2.1)$$

We will refer to the sample path $\boldsymbol{X}$ as a stochastic process.

With each reaction channel, $R_j$ (for $j = 1,\ldots,M$), we associate two quantities. The first is the stoichiometric or state-change vector,

$$\boldsymbol{\nu}_j = \begin{bmatrix} \nu_{1j} \\ \vdots \\ \nu_{Nj} \end{bmatrix}, \qquad (2.2)$$

where $\nu_{ij}$ is the change in the copy number of $S_i$ caused by reaction $R_j$ taking place. Thus, if the system is in state $\boldsymbol{X}$ and reaction $R_j$ 'fires', the system jumps to state $\boldsymbol{X} + \boldsymbol{\nu}_j$. The second quantity is the propensity function, $a_j$. This represents the average rate at which a reaction fires (i.e. takes place). Formally, for small $\mathrm{d}t$, and based on a state vector of $\boldsymbol{X}(t)$, we define $a_j(\boldsymbol{X}(t))$ as follows:

- the probability that reaction $R_j$ fires exactly once during the infinitesimal interval $[t, t + \mathrm{d}t)$ is $a_j(\boldsymbol{X}(t))\mathrm{d}t + \mathcal{O}(\mathrm{d}t^2)$;

- the probability of reaction $R_j$ firing more than once during this interval is $\mathcal{O}(\mathrm{d}t^2)$.

Since we have assumed that the system is well-stirred, it seems reasonable for the propensity function of reaction $R_j$, $a_j$, to be proportional to the number of possible combinations of reactant molecules in the system [9, 24].

**Example 2.1.** We expect that a reaction of the type $S_1 \rightarrow S_2$, where one $S_1$ molecule becomes one $S_2$ molecule, will broadly occur at a rate proportional to the abundance of $S_1$. In second-order reactions, such as $S_1 + S_2 \rightarrow S_3$, the rate should be proportional to the abundance of pairs of $(S_1, S_2)$ molecules. ∎

The volume of interest, $\Omega$, also affects the propensities: a larger volume would mean that the particles collide less frequently, thereby leading to fewer reactions taking place. We are therefore implementing *mass action kinetics* [9]; full details are given in Table 2.1.

We note that the propensity functions defined in Table 2.1 are time-independent, with any change in biological populations resulting in immediate updates to propensity functions. A range of methods have been developed to delay updating the propensity functions by a number of units of time for the cases where such methods can provide a better description of complex, multi-scale biochemical models [15, 54]. Following a small number of alterations, the methods developed throughout this thesis will be able to model the effects of delayed updates to propensity functions.

| Reaction | Example | Propensity |
| --- | --- | --- |
| Zero-order | $\emptyset \xrightarrow{c_1} S_1$ | $c_1 \cdot \Omega$ |
| First-order | $S_1 \xrightarrow{c_2} S_2$ | $c_2 \cdot X_1$ |
| Second-order | $S_1 + S_2 \xrightarrow{c_3} S_3$ | $c_3 \cdot X_1 \cdot X_2 / \Omega$ |
| Homo-dimer formation | $2S_1 \xrightarrow{c_4} S_4$ | $c_4 \cdot X_1 \cdot (X_1 - 1) / \Omega$ |
| Third-order | $3S_1 \xrightarrow{c_5} S_5$ | $c_5 \cdot X_1 \cdot (X_1 - 1) \cdot (X_1 - 2) / \Omega^2$ |

Table 2.1: Sample reaction propensities for a stochastic system embedded in a volume $\Omega$. Note for the propensity of homo-dimer formation we have adopted the common practice of absorbing the multiplier $1/2$ into $c_4$; we have also absorbed the multiplier $1/6$ into $c_5$. The values of $c_j$ (for $j = 1, \ldots, M$) are the rate constants.

Our approach to understanding the dynamics of the system comes from considering how the probability that the system is in a particular state changes through time. Define

$$\mathbb{P}[\boldsymbol{x}, t \mid \boldsymbol{x_0}, t_0] := \mathbb{P}\left[\boldsymbol{X}(t) = \boldsymbol{x}, \text{ given } \boldsymbol{X}(t_0) = \boldsymbol{x_0}\right]. \qquad (2.3)$$

By considering the possible changes in species numbers brought about by a single reaction taking place, it is possible to arrive at the aforementioned CME [55]:

$$\frac{\mathrm{d}\mathbb{P}[\boldsymbol{x}, t \mid \boldsymbol{x_0}, t_0]}{\mathrm{d}t} = \sum_{j=1}^{M} \left[ \mathbb{P}[\boldsymbol{x} - \boldsymbol{\nu}_j, t \mid \boldsymbol{x_0}, t_0] \cdot a_j(\boldsymbol{x} - \boldsymbol{\nu}_j) - \mathbb{P}[\boldsymbol{x}, t \mid \boldsymbol{x_0}, t_0] \cdot a_j(\boldsymbol{x}) \right]. \quad (2.4)$$

In very simple cases, Equation (2.4) can be directly integrated to determine $\mathbb{P}[\boldsymbol{x}, t \mid \boldsymbol{x_0}, t_0]$. Unfortunately, this is not possible in general, and different methods need to be used to analyse the dynamics of biochemical reaction networks of interest [9, 23, 24].

### 2.1.1 The random time change representation

An alternative mathematical description of the temporal evolution of the state vector, $\boldsymbol{X}(t)$, is given by the RTCR, which was first described by Kurtz [56]. When the RTCR is used, the dynamics of the biochemical network are described by a set of inhomogeneous Poisson processes. We first describe the Poisson process, and we then proceed to discussing the RTCR.

Let $\mathcal{Y}^\lambda$ describe a Poisson process, where the constant rate parameter $\lambda$ has been specified. A Poisson process is a counting process that tracks the occurrences of events (known as 'arrivals') during a time-period of interest. The number of 'arrivals' over the time-interval $[0, t)$ can be represented as $\mathcal{Y}^\lambda(0, t)$. The time between successive occurrences (known as the 'inter-arrival time') is an exponential random variate with rate $\lambda$, $\text{Exp}(\lambda)$, and the number of arrivals within the interval $[0, t)$ is Poisson random variate, $\mathcal{P}(\lambda \cdot t)$, with mean $\lambda \cdot t$ [27, 57, 58]. The Poisson process, $\mathcal{Y}^\lambda$, is homogeneous because the rate, $\lambda$, is constant.

By contrast, an inhomogeneous Poisson process is a Poisson process where the rate parameter, $\lambda$, is not constant in time. A mathematical representation of an inhomogeneous Poisson process is now discussed. First, consider a homogeneous Poisson process of fixed rate $\lambda$ that has been labelled as $\mathcal{Y}^\lambda$. Second, consider a Poisson process of unit rate, $\mathcal{Y}^1$. As mentioned, Poisson processes count the number of 'arrivals' over time, so we can compare Poisson processes by considering the distribution of the number of arrivals by some time $t$. If $\mathcal{Y}^\lambda(0, t)$ and $\mathcal{Y}^1(0, t)$ represent the number of arrivals by a time $t$ in the two processes, then there is an equality in distribution, that is $\mathcal{Y}^\lambda(0, t) \sim \mathcal{Y}^1(0, \lambda t)$. It is therefore possible to re-scale time to transform a unit rate Poisson process to one of arbitrary (but known) rate. The number of arrivals by time $t$ in an inhomogeneous Poisson process, $\overline{\mathcal{Y}}$, is given by:

$$\overline{\mathcal{Y}}(0,t) = \mathcal{Y}^1\left(0, \int_0^t \lambda(t', \{\overline{\mathcal{Y}}(s) : s < t'\})\mathrm{d}t'\right), \tag{2.5}$$

where $\lambda\left(t', \{\overline{\mathcal{Y}}(s) : s < t'\}\right)$ emphasises that $\lambda$ is a function of the particular path the process is taking.

The RTCR of the process $\boldsymbol{X}$ will now be set out. Note that the RTCR describes exactly the same stochastic process as the CME. The number of times reaction $R_j$ fires over the time-interval $(0,T]$ is given by a inhomogeneous Poisson counting process

$$\mathcal{Y}_j\left(0, \int_0^T a_j(\boldsymbol{X}(t))\mathrm{d}t\right),$$

where $\mathcal{Y}_j$ is a unit-rate Poisson process, and $\mathcal{Y}_j(0,\alpha)$ counts the number of times the unit-rate Poisson process fires over the interval $(0,\alpha]$. Every time reaction $R_j$ fires, the state vector (see Equation (2.1)), is updated by adding the appropriate stoichiometric vector, $\boldsymbol{\nu}_j$, to it. Therefore, by considering all possible reactions over the time-interval $(0,T]$, we can determine the state vector at time $T$ as

$$\boldsymbol{X}(T) = \boldsymbol{X}(0) + \sum_{j=1}^M \mathcal{Y}_j\left(0, \int_0^T a_j(\boldsymbol{X}(t))\mathrm{d}t\right) \cdot \boldsymbol{\nu}_j. \tag{2.6}$$

## 2.2 Monte Carlo simulation

As highlighted in Chapter 1, with the general exception of very simple systems, an analytic solution of the CME is unattainable. Therefore, stochastic simulation methods must be used to understand and explore physically-relevant biological models.

We generate an ensemble of $\mathcal{N}$ sample paths. Each sample path is a realisation of the stochastic process $\boldsymbol{X}$, and, as explained in Chapter 1, the sample paths are used to estimate suitable summary statistics. For example, we might estimate the expected population of species $S_i$ at time $T$. We estimate this quantity, $\mathcal{Q} = \mathbb{E}[X_i(T)]$, with an estimate $\widehat{\mathcal{Q}}$ given by

$$\widehat{\mathcal{Q}} := \frac{1}{\mathcal{N}} \sum_{r=1}^{\mathcal{N}} X_i^{(r)}(T), \qquad (2.7)$$

where the copy number of species $i$ at time $t$ in path $r$ is represented by $X_i^{(r)}(t)$. This is an example of a Monte Carlo estimate, and, as such, the estimate contains a statistical error [58]. This arises as we have generated only a subset of possible sample paths: with a different ensemble of realisations, the estimate, $\widehat{\mathcal{Q}}$, will be slightly different. The statistical error can be quantified by determining the estimator variance. If the variance of the sample $X_i(T)$ is given by $\sigma^2$, then it can be shown that the variance of the estimator $\widehat{\mathcal{Q}}$ is given by $\sigma^2/\mathcal{N}$. Throughout the remainder of this thesis, we will use the term 'sample variance' to refer to $\sigma^2$, and the term 'estimator variance' to refer to $\sigma^2/\mathcal{N}$. The estimator variance can then be used to construct a confidence interval around the estimate $\widehat{\mathcal{Q}}$ to illustrate how statistical errors affect the estimate. An *approximate* 95% confidence interval is typically provided by [44]

$$\left( \widehat{\mathcal{Q}} - 1.96 \sqrt{\frac{\sigma^2}{\mathcal{N}}}, \widehat{\mathcal{Q}} + 1.96 \sqrt{\frac{\sigma^2}{\mathcal{N}}} \right), \qquad (2.8)$$

where $\sigma^2$ is estimated using the $\mathcal{N}$ sample values of $X_i^{(r)}(T)$. To ensure a high degree of statistical accuracy the confidence interval provided by (2.8) must be small. The difficulty is that the total simulation time scales with the number of sample paths, i.e. as $\mathcal{O}(\mathcal{N})$, whilst the size of the confidence interval scales as $\mathcal{O}(1/\sqrt{N})$. Consequently, a confidence interval of length $\mathcal{O}(\varepsilon)$ requires a simulation time of $\mathcal{O}(\varepsilon^{-2})$, and naïve SSAs can therefore be very computationally expensive.

### 2.2.1 Two initial case studies

Before proceeding further, we provide two examples of a biochemical reaction network. These case studies will be used for testing and illustration purposes.

**Case Study 1.** We first consider a stochastic analogue of a logistic growth model:

$$R_1 : A \xrightarrow{10} 2A; \quad R_2 : 2A \xrightarrow{0.1} A. \tag{2.9}$$

Initially, the population[1] of $A$ is given as $X(0) = 10$. The stoichiometric matrix, $\boldsymbol{\nu}$, is given by

$$\boldsymbol{\nu} = \begin{bmatrix} 1 & -1 \end{bmatrix}.$$

The rate of constant of $R_1$, $c_1$, equals 10, and, in accordance with Table 2.1, the units of $c_1$ are $[t]^{-1}$, with $[t]$ referring to a unit of time. The rate constant of $R_2$, $c_2$, equals 0.1, and the units are $[\Omega] \cdot [t]^{-1}$, where $[\Omega]$ refers to a unit of volume. Reaction $R_1$ represents the birth of individuals, whilst reaction $R_2$ accounts for the death of individuals in the model.

**Case Study 2.** We consider a model of gene transcription and translation, as introduced by Anderson and Higham [47]:

$$R_1 : G \xrightarrow{25} G + M; \quad R_2 : M \xrightarrow{1000} M + P; \quad R_3 : P + P \xrightarrow{0.001} D; \tag{2.10}$$

$$R_4 : M \xrightarrow{0.1} \emptyset; \quad R_5 : P \xrightarrow{1} \emptyset.$$

In this model, molecules of mRNA ($M$) are transcribed from a single gene ($G$); these mRNA molecules are then used in the translation of protein molecules ($P$). Two protein molecules may combine to produce stable homodimers ($D$), whilst both the mRNA and protein molecules decay linearly. We assume that the system contains a single copy of the gene throughout, and that initially there are no copies of $M$, $P$ or $D$. We write the numbers of mRNA, protein and dimer molecules at time $t$, respectively, as $\boldsymbol{X}(t) = [X_1(t), X_2(t), X_3(t)]^\top$ and consequently the initial condition can be expressed as $\boldsymbol{X}(0) = [0, 0, 0]^\top$. The stoichiometric matrix is

$$\boldsymbol{\nu} = \begin{bmatrix} 1 & 0 & 0 & -1 & 0 \\ 0 & 1 & -2 & 0 & -1 \\ 0 & 0 & 1 & 0 & 0 \end{bmatrix}. \tag{2.11}$$

---
[1] As there is only one species, we will suppress the subscript and work with $X(t)$ instead of $X_1(t)$.

In order to reliably assess and interpret the computational performance of different SSA methods, values for the rate constants shown above are taken from Anderson and Higham [47], and we will therefore be able to compare our findings with the existing literature. Note that, for this, as well as all subsequent case studies, we assume that time has been non-dimensionalised.

We now discuss how one might produce sample paths to understand the dynamics of Case Studies 1 and 2.

## 2.3 Gillespie Direct method

In this section, we describe methods for generating sample paths of a biochemical reaction network. A wide variety of methods have been developed to generate such sample paths, but the Direct method (DM) due to Gillespie [59] is certainly the most commonly-used algorithm [24, 33, 59]. The DM can be derived from the same fundamental hypotheses as the CME; therefore, the sample paths the DM generates are associated with the same probability distributions as the CME. Numerous variations of Gillespie's exact DM algorithm have since been developed [60, 61, 62]. We will call the aforementioned methods 'exact' stochastic simulation algorithms ('exact SSAs').

The DM generates sample paths by simulating each and every reaction that takes place in the system. If the state vector at time $t$ is given by $\boldsymbol{X}(t)$, then a reaction is simulated by considering the following two questions:

1. How long passes until the next reaction occurs? Label this quantity as $\Delta$.

2. Which type of reaction fires next? Label the reaction as $R_k$.

If these two questions are asked, and then answered, a sufficient number of times, a sample path is generated. The DM proceeds as follows:

1. Write $a_0(\boldsymbol{X}(t)) := \sum_{j=1}^{M} a_j(\boldsymbol{X}(t))$ for the total propensity (i.e. the average rate at which *any* reaction fires). The value of $\Delta$ is simulated according to an exponential distribution with rate $a_0(\boldsymbol{X}(t))$:

$$\Delta \sim \text{Exp}(a_0(\boldsymbol{X}(t))).$$

2. The probability that reaction $R_j$ is the next reaction to fire is given by its fractional share of the total rate:

$$a_j(\boldsymbol{X}(t))/a_0(\boldsymbol{X}(t)).$$

$R_j$ can be stochastically chosen by using an inverse transform method. First, a random number, $u$, is uniformly generated on $(0,1)$. Then, we take $j$ to be the index that satisfies

$$\sum_{j'=1}^{j-1} a_{j'}(\boldsymbol{X}(t)) < u \cdot \sum_{j'=1}^{M} a_{j'}(\boldsymbol{X}(t)) < \sum_{j'=1}^{j} a_{j'}(\boldsymbol{X}(t)).$$

We provide a pseudo-code implementation of the DM as Algorithm 2.1.

---

Algorithm 2.1: The DM. This simulates a single sample path.

---

**Require:** initial conditions, $\boldsymbol{X}(0)$, and terminal time, $T$.
1: set $\boldsymbol{X} \leftarrow \boldsymbol{X}(0)$ and set $t \leftarrow 0$
2: **loop**
3:     for each $R_j$, calculate propensity values $a_j$ and set $a_0 \leftarrow \sum_{j=1}^{M} a_j$
4:     set $\Delta \leftarrow \text{Exp}(a_0)$
5:     **if** $t + \Delta > T$ **then**
6:         break
7:     **end if**
8:     choose reaction $R_k$ to fire next: $R_k$ fires with probability $a_k/a_0$
9:     set $\boldsymbol{X} \leftarrow \boldsymbol{X} + \boldsymbol{\nu}_k$, and set $t \leftarrow t + \Delta$
10: **end loop**

---

We now turn to considering the Modified next reaction method (the 'MNRM') [63]. Like the DM, the MNRM is an example of an SSA, and we will therefore compare these two methods. The MNRM will be derived by considering the RTCR.

The MNRM is equivalent to the DM, in that the sample paths it generates will share the same probability distribution. Specifically, if $\boldsymbol{x_0}$ and $t_0$ are given, then the probability $\mathbb{P}[\boldsymbol{x}, t \mid \boldsymbol{x_0}, t_0]$ is the same for sample paths generated with the DM and the MNRM.

### 2.3.1 Modified next reaction method

As outlined in Section 2.1.1, the stochastic process $\boldsymbol{X}$ evolves according to Equation (2.6), i.e.

$$\boldsymbol{X}(T) = \boldsymbol{X}(0) + \sum_{j=1}^{M} \mathcal{Y}_j \left(0, \int_0^T a_j(\boldsymbol{X}(t))\mathrm{d}t\right) \cdot \boldsymbol{\nu}_j.$$

Thus, we can determine the value of $\boldsymbol{X}(T)$ by determining the value of each term of Equation (2.6). As with the DM, a sample path will be generated by firing each reaction individually.

We now explain how Equation (2.6) can be used to simulate a sample path with the MNRM [63]. If, at time $t$, the system is in state $\boldsymbol{X}(t)$, we need to work out the time until the next event $\Delta$, as well as the particular reaction event that occurs, $R_k$. We determine $\Delta$ by repeatedly asking the following question: suppose a reaction $R_j$ fires next, then what would the putative value of $\Delta$ be? We exhaustively loop through all possible reactions $R_j$ (for $j = 1, \ldots, M$) to determine putative values for $\Delta$. The reaction that gives rise to the smallest putative value for $\Delta$, $R_k$, is the one that will fire next. To calculate the value of $\Delta$, for each reaction $R_j$ we define the following two quantities:

- $A_j = \int_0^t a_i(\boldsymbol{X}(t'))\mathrm{d}t'$, the internal or natural time of the reaction channel;

- $T_j$, the time of the next arrival in the associated unit-rate Poisson process $\mathcal{Y}_j$.

We exhaustively search for the value of $\Delta$ by taking

$$\Delta = \min_j \left( \frac{T_j - A_j}{a_j(\boldsymbol{X}(t))} \right), \tag{2.12}$$

and setting $k$ to be the index where this minimum is obtained (the 'argmin'). The reaction $R_k$ is fired; this simulation method is formally described in Algorithm 2.2.

### 2.3.2 Numerical experiments

We now illustrate the use of the DM (see Algorithm 2.1) and the MNRM (see Algorithm 2.2). The DM and MNRM are implemented in `C++` code that follows the 2011 standard (`C++11`).

The simulations were performed on an `AMD FX-4350` CPU, with an advertised clockspeed of 4.2 GHz.

Algorithm 2.2: The MNRM simulates a single sample path according to the CME. At each step of the loop, the next event is chosen. The population values and time are then updated. A new random number is generated to replace the one that has just been used to simulate an event, and the loop is repeated until the terminal time is reached.

**Require:** initial conditions, $\boldsymbol{X}(0)$ and terminal time, $T$.
1: set $\boldsymbol{X} \leftarrow \boldsymbol{X}(0)$
2: set $t \leftarrow 0$
3: for each $R_j$, set $A_j \leftarrow 0$, and generate $T_j \leftarrow \text{Exp}(1)$
4: **loop**
5:     calculate propensity values $a_j$ for each $R_j$
6:     calculate $\Delta_j$ as
$$\Delta_j = \frac{T_j - A_j}{a_j}$$
7:     set $\Delta \leftarrow \min_j \Delta_j$, and $k \leftarrow \text{argmin}\Delta_j$
8:     **if** $t + \Delta > T$ **then**
9:         break
10:     **end if**
11:     set $\boldsymbol{X} \leftarrow \boldsymbol{X} + \boldsymbol{\nu}_k$
12:     set $t \leftarrow t + \Delta$
13:     for each $R_j$, set $A_j \leftarrow A_j + a_j \cdot \Delta$
14:     set $T_k \leftarrow T_k + \text{Exp}(1)$
15: **end loop**

**Case Study 1.** We recall that Case Study 1 is a stochastic analogue of a logistic growth model, and its reaction channels can be summarised as:

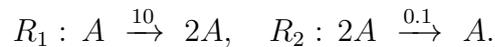

$$R_1 : A \xrightarrow{10} 2A, \quad R_2 : 2A \xrightarrow{0.1} A.$$

We use the DM to simulate $\mathcal{N} = 1000$ sample paths of this system. In Figure 2.1, we show five examples of such sample paths. We also show the mean population of $A$ during the time-interval $[0, T]$, $\mathbb{E}[X(t)]$, together with one and two standard deviations away from this mean value.

**Case Study 2.** In Case Study 2, a gene regulatory network proposed by Anderson

and Higham [47] is studied. The reaction channels are summarised as:

$$R_1 : G \xrightarrow{25} G + M, \quad R_2 : M \xrightarrow{1000} M + P, \quad R_3 : P + P \xrightarrow{0.001} D,$$

$$R_4 : M \xrightarrow{0.1} \emptyset, \quad R_5 : P \xrightarrow{1} \emptyset.$$

We first use the DM to simulate $\mathcal{N} = 1000$ sample paths of this system. Figure 2.2 shows the evolution of $\boldsymbol{X}(t) = [X_1(t), X_2(t), X_3(t)]^\top$ up until time $T = 1$. The solid black lines show the mean species numbers and the coloured bands one and two standard deviations from the mean. On average, approximately 17500 reactions must be fired to generate a single sample path; in our ensemble of $\mathcal{N} = 1000$ paths, we simulated between 6735 and 33256 reactions per sample path.

The DM and MNRM are tested and compared. We will estimate $\mathbb{E}[X_3(1)]$, the expected dimer population at time $T = 1$. Complete results are detailed in Figure 2.2. To compute $\mathbb{E}[X_3(1)]$ to within a single dimer, with a 95% confidence interval (see Equation (2.8)), requires the generation of approximately $4.8 \times 10^6$ sample paths. The DM calculation required $7223s$ (approximately 2 hours) of CPU time to run, whilst the MNRM was completed within $7513s$ (approximately 2 hours and 5 minutes). Although our DM and MNRM algorithms perform adequately, the fact remains that for this system, and many others, these exact simulation algorithms are too computationally demanding to be used for in-depth numerical studies.

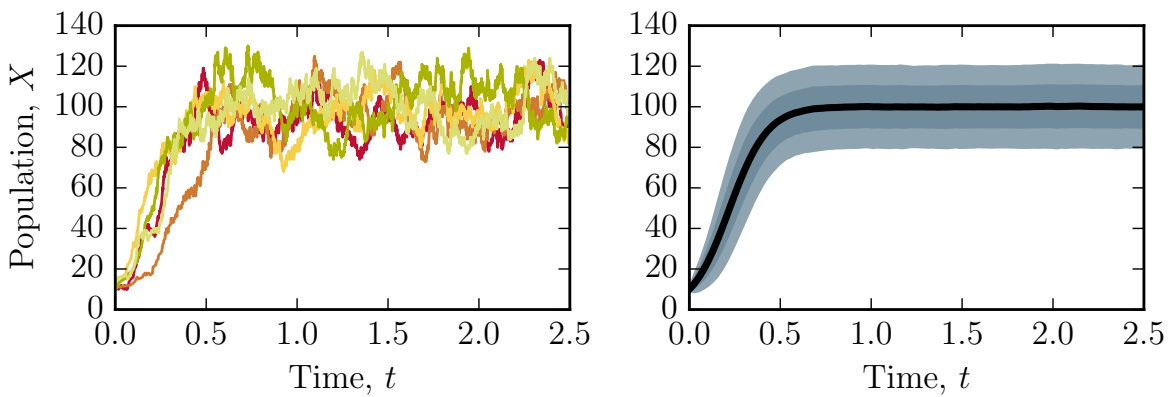

Figure 2.1: In the left pane, five sample paths of System (2.9) are shown. In the right pane, we show the expected value of $A$ over time, $\mathbb{E}[A(t)]$, in black. One standard deviation from the mean is illustrated with dark blue, whilst two standard deviations are shown in light blue. A total of $\mathcal{N} = 1000$ sample paths were used to generate this figure.

The DM and MNRM are computationally demanding because they both simulate each reaction individually. The large cost associated with firing reactions individually comes from two main sources: first is the computational overheads in generating the large quantity of random numbers required by the algorithm; and second is the search time involved in determining which reaction occurs at each step. In the next section, we discuss a number of approximate simulation algorithms that are able to overcome the aforementioned restrictions, thereby reducing the CPU time required for each sample path.

## 2.4 Approximate simulation algorithms

In this section, we describe approximate stochastic simulation algorithms ('approximate SSAs'). We will simulate a stochastic process, $\boldsymbol{Z}$, which is different from the stochastic process $\boldsymbol{X}$ that we described in Section 2.1. The *approximate* process $\boldsymbol{Z}$ will be a simplified version of the *exact* process, $\boldsymbol{X}$. The dynamics of $\boldsymbol{Z}$ will be chosen so that sample paths can be quickly simulated, but subject to the restriction that the sample paths of $\boldsymbol{Z}$ are able to roughly approximate the dynamics of $\boldsymbol{X}$.

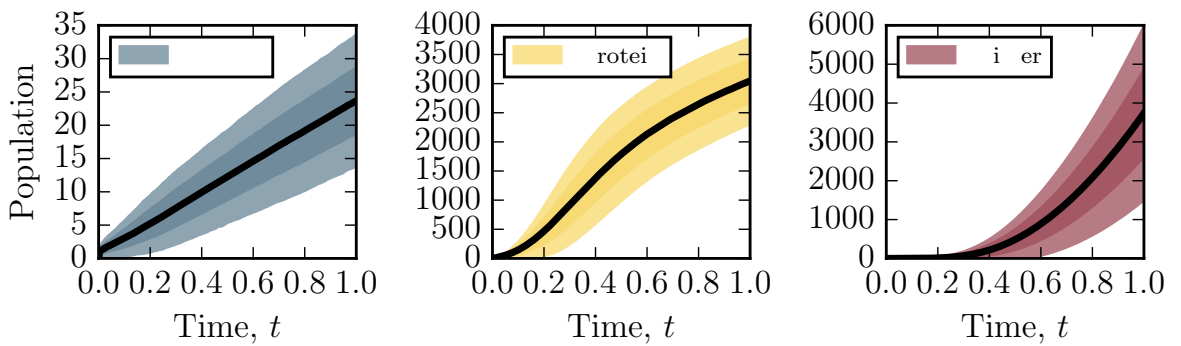

Figure 2.2: We show the expected value of each of the mRNA, dimer and protein populations of System (2.10) over time, in black. One standard deviation from the mean is illustrated in dark colour, whilst two standard deviations are shown in the lighter colour. A total of $\mathcal{N} = 1000$ sample paths were used to generate this figure.

| Estimator | DM | | MNRM | |
|---|---|---|---|---|
| | Estimate | CPU | Estimate | CPU |
| $\mathbb{E}[X_1]$ | $23.8 \pm 0.004$ | | $23.8 \pm 0.004$ | |
| $\mathbb{E}[X_2]$ | $3053.1 \pm 0.33$ | $7223s$ | $3052.9 \pm 0.33$ | $7513s$ |
| $\mathbb{E}[X_3]$ | $3714.3 \pm 0.99$ | | $3713.7 \pm 0.99$ | |

Table 2.2: Estimated populations of System (2.10) at time $T = 1$, as determined by the DM and MNRM. In each case, $4.8 \times 10^6$ sample paths were used. 95% confidence intervals have been constructed; these are indicated with the '$\pm$' terms.

We will estimate summary statistics of interest by using an ensemble of approximate sample paths, $\boldsymbol{Z}^{(1)}, \boldsymbol{Z}^{(2)}, \ldots, \boldsymbol{Z}^{(\mathcal{N})}$. The use of approximate sample paths will affect the estimated values of the summary statistics: this error is known as the 'bias'.

We now describe two simulation methods that admit a bias, but which, in exchange, reduce the CPU time taken to generate sample paths. We discuss the tau- and R-leap methods.

### 2.4.1 Tau-leap method

The tau-leap method, first proposed by Gillespie [45], generates approximate sample paths by taking time-steps of length $\tau$ to traverse the time-interval $[0, T]$. At the end of each time-step, *all* the reactions that are associated with the time-step are concurrently fired. The key, time-saving assumption we make is that the reaction propensities are constant during each time-step: this is known as the 'tau-leap assumption'. Thus, over each time-interval $[t, t + \tau)$, the number of times reaction $R_j$ fires will be given by a homogeneous Poisson process with fixed rate, $a_j(\boldsymbol{Z}(t))$. Based on the tau-leap assumption, the number of times reaction $R_j$ fires over the interval $[t, t+\tau)$, $K_j$ is given by a Poisson *random number* [45]:

$$K_j \sim \mathcal{P}(a_j(\boldsymbol{Z}(t)) \cdot \tau).$$

The state vector, $\boldsymbol{Z}$, is updated by setting

$$\boldsymbol{Z}(t+\tau) = \boldsymbol{Z}(t) + \sum_{j=1}^{M} K_j \cdot \boldsymbol{\nu}_j. \tag{2.13}$$

Appropriate choices of $\tau$ must be used to ensure that the approximate sample paths, $\boldsymbol{Z}$, can successfully approximate the exact stochastic process, $\boldsymbol{X}$. The value of $\tau$ can be fixed from the outset, or $\tau$ can dynamically change, as required. As we shall see, smaller values of $\tau$ will lead to summary statistics with a lower bias [45], but will require more steps to simulate a path and therefore a higher run-time. A sample implementation is provided as Algorithm 2.3. In this implementation, the value of $\tau$ is fixed, and we assume that $\tau$ divides the interval $[0, T]$ into equally-sized time-steps.

---

Algorithm 2.3: Tau-leap method. This simulates a single sample path, using fixed time-step $\tau$.

---

**Require:** initial conditions, $\boldsymbol{Z}(0)$, time-step $\tau$, and terminal time, $T$.
1: set $\boldsymbol{Z} \leftarrow \boldsymbol{Z}(0)$ and set $t \leftarrow 0$
2: **while** $t < T$ **do**
3:     **for** each $R_j$ **do**
4:         calculate propensity value $a_j(\boldsymbol{Z})$
5:         generate $K_j \sim \mathcal{P}(a_j(\boldsymbol{Z}) \cdot \tau)$
6:     **end for**
7:     set $\boldsymbol{Z} \leftarrow \boldsymbol{Z} + \sum_{j=1}^{M} K_j \cdot \boldsymbol{\nu}_j$
8:     set $t \leftarrow t + \tau$
9: **end while**

---

Under reasonably general circumstances, the tau-leap method can generate sample paths more quickly than the DM [23, 24, 45]. This is because, during every time-step of the algorithm (where we leap over time and advance $t \to t + \tau$), multiple reactions are simulated during that step. When the DM advances through time (where $t \to t + \Delta$), a single reaction is fired at that step. In general, each step of the tau-leap method takes more CPU time than the equivalent DM step; however, unless $\tau$ is very small, the tau-leap method requires far fewer time-steps than the DM to generate a sample path. The value of $\tau$ is typically considered too small if very few reactions are expected to take place during each time-step. For example, Cao et al. [64] consider the tau-leap method inefficient if fewer than 10 reactions are expected to take place during each time-step; therefore, we consider the tau-leap method as efficient if $\tau \geq 10/a_0(\boldsymbol{Z})$.

Having explained how one might implement the tau-leap method, we turn to discussing its mathematical representation.

## 2.4.2 Representing the tau-leap method with the RTCR

In this section, we use the RTCR to motivate and represent the tau-leap method. We recall that the state $\boldsymbol{X}$ of our biochemical reaction network evolves according to Equation (2.6), which we now restate:

$$\boldsymbol{X}(T) = \boldsymbol{X}(0) + \sum_{j=1}^{M} \mathcal{Y}_j \left( 0, \int_0^T a_j(\boldsymbol{X}(t)) \mathrm{d}t \right) \cdot \nu_j.$$

We continue to represent the state of a tau-leap process at time $t$ as $\boldsymbol{Z}(t)$. Following from Algorithm 2.3, we suppose that $[0, T]$ is divided into $K$ equal time-steps of length $\tau$. The tau-leap assumption is that the propensities can change only at fixed times $t = k \cdot \tau$ (for $k = 1, 2, \ldots$). Therefore, when we implement the tau-leap method we use the following approximation:

$$\int_0^T a_j(\mathbf{Z}(t))\mathrm{d}t \approx \sum_{k=0}^{K-1} a_j(\mathbf{Z}(\tau \cdot k)) \cdot \tau. \tag{2.14}$$

If we insert assumption (2.14) into Equation (2.6), then the evolution of the state of the tau-leap process, $\mathbf{Z}(t)$, is described by

$$\mathbf{Z}(T) = \mathbf{Z}(0) + \sum_{j=1}^{M} \mathcal{Y}_j \left(0, \sum_{k=0}^{K-1} a_j(\mathbf{Z}(\tau \cdot k)) \cdot \tau\right) \cdot \boldsymbol{\nu}_j, \tag{2.15}$$

where the $\mathcal{Y}_j$ ($j = 1, \ldots, M$) are unit-rate Poisson processes. We now rearrange Equation (2.15) so that it is easier to work with. First we set[2]

$$P_{k,j} = \sum_{k'=0}^{k} a_j(\mathbf{Z}(\tau \cdot k')) \cdot \tau, \tag{2.16}$$

with the special case of $P_{-1,j} = 0$. Then, we re-arrange equation (2.15) to give

$$\mathbf{Z}(T) = \mathbf{Z}(0) + \sum_{k=0}^{K-1} \sum_{j=1}^{M} \mathcal{Y}_j \left(P_{k-1,j}, P_{k,j}\right) \cdot \boldsymbol{\nu}_j. \tag{2.17}$$

The tau-leap method can be seen as a method for iterating over $k$: for each reaction $R_j$, at each step we calculate the number of events in the Poisson process $\mathcal{Y}_j$ between positions (or internal times) $P_{k-1,j}$ and $P_{k,j}$ (given by $\mathcal{Y}_j(P_{k-1,j}, P_{k,j})$). Equation (2.17) can therefore be re-arranged into an update formula:

$$\mathbf{Z}(k \cdot \tau) = \mathbf{Z}((k-1) \cdot \tau) + \sum_{j=1}^{M} \mathcal{Y}_j \left(P_{k-1,j}, P_{k,j}\right) \cdot \boldsymbol{\nu}_j. \tag{2.18}$$

---

[2] Note that $j$ indexes the reaction, $R_j$; and $k$ indexes the time.

### 2.4.3 Extending the tau-leap method

The tau-leap method has been developed and modified to deal with a wide variety of biochemical reaction networks. We briefly describe a number of major adaptations to the method below:

**Adaptive time-stepping tau-leap method.** Methods have been developed for dynamically changing the time-step, $\tau$, according to the stochastic behaviour of each sample path [64, 65, 66, 67]. Such methods are particularly suitable for handling stiff systems. Whilst most methods choose $\tau$ at the start of each time-step, a post-leap check mechanism has also been described [68].

**Partially-implicit tau-leap method.** The numerical scheme of the tau-leap method has been altered and recast in a partially-implicit format, which can take larger time-steps [69, 70].

**Binomial leaping.** Occasionally, the tau-leap method generates non-physical, negative population values. The negative populations arise where, as a consequence of using approximate reaction propensity values, more reactants than the number available are consumed during a leap. This issue often arises where too large a value of $\tau$ has been used [71]. However, as the Poisson variates that determine how many reactions fire during each leap have semi-infinite support, even a cautious choice of $\tau$ cannot certainly prevent a negative populations from occurring [72]. Tian and Burrage [72] have developed an alternative method that uses suitable binomial variates to simulate the reactions that occur during each time-step.

In addition, the tau-leap method has spurred the development of alternative approximate simulation algorithms. We set out one particularly important example below.

## 2.4.4 The R-leap method

Auger et al. [73] developed the R-leap method that generates sample paths by simulating a user-specified number of reactions during each step of the method. For our purposes, this 'jump size' will be labelled as $\mathcal{K}$. The R-leap method differs from tau-leap method in that each step of the tau-leap method covers a fixed time-interval (with the number of reaction events determined by simulation), whereas the time-interval covered by each step of the R-leap method must be determined by simulating a random number (though the number of reaction events, $\mathcal{K}$, is fixed).

We will refer to the R-leaping algorithm in Chapter 5, and further information will be provided at that time. A pseudo-code implementation of R-leaping is provided in Algorithm 2.4. In this algorithm, we have used the conditional binomial method to sample from a multinomial distribution. At this stage, we simply note that this is one of many possible methods for sampling from a multinomial distribution (further details are provided in Chapter 5).

---

Algorithm 2.4: R-leaping. This simulates a single sample path.

---

**Require:** initial conditions, $\boldsymbol{Z}(0)$, jump size, $\mathcal{K}$, and terminal time, $T$.
1: set $\boldsymbol{Z} \leftarrow \boldsymbol{Z}(0)$ and set $t \leftarrow 0$
2: **while** $t < T$ **do**
3:     for each $R_j$, calculate propensity value $a_j(\boldsymbol{Z})$, and set $a_0 \leftarrow \sum_{j=1}^{M} a_j$
4:     generate $\Delta \sim \Gamma(\mathcal{K}, 1/a_0)$
5:     **if** $t + \Delta > T$ **then**
6:         generate $\mathcal{K} \sim \mathcal{B}(\mathcal{K} - 1, (T - t)/\Delta)$, and set $t \leftarrow T$
7:     **else**
8:         set $t \leftarrow t + \Delta$
9:     **end if**
10:     **for** $j = 1, \ldots, M$ **do**
11:         generate $K_j \sim \mathcal{B}\bigl(\mathcal{K}, a_j / \sum_{j'=j}^{M} a_{j'}\bigr)$ and set $\mathcal{K} \leftarrow \mathcal{K} - K_j$
12:     **end for**
13:     set $\boldsymbol{Z} \leftarrow \boldsymbol{Z} + \sum_{j=1}^{M} K_j \cdot \boldsymbol{\nu}_j$
14: **end while**

---

**2.4.5 Error analysis**

The error induced by using an approximate SSA to generate sample paths can be measured in a number of ways. We introduce the numerical analysis concepts of *weak* and *strong* error [44, 74, 75]. The weak and strong errors will reflect the effect of using an approximate SSA to generate sample paths. For a summary statistic, $f : \mathbb{N}^N \to \mathbb{R}$, where $f(\cdot)$ satisfies suitable regularity conditions[3], the weak error is defined to be

$$\left| \mathbb{E}\big[f(\boldsymbol{Z})\big] - \mathbb{E}\big[f(\boldsymbol{X})\big] \right|, \qquad (2.19)$$

where $\boldsymbol{X}$ represents the *exact* process, and $\boldsymbol{Z}$ the *approximate* process.

If the tau-leap method is run with a fixed time-step of $\tau$, then this approximate SSA is said to a have a *weak order of convergence* of $\gamma$ if $\exists\, C > 0$ such that, for all $f : \mathbb{N}^N \to \mathbb{R}$ that satisfy suitable regularity conditions,

$$\left| \mathbb{E}\big[f(\boldsymbol{Z})\big] - \mathbb{E}\big[f(\boldsymbol{X})\big] \right| \leq C\tau^\gamma.$$

Informally, the order $\gamma$ can be thought of as measuring the rate of decay in the 'error of the means' [75].

The strong error of an approximation scheme is similarly defined:

$$\mathbb{E}\big[\,|f(\boldsymbol{Z}) - f(\boldsymbol{X})|\,\big]. \qquad (2.20)$$

A *strong order of convergence* can also be calculated; the strong order of convergence can be thought of as providing the rate of decay in the 'mean value of the errors' [75]. For the tau-leap method, in the limit $\tau \downarrow 0$, it can be shown [76] that the strong order

---

[3]Typically, numerical analysts assume that $f$ is smooth and grows at most polynomially.

of convergence is 1/2, whilst the weak order of convergence is 1.

Before proceeding further, we pursue two technical points:

**Vectors of summary statistics.** If, instead, we let $\boldsymbol{f}(\cdot)$ represent a vector of summary statistics, that is $\boldsymbol{f} : \mathbb{N}^N \to \mathbb{R}^R$, then a weak and a strong error can be similarly defined. To define such errors, we replace the absolute values operators of Equations (2.19) and (2.20), $|\cdot|$, with a suitable norm, $\|\cdot\|$. For example, we might take $\|\cdot\|$ to be the 1-norm[4]. It is straightforward to show that the weak and strong orders of convergence of the tau-leap method remain at 1 and 1/2, respectively.

**Measurability.** We have yet to formally define the expectation of a random variable, $\mathbb{E}[\cdot]$. We start by considering a *probability space*, $(\Omega, \mathcal{F}, \mathbb{P})$. The components of the probability space are defined as follows: $\Omega$ is the sample space, and consists of all possible outcomes; $\mathcal{F}$ is the set of events, where each event is a set containing zero or more outcomes; and $\mathbb{P}$ is the probability function, a map $\mathcal{F} \to [0, 1]$ [77]. Formally, a random variable, $X$, is a map from the sample space to a state space (for example, $\mathbb{R}$) given by $X : \Omega \to S$. If $X$ is *integrable*, then the expectation of $X$ is then defined as

$$\mathbb{E}[X] = \int_\Omega X(\omega) \mathbb{P}(\{\mathrm{d}\omega\}). \tag{2.21}$$

If a second random variable, $Y : \Omega \to S$, is introduced and is integrable, then the integral $\mathbb{E}[X - Y]$ can also be computed. Note that $X$ and $Y$ must make use of the *same* sample space. For further information, including a derivation of the strong error result for a tau-leap method, we refer readers to Li [76].

Statisticians measure the effect of an approximate scheme on a *specific* summary statistic by considering the estimator bias [58]. If our summary statistic is estimated

---

[4]In $R$ dimensions, the taxicab or 1-norm, $\|\cdot\|$, is given by $\|\boldsymbol{X}\| = \sum_{r=1}^{R} |X_r|$.

as $\widehat{\mathcal{Q}}$, whereas the true value is $\mathcal{Q}$, then the bias is stated as

$$\text{Bias}(\mathcal{Q}, \widehat{\mathcal{Q}}) = \mathbb{E}[\widehat{\mathcal{Q}}] - \mathcal{Q}. \tag{2.22}$$

Therefore, any summary statistics estimated with approximate sample paths will be laden with both a bias (see above), and a statistical error (see Section 2.2).

It is often not necessary to distinguish between the statistical error and the bias of an estimate; rather, the overall error contained within the estimate, $\widehat{\mathcal{Q}}$, is the quantity of interest. The overall error can be mathematically measured by considering the mean-squared error (the 'MSE'). Suppose a summary statistic is estimated as $\widehat{\mathcal{Q}}$, whereas its actual value equals $\mathcal{Q}$. Then the MSE is given by combining the statistical error and bias as follows [58]:

$$\begin{aligned}\text{MSE} &= \mathbb{E}\left[\left(\mathcal{Q} - \widehat{\mathcal{Q}}\right)^2\right] \\ &= \underbrace{\mathbb{E}\left[\left(\widehat{\mathcal{Q}} - \mathbb{E}[\widehat{\mathcal{Q}}]\right)^2\right]}_{\text{Statistical error}} + \underbrace{\left(\mathbb{E}[\widehat{\mathcal{Q}}] - \mathcal{Q}\right)^2}_{\text{Square of bias}}.\end{aligned} \tag{2.23}$$

The *root mean square error* (RMSE) is the square-root of the MSE.

Having explained how one might measure the error associated with an estimate of a summary statistic, we now discuss the relationship between the CPU time expended by an algorithm and its error.

### 2.4.6 Time complexity

The time complexity of an algorithm relates the CPU time required by an algorithm to an user-provided algorithm parameter. We will quantify the time complexity in terms of its *leading order* behaviour, and we explain with an example.

**Example 2.2.** Consider an algorithm that is provided with an unsorted list of $N$ numbers, and which must return a smallest-to-largest list of those numbers. A very simple algorithm would involve searching through the input list, from start to finish, finding the smallest number, and moving that number to the start of the output list. Then there are $N-1$ input numbers left to sort and the procedure is repeated. A total $N \times (N+1)/2$ steps are required to sort the list, and the computational complexity of this procedure is therefore $\mathcal{O}(N^2)$. ∎

We now study the time complexity of a Monte Carlo method. In particular, we wish to describe the CPU time required by the method to achieve a user-specified level of accuracy. The accuracy of our Monte Carlo method will be delineated by the RMSE. Thus, if a RMSE of order $\mathcal{O}(\varepsilon)$ is required, we wish to determine the leading-order term that, in terms of $\varepsilon$, characterises the CPU cost.

We will demonstrate the complexity of the tau-leap method, with our findings motivating the need for new Monte Carlo approaches. If the tau-leap estimate, $\widehat{\mathcal{Q}}$, requires a RMSE of order $\mathcal{O}(\varepsilon)$, we require:

1. a statistical error of order $\mathcal{O}(\varepsilon^2)$; and,

2. a bias of order $\mathcal{O}(\varepsilon)$.

Thus, we require that the number of sample paths, $\mathcal{N}$, be given by

$$\mathcal{N} \sim \mathcal{O}(\varepsilon^{-2}),$$

and the time-step of the tau-leap method, $\tau$, to be given by

$$\tau \sim \mathcal{O}(\varepsilon).$$

If we make the precursory assumption (that we will soon refine) that the CPU time scales as $\mathcal{O}(\tau^{-1})$, then our initial analysis suggests that, for the tau-leap method, a RMSE error of $\mathcal{O}(\varepsilon)$ requires $\mathcal{O}(\varepsilon^{-3})$ units of CPU time.

Therefore, if our estimate of a summary statistic, $\widehat{\mathcal{Q}}$, is to have a relatively small RMSE, then the CPU time required to generate this estimate could be very large.

### 2.4.7 Numerical examples

We now generate sample paths for case studies 1 and 2 using the tau-leap method. We illustrate the effect of varying the time-step, $\tau$, on the dynamics of the sample paths, as well as the overall CPU time taken to produce the required sample paths. As before, our algorithms have been implemented in `C++`. Where possible, the `C++` Standard Template Library has been used, and all calculations have been performed with double precision.

**Case Study 1.** To qualitatively illustrate the tau-leap method, and to compare the effect of varying $\tau$ on the system dynamics, two sample paths of the logistic growth model (2.9) are generated. We first take $\tau = 0.01$ and then we take $\tau = 0.10$. To compare the effect of changing $\tau$, the same random inputs are used to simulate each sample path. By 'the same random inputs' we mean that the reaction network has been expressed in the RTCR framework discussed in Section 2.4.2, and the same sets of Poisson processes have been used to generate the sample paths. The sample paths are exhibited in Figure 2.3.

Whilst it might be tempting to describe the sample path generated with $\tau = 0.1$ as a low-resolution version of the sample path generated with $\tau = 0.01$, we note that it is difficult to reliably draw any meaningful conclusions based on qualitative reasoning alone.

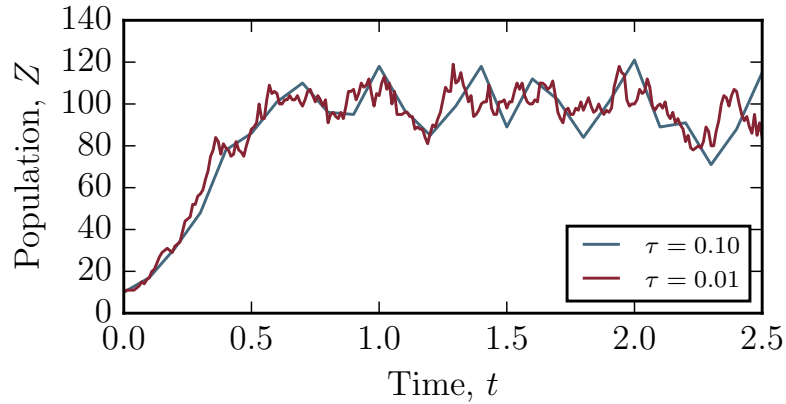

Figure 2.3: Two sample paths of System (2.9) are generated using the tau-leap method, with the time-steps as shown. The same random inputs have been used for both paths.

**Case Study 2.** To quantitatively illustrate the benefits of the tau-leap method, we consider Case Study 2 again. The average dimer population at terminal time $T = 1$, $\mathbb{E}[Z_3(1)]$, is estimated. A range of different values of $\tau$ can be tested, and the resultant estimates of the average dimer population can then be compared with the estimates given by the DM in Table 2.2.

For example, if we take $\tau = 3^{-5}$, then we estimate

$$\mathbb{E}[Z_3(T)] = 3694.5 \pm 1.0.$$

This calculation requires $4.8 \times 10^6$ sample paths, and takes approximately 1247 seconds of CPU time. In this case, the DM would take 5.8 times as long to estimate the same quantity. We estimate the bias of our estimate of $\mathbb{E}[Z_3(T)]$ to equal approximately 20 dimers, which is around 0.5% of the estimated value.

In Figure 2.4, we demonstrate the effect of changing $\tau$ on the bias of the tau-leap method. We compare the bias when $\mathbb{E}[Z_3(T)]$ is estimated with a time-step chosen from $\tau \in \{3^{-2}, \ldots, 3^{-7}\}$. Figure 2.4 numerically demonstrates the first order weak convergence[5] of the tau-leap method, as a least-squares regression[6] shows that $\gamma \approx 0.996$. We also show the computational cost of producing sample paths grows as $1/\tau$ increases; however, the relationship between simulation time and the number of time-steps is non-linear. Even though the number of steps carried out by the tau-leap method grows linearly with $1/\tau$, the relationship between simulation time and $1/\tau$ is non-linear, because the cost of each simulation step depends on $\tau$. This is because the CPU time taken to generate the required Poisson random variates is dependent on the parameters of those variates, with smaller parameters leading to faster random variate simulation. We will return to this point in Chapter 3.

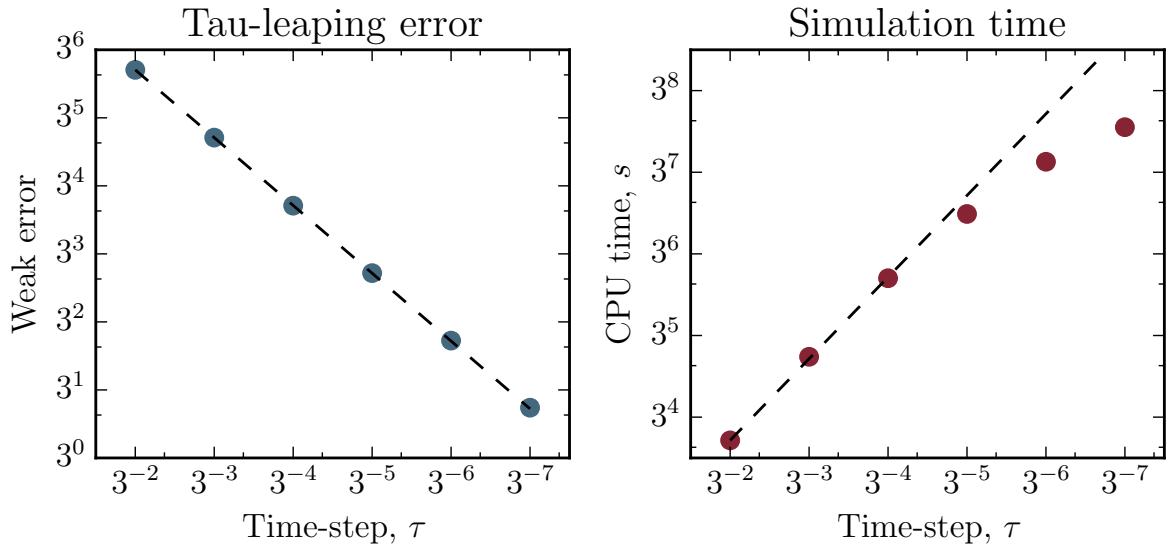

Figure 2.4: In the left pane, we show the weak error of the tau-leap method for different values of $\tau$. The dashed line shows a least-squares regression. In the right pane, we show the simulation time required to generate $\mathcal{N} = 4.8 \times 10^6$ sample paths with the tau-leap method. This choice of $\mathcal{N}$ is sufficient to estimate $\mathbb{E}[Z_3(T)]$ with a confidence interval semi-length of 1.0. The dashed line shows the CPU time predicted by a $\mathcal{O}(1/\tau)$ model.

We now shift focus, and describe how one might model biochemical networks that exhibit spatial heterogeneity.

---

[5] Recall that weak order is given by the largest $\gamma$ such that $\big|\mathbb{E}\big[f(\boldsymbol{Z})\big] - \mathbb{E}\big[f(\boldsymbol{X})\big]\big| \leq C\tau^\gamma$.

[6] For each choice of $\tau$, we estimated $\mathbb{E}[Z_3(T)]$, and we then used least-squares regression to fit a function of the form $\alpha + \beta \cdot \tau^\gamma$ to the estimated values.

## 2.5 Spatially-extended modelling

In Section 2.1 we explained the CME on the basis that the biochemical reaction network of interest is 'well-stirred'. However, many biochemical networks of interest are not 'well-stirred' [78, 79], and an alternative modelling approach is necessary. We will describe the RDME, which explicitly models spatial variability within a biochemical reaction network of interest.

The RDME framework is an example of a voxel-based or lattice-based stochastic model [8, 80]. The volume of interest, $\Omega$, is discretised into a finite number of voxels. Each particle of the system is located within a voxel, and is able to move ('diffuse') by transferring into a neighbouring voxel. We further assume that, within each voxel, the particles are 'well-stirred' and can react with one another. As with the CME, the RDME has a tractable, analytic solution in only a small number of special cases [80].

We now formally describe the RDME. In this thesis, $\Omega$ is assumed to be a volume of dimensions $L \times a \times a$, where $a \ll L$. In line with research presented elsewhere in the literature, our initial assumption is that the spatial variability in the distribution of the particles occurs in only the first dimension. We therefore discretise $\Omega$ into $K$ equally-sized voxels of dimensions $h \times a \times a$, where $h = L/K$. The choice of $h$ needs to be carefully considered [81, 82], and it might be necessary to use different voxel sizes for different biological species [83].

The voxels are labelled as $V^1, V^2, \ldots, V^K$. Then, $S_i^k$ is used to refer to particles of chemical species $S_i$ that are located in voxel $V^k$, whilst $X_i^k$ represents the population of $S_i^k$. The state matrix, $\boldsymbol{X}$, represents all the chemical populations, and is defined

as[7]:
$$\boldsymbol{X} = \begin{bmatrix} X_1^1 & \ldots & X_1^K \\ \vdots & & \vdots \\ X_N^1 & \ldots & X_N^K \end{bmatrix}. \quad (2.24)$$

The particles are able to diffuse (move) within $\Omega$, and can react to change the chemical populations of the system. Particles diffuse by jumping from one voxel into an adjacent voxel, and suitable boundary conditions can be implemented to handle the behaviour at the ends of the domain. A straightforward model might make use of fully-absorbing, reflective (also known as zero-flux), or periodic boundary conditions. Additionally, a partially-absorbing, or reactive boundary condition can also be used to represent phyiscal features of the model [84, 85]. Throughout this thesis, we work with zero-flux boundary conditions.

Each particle diffuses to each of its neighbouring voxels with an average rate of

$$d := \frac{D}{h^2},$$

where $D$ is the macroscale diffusion constant [8, 86]. The choice of $d$ can be justified in a number of ways, and we set out one such justification in the following paragraph. Thus, with zero-flux boundary conditions, the diffusion of each species $S_i$ can be represented by the collection of events

$$S_i^1 \underset{d}{\overset{d}{\rightleftharpoons}} S_i^2 \underset{d}{\overset{d}{\rightleftharpoons}} \ldots \underset{d}{\overset{d}{\rightleftharpoons}} S_i^{K-1} \underset{d}{\overset{d}{\rightleftharpoons}} S_i^K, \quad (2.25)$$

where $S_i^k \underset{d}{\overset{d}{\rightleftharpoons}} S_i^{k+1}$ denotes two events: firstly, diffusion of a $S_i$ particle from voxel $V^k$ to voxel $V^{k+1}$, and, secondly, diffusion of a $S_i$ particle from voxel $V^{k+1}$ to voxel $V^k$.

---

[7] For the well-stirred case, the populations were stored in *vector* (2.1), but where the particles are associated with voxels as well, it is more convenient to use a *matrix*.

Our choice of $d$ is justified as follows [80, 87]. We start by considering just a single species, and we let $M_i^k(t)$ be the mean population of that species in voxel $V^k$ at time $t$. We can manipulate the CME of System (2.25) to obtain a system of equations for $M_i^k(t)$:

$$\frac{\partial M_i^k}{\partial t} = d\big(M_i^{k+1} + M_i^{k-1} - 2M_i^k\big), \qquad \text{for } k = 2, \ldots, K-1; \qquad (2.26)$$

$$\frac{\partial M_i^1}{\partial t} = d\big(M_i^2 - M_i^1\big); \quad \text{and} \quad \frac{\partial M_i^K}{\partial t} = d\big(M_i^{K-1} - M_i^K\big). \qquad (2.27)$$

The concentration of $S_i$ in voxel $V^k$ can be approximated as $c_i(v^k, t) \approx M_i^k(t)/h$, where $v^k$ is the centre of voxel $V^k$. Dividing Equation (2.26) by $h$, we obtain

$$\frac{\partial c_i}{\partial t}(v^k, t) \approx d\Big(c_i(v^k + h, t) + c_i(v^k - h, t) - 2c_i(v^k, t)\Big). \qquad (2.28)$$

A Taylor expansion of Equation (2.28) reveals that

$$\frac{\partial c_i}{\partial t}(v^k, t) \approx dh^2\, \nabla^2 c_i(v^k, t). \qquad (2.29)$$

Macroscopically, Fick's Second Law describes the diffusion of a particle. We will choose $d$ so that the mean concentration within each voxel in the RDME model matches the concentration given by Fick's Second Law. Deterministically, $c_i(v, t)$ satisfies the following partial differential equation (PDE):

$$\frac{\partial c_i}{\partial t}(v, t) = D\nabla^2 c_i(v, t). \qquad (2.30)$$

Thus, choosing $d = D/h^2$ means that, as $h \downarrow 0$, the voxel-based description of diffusion matches the classical diffusion process.

Reactions describe the changes to the chemical populations of the model, and take place between reactants in the same voxel. As with the CME model, each reaction is associated with a propensity that describes the average rate at which the reaction takes place, and a stoichiometric vector that describes how the reaction changes the population levels of the particles. We label the $M$ reaction types of our model as $R_1, \ldots, R_M$. We will assume that each reaction can take place in each voxel, and so we refer to a reaction of type $R_j$ taking place in voxel $V^k$ as $R_j^k$. The propensity of $R_j^k$ depends on the number of particles inside voxel $V^k$, and, taking the voxel size into account, can be determined by referring to Table 2.1.

Having outlined the RDME, we now describe the simulation of such reaction-diffusion networks.

### 2.5.1 Simulating RDME kinetics

Mathematically, the CME (well-stirred) and RDME (spatially-extended) models are both continuous-time, discrete-state Markov chains. This means that, in principle, any simulation algorithm designed for the CME can be adjusted to simulate the dynamics of the RDME. We illustrate the simulation of the stochastic process, $\boldsymbol{X}$, with the DM.

As explained in Section 2.3, the DM algorithm records every change to the population matrix (2.24). The events that change the population matrix are enumerated as the set $\left(\zeta_j\right)_{j \in \{1,\ldots,J\}}$. There are $J$ events in total. The set $\left(\zeta_j\right)_{j \in \{1,\ldots,J\}}$ includes the diffusion of particles, and the various reactions that can take place (of which there are $M \cdot K$ possibilities). The propensity of each event $\zeta_j$ is given by $a_j$, and the stoichiometric matrix as $\boldsymbol{\nu}_j$. Should event $\zeta_j$ take place at time $t$, the population matrix $\boldsymbol{X}$ is updated by adding $\boldsymbol{\nu}_j$ to it. This approach is sometimes known as the 'all events method' [88], and Algorithm 2.1 is readily adjusted to produce such sample paths. We illustrate the method with an example.

**Case Study 3.** The third case study considers a stochastic model of the Fisher-KPP wave, which has been used to model the invasion of numerous biological populations [89]. We divide a volume $L \times a \times a$ into $K = 101$ equally-sized voxels along the first dimension. The particles, which are all of the same species, $A$, diffuse at a macroscale rate of $D$ throughout the domain. Within a voxel, the particles interact through the following two reaction channels:

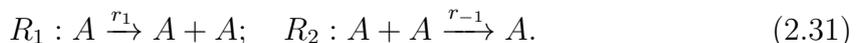

$$R_1 : A \xrightarrow{r_1} A + A; \quad R_2 : A + A \xrightarrow{r_{-1}} A. \qquad (2.31)$$

In order to study this system, we place $10^4$ particles in the left-most voxel (formally, $X^1 = 10^4$), with the remaining voxels left empty. We take $d = D/h^2 = 0.1$, $r_1 = 1$ and $r_{-1} = 0.01$; and generate paths until time $T = 100$. As before, we are working with non-dimensional time.

The Gillespie DM is employed to generate sample trajectories of this reaction network. Note that, within each individual voxel, the reaction dynamics are equivalent to the dynamics of Case Study 1. This model differs from Case Study 1 because the particles are able to diffuse between voxels. In Figure 2.5 we show a single sample path as an image matrix: the populations of each voxel, $X^k$, are shown over the time-interval $[0, 100]$.

## 2.6 Outlook

In this chapter, we described the widely-used CME framework for modelling and simulating biochemical reaction networks. We explained that the CME is generally analytically intractable, and that Monte Carlo methods are therefore used to characterise such stochastic models. We then described a number of generic SSAs, which can either be characterised as being exact, or approximate. Approximate simulation algorithms produce sample paths more efficiently than the exact SSAs, but the estimated summary statistics are biased by the use of an approximate SSA. On the other hand, with a constant level of computing resources, the more efficient approximate SSA can generate more sample paths, which leads to a lower statistical error. We concluded the chapter by introducing the RDME, which is widely used to model spatially-inhomogeneous reaction-diffusion systems.

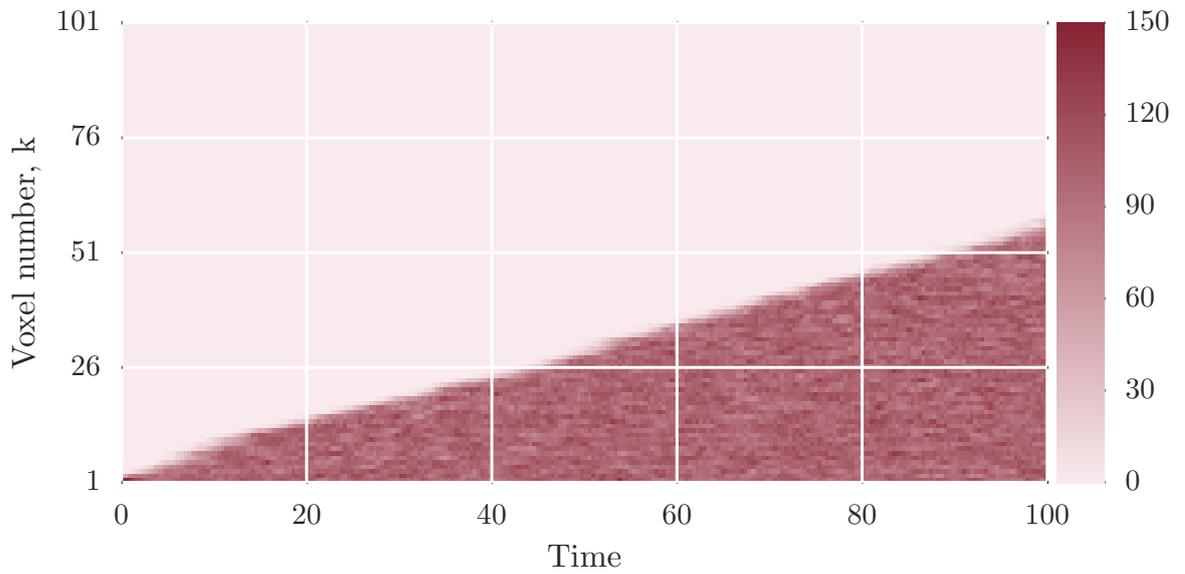

Figure 2.5: A single sample path of System (2.31) is shown as an image matrix. The population of each voxel, $X^k$, is shown over the time-interval $[0, 100]$. An indicated with the colour-bar, darker cells correspond with higher populations. The parameters are as described in the main text.

In the upcoming chapters, we present new and efficient simulation methods. We develop simulation algorithms that exploit particular characteristics of the biochemical reaction networks of interest, so that sample paths can be generated quickly. Our algorithms will also be tailored to estimating summary statistics of importance efficiently.

# Chapter 3

# Multi-level simulation

In this chapter, the multi-level method is introduced and implemented. The multi-level approach divides the computational work required to estimate a summary statistic of interest into parts, known as 'levels'. The chosen summary statistic is calculated by combining a hierarchy of estimators in a telescoping sum, with each term in the telescoping sum representing a different level. The multi-level method can dramatically reduce the CPU time taken to estimate a summary statistic of interest; as such, it has the potential to reshape the field of stochastic simulation [47].

## 3.1 Introduction

This chapter focuses on two distinct goals: firstly, we introduce and describe the multi-level method. Secondly, we explain how the multi-level method can be effectively implemented to study a wide range of reaction networks. As the multi-level method relies heavily on user input and careful configuration, fulfilling our second goal will ensure the efficiency of the multi-level method under a wide variety of circumstances. As part of explaining how to implement the multi-level method effectively, we suggest a number of refinements to the methodology that lead to general performance improvements, and we highlight a number of unusual implementation issues that need to be considered.

### 3.1.1 Outline

In Section 3.2, the multi-level method is introduced and described. A number of preliminary investigations are conducted in Section 3.3. The limitations of the multi-level method are summarised in Section 3.4, and the research goals of the remainder of this chapter are then set out. In Section 3.5, a dynamic calibration method is described to enhance the performance of the multi-level method. Then, in Section 3.6, methods for configuring the algorithm are discussed. Numerical results are set out in Sections 3.7 and 3.8, and our conclusions are drawn in Section 3.9.

## 3.2 Describing the multi-level method

The multi-level framework was first described by Giles [46] in 2008. The original multi-level scheme was designed to estimate summary statistics of a stochastic process that evolves according to a system of stochastic differential equations (SDEs). If sample paths of an SDE are generated with the standard Euler–Maruyama method[1], then Monte Carlo simulation has a computational complexity of $\mathcal{O}(\varepsilon^{-3})$ [44]. The multi-level method of Giles [46] can potentially reduce the computational complexity to $\mathcal{O}(\varepsilon^{-2}\ln(\varepsilon)^2)$. More nuanced implementations of the multi-level method can result in a computational complexity of $\mathcal{O}(\varepsilon^{-2})$ [46]. Anderson and Higham [47] then adapted the method for use with discrete-state, continuous-time Markov chains. In line with the seminal work of Anderson and Higham [47], an emphasis has been placed on rigorously understanding the computational complexity of the multi-level algorithm [90] and a wide range of improvements have been proposed [70, 91, 92, 93].

---

[1]The Euler-Maruyama method generates sample paths by using a Gaussian noise, but is otherwise very similar to the tau-leap method.

### 3.2.1 Aims of the multi-level method

The multi-level method estimates summary statistics of interest. Let $\mathcal{Q}$ be the quantity that will be estimated. For example, $\mathcal{Q}$ could represent the expected value of $X_i(T)$, the population of the $i$-th species at time $T$. Of course, alternative summary statistics can also be estimated, but we will initially focus on this straightforward example. Over the next few sections, we will define (sub) estimators, $\mathcal{Q}_0$, $\mathcal{Q}_1$, $\mathcal{Q}_2$, ..., $\mathcal{Q}_{\mathcal{L}+1}$, so that

$$\mathcal{Q} = \mathcal{Q}_0 + \mathcal{Q}_1 + \mathcal{Q}_2 + \cdots + \mathcal{Q}_{\mathcal{L}+1}. \tag{3.1}$$

We will independently estimate the values of each of $\mathcal{Q}_0, \mathcal{Q}_1, \ldots, \mathcal{Q}_{\mathcal{L}+1}$, which we then sum together to arrive at an estimate for $\mathcal{Q}$. The key savings provided by the multi-level method arise because each of the estimates for $\mathcal{Q}_1 \ldots, \mathcal{Q}_{\mathcal{L}+1}$ are calculated using a variance reduction technique. The benefit of using a variance reduction technique is that it can dramatically reduce the computational cost of estimating a particular statistic. It is therefore possible that, by repeatedly using a variance reduction technique, the multi-level method can estimate summary statistics more efficiently than traditional SSA methods.

The general variance reduction principle is now explained.

**Variance reduction.** Suppose we wish to estimate a summary statistic of a stochastic process, $\phi$, given by $\mathbb{E}[f(\phi)]$, where $f(\cdot)$ is a suitable function. Further, we suppose that estimating $\mathbb{E}[f(\phi)]$ is computationally intensive, because $f(\phi)$ has a relatively high variance, $\sigma^2_{f(\phi)}$. As outlined in Section 2.2, if we estimate $\mathbb{E}[f(\phi)]$ using $\mathcal{N}$ sample paths, then the estimator variance is given by $\sigma^2_{f(\phi)}/\mathcal{N}$.

If there is a different stochastic process, $\psi$, and function, $g(\cdot)$, such that

$$\mathbb{E}[f(\phi)] = \mathbb{E}[g(\psi)],$$

then we can instead estimate $\mathbb{E}[g(\psi)]$, and any resultant estimate is also an estimate for $\mathbb{E}[f(\phi)]$. The bias of the estimate remains unchanged. However, the estimator variance of the estimate is now given by $\sigma^2_{g(\psi)}/\mathcal{N}$, where $\sigma^2_{g(\psi)}$ is the variance of $g(\psi)$. Thus, if $\sigma^2_{g(\psi)} < \sigma^2_{f(\phi)}$, fewer sample paths are required to achieve a given confidence in the estimator when $\mathbb{E}[g(\psi)]$, and not $\mathbb{E}[f(\phi)]$, is estimated. Therefore, the overall simulation time can be reduced by estimating $\mathbb{E}[g(\psi)]$ instead of $\mathbb{E}[f(\phi)]$.

### 3.2.2 Setting out the multi-level method

The multi-level method, which is effectively a recursive control variate method, is now set out. On the *base level* (level $\ell = 0$), a tau-leap SSA with a large value of $\tau$ (which is denoted as $\tau_0$) is used to generate a large number (labelled as $\mathcal{N}_0$) of sample paths of the model. Let $\mathbf{Z}^{(r)}_{\tau_0}(t)$ represent the $r$-th sample path generated by the tau-leap method, where a time-step of $\tau_0$ is used[2]. We write the point statistic of interest as the scalar $Z^{(r)}_{\tau_0}$. The resulting summary statistic estimate is

$$\mathcal{Q}_0 := \mathbb{E}\left[Z_{\tau_0}\right] \approx \frac{1}{\mathcal{N}_0} \sum_{r=1}^{\mathcal{N}_0} Z^{(r)}_{\tau_0}. \tag{3.2}$$

As $\tau_0$ is large, this estimate is calculated cheaply, with the downside being that it is likely to contain considerable bias. The typical CPU time taken to generate a single sample path is recorded as $\mathcal{C}_0$.

The goal with the next level (level $\ell = 1$) is to introduce a *correction term* that begins to reduce this bias. In essence, in order to compute this correction term, *pairs* of sample paths are generated. For each pair, one sample path is generated using the tau-leap SSA with the same value of $\tau$ as on the base level (i.e. $\tau_0$). The second member of each pair is generated using the tau-leap SSA, but with a smaller value of $\tau$ (denoted as $\tau_1$). There are $\mathcal{N}_1$ pairs of sample paths generated on this level, $\ell = 1$. The correction term, $\mathcal{Q}_1$, is the difference between the point estimates, when considering the first and second sample paths of each of the $\mathcal{N}_1$ pairs of sample paths:

$$\mathcal{Q}_1 := \mathbb{E}\left[Z_{\tau_1} - Z_{\tau_0}\right] \approx \frac{1}{\mathcal{N}_1} \sum_{r=1}^{\mathcal{N}_1} \left[Z_{\tau_1}^{(r)} - Z_{\tau_0}^{(r)}\right].$$

---
[2]As indicated, $Z_i(T)$ might represent the population of $S_i$ at time $T$.

Adding this correction term to the estimator calculated on the base level gives an overall more accurate estimator. This can be seen by noting that $\mathcal{Q}_0 + \mathcal{Q}_1 = \mathbb{E}\left[Z_{\tau_0}\right] + \mathbb{E}\left[Z_{\tau_1} - Z_{\tau_0}\right] = \mathbb{E}\left[Z_{\tau_1}\right]$, so that the sum of the two estimators has a bias equivalent to that of the tau-leap method with $\tau = \tau_1$. The key to the efficiency of the multi-level method is to generate the pairs of sample paths,

$$\left\{\left[\mathbf{Z}_{\tau_1}^{(r)}, \mathbf{Z}_{\tau_0}^{(r)}\right] : r = 1, \ldots, \mathcal{N}_1\right\},$$

using a variance reduction technique, so that the variance of $\left[Z_{\tau_1}^{(r)} - Z_{\tau_0}^{(r)}\right]$ is minimised. If the sample variance is denoted as $\sigma_1^2$, then the estimator variance is given as $\mathcal{V}_1 := \sigma_1^2/\mathcal{N}_1$. The CPU time taken to generate each pair of such sample paths is stored as $\mathcal{C}_1$.

On the level 2, this process is repeated to give a second correction term. On this level, $\mathcal{N}_2$ pairs of sample paths are generated using the tau-leap algorithm: the first member of the pair is generated with $\tau = \tau_1$, and the second member is generated with $\tau = \tau_2 < \tau_1$. The $\ell = 2$ correction term estimate is

$$\mathcal{Q}_2 := \mathbb{E}\left[Z_{\tau_2} - Z_{\tau_1}\right] \approx \frac{1}{\mathcal{N}_2} \sum_{r=1}^{\mathcal{N}_2} \left[Z_{\tau_2}^{(r)} - Z_{\tau_1}^{(r)}\right],$$

and it is added to the combined estimators from levels 0 and 1 to give $\mathcal{Q}_0 + \mathcal{Q}_1 + \mathcal{Q}_2 = \mathbb{E}[Z_{\tau_2}]$. Carrying on in this way, the multi-level method forms a telescoping sum in the form of Equation (3.1),

$$\mathbb{E}[Z_{\tau_\mathcal{L}}] = \mathbb{E}[Z_{\tau_0}] + \sum_{\ell=1}^{\mathcal{L}} \mathbb{E}[Z_{\tau_\ell} - Z_{\tau_{\ell-1}}] = \sum_{\ell=0}^{\mathcal{L}} \mathcal{Q}_\ell.$$

With the addition of each subsequent level, the bias of the estimator is reduced further, until a desired level of accuracy is reached.

Finally, and optionally, by generating $\mathcal{N}_{\mathcal{L}+1}$ pairs of sample paths, with each pair comprising a sample path generated with the tau-leap method with $\tau = \tau_\mathcal{L}$, and a second sample path, generated with an exact SSA, we can efficiently compute a *final correction* term,

$$\mathcal{Q}_{\mathcal{L}+1} = \mathbb{E}[X - Z_{\tau_\mathcal{L}}] \approx \frac{1}{\mathcal{N}_{\mathcal{L}+1}} \sum_{r=1}^{\mathcal{N}_{\mathcal{L}+1}} \left[ X^{(r)} - Z_{\tau_\mathcal{L}}^{(r)} \right].$$

This can be added to the telescoping sum in order to make the estimator unbiased, and hence give

$$\begin{aligned}
\mathbb{E}[X(T)] &= \mathbb{E}[Z_{\tau_0}] + \sum_{\ell=1}^{\mathcal{L}} \mathbb{E}[Z_{\tau_\ell} - Z_{\tau_{\ell-1}}] + \mathbb{E}[X - Z_{\tau_\mathcal{L}}] \\
&= \mathcal{Q}_0 + \sum_{\ell=1}^{\mathcal{L}} \mathcal{Q}_\ell + \mathcal{Q}_{\mathcal{L}+1}.
\end{aligned} \quad (3.3)$$

Importantly, if variance reduction techniques are used, the total time taken to generate the sets of sample paths for the base level, $\mathcal{Q}_0$, each of the correction terms, $\mathcal{Q}_\ell$ (for $\ell = 1, \ldots, \mathcal{L}$), and the final correction term, $\mathcal{Q}_{\mathcal{L}+1}$, can be less than that taken to estimate $\mathbb{E}[X(T)]$ using an exact SSA (such as the DM or MNRM).

Note that the multi-level method can be used to estimate either a *biased* or an *unbi-*

*ased* estimator. The biased estimator is given by

$$\mathcal{Q}_b := \mathcal{Q}_0 + \sum_{\ell=1}^{\mathcal{L}} \mathcal{Q}_\ell. \tag{3.4}$$

As explained in Section 2.4.5, estimators are influenced by two distinct types of error: a *statistical error* and a *bias*. The bias of an estimate for $\mathcal{Q}_b$ is equivalent to that of a tau-leap method with time-step $\tau_\mathcal{L}$, because $\mathcal{Q}_b = \mathbb{E}[Z_{\tau_\mathcal{L}}]$. The statistical error can be controlled by bounding the associated estimator variance, $\mathcal{V}_b$. Each of the estimators, $\mathcal{Q}_\ell$, which make up $\mathcal{Q}_b$, has an estimator variance, $\mathcal{V}_\ell$, associated with it. As each level is estimated in an independent fashion, we can write

$$\mathcal{V}_b = \sum_{\ell=0}^{\mathcal{L}} \mathcal{V}_\ell = \sum_{\ell=0}^{\mathcal{L}} \frac{\sigma_\ell^2}{\mathcal{N}_\ell}.$$

Therefore, the values of $\mathcal{N}_\ell$ (for $\ell = 1, \ldots, \mathcal{L}$) can be chosen such that the overall estimator variance is below a given threshold. That is, $\mathcal{V}_b < \varepsilon^2$.

The unbiased estimator is given by

$$\mathcal{Q}_u := \mathcal{Q}_0 + \sum_{\ell=1}^{\mathcal{L}} \mathcal{Q}_\ell + \mathcal{Q}_{\mathcal{L}+1}. \tag{3.5}$$

Any unbiased estimate that we calculate will contain only a statistical error. The statistical error of the unbiased estimator, $\mathcal{V}_u = \sum_{\ell=0}^{\mathcal{L}+1} \sigma_\ell^2/\mathcal{N}_\ell$, is bounded by following the procedure outlined for the biased estimator, $\mathcal{Q}_b$.

### 3.2.3 Defining algorithm parameters

To use the multi-level method, a number of *algorithm parameters* will need to be specified. The algorithm parameters specify exactly how the estimator $\mathcal{Q}$ is broken into (sub)-estimators $\mathcal{Q}_0, \mathcal{Q}_1, \ldots$.

Therefore, we will need to specify $\mathcal{L}$, which controls the number of levels that the algorithm uses, and the time-steps, $\tau_0, \ldots, \tau_{\mathcal{L}}$. To avoid needless complexity, we will assume that $T/\tau_\ell$ is always an integer, thereby ensuring that the time-intervals are neatly divided into equally-sized steps. To implement the multi-level algorithm, we follow Giles [46] in making the following important decision about the time-steps. Let $\mathcal{M} \in \{2, 3, \ldots\}$ be a scaling factor and take $\tau_\ell = \tau_{(\ell-1)}/\mathcal{M}$ so that

$$\mathcal{Q}_0 := \mathbb{E}[Z_{\tau_0}], \tag{3.6}$$

$$\mathcal{Q}_1 := \mathbb{E}[Z_{\tau_0/\mathcal{M}} - Z_{\tau_0}], \tag{3.7}$$

$$\mathcal{Q}_2 := \mathbb{E}[Z_{\tau_0/\mathcal{M}^2} - Z_{\tau_0/\mathcal{M}}], \tag{3.8}$$

$$\vdots$$

$$\mathcal{Q}_{\mathcal{L}} := \mathbb{E}[Z_{\tau_0/\mathcal{M}^{\mathcal{L}}} - Z_{\tau_0/\mathcal{M}^{\mathcal{L}-1}}] \tag{3.9}$$

$$\left(\mathcal{Q}_{\mathcal{L}+1} := \mathbb{E}[X - Z_{\tau_0/\mathcal{M}^{\mathcal{L}}}]\right). \tag{3.10}$$

The final sub-estimate, $\mathcal{Q}_{\mathcal{L}+1}$ is shown in brackets, because it is only required if the overall estimate is unbiased.

Therefore, the time-steps are nested[3], with the same scaling factor between each, and this choice renders the multi-level method simpler to understand and implement. The implications of this restriction are discussed in Chapter 4.

We need to decide how many sample paths must be generated on each level, i.e. for $\ell = 0, \ldots, \mathcal{L}$ (+1). The (+1) term that appears in the preceding list indicates that the final level should be included as appropriate, i.e. where an unbiased estimator is sought. As in Section 3.2.2, the CPU time taken to generate a sample value on level $\ell$ is denoted as $\mathcal{C}_\ell$. Thus, the values of $\mathcal{N}_\ell$ should be chosen to minimise the total expected computational time, $\sum_{\ell=0}^{\mathcal{L}(+1)} \mathcal{C}_\ell \cdot \mathcal{N}_\ell$, subject to the overall estimator variance, $\sum_{\ell=0}^{\mathcal{L}(+1)} \sigma_\ell^2/\mathcal{N}_\ell$, being bounded by $\varepsilon^2$. The value of $\varepsilon^2$ will be user-specified, and can be chosen on a problem-specific basis.

---
[3]With the exception of level $\mathcal{L} + 1$.

Therefore, we choose $\mathcal{N}_\ell$ such that we minimise

$$\sum_{\ell=0}^{\mathcal{L}(+1)} \mathcal{N}_\ell \cdot \mathcal{C}_\ell \quad \text{subject to the constraint} \quad \sum_{\ell=0}^{\mathcal{L}(+1)} \frac{\sigma_\ell^2}{\mathcal{N}_\ell} < \varepsilon^2. \tag{3.11}$$

We perform the required constrained optimisation by using the method of Lagrange multipliers. We seek a $\lambda \in \mathbb{R}$ such that

$$\frac{\partial}{\partial \mathcal{N}_\ell} \left[ \sum_{m=0}^{\mathcal{L}(+1)} \mathcal{C}_m \cdot \mathcal{N}_m + \lambda \sum_{m=0}^{\mathcal{L}(+1)} \frac{\sigma_m^2}{\mathcal{N}_m} \right] = 0 \quad \text{for} \quad \ell = 0, 1, \ldots, \mathcal{L}\,(+1).$$

This implies $\mathcal{N}_\ell = \sqrt{\lambda \cdot \sigma_\ell^2/\mathcal{C}_\ell}$. As we require $\sum_{m=0}^{\mathcal{L}(+1)} \sigma_m^2/\mathcal{N}_m < \varepsilon^2$, it follows that $\sqrt{\lambda} = \sum_{m=0}^{\mathcal{L}(+1)} \sqrt{\sigma_m^2 \cdot \mathcal{C}_m}/\varepsilon^2$. Therefore, each $\mathcal{N}_\ell$ should be chosen to be

$$\mathcal{N}_\ell = \left\{ \frac{1}{\varepsilon^2} \sum_{m=0}^{\mathcal{L}(+1)} \sqrt{\sigma_m^2 \cdot \mathcal{C}_m} \right\} \sqrt{\sigma_\ell^2/\mathcal{C}_\ell}. \tag{3.12}$$

We will now substitute values for $\sigma_\ell^2$ and $\mathcal{C}_\ell$ into Equation (3.12). The exact values of $\sigma_\ell^2$ (for $\ell = 0, \ldots, \mathcal{L}\,(+1)$) are analytically unknown. Therefore, we must estimate the variances, $\sigma_\ell^2$. Section 3.5 contains a detailed discussion of how we should estimate $\sigma_\ell^2$. At this stage, we simply note that Anderson and Higham [47] generate 100 initial sample paths on each level as a basis for estimating the variances. The process of generating these simulations is timed to provide estimates for $\mathcal{C}_\ell$.

The method of Lagrange multipliers estimates that the total CPU time required by the multi-level method is

$$\frac{1}{\varepsilon^2} \left\{ \sum_{\ell=0}^{\mathcal{L}(+1)} \sqrt{\mathcal{C}_\ell \cdot \sigma_\ell^2} \right\}^2, \tag{3.13}$$

units of CPU time.

Equation (3.13) confirms that if we can reduce the variance on each level, $\sigma_\ell^2$, then the overall simulation time will decrease.

Based on the description of the multi-level method set out in Section 3.2.2, we will consider three types of level as part of the multi-level estimator.

- $\ell = 0$ represents the *base level*. We estimate $\mathcal{Q}_0$ by generating sample paths using the tau-leap method with time-step $\tau_0$;

- $\ell = 1, \ldots, \mathcal{L}$ represents the tau-leap *correction levels*. We estimate $\mathcal{Q}_\ell$ by comparing pairs of sample paths: for each pair, one path is simulated with time-step $\tau_\ell$, and the second path is simulated with time-step $\tau_{\ell-1}$. A detailed simulation method is stated below;

- for an unbiased estimator only, $\ell = \mathcal{L} + 1$ represents the *final correction level*. We estimate $\mathcal{Q}_{\mathcal{L}+1}$ by comparing pairs of sample paths: for each pair, one path is simulated exactly, and the second path is simulated with the tau-leap method with time-step $\tau_\mathcal{L}$. Again, a detailed simulation algorithm will be provided below.

In order for estimates on level $\ell = 1, \ldots, \mathcal{L}$ (+1) to be generated quickly, a variance reduction technique will be used.

### 3.2.4 Estimating the correction terms, $\mathcal{Q}_\ell$

We now explain how a variance reduction technique can be implemented for the correction levels, $\mathcal{Q}_\ell$. For $\ell = 1, \ldots, \mathcal{L}$, the estimate of $\mathcal{Q}_\ell$ is given by

$$\mathcal{Q}_\ell := \mathbb{E}\left[Z_{\tau_\ell} - Z_{\tau_{\ell-1}}\right] \approx \frac{1}{\mathcal{N}_\ell} \sum_{r=1}^{\mathcal{N}_\ell} \left[Z_{\tau_\ell}^{(r)} - Z_{\tau_{\ell-1}}^{(r)}\right]. \tag{3.14}$$

To generate the $r$-th sample value, $Z^{(r)}_{\tau_0/\mathcal{M}^\ell} - Z^{(r)}_{\tau_0/\mathcal{M}^{\ell-1}}$, we will need to simultaneously generate two sample paths using the tau-leap method, but with different time-steps. As we are constructing a Monte Carlo estimator, we require each of the sample values, $\left[Z^{(r)}_{\tau_0/\mathcal{M}^\ell} - Z^{(r)}_{\tau_0/\mathcal{M}^{\ell-1}}\right]$, to be independent of the other bracketed terms in Equation (3.14). The key point to note is that for each sample there is no need for $Z^{(r)}_{\tau_0/\mathcal{M}^\ell}$ and $Z^{(r)}_{\tau_0/\mathcal{M}^{\ell-1}}$ to be independent of one another. This is because the estimator $\mathcal{Q}_\ell$ is not dependent on the actual copy numbers within each sample path, but merely their difference. Therefore, we have a choice of two suitable estimators for $\mathcal{Q}_\ell$:

1. for each $r$, let $Z^{(r)}_{\tau_0/\mathcal{M}^\ell}$ and $Z^{(r)}_{\tau_0/\mathcal{M}^{\ell-1}}$ be dependent;

2. for each $r$, let $Z^{(r)}_{\tau_0/\mathcal{M}^\ell}$ and $Z^{(r)}_{\tau_0/\mathcal{M}^{\ell-1}}$ be independent.

The variance reduction principle says that we should choose the estimator with the lower variance. By recalling that

$$\mathrm{Var}\left[Z_{\tau_0/\mathcal{M}^\ell} - Z_{\tau_0/\mathcal{M}^{\ell-1}}\right] = \mathrm{Var}\left[Z_{\tau_0/\mathcal{M}^\ell}\right] + \mathrm{Var}\left[Z_{\tau_0/\mathcal{M}^{\ell-1}}\right] - 2\,\mathrm{Cov}\left[Z_{\tau_0/\mathcal{M}^\ell}, Z_{\tau_0/\mathcal{M}^{\ell-1}}\right],$$

we note it is in our interests for $Z^{(r)}_{\tau_0/\mathcal{M}^\ell}$ and $Z^{(r)}_{\tau_0/\mathcal{M}^{\ell-1}}$ to depend on one another. In particular, a strong positive correlation will give rise to a lower estimator variance. We achieve this positive correlation by keeping the $r$-th sample paths of the approximate processes with time-steps $\tau_0/\mathcal{M}^\ell$ and $\tau_0/\mathcal{M}^{\ell-1}$ as similar to each other as possible, throughout the time-period of interest.

For each pair of sample paths generated on level $\ell$, we call the sample path with time-step $\tau_\ell = \tau_0/\mathcal{M}^\ell$ the 'fine' path, and the path with time-step $\tau_{\ell-1} = \tau_0/\mathcal{M}^{\ell-1}$ the 'coarse' path. We let $\boldsymbol{Z}^C$ and $\boldsymbol{Z}^F$ be the copy numbers in the coarse and fine resolution sample paths, respectively. For each reaction channel, $R_j$, we define $a_j^C$ to be its propensity function in the coarse resolution sample path, and $a_j^F$ to be its propensity in the fine resolution sample path.

Since both paths have the same initial conditions, one approach to achieving strong positive correlation between the two paths is to simultaneously simulate each sample path, and aim to have each reaction channel fire a similar number of times in both sample paths.

The simulation method will rely on the thickening property of the Poisson distribution. We state the following fact [57]:

**Fact 1.** Suppose $\mathcal{P}_1$, $\mathcal{P}_2$, and $\mathcal{P}_3$ are independent Poisson random variates. Then, for parameters $\alpha > 0$, $\beta > 0$,

$$\mathcal{P}_1(\alpha + \beta) \sim \mathcal{P}_2(\alpha) + \mathcal{P}_3(\beta), \qquad (3.15)$$

where $\sim$ implies equality in distribution. ∎

This means that a Poisson random variate with parameter $\alpha + \beta$ can be generated by generating two Poisson variates, one with parameter $\alpha$ and the other with parameter $\beta$, and then adding them.

In terms of the sample paths, the thickening property implies that we can use one Poisson random variate to determine how many of a particular type of reaction happen in *both* the coarse and fine resolution sample paths during a time-step and then 'top up' any further reactions that happen in only one of the sample paths using further Poisson random variates.

In practice, this can be achieved by re-formulating each reaction channel, $R_j$, into three virtual reaction channels. We call these virtual channels $R_j^1$, $R_j^2$ and $R_j^3$ and define the virtual channels such that:

- $R_j^1$ : reaction $R_j$ occurs in both the coarse and fine paths;

- $R_j^2$ : reaction $R_j$ occurs only in the coarse path;

- $R_j^3$ : reaction $R_j$ occurs only in the fine path.

We assign the following values as propensities for $R_j^1$, $R_j^2$, and $R_j^3$:

$$\begin{aligned} b_j^1 &= \min\{a_j^F, a_j^C\}; \\ b_j^2 &= a_j^C - b_j^1; \\ b_j^3 &= a_j^F - b_j^1. \end{aligned} \qquad (3.16)$$

Note that, for each $j$, at least one of $b_j^2$ and $b_j^3$ will equal 0. To generate coupled sample paths, we use an algorithm that steps forward, in time, with fine-resolution time-steps, $\tau_\ell$.

We update the propensity functions of the fine resolution sample path at each time-step, but only update the propensity functions of the coarse resolution sample path every $\mathcal{M}$ steps. In other words, we only update the propensity functions of the coarse resolution sample path every $\tau_{\ell-1}$ units of time. Following this procedure ensures that both sample paths exhibit the appropriate dynamics. The full method is stated in Algorithm 3.1.

We note that by Equation (3.15), with a time-step of $\tau_\ell = \tau_0/\mathcal{M}^\ell$,

$$\begin{aligned} \mathcal{P}(a_j^C \cdot \tau_\ell) &\sim \mathcal{P}(b_j^1 \cdot \tau_\ell) + \mathcal{P}(b_j^2 \cdot \tau_\ell), \\ \mathcal{P}(a_j^F \cdot \tau_\ell) &\sim \mathcal{P}(b_j^1 \cdot \tau_\ell) + \mathcal{P}(b_j^3 \cdot \tau_\ell), \end{aligned} \qquad (3.17)$$

so that each sample path is updated using the correct propensity functions. In Figure 3.1 we illustrate how the time-steps are arranged on the time axis. We have shown

---

Algorithm 3.1: Coupled tau-leap method. This simulates a pair of sample paths.

**Require:** initial conditions, $\boldsymbol{Z}(0)$, time-steps $\tau_C$, $\tau_F$, and terminal time, $T$.
 1: set $\boldsymbol{Z}^C \leftarrow \boldsymbol{Z}(0)$, $\boldsymbol{Z}^F \leftarrow \boldsymbol{Z}(0)$ and $t \leftarrow 0$. Set `constant` $\mathcal{M} \leftarrow \tau_C/\tau_F$
 2: **for** each $t \in \{0, \tau_{\ell-1}, 2\cdot\tau_{\ell-1}, \cdots, T - \tau_{\ell-1}\}$ **do**
 3:     for each $R_j$, calculate propensity value $a_j^C(\boldsymbol{Z}^C)$
 4:     **for** each $s \in \{t, t + \tau_\ell, \ldots, t + (\mathcal{M}-1)\cdot\tau_\ell\}$ **do**
 5:         for each $R_j$, calculate propensity value $a_j^F(\boldsymbol{Z}^F)$
 6:         for each $R_j$, calculate virtual propensities $b_j^1$, $b_j^2$ and $b_j^3$
 7:         **for** each $R_j$ and **for** each $k \in \{1, 2, 3\}$ **do**
 8:             generate $K_{jk} \sim \mathcal{P}(b_j^k \cdot \tau_\ell)$
 9:         **end for**
10:         set $\boldsymbol{Z}^C \leftarrow \boldsymbol{Z}^C + \sum_{j=1}^M (K_{j1} + K_{j2}) \cdot \boldsymbol{\nu}_j$
11:         set $\boldsymbol{Z}^F \leftarrow \boldsymbol{Z}^F + \sum_{j=1}^M (K_{j1} + K_{j3}) \cdot \boldsymbol{\nu}_j$
12:     **end for**
13: **end for**

---

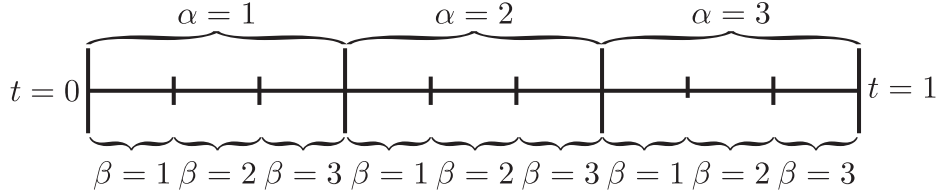

Figure 3.1: A diagrammatic representation of the steps in the algorithm, shown on a time axis, from time $t = 0$ to $t = 1$. The vertical lines represent the discretisation of time. The values of $\alpha$ represent successive runs of the outer for loop in Algorithm 3.1, whilst $\beta$ represents runs of the inner for loop of Algorithm 3.1.

---

a coarse time-step of $\tau_{\ell-1} = 1/3$, and a fine time-step of $\tau_\ell = 1/9$. In this case, the scaling factor is $\mathcal{M} = 3$, and so we have three steps of the fine sample path for every step of the coarse sample path.

Using the same Poisson random variates, $K_{j1}$ (for $j = 1, \ldots, M$) to update both the coarse and fine sample path populations is crucial to the success of the method, and has the effect of introducing a strong path-wise correlation between the coarse and fine resolution sample paths. The premise is as follows: if the state vectors $\boldsymbol{Z}^C(t)$ and $\boldsymbol{Z}^F(t)$ show similar populations for each species, then we expect $a_j^C$ and $a_j^F$ to be similar for all $j$, as these are continuous functions of the underlying populations. If this is the case then the $b_j^k$, as defined in Equation (3.15), are such that for all $j$, $b_j^1 \gg \max\{b_j^2, b_j^3\}$. To ensure compliance with the tau-leap method, we introduce a further $K_{j2}$ and $K_{j3}$ reactions in the coarse and fine sample paths, respectively. Note that at least one of $K_{j2}$ and $K_{j3}$ will be zero so that we 'top up' at most one of the sample paths. As we expect $K_{j1}$ to be significantly larger than both $K_{j2}$ and $K_{j3}$, the main part of the fluctuation is common to both sample paths. Here, the aim is to ensure that the state vectors in both sample paths remain comparable. The argument then repeats itself for each time-step, and the population differences between equivalent species at the terminal time are therefore likely to be small.

### 3.2.5  Estimating the final correction term, $\mathcal{Q}_{\mathcal{L}+1}$

We now provide a new technique for estimating $\mathcal{Q}_{\mathcal{L}+1}$, the final correction term that is needed to produce an unbiased summary statistic, $\mathcal{Q}_u$. $\mathcal{Q}_{\mathcal{L}+1}$ is the expected difference between the summary statistic generated from a tau-leap approximation with $\tau_{\mathcal{L}} = \tau_0/\mathcal{M}^{\mathcal{L}}$ and that generated using an exact SSA. The benefit of including this final correction term into the multi-level estimator is that it allows us to produce an overall unbiased estimator, and therefore provides an output comparable to that of the DM. Recall that

$$\mathcal{Q}_{\mathcal{L}+1} = \mathbb{E}\left[X - Z_{\tau_0/\mathcal{M}^{\mathcal{L}}}\right] \approx \frac{1}{\mathcal{N}_{\mathcal{L}+1}} \sum_{r=1}^{\mathcal{N}_{\mathcal{L}+1}} \left[X^{(r)} - Z_{\tau_{\mathcal{L}}}^{(r)}\right],$$

where $X^{(r)}$ and $Z_{\tau_{\mathcal{L}}}^{(r)}$ represent the copy numbers of the species of interest at time $T$ in the $r$-th sample paths, generated by the DM, and tau-leap SSA with time-step $\tau_{\mathcal{L}}$, respectively. As for levels $\ell = 1, \ldots, \mathcal{L}$, we aim to correlate the sample paths $X^{(r)}$ and $Z_{\tau_{\mathcal{L}}}^{(r)}$ in order to reduce the variance in $\mathcal{Q}_{\mathcal{L}+1}$.

The difficulty in coupling the sample paths for this estimator arises because the tau-leap sample path has its reaction propensities updated after a fixed period of time, and not after a fixed number of reactions. The DM is not equipped to provide sample paths that exhibit this time-dependent behaviour. One approach to handling this situation is to use a form of the MNRM (see Algorithm 2.2) to simulate the required sample paths [47]. We do not explore this approach further, but rather present our own, simpler method. Our technique is mathematically equivalent to the MNRM, and therefore generates equivalent statistics (detailed results not shown).

In order to couple the same paths, we reformulate the tau-leap method so that it can be implemented in the same way as the DM, in the sense that we will individually simulate each reaction of the tau-leap sample path. We start by recalling that the tau-leap process $\boldsymbol{Z}(t)$ can be expressed in the form of Equation (2.18):

$$\boldsymbol{Z}(k \cdot \tau) = \boldsymbol{Z}((k-1) \cdot \tau) + \sum_{j=1}^{M} \mathcal{Y}_j\left(P_{k-1,j}, P_{k,j}\right) \cdot \boldsymbol{\nu}_j.$$

The tau-leap method, as stated in Algorithm 2.3, uses Poisson *random numbers* to generate the values of $\mathcal{Y}_j(P_{k-1,j}, P_{k,j})$ (for $j = 1, \ldots, M$). The values of $\mathcal{Y}_j(P_{k-1,j}, P_{k,j})$ are Poisson distributed, i.e.

$$\mathcal{Y}_j(P_{k-1,j}, P_{k,j}) \sim \mathcal{P}(a_j(\boldsymbol{Z}((k-1) \cdot \tau))).$$

Instead of using a Poisson variate, the reactions could simply be fired by running the DM over each time-interval, $((k-1) \cdot \tau, k \cdot \tau)$. The trick is that the propensities are to remain constant: the propensity of $R_j$ is fixed at $a_j(\boldsymbol{Z}((k-1) \cdot \tau))$ for the time-interval. In line with the regular tau-leap method, the propensity functions are only updated at the end of each time-step.

As before, let $\boldsymbol{X}$ and $\boldsymbol{Z}$ represent the exact and approximate stochastic processes,

respectively. For each reaction, $R_j$, let the respective propensity values be $a_j^X$ and $a_j^Z$. Following Anderson and Higham [47], for each $R_j$ we define virtual channels, $R_j^1$, $R_j^2$ and $R_j^3$, such that:

- $R_j^1$ : reaction $R_j$ occurs in both paths; the propensity is $\min\{a_j^X, a_j^Z\}$;

- $R_j^2$ : reaction $R_j$ occurs only in the exact sample path; the propensity is $a_j^X - \min\{a_j^X, a_j^Z\}$;

- $R_j^3$ : reaction $R_j$ occurs only in the approximate, tau-leap sample path; the propensity is $a_j^Z - \min\{a_j^X, a_j^Z\}$.

In Algorithm 3.2, we present a new, time-saving method. When we implement the simulation algorithm, we will follow this procedure. For each $j = 1, \ldots, M$, we place virtual reactions $R_j^1$, $R_j^2$ and $R_j^3$ into a group, which we label $R_j^{1,2,3}$. The total propensity of the group $R_j^{1,2,3}$ is $\max\{a_j^X, a_j^Z\}$[4], and at least one of $R_j^2$ and $R_j^3$ has a propensity of zero. Each reaction is simulated by following a two-step procedure:

1. we first pick a group, $R_j^{1,2,3}$, where $R_j^{1,2,3}$ has propensity $\max\{a_j^X, a_j^Z\}$. Exactly one of $R_j^1$, $R_j^2$ and $R_j^3$ will now take place, and we need to choose which one;

2. if $a_j^X > a_j^Z$, then the propensity of $R_j^3$ is zero, and we must choose between $R_j^1$ and $R_j^2$. This means that $R_j$ certainly fires in the exact sample path (as both $R_j^1$ and $R_j^2$ result in this outcome). Reaction $R_j$ fires in the tau-leap sample path with probability $a_j^Z / a_j^X$ (which is the probability that $R_j^2$, and not $R_j^1$, fires). An equivalent result applies if $a_j^Z > a_j^X$.

Algorithm 3.2 provides computational savings because it performs the second step by using a rejection sampling approach (details not shown).

---

[4]Note that $\min\{a_j^X, a_j^Z\} + \left(a_j^X - \min\{a_j^X, a_j^Z\}\right) + \left(a_j^Z - \min\{a_j^X, a_j^Z\}\right) = \max\{a_j^X, a_j^Z\}$.

Algorithm 3.2: Simulates a pair of sample paths: an exact path and a path with the tau-leap method using time-step $\tau$.

---

**Require:** initial conditions, $\boldsymbol{X}(0)$, time-step, $\tau$, and terminal time, $T$.
1: set $\boldsymbol{X} \leftarrow \boldsymbol{X}(0)$, $\boldsymbol{Z} \leftarrow \boldsymbol{X}(0)$, $t \leftarrow 0$, and $t^* \leftarrow \tau$
2: for each $R_j$, calculate propensity values $a_j^X(\boldsymbol{X})$ and $a_j^Z(\boldsymbol{Z})$
3: **loop**
4:     for each $R_j$, set $a_j^* = \max\{a_j^X, a_j^Z\}$, and set $a_0^* = \sum_{j=1}^M a_j^*$
5:     set $\Delta \leftarrow \mathrm{Exp}(a_0^*)$
6:     **if** $t + \Delta > T$ **then**
7:         **break**
8:     **else if** $t + \Delta > t^*$ **then**
9:         set $t \leftarrow t^*$, and $t^* \leftarrow t^* + \tau$
10:        for each $R_j$, calculate propensity values $a_j^Z(\boldsymbol{Z})$
11:     **else**
12:        choose index $k$, where $k$ has probability $a_k^*/a_0^*$ of being chosen
13:        with probability $a_k^X/a_k^*$, set $\boldsymbol{X} \leftarrow \boldsymbol{X} + \boldsymbol{\nu}_k$
14:        with probability $a_k^Z/a_k^*$, set $\boldsymbol{Z} \leftarrow \boldsymbol{Z} + \boldsymbol{\nu}_k$
15:        set $t \leftarrow t + \Delta$
16:        for each $R_j$, calculate propensity values $a_j^X(\boldsymbol{X})$
17:     **end if**
18: **end loop**

---

In the context of computational efficiency, we expect the revised algorithm to differ from the original MNRM [47] in two significant ways. Firstly, in line 9 of Algorithm 3.2, we 'reject' an algorithm step, as the random number generated in line 6 is discarded. This step is justified by the memory-less property [57]. In contrast, one of the main attractions of the MNRM is that it discards effectively no random numbers. Secondly, in line 13 of Algorithm 3.2, an index $k$ is chosen. There are $M$ possibilities to search through at this step; this step is amenable to optimisation in much the same way as the DM [60, 61, 62]. The downside of the MNRM in this context is that a substantial amount of complex maintenance work needs to be carried out within the algorithm, and that choice of the next reaction involves an unavoidably time-consuming search to find the minimum within a matrix of 'next reaction times' on a substantially enlarged state space (as there are $3 \cdot M$ reactions, i.e. three virtual channels for every reaction channel, to search through). Optimisation of the MNRM is somewhat less straightforward, but the method of Gibson and Bruck [94] can be adapted.

In the next section we test and evaluate the multi-level method. We will show how the multi-level method provides superior computational performance when estimating both biased and unbiased summary statistics of interest.

## 3.3 Preliminary investigations

**Case Study 2.** To demonstrate the benefits of the multi-level method, we return to Case Study 2. As before, we estimate the expected dimer population of the gene regulatory network (2.10), which we recall comprises the following reaction channels:

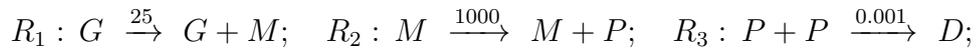

$$R_1 : G \xrightarrow{25} G + M; \quad R_2 : M \xrightarrow{1000} M + P; \quad R_3 : P + P \xrightarrow{0.001} D;$$

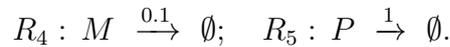

$$R_4 : M \xrightarrow{0.1} \emptyset; \quad R_5 : P \xrightarrow{1} \emptyset.$$

To implement the multi-level method, sensible algorithm parameters need to be chosen to minimise the CPU time required. To demonstrate the multi-level method, we take $\mathcal{M} = 3$, $\tau_0 = 1/9$ and $\mathcal{L} = 5$ as the algorithm parameters. In the upcoming sections, we show that our choices of $\mathcal{M}$, $\mathcal{L}$ and $\tau_0$ are appropriate for the chosen reaction network. The value of the mean dimer population is estimated in two ways:

1. using an unbiased estimator, $\mathcal{Q}_u$. In this case, we compare our results with DM- and MNRM-derived estimates;

2. using a biased estimator, $\mathcal{Q}_b$. In this case, our results are compared with a tau-leap estimate, where $\tau = \tau_0/\mathcal{M}^{\mathcal{L}}$. With our choice of algorithm parameters, $\tau_{\mathcal{L}} = 1/2187$.

| Level | $\tau_{\ell-1}$ | $\tau_\ell$ | Estimate | Variance | Paths | Time |
|---|---|---|---|---|---|---|
| $\mathcal{Q}_0$ | - | $3^{-2}$ | 3187.47 | $1.03 \times 10^6$ | $7.11 \times 10^6$ | 89.9s |
| $\mathcal{Q}_1$ | $3^{-2}$ | $3^{-3}$ | 350.52 | 16287.10 | $4.21 \times 10^5$ | 24.9s |
| $\mathcal{Q}_2$ | $3^{-3}$ | $3^{-4}$ | 117.48 | 2666.80 | $1.14 \times 10^5$ | 15.5s |
| $\mathcal{Q}_3$ | $3^{-4}$ | $3^{-5}$ | 39.15 | 658.14 | $4.10 \times 10^4$ | 12.0s |
| $\mathcal{Q}_4$ | $3^{-5}$ | $3^{-6}$ | 13.00 | 196.09 | $1.75 \times 10^4$ | 10.7s |
| $\mathcal{Q}_5$ | $3^{-6}$ | $3^{-7}$ | 4.42 | 48.28 | $6.41 \times 10^3$ | 7.0s |
| $\mathcal{Q}_6$ | $3^{-7}$ | DM | 2.19 | 38.75 | $2.87 \times 10^3$ | 5.8s |
| **Total** | | | **3714.23 ± 0.99** | | - | **165.8s** |

Table 3.1: The contribution from each level in producing an unbiased overall estimator, $\mathcal{Q}_u$, for $\mathbb{E}[X_3]$ in system (2.10) at $T = 1$. We have taken $\tau_0 = 1/9$, $\mathcal{M} = 3$, and $\mathcal{L} = 5$.

We estimate $\mathcal{Q}_u = 3714.23 \pm 0.99$ within 165.8 seconds by using the multi-level method. The error tolerance refers to a 95% confidence interval; we took $\varepsilon^2 = 0.2603$. A detailed breakdown of the contribution to the estimate $\mathcal{Q}_u$ by each level is provided in Table 3.1. Compared with the 7223 seconds taken for the DM (see Section 2.3.2; the corresponding figure for the MNRM is 7513 seconds), the multi-level approach is nearly 44 times faster for this case study (45 times faster than the MNRM).

We also provide an estimate for the biased estimator, $\mathcal{Q}_b$. We estimate that $\mathbb{E}[X_3(1)]$ = 3711.81 ± 1.00 by implementing the multi-level method. This calculation takes 156.5 seconds; this duration represents a simulation time that is nearly 26 times smaller than that taken by a regular tau-leap method with the same bias (which takes 4010 seconds; see Figure 2.4 for further information). A detailed breakdown of the contribution to the estimate $\mathcal{Q}_u$ by each level is provided in Table 3.2.

Therefore, under suitable conditions, the multi-level method has the capability to reduce the computational demands of Monte Carlo simulation dramatically. However, the results presented in Tables 3.1 and 3.2 are only achievable after careful fine-tuning of the algorithm parameters. In the following sections, we provide a systematic approach towards configuring the algorithm. We also highlight where the fine-tuning procedure can be automated, meaning that the multi-level can be relied on to provide superior computational performance.

| Level | $\tau_{\ell-1}$ | $\tau_\ell$ | Estimate | Variance | Paths | Time |
|---|---|---|---|---|---|---|
| $\mathcal{Q}_0$ | - | $3^{-2}$ | 3187.19 | $1.03 \times 10^6$ | $7.11 \times 10^6$ | 87.1s |
| $\mathcal{Q}_1$ | $3^{-2}$ | $3^{-3}$ | 350.52 | 16263.66 | $4.17 \times 10^5$ | 24.8s |
| $\mathcal{Q}_2$ | $3^{-3}$ | $3^{-4}$ | 117.67 | 2618.71 | $1.11 \times 10^5$ | 15.8s |
| $\mathcal{Q}_3$ | $3^{-4}$ | $3^{-5}$ | 39.21 | 692.05 | $4.72 \times 10^4$ | 14.4s |
| $\mathcal{Q}_4$ | $3^{-5}$ | $3^{-6}$ | 12.98 | 206.63 | $17.8 \times 10^4$ | 11.4s |
| $\mathcal{Q}_5$ | $3^{-6}$ | $3^{-7}$ | 4.24 | 74.72 | $2.72 \times 10^3$ | 3.1s |
| Total | | | $3711.81 \pm 1.00$ | | - | 156.5s |

Table 3.2: The contribution from each level in producing a biased overall estimator, $\mathcal{Q}_b$, for $\mathbb{E}[X_3]$ in system (2.10) at $T = 1$. We have taken $\tau_0 = 1/9$, $\mathcal{M} = 3$, and $\mathcal{L} = 5$.

## 3.4 Overviewing the research aims of this chapter

We now state our research aims for the remainder of this chapter, and we highlight a number of our enhancements to the multi-level method. We will aim to develop the multi-level method so that it can be reliably applied to study a range of reaction networks and summary statistics of interest.

To recap: a simple, straightforward multi-level implementation might proceed with three major steps:

1. the parameter values for $\tau_0$, $\mathcal{M}$, and $\mathcal{L}$ are chosen;

2. we decide how many simulations to perform on each level of the multi-level method by considering Equation (3.12). To use Equation (3.12), the variance, $\sigma_\ell^2$, and simulation cost, $\mathcal{C}_\ell$, of each level, must be estimated;

3. finally, for each level $\ell$, we generate the $\mathcal{N}_\ell$ simulations required.

Having completed these three steps, we collate our results to arrive at an estimate,

$\widehat{\mathcal{Q}}$. We now discuss, in detail, how the multi-level method can be streamlined.

As the sample paths of the multi-level method are generated during steps (2) and (3), we first evaluate and optimise these steps. Once we have optimised steps (2) and (3), we can return to step (1) and then, taking into account our improvements to the multi-level method, the appropriate algorithm parameters can be chosen.

In Section 3.5, we develop a dynamic calibration method that repeatedly recalculates the number of simulations required on each level. We start with an initial estimate of the number of simulations required on level $\ell$, $\mathcal{N}_\ell$. As more sample paths are generated, we can update and improve the estimated values of $\sigma_\ell^2$, and then, by considering the updated estimate, revise the number of samples required for each level. This recalculation procedure leads to a multi-level method that is fast and reliable.

In Section 3.6, we return to step (1), and discuss how the values of the algorithm parameters, $\tau_0$, $\mathcal{M}$ and $\mathcal{L}$, can be chosen to optimise computational performance. The value of $\tau_0$ is chosen through a recursive search procedure. A distinction is then drawn between a biased and an unbiased estimator, and this distinction allows us to assign a value for $\mathcal{L}$. We argue that for a biased estimator, the mean-squared error should be used to quantify the accuracy of the estimate, and a suitable simulation algorithm is then described. In addition, the value of the scaling parameter, $\mathcal{M}$, is discussed.

In Sections 3.7 and 3.8 we will demonstrate the numerical benefits of our developments.

## 3.5 Dynamic calibration of the multi-level method

In this section, we investigate the method for choosing $\mathcal{N}_\ell$, the number of simulations on each level, $\ell$ (for $\ell = 0, \ldots, \mathcal{L}\,(+1)$). From Equation (3.12), we see that the number of simulations required on level $\ell$ is given by

$$\mathcal{N}_\ell = \left\{ \frac{1}{\varepsilon^2} \sum_{m=0}^{\mathcal{L}(+1)} \sqrt{\sigma_m^2 \cdot \mathcal{C}_m} \right\} \sqrt{\sigma_\ell^2 / \mathcal{C}_\ell},$$

where $\mathcal{C}_\ell$ records the per-sample computational cost on each level, and $\sigma_\ell^2$, the sample variance. This section will be concerned with the competing approaches for estimating the unknown quantities, $\sigma_\ell^2$.

### 3.5.1 Investigating one-step calibration

We start by investigating the feasibility of using a small number of 'initial simulations' to estimate $\sigma_\ell^2$ in a single step. In Anderson and Higham [47], $\mathcal{N} = 10^2$ initial simulations were performed on each level, so that $\sigma_\ell^2$ could be estimated, and the number of simulations required for each level subsequently calculated. We show that, whilst this estimation procedure is quick and straightforward, the resulting accuracy and computational performance of the multi-level method is somewhat lacklustre.

To investigate the suitability of the initial simulations or one-step calibration approach, we return to Case Study 2 where the reaction network (2.10) is used to describe a gene regulatory network. As before, we take $\tau_0 = 1/9$, $\mathcal{M} = 3$ and $\mathcal{L} = 5$ as the multi-level algorithm parameters. We calculate biased and unbiased estimators for $\mathcal{Q} = \mathbb{E}[X_3(1)]$, the expected dimer population, and we seek a statistical error commensurate with a confidence interval of semi-length 1.0. To understand the behaviour of the multi-level method, the full algorithm is run, from start-to-finish, 1000 times over. For each of the 1000 runs, we will first generate the initial simulations (using $\mathcal{N} = 10^2$ trials), then decide how many simulations are required to determine each estimator (as per Section 3.2.3), and finally perform the remaining required simulations. Each run is associated with a different (pseudo)-random number seed, and therefore, uses different random numbers. This procedure is then repeated with $\mathcal{N} = 10^3$ initial simulations and the results are compared. We demonstrate that a fixed number of initial simulations (be that $\mathcal{N} = 10^2$ or $\mathcal{N} = 10^3$) leads to unsatisfactory outcomes.

In Figure 3.2 we show the effect of using $\mathcal{N} = 10^2$ and $\mathcal{N} = 10^3$ initial simulations on the multi-level estimator. For each of the 1000 runs of the multi-level method, the absolute simulation time is plotted against the resultant confidence interval semi-length. When $\mathcal{N} = 10^2$, the average simulation time of the unbiased multi-level method is 136.7 seconds, and the average confidence interval semi-length is 1.18. The biased multi-level method takes on average 132.4 seconds to run, and generates a confidence interval of semi-length 1.13. When $\mathcal{N}$ is increased to equal $10^3$, the unbiased method has a mean run-time of 151.2 seconds, and generates an average confidence interval of semi-length 1.06. On average, the biased method runs in 143.6 seconds, and generates a confidence interval of semi-length 1.03.

The simulation results shown in Figure 3.2 are disappointing. In fact, when considering the unbiased estimator, $\mathcal{Q}_u$, if $\mathcal{N} = 10^2$ initial simulations are performed, only 8% of the estimates have a confidence interval of semi-length 1.0 or smaller. If $\mathcal{N} = 10^3$ initial simulations are performed, then approximately 33% of estimates achieve the target confidence interval. If the biased estimator, $\mathcal{Q}_b$, is considered, then the proportion of runs with the required confidence interval (or better) are 11% and 41%, respectively.

### 3.5.2 Setting out dynamic, multi-step calibration

We will therefore develop a new, dynamic simulation method to improve the reliability of the multi-level method. We start by supposing that the multi-level estimator comprises only the level $\ell = 0$. In the next few paragraphs, we will gradually add more levels into the estimator. We generate $\mathcal{N} = 10^2$ trial simulations of this single level. From Section 3.2, we recall this means simulating $10^2$ tau-leap sample paths, with $\tau = \tau_0$. Then, we estimate $\sigma_0^2$ and $\mathcal{C}_0$. If, with only level $\ell = 0$, we are to achieve an estimator variance of $\varepsilon^2$, we require[5]

$$\widehat{\mathcal{N}_0^{(0)}} = \frac{1}{\varepsilon^2} \left( \sqrt{\sigma_0^2 \cdot \mathcal{C}_0} \right) \sqrt{\sigma_0^2/\mathcal{C}_0}.$$

simulations on this level. The superscript (0) indicates that this is the first step of the dynamic calibration programme, and the circumflex denotes that this quantity indicates the number of simulations required (rather than the quantity that has been generated). As $\mathcal{N}_0 = 10^2$, a further $\widehat{\mathcal{N}_0^{(0)}} - 10^2$ simulations are now generated.

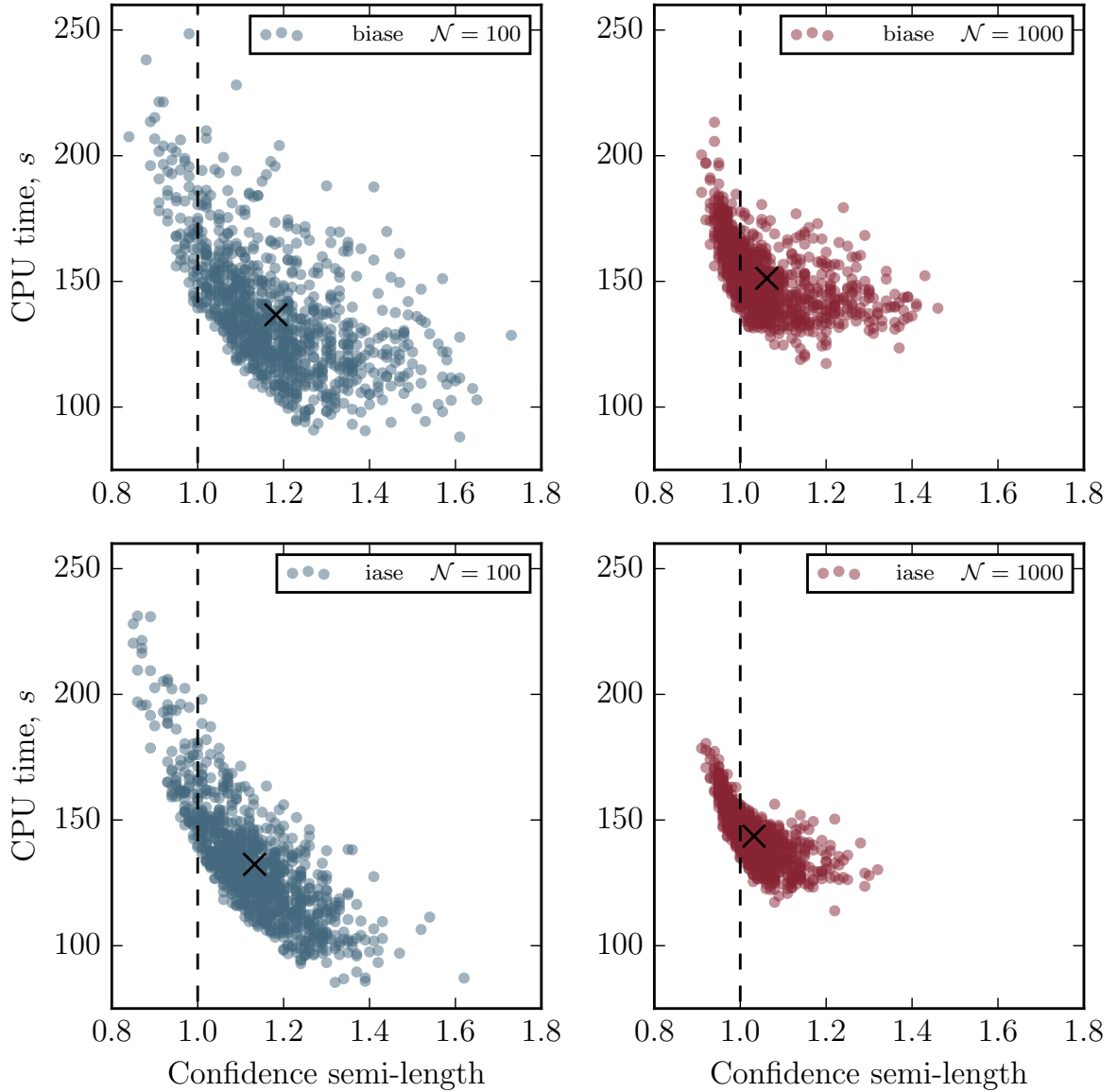

Figure 3.2: We compare the unbiased and biased multi-level methods, where either $\mathcal{N} = 10^2$ or $\mathcal{N} = 10^3$ initial simulations are performed. For each figure, the full multi-level method is run 1000 times to estimate the dimer population of system (2.10). The simulation times are plotted against the semi-length of the confidence intervals attained. The black crosses represent the mean values of the data. The target confidence interval is shown with a dashed line. As implemented, the method is unlikely to attain the required confidence interval.

Our next step is to assume that there are two levels, denoted $\ell = 0$ and $\ell = 1$. We have already generated $\mathcal{N}_0$ simulations for level $\ell = 0$. We generate an initial $\mathcal{N}_1 = 10^2$ trial simulations for level $\ell = 1$. The variances, $\sigma_0^2$ and $\sigma_1^2$, are now calculated based on *all* the simulations generated thus far. Then, assuming we have levels $\ell = 0$ and $\ell = 1$ only, an estimator variance of $\varepsilon^2$ is achieved if $\mathcal{N}_0$ and $\mathcal{N}_1$ are chosen to be:

$$\widehat{\mathcal{N}_0^{(1)}} = \frac{1}{\varepsilon^2} \left( \sqrt{\sigma_0^2 \cdot \mathcal{C}_0} + \sqrt{\sigma_1^2 \cdot \mathcal{C}_1} \right) \sqrt{\sigma_0^2/\mathcal{C}_0};$$
$$\widehat{\mathcal{N}_1^{(1)}} = \frac{1}{\varepsilon^2} \left( \sqrt{\sigma_0^2 \cdot \mathcal{C}_0} + \sqrt{\sigma_1^2 \cdot \mathcal{C}_1} \right) \sqrt{\sigma_1^2/\mathcal{C}_1}.$$

Note that the value of $\widehat{\mathcal{N}_0^{(0)}}$ was calculated based on the initial estimate of $\sigma_0^2$, whilst $\widehat{\mathcal{N}_0^{(1)}}$ was based on a refined, and more accurate, estimate of $\sigma_0^2$. Unless the estimate for $\sigma_0^2$ *decreases* substantially, then $\widehat{\mathcal{N}_0^{(1)}} > \widehat{\mathcal{N}_0^{(0)}}$, and so additional simulations will need to be performed on level $\ell = 0$. The required simulations on levels $\ell = 0$ and $\ell = 1$ are generated (for the circumstances where $\widehat{\mathcal{N}_0^{(1)}} < \widehat{\mathcal{N}_0^{(0)}}$, please see the next

---
[5]This equation is written in a clunky format for later comparison.

page).

We now add another level, $\ell = 2$, and we repeat the argument from above. This procedure is followed until all the required levels are included in the estimator. If the estimated sample variances prove to be inaccurate, further simulations can be run to ensure that the required estimator variance is attained, i.e. the values of $\mathcal{N}_\ell$ can be increased until $\sum_{\ell=0}^{\mathcal{L}(+1)} \sigma_\ell^2 / \mathcal{N}_\ell < \varepsilon^2$.

The incremental approach outlined above therefore provides the opportunity to correct the errors inherited from the use of (initially poorly) approximated values for $\sigma_\ell^2$. However, the use of updated values of $\sigma_\ell^2$ means that there may be a set of levels, $\Omega$, where more sample paths have already been simulated than the number required by (3.12). In the case where $\widehat{\mathcal{N}_\ell} < \mathcal{N}_\ell$, we see that the estimator variance (that takes into account all samples generated on level $\ell$), $\mathcal{V}_\ell$, is lower than the estimator variance implied by the method of Lagrange multipliers optimisation, $\widehat{\mathcal{V}_\ell}$, i.e.

$$\mathcal{V}_\ell = \sigma_\ell^2 / \mathcal{N}_\ell < \sigma_\ell^2 / \widehat{\mathcal{N}_\ell} = \widehat{\mathcal{V}_\ell}. \tag{3.18}$$

This situation arises where the estimated values of $\sigma_\ell^2$ are significantly revised.

As the estimator variance is lower than it needs to be on levels $\ell \in \Omega$, the estimator variances on levels $\ell \notin \Omega$ are permitted to be higher than originally anticipated. If we set

$$\varepsilon_*^2 := \varepsilon^2 - \sum_{\ell \in \Omega} \sigma_\ell^2 / \mathcal{N}_\ell, \tag{3.19}$$

then we can still satisfy the overall variance target by achieving the variance target of $\varepsilon_*^2$ for the combined levels $\ell \in \{0, \ldots \mathcal{L}(+1)\} \setminus \Omega$. Note that if we re-calculate the target $\widehat{\mathcal{N}_\ell}$ for $\ell \in \{0, \ldots \mathcal{L}(+1)\} \setminus \Omega$, we now require fewer sample paths for each level. It is therefore now possible that more sample paths than required have already been generated for some $\ell \in \{0, \ldots \mathcal{L}\, (+1)\} \setminus \Omega$. These levels can then be added to the set $\Omega$ and $\varepsilon_*^2$ can be recalculated. This argument can be repeated until no more levels can be added to $\Omega$.

We now provide pseudo-code to control the implementation of the multi-level method. Please see Algorithm 3.3 for further information.

---

Algorithm 3.3: This dynamic calibration method runs the multi-level algorithm.

**Require:** algorithm parameters: base time-step, $\tau_0$, refinement factor, $\mathcal{M}$, level indicator, $\mathcal{L}$, and whether the estimator is unbiased.
**Require:** target estimator variance, $\varepsilon^2$.
 1: set $\ell \leftarrow 0$
 2: **loop**
 3:    if $\mathcal{N}_\ell = 0$, generate $10^2$ simulations on level $\ell$
 4:    for $\ell' = 0, \ldots, \ell$, use $\sigma_{\ell'}^2$ and $\mathcal{C}_{\ell'}$ to determine $\widehat{\mathcal{N}_{\ell'}}$
 5:    let $\Omega \leftarrow \{\ell' \in \{0, \ldots, \ell\} : \mathcal{N}_{\ell'} > \widehat{\mathcal{N}_{\ell'}}\}$
 6:    **loop**
 7:       set $\varepsilon_*^2 \leftarrow \varepsilon^2 - \sum_{\ell' \in \Omega} \sigma_{\ell'}^2 / \mathcal{N}_{\ell'}$
 8:       for $\ell' \in \{0, \ldots, \ell\} \setminus \Omega$, use $\varepsilon_*^2$ to recalculate $\widehat{\mathcal{N}_{\ell'}}$
 9:       let $\Omega' \leftarrow \{\ell' \in \{0, \ldots, \ell\} : \mathcal{N}_{\ell'} > \widehat{\mathcal{N}_{\ell'}}\}$
10:       **if** $\Omega' = \Omega$ **then break**
11:       **else** set $\Omega \leftarrow \Omega'$
12:       **end if**
13:    **end loop**
14:    for $\ell' \notin \Omega$, generate $\widehat{\mathcal{N}_{\ell'}} - \mathcal{N}_{\ell'}$ simulations on level $\ell'$
15:    **if** $\ell = \mathcal{L}\,(+1)$ **and** $\sum_{\ell'=0}^{\mathcal{L}(+1)} \sigma_{\ell'}^2 / \mathcal{N}_{\ell'} < \varepsilon^2$ **then**
16:       **break**
17:    **end if**
18:    **if** $\ell < \mathcal{L}\,(+1)$ **then**
19:       set $\ell \leftarrow \ell + 1$
20:    **end if**
21: **end loop**

---

### 3.5.3 Sample implementation

We return to the gene expression model of Case Study 2. As before, we set $\tau_0 = 1/9$, $\mathcal{M} = 3$, and $\mathcal{L} = 5$. The dynamic calibration multi-level method, as detailed in

| Level | $\tau_{\ell-1}$ | $\tau_\ell$ | Estimate | Variance | Paths | Time |
|---|---|---|---|---|---|---|
| $\mathcal{Q}_0$ | - | $3^{-2}$ | 3187.06 | $1.03 \times 10^6$ | $7.21 \times 10^6$ | 86.8s |
| $\mathcal{Q}_1$ | $3^{-2}$ | $3^{-3}$ | 350.60 | 16210.70 | $4.22 \times 10^5$ | 23.4s |
| $\mathcal{Q}_2$ | $3^{-3}$ | $3^{-4}$ | 117.462 | 2675.21 | $1.64 \times 10^5$ | 14.8s |
| $\mathcal{Q}_3$ | $3^{-4}$ | $3^{-5}$ | 39.31 | 667.88 | $3.74 \times 10^4$ | 11.0s |
| $\mathcal{Q}_4$ | $3^{-5}$ | $3^{-6}$ | 13.07 | 182.68 | $1.33 \times 10^4$ | 8.4s |
| $\mathcal{Q}_5$ | $3^{-6}$ | $3^{-7}$ | 4.54 | 17.64 | $3.55 \times 10^3$ | 3.9s |
| $\mathcal{Q}_6$ | $3^{-7}$ | DM | 2.11 | 89.94 | $6.53 \times 10^3$ | 11.6s |
| **Total** | | | **$3714.32 \pm 0.99$** | | - | **159.9s** |

Table 3.3: The contribution from each level in producing an unbiased overall estimator, $\mathcal{Q}$, for $\mathbb{E}[X_3]$ in system (2.10) at $T = 1$. The variance is repeatedly recalculated in line with Algorithm 3.3. We have taken $\tau_0 = 1/9$, $\mathcal{M} = 3$, and $\mathcal{L} = 5$.

Algorithm 3.3, is implemented to provide an unbiased estimate of $\mathbb{E}[X_3(T)]$. We take $\varepsilon^2 = 0.2603$, and in Table 3.3 we show how the dimer population is estimated as $3714.32 \pm 0.99$ within 159.9 seconds.

To illustrate the dynamic calibration method, Figure 3.3 shows the number of simulations performed on each level during each iteration of the main loop of Algorithm 3.3 (see line 9). The data from these simulations are contained in the previously-mentioned Table 3.3.

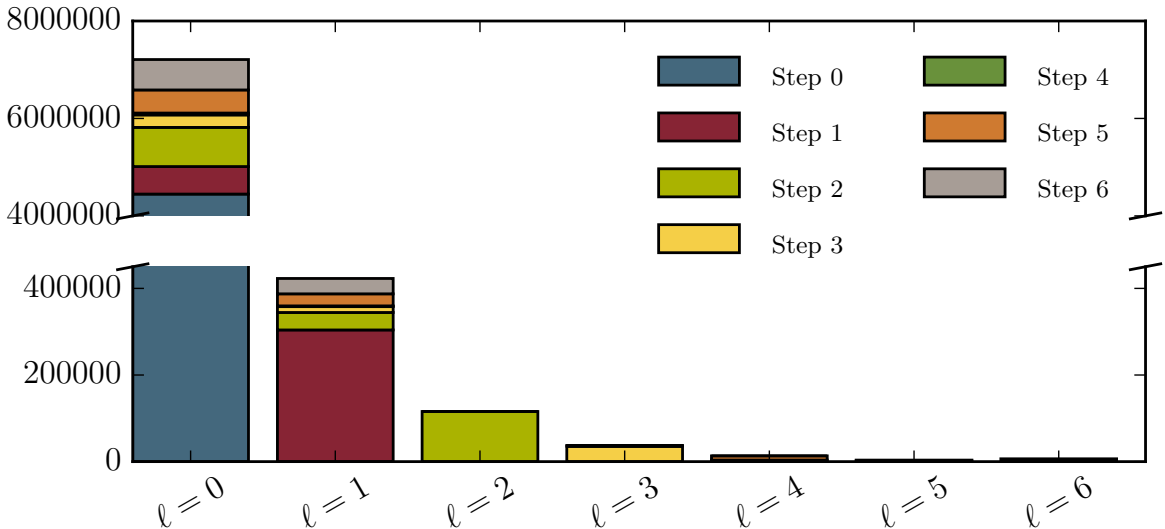

Figure 3.3: We show the number of simulations generated for each level, $\ell$, at each iteration, $m$, of the dynamic algorithm 3.3, given by $\mathcal{N}_\ell^{(m)}$. We have studied system (2.10) and taken $\tau_0 = 1/9$, $\mathcal{M} = 3$, and $\mathcal{L} = 5$.

The dynamic calibration multi-level method is now fully implemented. The algorithm is run, from start until finish, a total of 1000 times. These simulations take on average 158.05 seconds to complete, with the 10-th and 90-th percentiles of the CPU time corresponding to 142 and 176 seconds, respectively. The average estimate of $\mathbb{E}[X_3(1)]$ is 3714.30. In Figure 3.4, a histogram details the CPU times required by the dynamic calibration method. Whilst the average simulation duration might be slightly longer than that of the one-step calibration method, with the dynamic calibration method all runs of the algorithm attain the required confidence interval semi-length (and, further, the average confidence interval semi-length is 0.99).

In Figure 3.5, we plot the estimated values of $\mathbb{E}[X_3(1)]$. A histogram is plotted (after placing the simulated values into bins), together with an empirical cumulative density function (a CDF, which is not binned). A normal distribution is then fitted to the estimated values: the mean is chosen as $\mu = 3714.30$ and the variance as $\sigma^2 = 0.2660$. Therefore, an appropriate confidence interval for the statistical error of the estimate is given by the interval (2.8).

A more comprehensive assessment of Algorithm 3.3, that will vary the algorithm parameters, is postponed until all our method enhancements have been discussed.

## 3.6 Setting algorithm parameters

In this section, we discuss how one might systematically choose values for $\tau_0$, $\mathcal{M}$ and $\mathcal{L}$. We also discuss whether a biased or unbiased estimator should be preferred.

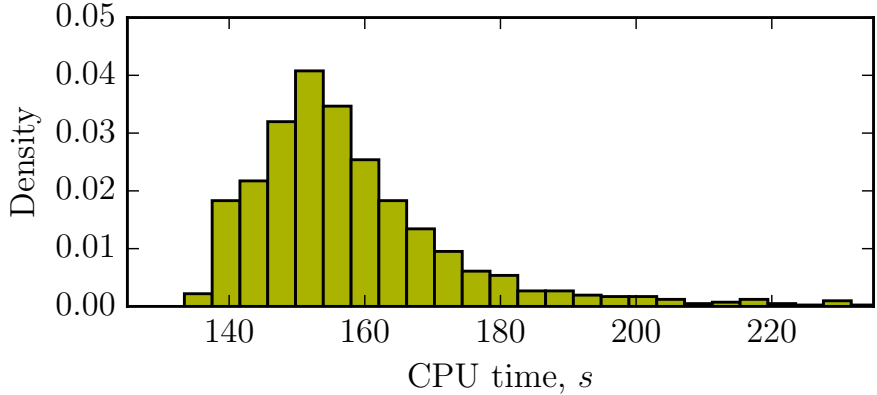

Figure 3.4: A histogram showing the distribution of the total duration taken by the multi-level algorithm to estimate $\mathbb{E}[X_3(1)]$ of reaction network (2.10). The dynamic calibration method stated in Algorithm 3.3 has been run a total of 1000 times. The algorithm parameters are as stated in the main text.

### 3.6.1 Choosing $\tau_0$, the base level time-step

It is tempting to assume that, since the multi-level method benefits from using many low quality population estimates that are simulated quickly, a large choice of $\tau_0$ would be prudent. However, the impact of choosing too large a value of $\tau_0$ is that, whilst the base level estimate, $\mathcal{Q}_0$, may be calculated quickly, the bias of $\mathcal{Q}_0$ will be large. This large bias needs to be corrected in subsequent levels. The process of eliminating a large bias might require a significant number of samples for the more expensive correction levels, thereby outweighing the time-savings achieved on the base level.

We take a structured approach towards choosing $\tau_0$. The optimal value for $\tau_0$ may well depend on the particular choice of $\mathcal{M}$, the scaling constant. For the purposes of this investigation, however, we fix the value of $\mathcal{M}$. Our discussion of the merits of different choices of $\mathcal{M}$ is left to Section 3.6.2. We will also consider the time-step on the finest level, $\tau_{\mathcal{L}}$, as a fixed but unknown value. Based on this, we choose a value for $\tau_0$ and, subsequently, $\mathcal{L}$.

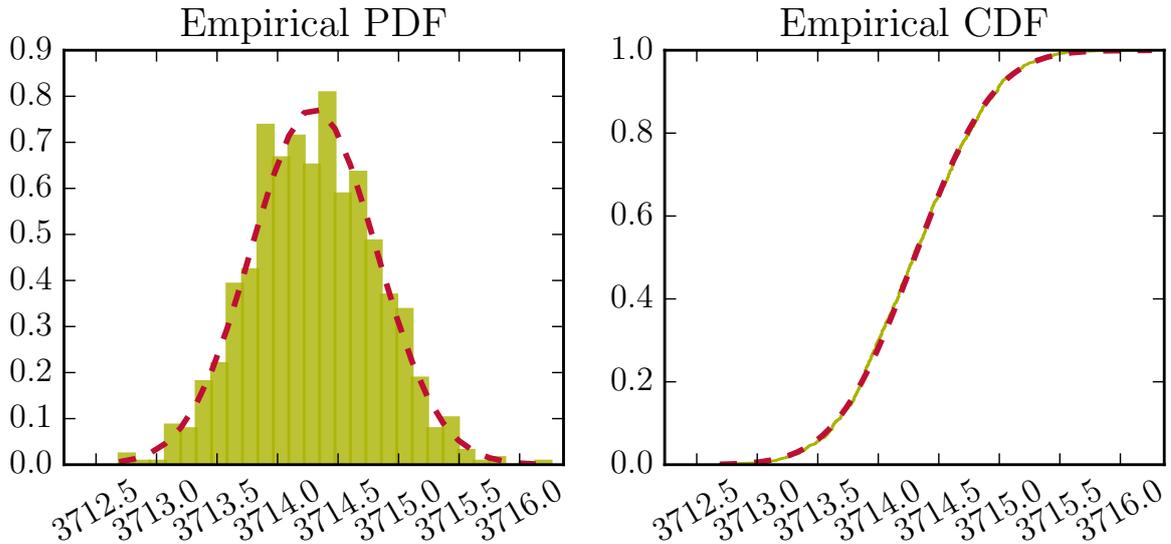

Figure 3.5: We plot the empirical distribution of the estimates of $\mathbb{E}[X_3(1)]$ of reaction network (2.10). The dynamic calibration method stated in Algorithm 3.3 has been run a total of 1000 times. A histogram is shown on the left, whilst an empirical cumulative density function is shown on the right. A normal distributed has been fitted and is shown in red. The algorithm parameters are as stated in the main text.

From Equation (3.13) we recall that

$$\frac{1}{\varepsilon^2} \left\{ \sum_{\ell=0}^{\mathcal{L}} \sqrt{\mathcal{C}_\ell \cdot \sigma_\ell^2} \right\}^2,$$

units of CPU time are required to attain an estimator variance of $\varepsilon^2$. To simplify notation, we introduce $k_\ell$, where

$$k_\ell \coloneqq \mathcal{C}_\ell \cdot \sigma_\ell^2. \tag{3.20}$$

We will call $k_\ell$ the *relative cost* of producing simulated paths for level $\ell$. The relative cost of a level therefore indicates the contribution of the level to the total computational time given by Equation (3.13)

We take an iterative approach to optimising the choice of $\tau_0$, beginning with an initial guess, and improving on it in subsequent iterations. Given an initial choice of $\tau_0$, $\tau_0^{(1)}$, we propose two candidates for an improved choice, $\tau_0^{(2)}$:

- a smaller choice, $\tau_0^{(2,1)} = \tau_0^{(1)}/\mathcal{M}$;

- a larger choice, $\tau_0^{(2,2)} = \tau_0^{(1)}\mathcal{M}$.

Making the reasonable assumption that there will be at least one level in addition to the *base* level, we can calculate the difference in expected overall simulation times using $\tau_0^{(2,1)}$ or $\tau_0^{(2,2)}$. If using $\tau_0^{(2,1)}$ or $\tau_0^{(2,2)}$ results in a time saving compared with using $\tau_0^{(1)}$, we set the improved guess $\tau_0^{(2)}$ to equal the better value. We can repeat this algorithm until we reach a choice of $\tau_0$ for which no further improvement is gained. This corresponds to a local minimum of the overall simulation time[6]. If, by chance, we begin at a local maximum we follow the refinement process in both directions (both increasing and decreasing $\tau_0$).

In general, our iterative algorithm will require comparison of the computational complexity of generating an estimator with coarse base level time-step, $\tau_0^C$, with the computational complexity of generating an estimator with a fine base level time-step, $\tau_0^F = \tau_0^C/\mathcal{M}$. The estimator for the coarse base level, given a desired level of accuracy, will be given by

$$\mathcal{Q} = \mathbb{E}\left[Z_{\tau_0^C}\right] + \mathbb{E}\left[Z_{\tau_0^C/\mathcal{M}} - Z_{\tau_0^C}\right] + \sum_{\ell=2}^{\mathcal{L}(+1)} \mathbb{E}\left[Z_{\tau_0^C/\mathcal{M}^\ell} - Z_{\tau_0^C/\mathcal{M}^{\ell-1}}\right], \qquad (3.21)$$

and the estimator for the fine base level will be given by

$$\mathcal{Q} = \mathbb{E}\left[Z_{\tau_0^C/\mathcal{M}}\right] + \sum_{\ell=2}^{\mathcal{L}(+1)} \mathbb{E}\left[Z_{\tau_0^C/\mathcal{M}^\ell} - Z_{\tau_0^C/\mathcal{M}^{\ell-1}}\right]. \qquad (3.22)$$

The majority of the levels are simulated for both choices of base level and, as such,

---

[6] We have found that restricting our search to the set $\{\tau : \tau = \tau_0^{(1)}\mathcal{M}^{-\alpha}\}$ still produces acceptable computational performance. We aim to ensure that particularly poor choices of $\tau_0$ are avoided.

will have the same relative cost, $k_\ell$. The terms that will have different relative costs will be $\mathbb{E}\left[Z_{\tau_0^C}\right]$ and $\mathbb{E}\left[Z_{\tau_0^C/\mathcal{M}} - Z_{\tau_0^C}\right]$ on the coarse level (for which we will denote the relative costs as $k_0^C$ and $k_1^C$, respectively), and $\mathbb{E}\left[Z_{\tau_0^C/\mathcal{M}}\right]$ on the fine level (for which we will denote the relative cost as $k_0^F$). We now state rules to decide whether a proposed base level time-step should be accepted or rejected by using a simple comparison of these three proportionality constants.

**Fact 2.** The configuration with the fine base level time-step, $\tau_0^F = \tau_0^C/\mathcal{M}$, should be preferred over that with the coarse base level time-step, $\tau_0^C$, if

$$\sqrt{k_0^F} < \sqrt{k_0^C} + \sqrt{k_1^C}, \qquad (3.23)$$

where we recall that $k_\ell$ represents the relative cost of generating a single sample value for level $\ell$, and is given by $k_\ell = \mathcal{C}_\ell \cdot \sigma_\ell^2$. ∎

**Proof.** In order to see where this inequality comes from proceed as follows: without loss of generality set the variance target at $\varepsilon^2 = 1$. Then, the expected difference in simulation time between the estimator with the fine base level time-step and the estimator with the coarse base level time-step is given by

$$\left\{\sum_{\ell=0}^{\mathcal{L}^F} \sqrt{k_\ell^F}\right\}^2 - \left\{\sum_{\ell=0}^{\mathcal{L}^C} \sqrt{k_\ell^C}\right\}^2,$$

where $\mathcal{L}^F$ and $\mathcal{L}^C$ indicate the (unknown) number of levels between the two competing configurations. Using the fact that, for $i \geq 1$, $k_{i+1}^C = k_i^F$, and that $\mathcal{L}^F + 1 = \mathcal{L}^C$, we can rewrite this as

$$\left\{\sqrt{k_0^F} - \sqrt{k_1^C} + \sum_{\ell=1}^{\mathcal{L}^F+1} \sqrt{k_\ell^C}\right\}^2 - \left\{\sqrt{k_0^C} + \sum_{\ell=1}^{\mathcal{L}^F+1} \sqrt{k_\ell^C}\right\}^2.$$

Thus, after rearrangement, the net change in simulation time is

$$\left[\sqrt{k_0^F} - \sqrt{k_0^C} - \sqrt{k_1^C}\right]\left\{\sqrt{k_0^F} + \sqrt{k_0^C} + \sqrt{k_1^C} + 2\sum_{\ell=2}^{\mathcal{L}^F+1}\sqrt{k_\ell^C}\right\}. \qquad (3.24)$$

As the terms within the braced brackets are positive, we have the required condition.

**Example 3.1.** We again consider the gene expression system (2.10) of Case Study 2, and use our algorithm to choose $\tau_0$. First impose the choice of $\mathcal{M} = 3$. If we take $\tau_0^{(1)} = 1/9$, then there are two alternatives to consider, $\tau_0^{(2,1)} = 1/27$ and $\tau_0^{(2,2)} = 1/3$. With $10^5$ samples (which took a total of 10.5 seconds to generate), we calculate estimates for the relevant proportionality constants and present the results in Table 3.4. We then use Fact 2 to decide on the appropriate choice of $\tau_0$. The initial base level time-step $\tau_0^{(1)}$ is coarse in comparison to the proposed base level time-step $\tau_0^{(2,1)}$. Since we have $\sqrt{k_0^{(1)}} + \sqrt{k_1^{(1)}} = 4.5233 < 6.5558 = \sqrt{k_0^{(2,1)}}$, by Fact 2 $\tau_0^{(2,1)} = 1/27$ is an inferior choice to $\tau_0^{(1)} = 1/9$. Similarly, as $\sqrt{k_0^{(2,2)}} + \sqrt{k_1^{(2,2)}} = 8.1182 > 3.5452 = \sqrt{k_0^{(1)}}$, Fact 2 implies that $\tau_0^{(2,2)} = 1/3$ is also an inferior choice. We therefore take $\tau_0 = 1/9$. ∎

| Guess | Estimates | |
|---|---|---|
| $\tau_0^{(1)} = 1/9$ | $\sqrt{k_0^{(1)}} = 3.5452,$ | $\sqrt{k_1^{(1)}} = 0.9781.$ |
| $\tau_0^{(2,1)} = 1/27$ | $\sqrt{k_0^{(2,1)}} = 6.5558,$ | N/A. |
| $\tau_0^{(2,2)} = 1/3$ | $\sqrt{k_0^{(2,2)}} = 3.0751,$ | $\sqrt{k_1^{(2,2)}} = 5.0431.$ |

Table 3.4: Details of the cost measure for each potential ensemble of estimators for the gene expression system (2.10) with different choices of $\tau_0$, the time-step on the base level.

Our iterative procedure therefore provides a mechanism by which a value of the base level time-step, $\tau_0$, can be selected, given a value of $\mathcal{M}$. A further benefit of the algorithm is that it does not require that $\tau_\mathcal{L}$ be chosen at the outset.

Within this framework for choosing $\tau_0$, the value of $\mathcal{M}$ leads to different possibilities for $\tau_0$.

### 3.6.2 Choosing the refinement factor, $\mathcal{M}$

In this section, we outline the approach that Giles [46] follows to choose the refinement factor, $\mathcal{M}$. To choose the value of $\mathcal{M}$, we again return to considering the total simulation cost of the multi-level method given by Equation (3.13):

$$\frac{1}{\varepsilon^2} \left\{ \sum_{\ell=0}^{\mathcal{L}(+1)} \sqrt{\mathcal{C}_\ell \cdot \sigma_\ell^2} \right\}^2 .$$

The following three assumptions can be made:

1. if the biased multi-level estimator is used, then, for a fixed bias[7], we require $\mathcal{L} = \mathcal{O}(\log(1/\mathcal{M}))$;

2. the values of $\sigma_\ell^2$ can be estimated in terms of $\mathcal{M}$;

3. the values of $\mathcal{C}_\ell$ can be determined through simulation (see Section 2.4.7) or otherwise (poorly) approximated as order $\mathcal{O}(\mathcal{M}^{-\ell})$.

The values of $\sigma_\ell^2$ can be estimated as per Giles [46]. These three assumptions can be used to construct a closed-form estimate for the total computational cost (given by Equation (3.13)) in terms of $\mathcal{M}$, and an optimisation algorithm can then select a value of $\mathcal{M}$.

In Section 3.7, we investigate the effect of variations in $\mathcal{M}$. On the basis of our numerical investigations, our view is that the performance of the multi-level method is not particularly sensitive to the choice of $\mathcal{M}$.

---

[7]Recall that the bias is $\mathcal{O}(\tau_\mathcal{L})$, and $\tau_\mathcal{L} = \tau_0 \cdot \mathcal{M}^{-\mathcal{L}}$.

We now discuss how many levels to include in the estimator, as indicated by $\mathcal{L}$. The biased and unbiased cases are considered separately.

### 3.6.3 Levels of a biased estimator

For the biased estimator, $\mathcal{Q}_b$, the choices of $\tau_0$ and $\mathcal{L}$ determine the final time-step, $\tau_{\mathcal{L}}$, and therefore the overall bias of the estimator. A larger value of $\mathcal{L}$ will result in a lower bias, but also to increased simulation time. Additionally, we must consider the statistical error induced by Monte Carlo simulation.

From Section 2.4.5, we note that it is often useful to consider the MSE (mean squared error) of our estimates. We recall that:

$$\text{MSE} = \underbrace{\mathbb{E}\left[\left(\widehat{\mathcal{Q}} - \mathbb{E}[\widehat{\mathcal{Q}}]\right)^2\right]}_{\text{Statistical error}} + \underbrace{\left(\mathbb{E}[\widehat{\mathcal{Q}}] - \mathcal{Q}\right)^2}_{\text{Square of bias}}.$$

As before, the statistical error is given by $\mathcal{V} = \sum_{\ell=0}^{\mathcal{L}} \mathcal{V}_\ell$. To evaluate the bias we note, for large $\ell$ (or, equivalently, small values of $\tau_{\mathcal{L}}$),

$$\begin{aligned}
Q_\ell = \mathbb{E}\left[Z_{\tau_0/\mathcal{M}^\ell} - Z_{\tau_0/\mathcal{M}^{\ell-1}}\right] &= \mathbb{E}\left[Z_{\tau_0/\mathcal{M}^\ell} - X\right] - \mathbb{E}\left[Z_{\tau_0/\mathcal{M}^{\ell-1}} - X\right] \\
&\approx C\tau_0/\mathcal{M}^\ell - C\tau_0/\mathcal{M}^{\ell-1} \\
&\approx (\mathcal{M} - 1)\mathbb{E}\left[X - Z_{\tau_0/\mathcal{M}^\ell}\right],
\end{aligned}$$

where $C$ is a constant (that does not need to be determined explicitly). Therefore, the bias can be estimated as,

$$\mathbb{E}\left[X - Z_{\tau_0/\mathcal{M}^\ell}\right] \approx \frac{Q_\ell}{\mathcal{M} - 1}. \qquad (3.25)$$

Suppose we are given a MSE target, $\varepsilon^2$, and have to ascribe a portion of this to the square of the bias, and the remainder to the variance. As a first attempt at a solution, we pre-assign a proportion, $\lambda \in (0, 1)$, of the MSE target to the square of the bias, and leave $1 - \lambda$ to the variance. Previous work [46] has made the simple choice $\lambda = 1/2$, that is, assigning half the MSE to the square of the bias, and the other half to the estimator variance. However, it is not clear how best to choose $\lambda$ for a particular problem. We demonstrate the effects of varying $\lambda$ later in this chapter.

Following Equation (3.25) above, to obtain a MSE $= \varepsilon^2$, the value of $\mathcal{L}$ should be chosen to ensure that

$$|Q_\mathcal{L}| \leq \sqrt{\lambda}(\mathcal{M} - 1)\varepsilon. \qquad (3.26)$$

Note that to improve the reliability of this approach, we follow Giles [46] and estimate the bias using the last two levels (i.e. $\mathcal{L} - 1$ and $\mathcal{L}$). Therefore we choose $\mathcal{L}$ such that

$$\max\{\mathcal{M}^{-1}|Q_{\mathcal{L}-1}|, |Q_\mathcal{L}|\} \leq \sqrt{\lambda}(\mathcal{M} - 1)\varepsilon. \qquad (3.27)$$

Our implementation of the biased multi-level method is now set out. We will combine our dynamic calibration method, as stated in Algorithm 3.3, with the *stopping criterion* described by Equation (3.27). Our procedure will be to repeatedly add levels to the estimator (i.e. to increase the value of $\mathcal{L}$) until Equation (3.27) is satisfied. At each iteration of the dynamic calibration method, we will check whether the stopping criterion is satisfied. If so, then the algorithm terminates. Otherwise, an additional level is generated and the algorithm proceeds. This combined approach has the added advantage of carefully controlling the statistical error. The full method is stated as Algorithm 3.4.

---

Algorithm 3.4: This dynamic calibration method runs the biased multi-level algorithm, and attains a given MSE.

---

**Require:** Mean Squared Error (MSE) $\varepsilon^2$, algorithm parameter, $\lambda$, and, refinement factor, $\mathcal{M}$.
1: choose $\tau_0$ using Fact 2
2: set $\ell \leftarrow 0$
3: **loop**
4:     if $\mathcal{N}_\ell = 0$, generate $10^2$ simulations on level $\ell$
5:     for $\ell' = 0, \ldots, \ell$, use $\sigma^2_{\ell'}$ and $\mathcal{C}_{\ell'}$ to determine $\widehat{\mathcal{N}_{\ell'}}$
6:     let $\Omega \leftarrow \{\ell' \in \{0, \ldots, \ell\} : \mathcal{N}_{\ell'} > \widehat{\mathcal{N}_{\ell'}}\}$
7:     **loop**
8:         set $\varepsilon^2_* \leftarrow \varepsilon^2 - \sum_{\ell' \in \Omega} \sigma^2_{\ell'}/\mathcal{N}_{\ell'}$
9:         for $\ell' \in \{0, \ldots, \ell\} \setminus \Omega$, use $\varepsilon^2_*$ to recalculate $\widehat{\mathcal{N}_{\ell'}}$
10:        let $\Omega' \leftarrow \{\ell' \in \{0, \ldots, \ell\} : \mathcal{N}_{\ell'} > \widehat{\mathcal{N}_{\ell'}}\}$
11:        **if** $\Omega' = \Omega$ **then break**
12:        **else** set $\Omega \leftarrow \Omega'$
13:        **end if**
14:     **end loop**
15:     for $\ell' \notin \Omega$, generate $\widehat{\mathcal{N}_{\ell'}} - \mathcal{N}_{\ell'}$ simulations on level $\ell'$
16:     **if** Equation (3.27) is not satisfied **then**
17:        set $\ell \leftarrow \ell + 1$
18:     **else if** $\sum_{\ell'=0}^{\ell} \sigma^2_{\ell'}/\mathcal{N}_{\ell'} < (1-\lambda) \cdot \varepsilon^2$ **then**
19:        break
20:     **end if**
21: **end loop**

---

### 3.6.4 Levels of an unbiased estimator

In this section, we discuss how one might decide on the number of levels that comprise an unbiased estimator. When an unbiased estimator is sought, a change in the value of $\mathcal{L}$ does not change the bias of the estimator; rather, it adjusts the simulation time.

The estimators, $\mathcal{Q}_1, \ldots, \mathcal{Q}_\mathcal{L}$ will be determined through simulations conducted by Algorithm 3.1, and the estimator $\mathcal{Q}_{\mathcal{L}+1}$ will be determined by using simulations generated with Algorithm 3.2. The estimate given for $\mathcal{Q}_b = \sum_{\ell=0}^{\mathcal{L}} \mathcal{Q}_\ell$ has a bias of order $\mathcal{O}(\tau_\mathcal{L})$. Following the notation from Section 3.6.1, the aim is to choose the $\mathcal{L}$ that

| Level indicator, $\mathcal{L}$ | 0 | 1 | 2 | 3 | 4 | 5 | 6 | 7 | 8 |
|---|---|---|---|---|---|---|---|---|---|
| Duration (s) | 455 | 209 | 164 | 154 | 155 | 168 | 178 | 178 | 208 |

Table 3.5: The expected run-time of the multi-level method for system (2.10), when different values of $\mathcal{L}$ are employed. We take $\mathcal{M} = 3$. Note that these quantities are expected values only.

minimises the total computational cost, which can be restated as

$$\mathcal{C} \propto \left( k_0 + \sum_{\ell=0}^{\mathcal{L}} k_\ell + k_{\mathcal{L}+1} \right)^2. \qquad (3.28)$$

On first glance, it seems that a small value of $\mathcal{L}$ will reduce the number of terms in Equation (3.28), and will therefore lead to a small total simulation time. However, changing the value of $\mathcal{L}$ also affects $k_{\mathcal{L}+1}$[8]. There is no straightforward way to assign a value to $\mathcal{L}$, and a value should be chosen on a case-by-case basis.

**Example 3.2.** Again, we return to considering Case Study 2. In Section 3.6.1, we concluded that taking $\tau_0 = 1/9$ as the base level time-step is reasonable. We work with $\mathcal{M} = 3$ and investigate the expected effect of varying $\mathcal{L}$. The expected run-time will be calculated according to Equation (3.28). Therefore, the values of $k_\ell$ are estimated (in each case, using $10^5$ simulations), so that they can be substituted into Equation (3.28). In Table 3.5, we show the effect of varying $\mathcal{L}$ on the total simulation time. This table suggests that, where $\mathcal{L} = 2, \ldots, 6$, the expected simulation time is not particularly sensitive to the choice of $\mathcal{L}$. ∎

## 3.7 Comparing variations in algorithm parameters

The full multi-level method, as stated in Algorithms 3.3 and 3.4, will now be implemented and tested on Case Study 2 (again, the dimer population is estimated). We ask what the effect of various combinations of $\mathcal{L}$, the number of levels, $\tau_0$, the size of the time-step on the base level, and $\mathcal{M}$, the scaling factor, will be on the CPU time.

---

[8]Recall that $k_{\mathcal{L}+1} = \mathcal{C}_{\mathcal{L}+1} \cdot \sigma^2_{\mathcal{L}+1}$, and both $\sigma^2_{\mathcal{L}+1}$ and $\mathcal{C}_{\mathcal{L}+1}$ change when $\tau_{\mathcal{L}}$ changes.

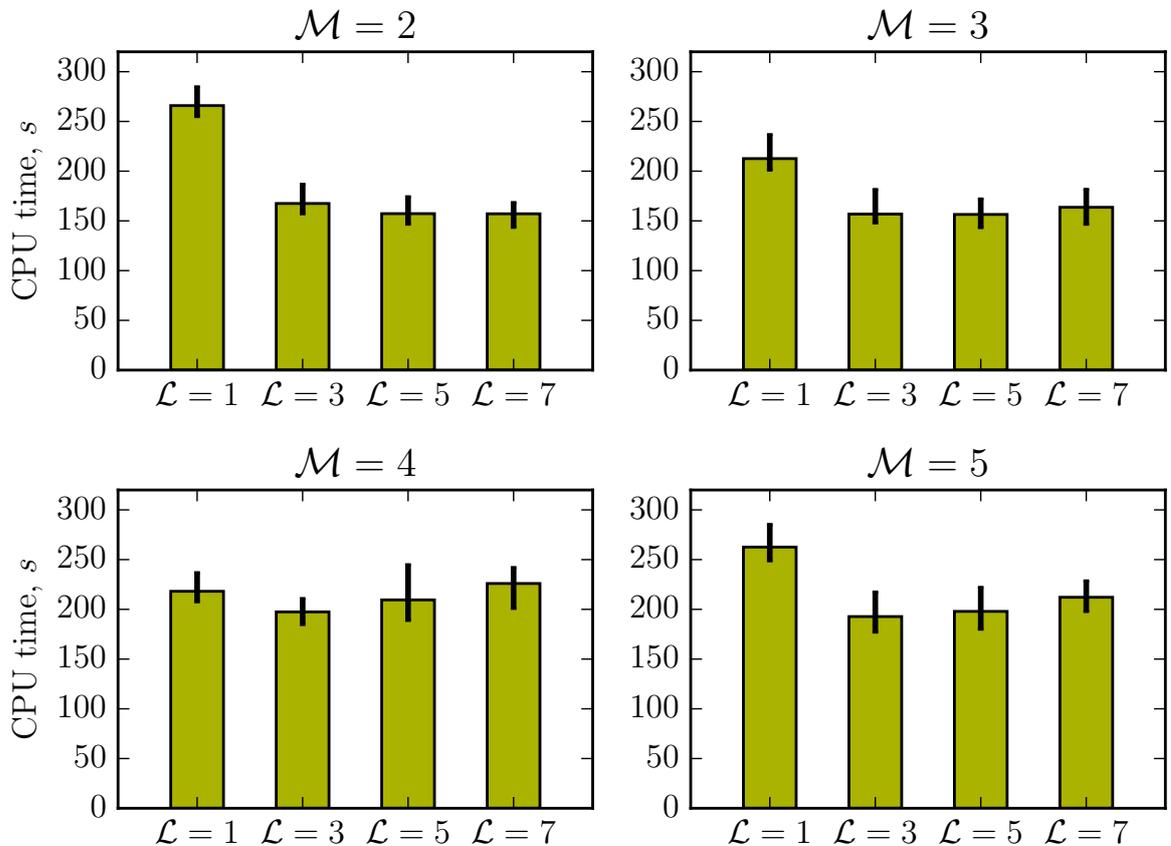

Figure 3.6: The average CPU time required by the multi-level method to estimate $\mathbb{E}[X_3(1)]$ in Case Study 2. We vary $\mathcal{M}$ and $\mathcal{L}$; the estimator is unbiased. The black bars indicate the range occupied by the 10-th to 90-th percentiles of the data. The values of $\tau_0$ are: $\mathcal{M} = 2 \Rightarrow \tau_0 = 1/8$, $\mathcal{M} = 3 \Rightarrow \tau_0 = 1/9$, $\mathcal{M} = 4 \Rightarrow \tau_0 = 1/16$, and $\mathcal{M} = 5 \Rightarrow \tau_0 = 1/5$.

**Unbiased estimation.** The multi-level method is now run with different choices of $\mathcal{M}$ and $\mathcal{L}$. For each choice of $\mathcal{M}$ and $\mathcal{L}$, the choice of $\tau_0$ is optimised according to Fact 2. The CPU times taken by the unbiased algorithm are represented in Figure 3.6. For each choice of algorithm parameters, the entire multi-level algorithm is run 100 times, and, in each case, Algorithm 3.3 is used to determine an unbiased estimator. The mean CPU time is then calculated, together with various percentiles. The average simulation durations are largely in accordance with Table 3.5. When $\mathcal{M} = 3$ and $\mathcal{L} = 5$, the best average simulation time is recorded; it is 156.4 seconds. Very similar average CPU times (within 0.5%) are observed with $\mathcal{M} = 3$, $\mathcal{L} = 3$, and with ($\mathcal{M} = 2$, $\mathcal{L} = 5$ or 7). Please note that a larger sample is represented in Figure 3.4, and therefore the statistics of Figure 3.4 are slightly different to Figure 3.6 (these differences apply where $\mathcal{M} = 3$ and $\mathcal{L} = 5$).

**Biased estimation.** As previously noted, for the canonical parameter values, Table 3.1 suggests that an unbiased estimator comes at little additional cost to a biased estimator, and should, therefore, be preferred. However, for completeness in Figure 3.7 we show box-plots of the simulation times of the biased multi-level algorithm 3.3, where $\mathcal{M}$ and $\mathcal{L}$ have been *fixed* in advance. As expected, lower values of $\mathcal{L}$ are associated with lower run-times. In addition, larger values of $\mathcal{L}$ are associated with more variability in the CPU time: we explain why this happens in Section 3.9.

In Section 3.6.3 we argued that it might be preferable to insist that the estimator meets a specified MSE. If this approach is followed, the value of $\mathcal{L}$ is dynamically chosen according to Algorithm 3.4. To compare our results with the unbiased estimators presented above, we take the MSE as $\varepsilon^2 = 0.2603$. To run Algorithm 3.4, algorithm parameter $\lambda$ is chosen: $\lambda$ ascribes a portion of the MSE to the square of the bias, with the remaining proportion $(1 - \lambda)$ being reserved for the estimator variance. A range of values for $\lambda$ are chosen, and the algorithm is implemented. In Figure 3.8, the total run-time for Algorithm 3.4 is shown for a range of choices of $\lambda$. For each choice of $\lambda$, the complete algorithm was run 100 times. The best average run-time is achieved is 169.2 seconds, which, for this particular model problem, is higher than the average CPU time of the unbiased estimation scheme. The best average run-time is achieved where $\lambda = 1/64$, though this CPU time does not differ significantly from the CPU time where $\lambda$ is instead chosen from the set $\{1/8, 1/16, 1/32\}$.

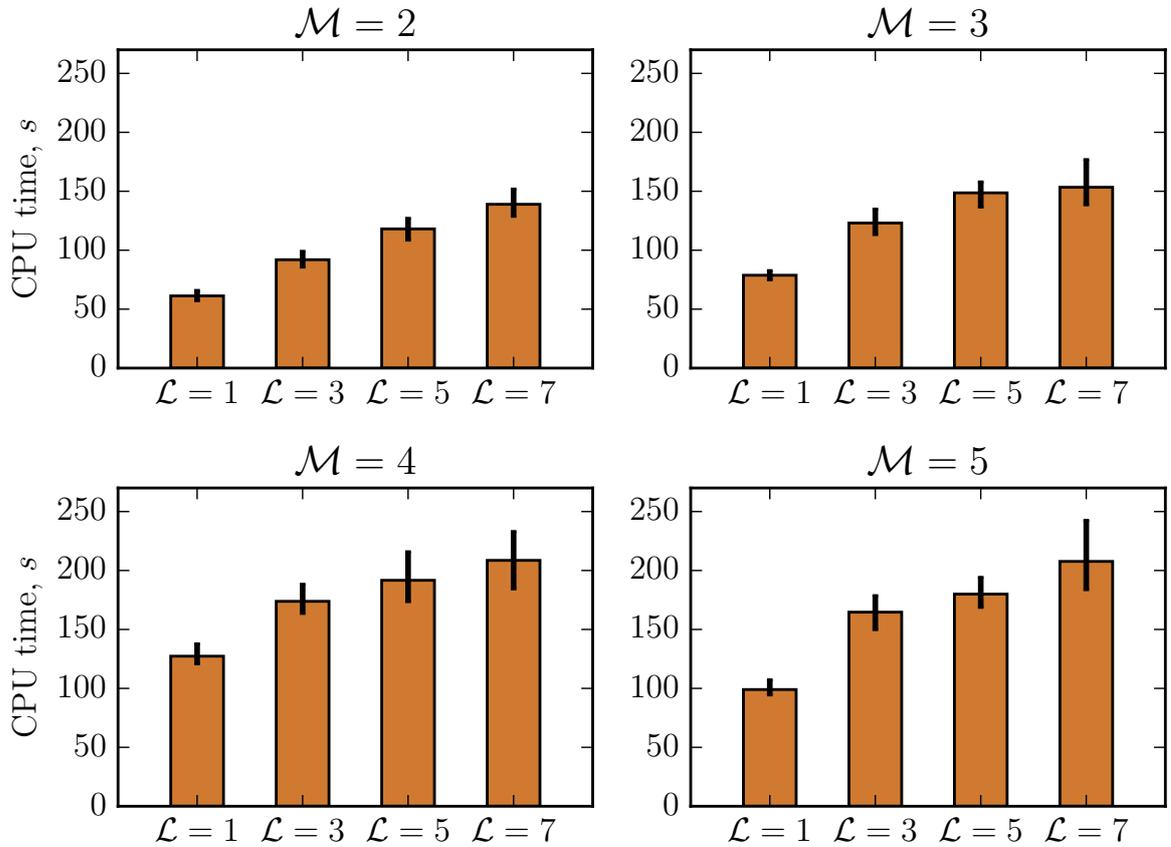

Figure 3.7: The average CPU time required by the multi-level method to estimate $\mathbb{E}[X_3(1)]$ in Case Study 2. We vary $\mathcal{M}$ and $\mathcal{L}$; the estimator is biased. The black bars indicate the range occupied by the 10-th to 90-th percentiles of the data. The values of $\tau_0$ are: $\mathcal{M} = 2 \Rightarrow \tau_0 = 1/8$, $\mathcal{M} = 3 \Rightarrow \tau_0 = 1/9$, $\mathcal{M} = 4 \Rightarrow \tau_0 = 1/16$, and $\mathcal{M} = 5 \Rightarrow \tau_0 = 1/5$.

In Figure 3.9, the numerical properties of the MSE estimation scheme are demonstrated. For each choice of $\lambda$, a histogram of the estimated dimer populations is provided. In particular, a smaller value of $\lambda$ means a lower bias in the estimator, and consequently, a larger value of $\mathcal{L}$. Larger values of $\lambda$ lead to a higher bias, but greater statistical accuracy. Figure 3.9 confirms that as $\lambda$ decreases, the mean estimate (where the mean is taken over independent runs of the multi-level scheme) increases towards the true dimer population (corresponding with decreasing bias, and a distribution shifting to the right), whilst the variance increases (corresponding with an increasing statistical error, and a wider distribution in Figure 3.9).

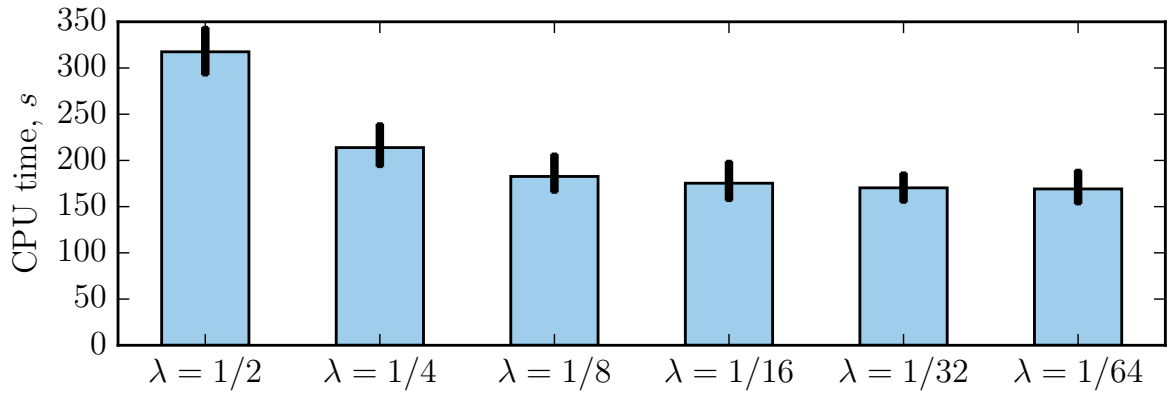

Figure 3.8: The average CPU time required by the multi-level method to estimate $\mathbb{E}[X_3(1)]$ in Case Study 2. The estimator is biased, and the MSE is bounded by $\varepsilon^2 = 0.2603$. We have fixed $\mathcal{M} = 3$ and $\tau_0 = 1/9$, with the parameter $\lambda$ varied as per the text. For each choice of $\lambda$, the multi-level method was run 100 times. The black bars indicate the range occupied by the 10-th to 90-th percentiles of the data.

The gene regulatory network case study is now exhausted, and we turn to considering a range of other model problems and summary statistics.

## 3.8 Further numerical examples

The multi-level method is now explored in more detail by considering two further case studies. The first case study was previously discussed in Chapter 2, and is a stochastic logistic growth model. In this case, we will estimate the expected time-average population of a species.

The second case study is a model concerned with the mitogen-activated protein kinase (MAPK) cascade, which is a relatively complicated cell signalling pathway [95]. Our MAPK cascade model will incorporate Michaelis-Menten kinetics.

---

[9]Note that, for reasons of comparability, Figure 3.7 uses only 100 runs for each choice of $\lambda$. Additional simulations were generated so that Figure 3.8 could be based on 300 runs of the multi-level method.

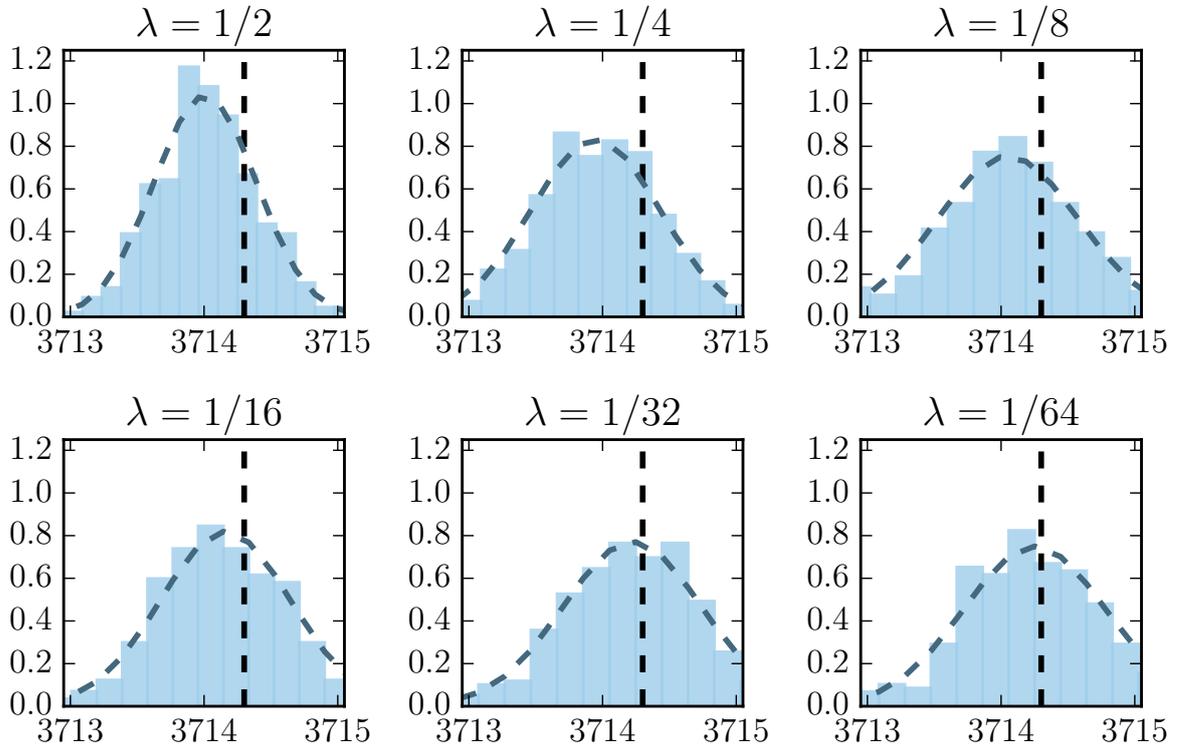

Figure 3.9: For each choice of $\lambda$, the multi-level method is run 300 times[9]. The estimates of the dimer population of network (2.10) are plotted as a histogram, and a normal distribution is fitted. A MSE of $\varepsilon^2 = 0.2603$ is attained. The dashed black line indicates the true dimer population. As $\lambda$ decreases, so does the bias, but the estimator variance increases.

### 3.8.1 Estimating different summary statistics

This section illustrates the use of the multi-level method with a more general summary statistic of interest.

**Case Study 1.** We return to Case Study 1, which involves a stochastic logistic growth model. The model comprises a single species, with the following two reaction channels[10]:

$$R_1 : A \xrightarrow{10} 2A; \quad R_2 : 2A \xrightarrow{0.01} A.$$

We will estimate the average population over the time-interval $[0, T]$, where $T = 3$.

---

[10]In comparison to our original description in Chapter 2, the parameters have been adjusted.

Therefore, we set

$$\mathcal{Q} = \mathbb{E}\left[\frac{1}{T}\int_0^T X(t)\,dt\right], \tag{3.29}$$

where $X(t)$ represents the population of $A$ at time $t$. Initially, $X(0) = 50$.

The DM estimates $\mathcal{Q} = 899.91 \pm 0.10$ using $3.3 \times 10^4$ simulations in 136.9 seconds (as before, $\pm$ refers to a 95% confidence interval).

To improve on the CPU time provided by the DM, the complete multi-level algorithm is run with a wide variety of algorithm parameters and the results are compared. To structure our investigation, we first choose $\mathcal{M}$, and then follow Section 3.6.1 to choose $\tau_0$: for $\mathcal{M} = 2$, $\tau_0 = 3/256$; for $\mathcal{M} = 3$, $\tau_0 = 3/243$; and for $\mathcal{M} = 5$, $\tau_0 = 3/256$. A variety of values for $\mathcal{L}$ are then tested. In Figure 3.10 the computational performance of the unbiased multi-level method is demonstrated. For each choice of algorithm parameters, the entire multi-level algorithm is run 100 times.

Figure 3.10 shows that an average multi-level simulation time of 29.6 seconds is attainable. Compared with the DM, this figure represents a speed-up ratio of approximately 4.6. There is not much to be said about $\mathcal{M}$, but for this example, smaller values of $\mathcal{L}$ might be preferred. Whilst not as impressive as the results presented in Section 3.7, the time-savings are nevertheless substantial.

### 3.8.2 A model of the MAPK cascade

In this section, we consider a fourth case study. This model describes the MAPK, which is involved in a wide variety of signalling processes that govern transitions relating to the phenotype of a cell, and has previously been used as a test case for various SSAs [96].

**Case Study 4.** This model of a MAPK cascade comprises ten coupled Michaelis-

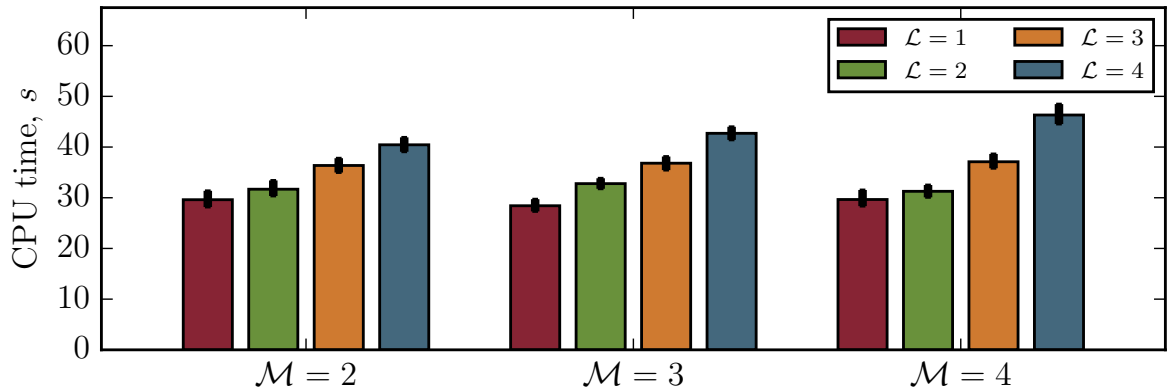

Figure 3.10: The average CPU time required by the multi-level method to estimate $\mathbb{E}\left[\int_0^3 X(t)\mathrm{d}t/3\right]$ in Case Study 3. We vary $\mathcal{M}$ and $\mathcal{L}$; the estimator is biased. The black bars indicate the range occupied by the 10-th to 90-th percentiles of the data. The values of $\tau_0$ are: $\mathcal{M} = 2 \Rightarrow \tau_0 = 3 \times 2^{-8}$, $\mathcal{M} = 3 \Rightarrow \tau_0 = 3^{-4}$, and $\mathcal{M} = 4 \Rightarrow \tau_0 = 3 \times 4^{-4}$.

Menten schemes [97], and has $N = 22$ species and $M = 30$ reactions.

A Michaelis-Menten scheme is constructed as follows [96]: there are four species and three reaction channels within the scheme. The species are substrate ('S'), enzyme ('E'), complex ('ES') and product ('P'). The reaction channels are as follows:

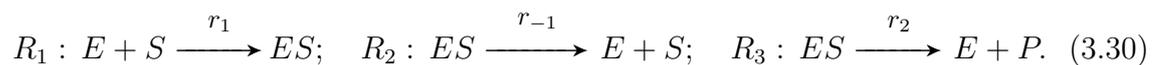

$$R_1 : E + S \xrightarrow{r_1} ES; \quad R_2 : ES \xrightarrow{r_{-1}} E + S; \quad R_3 : ES \xrightarrow{r_2} E + P. \quad (3.30)$$

A quasi-steady state assumption can be applied to reduce the computational complexity associated with simulating the reaction network. This reduces the scheme to two species: substrate ('S') and product ('P') particles. The three reaction channels described by (3.30) are reduced into a single reaction channel, which is given as

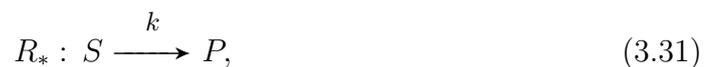

$$R_* : S \xrightarrow{k} P, \quad (3.31)$$

where the propensity function follows Michaelis-Menten kinetics, which are given by

$$k = \frac{k_2 E_0 S}{S + \frac{r_{-1}+r_2}{r_1}}, \quad (3.32)$$

where $E_0$ represents the initial enzyme population. As explained, the MAPK cascade comprises ten coupled Michaelis-Menten schemes: we provide a diagrammatic representation in Figure 3.11. We will now simulate the model using the quasi-steady state assumption; therefore, the reduced model comprises ten reaction channels. The rate constant of each reaction channel is given by the Michaelis-Menten kinetic formula detailed in Equation (3.32). The substrate, enzyme and product particles for each channel are as shown in Figure 3.11. The reaction channels are therefore as follows:

$$\begin{aligned}
R_1 &: KKK \xrightarrow{k_1} KKK\text{-}P; & R_2 &: KKK\text{-}P \xrightarrow{k_2} KKK; \\
R_3 &: KK \xrightarrow{k_3} KK\text{-}P; & R_4 &: KK\text{-}P \xrightarrow{k_4} KK; \\
R_5 &: KK\text{-}P \xrightarrow{k_5} KK\text{-}PP; & R_6 &: KK\text{-}PP \xrightarrow{k_6} KK\text{-}P; \\
R_7 &: K \xrightarrow{k_7} K\text{-}P; & R_8 &: K\text{-}P \xrightarrow{k_8} K; \\
R_9 &: K\text{-}P \xrightarrow{k_9} K\text{-}PP; & R_{10} &: K\text{-}PP \xrightarrow{k_{10}} K\text{-}P.
\end{aligned} \quad (3.33)$$

where the $k_j$ are rate functions given in Equation (3.32).

We will estimate the mean MAPK population (indicated by 'K-P' in System (3.33) and Figure 3.11) at a terminal time $T$. The initial conditions are detailed in Table 3.6, and we take $T = 250$. We now provide the model parameters. Each Michaelis-Menten reaction is of the form $R_j : X \xrightarrow{k_j} Y$, and the function $k_j$ is expressed as $k_j = \alpha_j \cdot X/(X + \beta_j)$. For each reaction $R_j$, the initial enzyme populations give

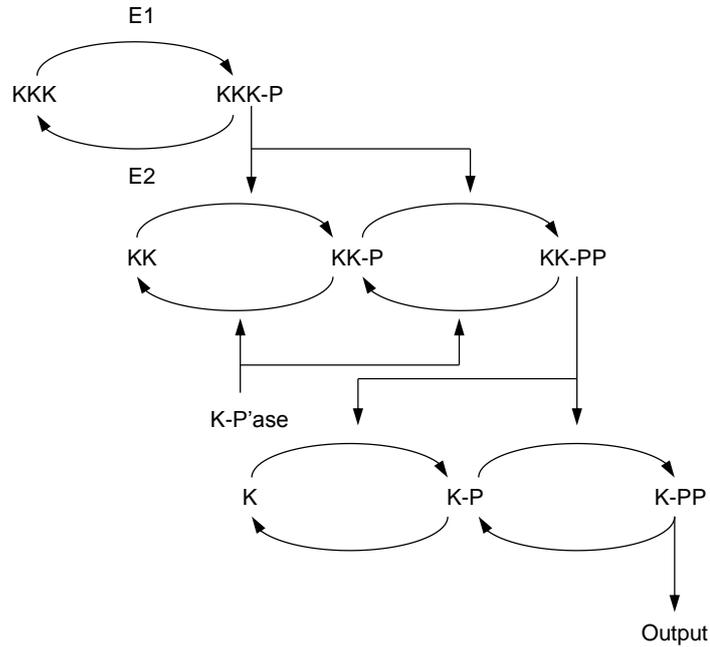

Figure 3.11: A diagrammatic representation of the MAPK cascade. The text refers to chemical species; whilst the arrows represent Michaelis-Menten schemes. The arrow points from the substrate towards the product; the species on top of the arc indicates the enzyme. This diagram has been adapted from Huang and Ferrell [97].

$\alpha_j$ and $\beta_j$ their values. The values that we use for $\alpha_j$ and $\beta_j$ (for $j = 1, \ldots, 10$) are stated in Table 3.7.

| Species | Initial value | Species | Initial value |
|---------|---------------|---------|---------------|
| $KKK$   | 90            | $KKK\text{-}P$  | 10 |
| $KK$    | 280           | $KK\text{-}P$   | 10 |
| $KK\text{-}PP$ | 10     | $K$     | 280 |
| $K\text{-}P$   | 10     | $K\text{-}PP$ | 10 |

Table 3.6: The initial values for the MAPK cascade model detailed in (3.33).

If the DM is used, it takes approximately 64.8 seconds to estimate the mean MAPK population at time $T = 250$ (with a confidence interval of semi-length 1.0). In Table 3.8 we show a multi-level configuration that estimates the mean MAPK population as $2683.16 \pm 0.99$; this calculation takes approximately 13.1 seconds of CPU time. This demonstrates that, even with a relatively complicated reaction network, a significant

| Reaction | Parameters | Reaction | Parameters |
|----------|------------|----------|------------|
| $R_1$ | $\alpha_1 = 2.5, \beta_1 = 10$ | $R_2$ | $\alpha_2 = 0.25, \beta_2 = 8$ |
| $R_3$ | $\alpha_3 = 0.025, \beta_3 = 15$ | $R_4$ | $\alpha_4 = 0.75, \beta_4 = 15$ |
| $R_5$ | $\alpha_5 = 0.025, \beta_5 = 10$ | $R_6$ | $\alpha_6 = 0.75, \beta_6 = 15$ |
| $R_7$ | $\alpha_7 = 0.025, \beta_7 = 10$ | $R_8$ | $\alpha_8 = 0.5, \beta_8 = 15$ |
| $R_9$ | $\alpha_9 = 0.025, \beta_9 = 10$ | $R_{10}$ | $\alpha_{10} = 0.5, \beta_{10} = 15$ |

Table 3.7: The parameters for the MAPK cascade model (3.33).

reduction in simulation time can be achieved with the multi-level method.

We now repeat the multi-level procedure, 1000 times over, to determine how the simulation time varies. We take $\mathcal{M} = 4$, $\tau_0 = 1/16$ and $\mathcal{L} = 3$. In Figure 3.12 the estimates of the MAPK population (indicated as 'K-P') are placed into a histogram, and a normal distribution is fitted. An empirical cumulative density function is also shown. As expected, the distribution of estimates is asymptotically normal. The total simulation durations are plotted in Figure 3.13. This figure confirms that the average simulation duration is approximately 13.8 seconds (with the 10-th percentile corresponding to 13.3 seconds, and the 90-th percentile 14.4 seconds), which represents a factor 4.7 improvement over the SSA.

| Level | $\tau_{\ell-1}$ | $\tau_\ell$ | Estimate | Sample variance | Paths | Time |
|-------|-----------------|-------------|----------|-----------------|-------|------|
| $\mathcal{Q}_0$ | - | $4^{-2}$ | 2331.92 | 15805.69 | 124237 | $6.5s$ |
| $\mathcal{Q}_1$ | $4^{-2}$ | $4^{-3}$ | 276.28 | 535.58 | 11346 | $2.4s$ |
| $\mathcal{Q}_2$ | $4^{-3}$ | $4^{-4}$ | 57.50 | 160.54 | 4103 | $2.0s$ |
| $\mathcal{Q}_3$ | $4^{-4}$ | $4^{-5}$ | 13.32 | 32.54 | 1170 | $1.4s$ |
| $\mathcal{Q}_4$ | $4^{-6}$ | DM | 4.22 | 10.61 | 703 | $1.0s$ |
| **Total** | | | **2683.16 ± 0.99** | | - | **13.42s** |

Table 3.8: The contribution from each level estimator $\mathcal{Q}_\ell$ in producing an unbiased overall estimator for the mean MAPK population at time $T = 250$. We have taken $\tau_0 = 1/16$, $\mathcal{M} = 4$, and $\mathcal{L} = 3$.

## 3.9 Discussion

In this final section we discuss some remaining computational challenges encountered whilst implementing the multi-level method.

### 3.9.1 Comparing MATLAB and C++

In this section, we briefly discuss the merits of using `C++` to generate sample paths, and we compare `C++` with alternative computing approaches.

**Benefits of C++.** We used `C++` due to its reliability and well-documented performance characteristics. We used `g++` from the GNU compiler collection to compile the simulation routines, and a compiler flag of `-Ofast` was set to ensure that the code will execute quickly. The `2011` coding-standard was used, as the built-in libraries include high-quality algorithms for random variate generation. As we sought to focus on the mathematical properties of the multi-level method, we proceeded with caution and used the well-established Mersenne Twister method for generating random numbers. We did not attempt to exploit any particular hardware characteristics.

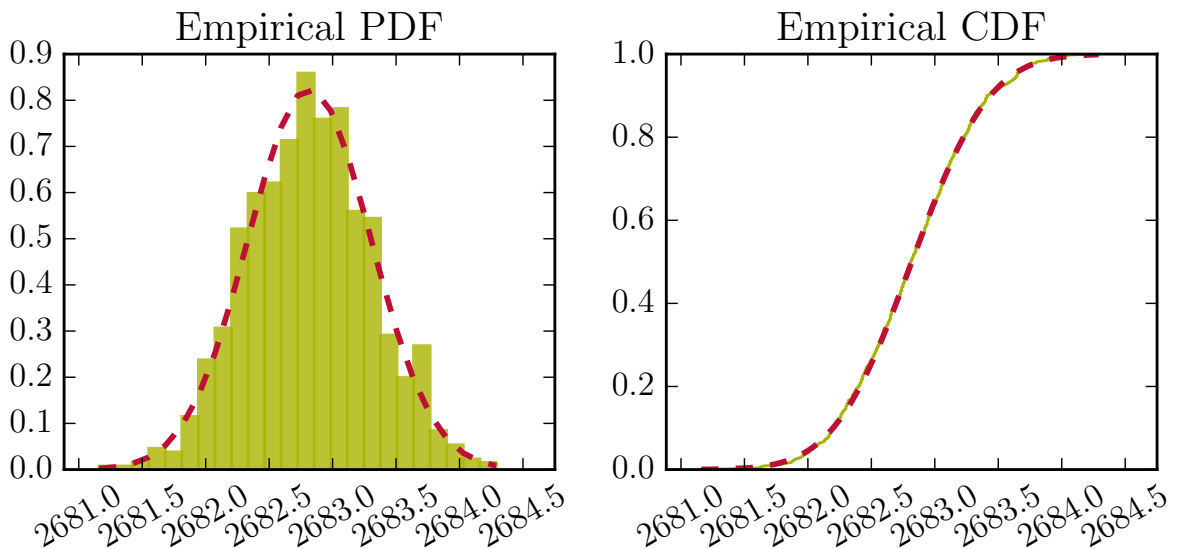

Figure 3.12: We plot the empirical distribution of the estimates of the MAPK population of reaction network (3.33). The dynamic calibration method stated in Algorithm 3.3 has been run a total of 1000 times. A histogram is shown on the left, whilst an empirical cumulative density function is shown on the right. A normal distributed has been fitted and is shown in red. The algorithm parameters and parameters of the normal distribution are as stated in the main text.

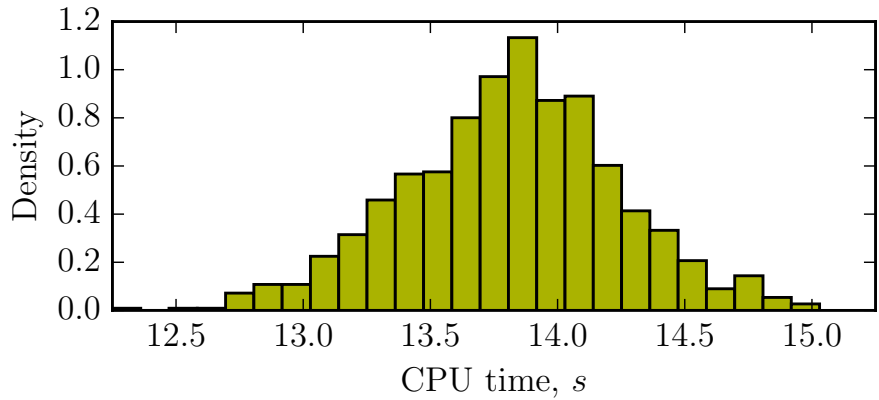

Figure 3.13: This histogram depicts the total duration taken by the multi-level algorithm to estimate the MAPK population of reaction network (3.33). The dynamic calibration method stated in Algorithm 3.3 has been run a total of 1000 times.

**Drawbacks of C++.** Arguably, `C++` cannot be learned quickly, and there are many pitfalls for a novice `C++` coder to avoid. As such, `MATLAB` is often seen as a more natural programming framework, and it has been widely used throughout the literature. Below, a number of `MATLAB`-specific issues are discussed.

**Considering MATLAB.** The preparatory simulation work for this thesis was undertaken in `MATLAB`. Whilst writing a `MATLAB` script is relatively straightforward, understanding the intricacies of `MATLAB` is more challenging. For `MATLAB` code to run quickly, it needs to be 'vectorized'. A piece of code is vectorized if it performs operations on multiple components of a vector at the same time.

**Example 3.3.** The results shown in Table 2.2 can also be generated with `MATLAB`. With vectorized code, it takes around three-and-a-half hours $(12,392$ seconds$)$ to regenerate the $4.8 \times 10^6$ sample paths required to produce this table. A non-vectorized implementation requires approximately 162 hours (nearly a week) to run. ∎

Other researchers have attempted to understand and quantify the effect of vectorization on the overall simulation time [91]. We merely make the following remarks regarding this problem. Throughout this section, we argued that the simulation time devoted to a level $\ell$ is given as $\mathcal{C}_\ell \cdot \mathcal{N}_\ell$. In MATLAB, code vectorization means that, for example, the CPU time per path when generating $10^2$ sample paths is different (and usually much greater) than the CPU time per path when generating $10^3$ paths. It could therefore be unwise to assume that the CPU time taken to generate the sample paths on a given level will scale linearly with the number of such paths.

### 3.9.2 Catastrophic decoupling

Consider the contribution of each term in $\mathcal{Q} = \mathcal{Q}_0 + \mathcal{Q}_1 + \cdots + \mathcal{Q}_\mathcal{L}\ (+\mathcal{Q}_{\mathcal{L}+1})$ to the multi-level estimator. In the course of our exploration of the multi-level method, we have noticed that occasionally pairs of sample paths on a level undergo what we will call a 'catastrophic decoupling' so that species populations in these sample paths become very different from one another. This can have a dramatic effect on the sample variance on that level, and hence on the results of the multi-level method. We illustrate with an example:

**Example 3.4.** For the example gene expression model (Case Study 2), we take $\tau_0 = 1/9$ and $\mathcal{M} = 3$. In Table 3.9 we show percentile data for distributions of $\mathcal{Q}_3$, $\mathcal{Q}_4$ and $\mathcal{Q}_5$. It is clear that the sample values contributing to $\mathcal{Q}_3$, $\mathcal{Q}_4$ and $\mathcal{Q}_5$ all possess extreme tails to their distributions as a result of one or more catastrophic decoupling events. For example, over 90% of the sample values for $\mathcal{Q}_5$ lie in the interval $[1, 9]$, but approximately 1 in 1150 sample paths provide sample values of $-100$ or less. This makes catastrophic decoupling events appear deceptively unlikely; however, if $10^2$ sample paths are generated, there is an 8% chance that such an event is encountered.

If $10^3$ sample paths are performed, this rises to approximately 58%.

| Estimator | 0.01 | 0.1 | 1 | 5 | 50 | 95 | 99 | 99.9 | 99.99 |
|---|---|---|---|---|---|---|---|---|---|
| $\mathcal{Q}_3$ | $-360.5$ | $-311$ | 10 | 21 | 40 | 63 | 74 | 87 | 198.5 |
| $\mathcal{Q}_4$ | $-379$ | $-256.5$ | 3 | 6 | 13 | 22 | 27 | 32 | 43 |
| $\mathcal{Q}_5$ | $-349.5$ | $-58.5$ | 0 | 1 | 4 | 9 | 11 | 13 | 15 |

Table 3.9: Various percentiles of the samples used to estimate $\mathcal{Q}_3$, $\mathcal{Q}_4$ and $\mathcal{Q}_5$ of System (2.10). In each case, $10^5$ samples were used, with $\tau_0 = 1/9$ and $\mathcal{M} = 3$.

We now explain the cause of this problem, and then discuss its consequences. In effect, a decoupling is possible each time a new mRNA molecule is introduced into the system. The coupling technique ensures it is introduced into *both* the coarse and fine sample paths. In the fine system, the decay process of this mRNA starts immediately. However, in the coarse system this is not always the case: this is because decay of the mRNA cannot take place in the coarse system until the reaction propensities are updated. Hence, during this interim period, it is possible for the new mRNA particle to decay in the fine system but not in the coarse system.

It is clear that the scaling of the system is then what causes problems with the variance. At time $T = 1$, there are approximately 24 mRNA molecules, compared with over 3000 protein molecules. If the decoupling in mRNA species counts occurs at an early time, the extra mRNA molecule in the coarse system leads to increased protein generation which, in turn, leads to increased dimer generation. This difference in generation rates remains until the mRNA populations converge again (if at all). As the dimer population is monotonically increasing, the population difference is 'locked in' for all subsequent times, and the difference in sample values of $X_3(T)$ is large. ■

A novel coupling method that avoids the catastrophic decoupling is presented in Chapter 5. At this stage, in Table 3.10 we provide sample means, variances and kurtoses of the different levels in the gene expression system (2.10), and we note that the kurtosis of $\mathcal{Q}_\ell$ increases with $\ell$, which might suggest that algorithms with many levels are more likely to be affected. We also note that even when the dynamic calibration method is used, a catastrophic decoupling will still affect the simulation time.

| Estimator | Mean | Variance | Kurtosis |
|---|---|---|---|
| $\mathcal{Q}_3$ | 39.25 | 665.96 | 107.05 |
| $\mathcal{Q}_4$ | 13.09 | 204.50 | 420.54 |
| $\mathcal{Q}_5$ | 4.34 | 57.54 | 1628.33 |

Table 3.10: Statistics describing the samples for $\mathcal{Q}_3$, $\mathcal{Q}_4$ and $\mathcal{Q}_5$ for system (2.10) using $\tau_0 = 1/9$ and $\mathcal{M} = 3$. Further information is available in Figure 5.2 in Chapter 5.

### 3.9.3 Conclusion

The multi-level method provides the potential for substantial time-savings to be made in the field of stochastic simulation of chemical systems. Although there are many intricacies associated with the method, many of them software- and system-dependent, the benefits of using multi-level approaches are enormous, and as such, they open up the range of problems that can be fully explored using stochastic simulation.

In this chapter, we have introduced a number of novel enhancements to the multi-level method, which, we feel, make it easier to understand and implement, as well more computationally efficient. We will now describe the *adaptive* multi-level scheme. The *adaptive* method that we develop will be a new and generalised multi-level approach that is particularly suited to studying the summary statistics of stiff systems.

# Chapter 4

# Adaptive Multi-Level Monte Carlo

The focus of this chapter is on extending the range of biochemical reaction networks for which the multi-level method can efficiently estimate summary statistics. When compared with the Gillespie DM, the use of a suitably implemented multi-level method can lead to substantially lower simulation times. However, the formulation of the multi-level method provided in Chapter 3 places restrictions on the time-steps used, which can limit the efficiency of the algorithm. In this chapter we work towards overcoming these constraints, and, in particular, we describe a multi-level method that is particularly suitable for the simulation of stiff systems. We demonstrate the efficiency of our method using a number of examples.

## 4.1 Introduction

In Chapter 3, we explained that the multi-level method can be used to estimate a summary statistic of interest, $\mathcal{Q}$, by independently estimating a number of (sub)- estimators, $\mathcal{Q}_0, \ldots, \mathcal{Q}_{\mathcal{L}+1}$, and adding the estimates together[1]:

$$\mathcal{Q} = \mathcal{Q}_0 + \sum_{\ell=1}^{\mathcal{L}} \mathcal{Q}_\ell + \mathcal{Q}_{\mathcal{L}+1}.$$

In Chapter 3, the estimators, $\mathcal{Q}_0, \ldots \mathcal{Q}_{\mathcal{L}}$, are generated using tau-leap methods, where the time-step, $\tau$, is fixed in advance. This restriction can result in poor performance when the reaction activity of a system changes substantially over the time-scale of interest.

In order to generalise and reformulate the multi-level method, in this chapter we will relax the following two restrictions that have been built into the method:

- that the time-steps are uniformly-sized;

- that the time-steps for sample paths with different accuracies are *nested*. In other words, the time-step is reduced by some integer factor $\mathcal{M} \in \{2, 3, \ldots\}$ as the value of $\ell$ increases.

Under certain conditions, these unnecessary restrictions can cause the multi-level to be inefficient. For example, difficulties are likely to arise where a system displays stiff behaviour, with markedly different propensities observed on different time-scales. Our new method will be called the *adaptive* multi-level method, because the time-steps for the approximate simulation algorithms will adapt to the stochastic behaviour of each sample path. Our focus will be on describing a method that is straightforward to implement.

---

[1] We have assumed the estimator is unbiased; the analysis is nearly identical for a biased estimator.

### 4.1.1 Outline

This chapter is arranged as follows: the remainder of Section 4.1 highlights the limitations of the fixed time-step multi-level method by presenting two case studies for which the method, when compared with the DM, provides only nominal improvements in CPU time. In Sections 4.2 and 4.3, a novel *adaptive* multi-level method is presented as a solution to the problem. The benefits of a new method are fully explored with reference to the motivating examples in Section 4.4. An automated procedure, that can quickly implement the adaptive multi-level scheme, is described in Section 4.5. We conclude by discussing our results in Section 4.6.

### 4.1.2 Two motivating examples

In this section we introduce two motivating examples that highlight the potential limitations of the fixed time-step multi-level method.

**Case Study 5.** This dimerisation model has been employed widely as a test of stochastic simulation algorithms [45, 72] as it exhibits distinct behaviours on multiple time-scales. The reaction network is given by:

$$
\begin{aligned}
&R_1 : S_1 \xrightarrow{1} \emptyset; &\qquad &R_2 : S_2 \xrightarrow{1/25} S_3; \\
&R_3 : S_1 + S_1 \xrightarrow{1/500} S_2; &\qquad &R_4 : S_2 \xrightarrow{1/2} S_1 + S_1.
\end{aligned}
\tag{4.1}
$$

We take the initial conditions to be $[X_1, X_2, X_3]^\top = [10^5, 0, 0]^\top$. Using the Gillespie DM (see Section 2.3), we calculate that the expected population of $S_3$ at time $T = 30$ is

$$\mathcal{Q} := \mathbb{E}[X_3(30)] = 20591.6 \pm 1.0.$$

As before, the '$\pm$' term provides a 95% confidence interval for the estimator (see Section 2.2 for further information). This calculation requires $3.6 \times 10^4$ sample paths,

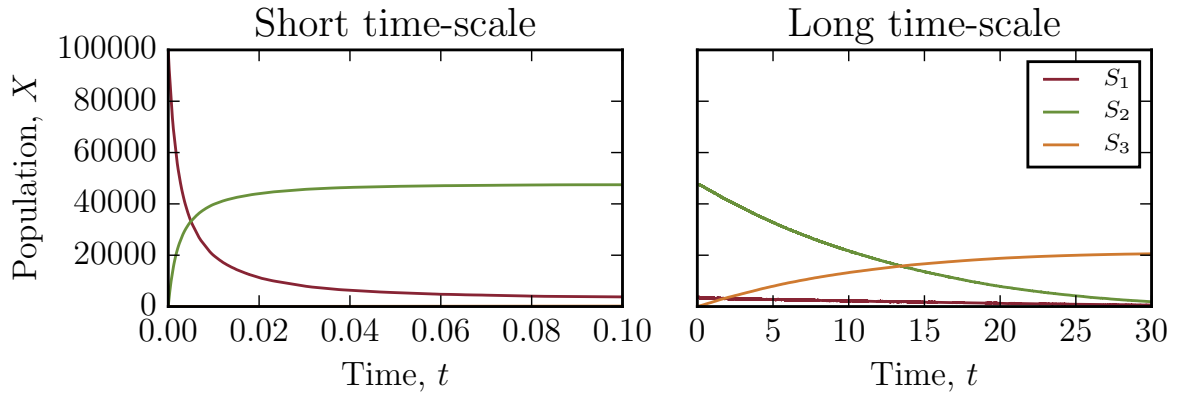

Figure 4.1: The temporal evolution of a single sample path of System (4.1) on two different time-scales.

and the algorithm takes a total of 1914 seconds (approximately 32 minutes) to run.

In order to better understand the dynamics of System (4.1), we consider a typical sample path. In Figure 4.1, the temporal evolution of a single sample path of the system, generated with the Gillespie DM, is shown on two distinct time scales. For this sample path, $X_3(30) = 20547$, and there are therefore approximately 45 fewer $S_3$ particles than average. In this case study, stochastic effects appear to play a more limited role than that seen, for example, in Case Study 2, and the computational cost of the DM, when compared with an ODE-modelling approach, might therefore be perceived as high. However, it is important to note that, to some extent, stochasticity still affects the temporal dynamics of the system, and a closer inspection will demonstrate that the sample paths are not smooth. By using a substantially lower level of computational resources, our adaptive multi-level method justifies the use of a stochastic model for this case study.

A detailed examination of the trajectory represented in Figure 4.1 shows that the initial phase is marked by a rush of reaction activity, but, once this phase has passed, reaction activity slows dramatically. For this particular realisation (statistics are broadly similar across all repeats), approximately $6 \times 10^5$ individual reactions are simulated over the time-interval $[0, 30]$. Of these, roughly $5 \times 10^4$ are in the first $0.05$ seconds. This is equivalent to $10^6$ reactions per unit time. For the remaining $29.95$ units of time, reactions fired at a rate of roughly $2 \times 10^4$ reactions per unit time, which is 50 times slower.

We now display results of our attempts at applying the unbiased, fixed time-step multi-level method to this problem. Adopting a trial-and-error approach, we vary the choices of $\mathcal{M}$ and $\mathcal{L}$, whilst choosing $\tau_0$ automatically by following the procedure set out in Section 3.6.1. With each choice of $\mathcal{M}$ and $\mathcal{L}$, the full dynamic-calibration method (as described in Algorithm 3.3) is run 100 times to test the multi-level method. As before, the algorithm produces an estimate of $\mathbb{E}[X_3(30)]$ with a 95% confidence interval of semi-length of $1.0$.

In Figure 4.2, we show the average simulation time of the multi-level method, where different choices of $\mathcal{M}$ and $\mathcal{L}$ are used. For each choice of $\mathcal{M}$ and $\mathcal{L}$, the black lines in Figure 4.2 indicate the range occupied by the 10-th to 90-th percentiles of the total simulation time; this demonstrates a tight clustering around the average total simulation time. In particular, with $\mathcal{M} = 2$, $\mathcal{L} = 1$ and $\tau_0 = 30 \times 2^{-14}$, the multi-level method can produce an estimate within approximately 676.5 seconds. This represents a factor 2.8 time saving over the Gillespie DM. We note that $\mathcal{L}$ is relatively small: in Section 4.1.3 we show that this is a consequence of the relatively small value $\tau_0$ necessarily takes on. Whilst significant, this time saving is substantially lower than those that have been demonstrated elsewhere in the literature for other reaction networks.

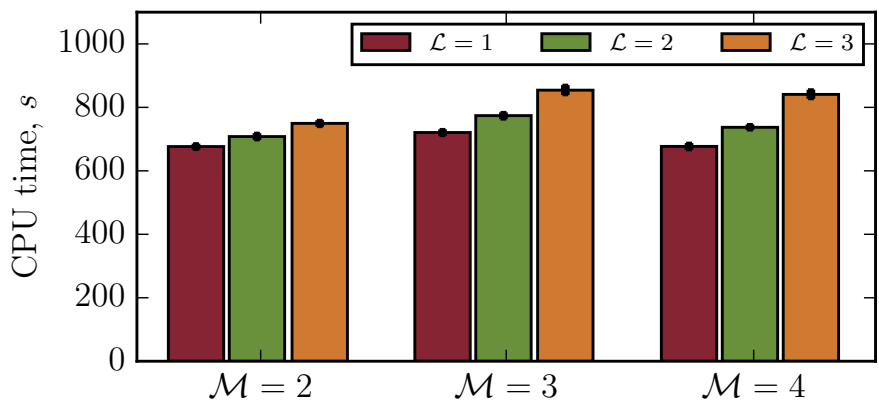

Figure 4.2: The simulation time for a range of system configurations for the fixed time-step multi-level method used to estimate $\mathbb{E}[X_3(30)]$ for System (4.1). The black bars indicate the range occupied by the 10-th to 90-th percentiles of the data. The estimator is unbiased; each case uses a different choice of $\mathcal{M}$ and $\mathcal{L}$. With $T = 30$, the values of $\tau_0$ are: $\mathcal{M} = 2 \Rightarrow \tau_0 = T \times 2^{-14}$; $\mathcal{M} = 3 \Rightarrow \tau_0 = T \times 3^{-9}$; and, $\mathcal{M} = 4 \Rightarrow \tau_0 = T \times 4^{-7}$.

**Case Study 6.** We next consider a synthetic test model. This model involves three species, $S_1$, $S_2$ and $S_3$. and four reactions. The reactions are specified as:

$$R_1 : \emptyset \xrightarrow{1/4} S_1; \qquad R_2 : S_1 + S_2 \xrightarrow{1/2} S_1 + 2S_2; \\ R_3 : 2S_2 \xrightarrow{1/50} S_2; \qquad R_4 : 2S_2 \xrightarrow{1/10000} 2S_2 + S_3. \tag{4.2}$$

We take the initial conditions to be $[X_1, X_2, X_3]^\top = [1, 5, 0]^\top$. In Figure 4.3, we present a sample trajectory of this system for $t \in [0, 100]$. This time, reaction activity increases dramatically over the course of the simulation.

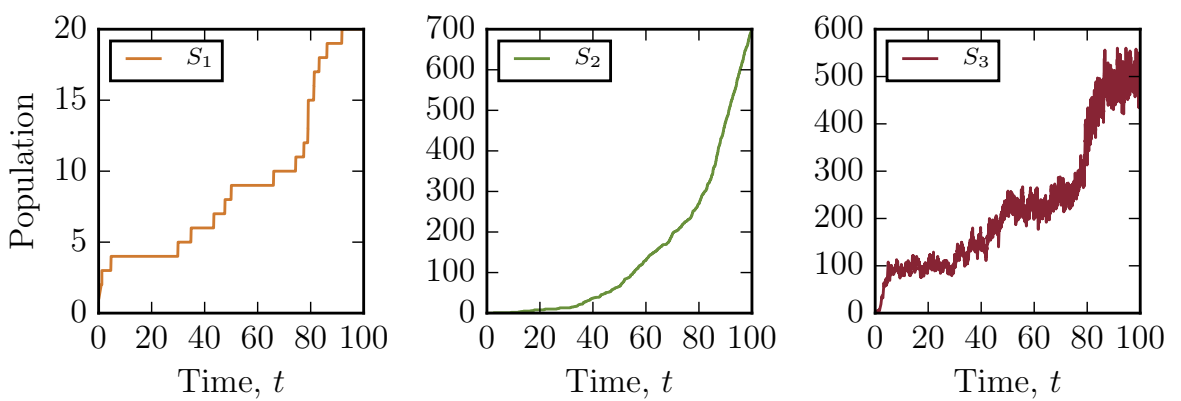

Figure 4.3: The temporal evolution of species $S_1$, $S_2$ and $S_3$ for one realisation of System (4.2).

Using the Gillespie DM (see Section 2.3), we calculate that the expected population of $S_3$ at time $T = 100$ is

$$\mathcal{Q} := \mathbb{E}[X_3(100)] = 1535.9 \pm 2.50.$$

This calculation takes 14235 seconds (nearly 4 hours) and requires $2.6 \times 10^5$ sample paths.

The full, dynamic-calibration method, as described in Algorithm 3.3, is run 100 times to test the fixed-step multi-level method with different choices of $\mathcal{M}$ and $\mathcal{L}$. For each choice of $\mathcal{M}$ and $\mathcal{L}$, Figure 4.4 indicates the mean total simulation time required by the algorithm; the black line indicates the range occupied by the 10-th to 90-th percentiles.

In particular, Figure 4.4 shows that the fixed time-step multi-level method provides at least a factor 50 time saving over the Gillespie DM, as the multi-level method can estimate $\mathcal{Q}$ within an average of 279.2 seconds (by taking $\mathcal{M} = 3$, $\mathcal{L} = 1$ and $\tau_0 = 100 \times 3^{-5}$). As above, a 95% confidence interval of semi-length 2.5 is obtained.

We now demonstrate that, through the use of an adaptive multi-level algorithm, even this significant saving can be improved upon. We start by discussing the disadvantages of using a fixed time-step tau-leap algorithm as part of the multi-level method.

### 4.1.3 Disadvantages of fixed time-step multi-level

Generating sample paths using the tau-leap method with a fixed choice of $\tau$ throughout a simulation poses inherent difficulties. Firstly, for temporal regions in which species numbers are changing rapidly we need to be careful not to choose $\tau$ too large that the propensity functions change considerably over the course of a leap [45]. At its worst, too large a $\tau$ can render the tau-leap method numerically unstable and therefore non-convergent. With a fixed choice of $\tau$ this means that the temporal region of the path that requires the most stringent bound on $\tau$ determines the maximum value that $\tau$ can feasibly take. Therefore, the values of $\tau_0, \tau_1, \ldots$ are restricted. In particular, it is possible that the time taken to generate a single tau-leap sample path – with even the largest feasible choice of $\tau$ – can be greater than that required generate a single sample path using the DM. Under these circumstances, the fixed time-step multi-level method is rendered redundant.

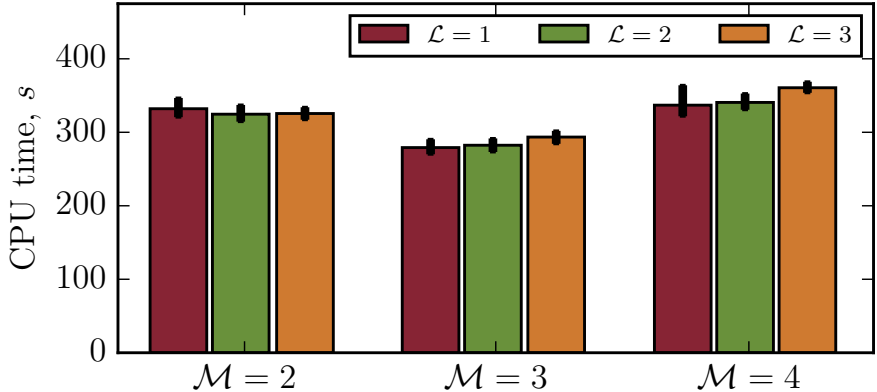

Figure 4.4: The simulation time for a range of system configurations for the fixed time-step multi-level method used to estimate $\mathbb{E}[X_3(100)]$ for System (4.2). The black bars indicate the range occupied by the 10-th to 90-th percentiles of the data. The estimator is unbiased; each case uses a different choice of $\mathcal{M}$ and $\mathcal{L}$. With $T = 100$, the values of $\tau_0$ are: $\mathcal{M} = 2 \Rightarrow \tau_0 = T \times 2^{-9}$; $\mathcal{M} = 3 \Rightarrow \tau_0 = T \times 3^{-5}$; and, $\mathcal{M} = 4 \Rightarrow \tau_0 = T \times 4^{-5}$.

Secondly, at different times during the evolution of a sample path, the propensities will change at different rates. In System (4.1), within the initial transient phase of a simulation, the propensity functions change quickly with respect to time and therefore must be updated frequently in order to maintain accuracy of the tau-leap method. However, in the slower phase, propensity functions change more slowly and hence larger time-steps can be tolerated between recalculation. Similarly, in System (4.2), the propensity of $R_4$, the reaction governing the production of $S_3$, is more sensitive to population changes at higher populations of $S_2$ than at lower populations (as it is proportional to $X_2(X_2 - 1)$).

This means that, if a constant level of CPU resources is dedicated to generating each sample path, then varying the lengths of the leaps over the course of each individual sample path (rather than maintaining constant leap sizes) may enable the generation of a more accurate, lower bias estimator. Therefore, the use of an adaptive time-step selection method can lead to a better use of resources. As such, we next present a generalisation of the multi-level method with an adaptive choice of $\tau$. Although our revised method introduces some additional computational overheads, we demonstrate that it can give rise to significantly reduced simulation times.

## 4.2 A new multi-level approach

The key difference between the original multi-level method described in Chapter 3 and the improved approach presented here is that the sample paths generated on each level will no longer be generated using the tau-leap method with constant values of $\tau$. Instead of relating each set of sample paths with a choice of $\tau_\ell$ (i.e. associating $\boldsymbol{Z}_{\tau_\ell}$ with time-step $\tau_\ell$) we will work with a *control parameter*, $\xi_\ell$, and the time-steps for each set of sample paths will be determined with a formula parametrised by $\xi_\ell$.

As such, we will use a *control parameter ensemble*, $\boldsymbol{\xi} := (\xi_0, \xi_1, \xi_2, \ldots, \xi_\mathcal{L})$, to generate our multi-level estimator. As before, suppose the summary statistic $\mathcal{Q} = \mathbb{E}[f(\boldsymbol{X})]$ must be estimated. We let $\boldsymbol{Z}_{h(\xi_\ell, \cdot)}$ represent a sample path of the tau-leap method, where $\tau$ is chosen according to the rule $h(\xi_\ell, \boldsymbol{Z})$. The first variable of $h(\xi_\ell, \boldsymbol{Z})$, $\xi_\ell$, is the previously-mentioned control parameter, and the second variable, $\boldsymbol{Z}$, indicates that the time-steps depend on the state vector (2.1). Further, let the scalar $Z_{h(\xi_\ell, \cdot)} := f(\boldsymbol{Z}_{h(\xi_\ell, \cdot)})$ be the quantity of interest.

Again, there is a choice between a biased and an unbiased estimator. If we wish to use the biased multi-level estimator to estimate a quantity of interest, we write the following telescoping sum of $\mathcal{L}+1$ components

$$\mathcal{Q}_b = \mathbb{E}[Z_{h(\xi_\mathcal{L},\cdot)}] = \mathbb{E}[Z_{h(\xi_0,\cdot)}] + \mathbb{E}[Z_{h(\xi_1,\cdot)} - Z_{h(\xi_0,\cdot)}] + \cdots + \mathbb{E}[Z_{h(\xi_\mathcal{L},\cdot)} - Z_{h(\xi_{\mathcal{L}-1},\cdot)}]. \quad (4.3)$$

In this equation, $h(\xi, \cdot)$ is the function that selects the value of $\tau$ for each time-step. We will assume that if $\xi$ is *decreased*, then the accuracy of our tau-leap method *increases*, and therefore the bias of a chosen summary statistics *decreases*.

An unbiased estimator of $\mathcal{L}+2$ terms is given by

$$\mathcal{Q}_u = \mathbb{E}[Z_{h(\xi_\mathcal{L},\cdot)}] = \mathbb{E}[Z_{h(\xi_0,\cdot)}] + \cdots + \mathbb{E}[Z_{h(\xi_\mathcal{L},\cdot)} - Z_{h(\xi_{\mathcal{L}-1},\cdot)}] + \mathbb{E}[X - Z_{h(\xi_\mathcal{L},\cdot)}], \quad (4.4)$$

where $X$ is shorthand for the (exact) quantity of interest, $f(\boldsymbol{X})$.

We now turn to discussing how each term of Equation (4.3) (in the case of a biased estimate) or Equation (4.4) (in the case of an unbiased estimate) can be estimated.

### 4.2.1 The base level, $\mathcal{Q}_0$

The base level estimator

$$\mathcal{Q}_0 := \mathbb{E}\left[Z_{h(\xi_0,\cdot)}\right], \quad (4.5)$$

can be estimated with sample paths generated with an adaptive tau-leap method. In contrast with the tau-leap method presented in Section 2.4, at each time-step of the tau-leap method, a new value for $\tau$ must be chosen. A variety of methods have been developed to choose the value of $\tau$, each of which has different aims and purposes [45, 64, 66, 67]. Details of the algorithm we use to choose $\tau$ in this chapter are discussed in Section 4.3. The adaptive tau-leap method is stated as pseudo-code in Algorithm 4.1.

---

Algorithm 4.1: Adaptive tau-leap method. This simulates a single sample path, using time-step chosen with rule $h(\xi, \cdot)$.

---

**Require:** initial conditions, $\bm{Z}(0)$, rule $h(\xi, \cdot)$, parameter, $\xi$, and terminal time, $T$.
1: set $\bm{Z} \leftarrow \bm{Z}(0)$ and set $t \leftarrow 0$
2: **while** $t < T$ **do**
3:     set $\tau \leftarrow \min\{h(\xi, \bm{Z}), T - t\}$
4:     **for** each $R_j$ **do**
5:         calculate propensity value $a_j(\bm{Z})$
6:         generate $K_j \sim \mathcal{P}(a_j(\bm{Z}) \cdot \tau)$
7:     **end for**
8:     set $\bm{Z} \leftarrow \bm{Z} + \sum_{j=1}^{M} K_j \cdot \bm{\nu}_j$, and $t \leftarrow t + \tau$
9: **end while**

---

### 4.2.2 The tau-leap correction terms, $\mathcal{Q}_\ell$

We now describe an approach for estimating terms of the form

$$\mathcal{Q}_\ell := \mathbb{E}\left[ Z_{h(\xi_\ell, \cdot)} - Z_{h(\xi_{\ell-1}, \cdot)} \right], \tag{4.6}$$

where $Z_{h(\xi_\ell, \cdot)}$ represents the point statistic of interest, and sample paths are generated using the tau-leap method with time-steps determined according to rule $h(\xi_\ell, \cdot)$.

We follow the same variance reduction argument as presented in Chapter 3: if we are able to generate sample paths so that $\mathcal{Q}_\ell$ can be estimated using a low sample variance, then few sample paths will be required to attain a desired statistical error.

We now state our approach to simultaneously generating the sample paths $\bm{Z}^{(r)}_{h(\xi_{\ell-1}, \cdot)}$ and $\bm{Z}^{(r)}_{h(\xi_\ell, \cdot)}$, which we will denote as the 'coarse' and 'fine' paths, respectively. At time $t$, let $\bm{Z}^C(t)$ and $\bm{Z}^F(t)$ denote the state vectors of the approximate coarse and fine paths, where the time-steps are determined according to rules $h(\xi_{\ell-1}, \cdot)$ and $h(\xi_\ell, \cdot)$, respectively. Following the tau-leap assumption, the propensities of the coarse and fine paths are updated only at the end of each of their respective time-steps. We let $T^C$ and $T^F$ record the 'next update times' (NUTs) for the propensities of the coarse and fine paths, respectively.

For each reaction $R_j$, define $a_j^C$ to be its propensity function when considering the coarse resolution path, and, similarly, define $a_j^F$ to be the propensity function for the fine path. Based on $\boldsymbol{Z}^C(t)$ and $\boldsymbol{Z}^F(t)$, calculate reaction propensities $a_j^C$ and $a_j^F$ for each reaction channel, $R_j$. As with the fixed time-step method presented in Chapter 3, for each reaction $R_j$, we create three 'virtual reactions', with propensities given as:

$$\begin{aligned} b_j^1 &= \min\{a_j^F, a_j^C\}; \\ b_j^2 &= a_j^C - b_j^1; \\ b_j^3 &= a_j^F - b_j^1. \end{aligned} \quad (4.7)$$

To emulate our previous approach, we will simulate the stochastic system with propensities $b_j^1$, $b_j^2$, and $b_j^3$ (for $j = 1, \ldots, M$). The propensities $b_j^1, b_j^2, b_j^3$ (for $j = 1, \ldots, M$), will need to be recalculated when either $a_j^C$ or $a_j^F$ (for $j = 1, \ldots, M$) are updated. As the values of $a_j^C$ (for $j = 1, \ldots, M$) are updated at time $T^C$, and the values of $a_j^F$ at time $T^F$, the values of $b_j^1$, $b_j^2$, and $b_j^3$ (for $j = 1, \ldots, M$) are next updated at time $\min\{T^C, T^F\}$. Therefore, we take a time-step of $\eta := \min\{T^C, T^F, T\} - t$, to simulate reactions that occur before some of the propensity functions need to be updated again[2]. The values of $T^C$ and $T^F$ are updated as required: we either set $T^C := t + h\bigl(\xi_{\ell-1}, \boldsymbol{Z}^F(t)\bigr)$ or $T^F := t + h\bigl(\xi_\ell, \boldsymbol{Z}^C(t)\bigr)$. Our full method is now stated as pseudo-code in Algorithm 4.2.

Figure 4.5 provides a visual representation of how the first four iterations of the algorithm might unfold. Our algorithm shows that it is possible to decide on the time-steps of the coarse and fine paths independently, and then to generate the sample paths simultaneously. Perhaps most importantly, due to the Poisson thickening

---

[2]The presence of $T$ ensures that the algorithm terminates at the correct time.

---

Algorithm 4.2: Adaptively-coupled tau-leap method. This simulates a correlated pair of sample paths with dynamically-chosen time-steps.

---

**Require:** initial conditions, $\mathbf{Z}(0)$, rule $h(\xi, \cdot)$, control parameters, $\xi_C$ and $\xi_F$, and terminal time, $T$.
1: set $\mathbf{Z}^C \leftarrow \mathbf{Z}(0)$, $\mathbf{Z}^F \leftarrow \mathbf{Z}(0)$ and $t \leftarrow 0$.
2: for each $R_j$, calculate propensities $a_j^C(\mathbf{Z}^C)$, and $a_j^F(\mathbf{Z}^F)$
3: for each $R_j$, calculate virtual propensities $b_j^1$, $b_j^2$ and $b_j^3$
4: set $T^C \leftarrow h(\xi_C, \mathbf{Z}^C)$ and $T^F \leftarrow h(\xi_F, \mathbf{Z}^F)$
5: **loop**
6:    **if** $t = T$ **then break**
7:    **end if**
8:    set $\Delta \leftarrow \min\{T, T^C, T^F\} - t$, and then set $t \leftarrow t + \Delta$
9:    **for** each $R_j$ and each $k \in \{1, 2, 3\}$ **do**
10:      generate $K_{jk} \sim \mathcal{P}(b_j^k \cdot \Delta)$
11:    **end for**
12:    set $\mathbf{Z}^C \leftarrow \mathbf{Z}^C + \sum_{j=1}^{M}(K_{j1} + K_{j2}) \cdot \boldsymbol{\nu}_j$
13:    set $\mathbf{Z}^F \leftarrow \mathbf{Z}^F + \sum_{j=1}^{M}(K_{j1} + K_{j3}) \cdot \boldsymbol{\nu}_j$
14:    set $S \leftarrow \operatorname{argmin}\{T^C, T^F\}$
15:    for each $R_j$, recalculate $a_j^S(\mathbf{Z}^S)$, and virtual propensities $b_j^1$, $b_j^2$ and $b_j^3$
16:    set $T_S \leftarrow T_S + h(\xi_S, \mathbf{Z}_S)$
17: **end loop**

---

property (see Fact 1), there is no need for the time increments of the fine path to be nested within (or, indeed, even to be smaller than) those of the coarse path.

### 4.2.3 The final level, $\mathcal{Q}_{\mathcal{L}+1}$

Lastly, we describe an approach for the final term of the unbiased estimator (Equation (4.4)), which is given by

$$\mathcal{Q}_{\mathcal{L}+1} := \mathbb{E}\big[X - Z_{h(\xi_\mathcal{L}, \cdot)}\big]. \tag{4.8}$$

In Equation (4.8), $Z_{h(\xi_\mathcal{L}, \cdot)}$ represents the point statistic of interest, with the tau-leap method time-steps determined according to rule $h(\xi_\mathcal{L}, \cdot)$, and $X$ represents the exact quantity of interest.

Sample values for estimator (4.8) can be generated using Algorithm 3.2 in Section

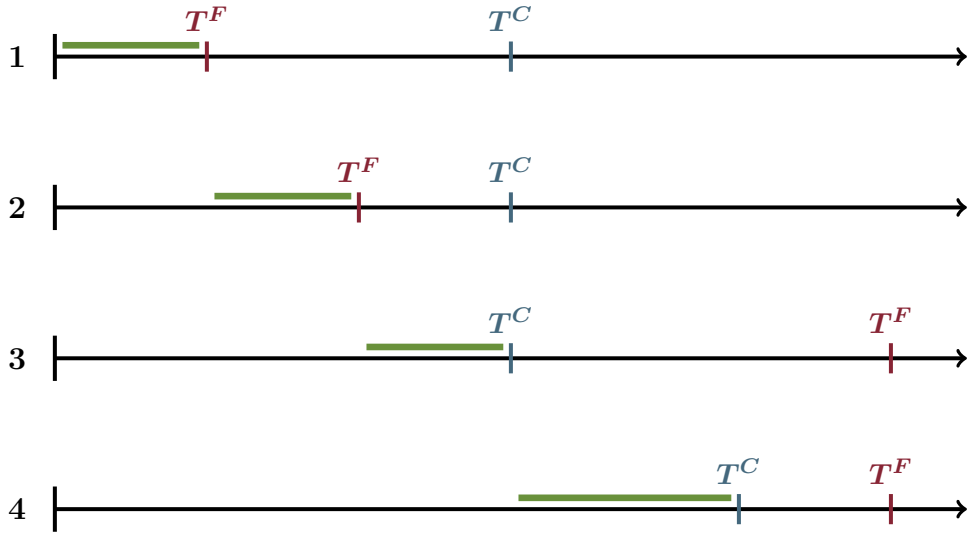

Figure 4.5: A diagrammatic representation of a possible first four steps of Algorithm 4.2, shown on a time axis. The vertical lines represent the discretisation of time: the NUT of the fine system is shown in red, and the corresponding NUT of the coarse path is in blue. The green bars show the time-steps that are used for each step. For the first step, the NUT of the fine path, $T^F$, is reached before the NUT of the coarse path, $T^C$. Consequently, both paths are advanced to the fine path NUT. We update the propensities of the fine path, and revise the NUT of the fine path. The second step starts by noting that the NUT of the fine path again occurs sooner than the NUT of the coarse path, and so a jump to reach the fine path NUT is implemented. The propensities and NUT of the fine path are updated. For step three, $T^F$ is larger than $T^C$, and so a jump is taken to the coarse path NUT: a new set of propensities and NUT are calculated for the coarse path. The fourth step progresses the system to time $T^C$ (as, again, $T^C < T^F$) and the appropriate updates are performed.

3.2.5. We simply insert the appropriate value of $\tau$ at lines `1` and `10` of Algorithm 3.2.

We have now described methods for determining the base estimate, $\mathcal{Q}_0$, the correction estimates, $\mathcal{Q}_1, \ldots, \mathcal{Q}_\mathcal{L}$, and the final estimate, $\mathcal{Q}_{\mathcal{L}+1}$. We now discuss a possible method for dynamically choosing $\tau$.

## 4.3 Adaptively choosing $\tau$

The method of adaptive $\tau$ choice we employ in this chapter is that of Cao, Gillespie and Petzold [64] (the 'CGP' method). This method has been chosen as it has undergone extensive testing and revision in the literature. However, our approach is modular, and in principal, any sensible method for choosing $\tau$ can be included in a multi-level algorithm. For example, any of the methods noted in Chapter 2 [66, 67] could be used to choose $\tau$.

The CGP method is predicated on the 'leap condition', that is, over a time-step $\tau$, the propensity functions within a system should remain 'approximately constant'. This can be taken to mean that the change in each propensity function should be small in relation to its magnitude. We start by outlining the approach presented in the original tau-leap method [45], and its immediate improvements [65]. Then, we set out the widely used CGP method [64].

### 4.3.1 Traditional $\tau$ selection schemes

If the state vector at a time $t$ is given by $\boldsymbol{Z}$, then denote the change in the propensity function for channel $R_j$ from time $t$ to $t+\tau$ as $\Delta_\tau a_j(\boldsymbol{Z})$. Earlier works [45, 65] enforced a 'leap condition' by attempting to ensure that $\Delta_\tau a_j(\boldsymbol{Z})$ does not change 'too much' over a time-step. It was suggested that choosing the maximal (and therefore, the least computationally demanding) value of $\tau$ such that

$$|\Delta_\tau a_j(\boldsymbol{Z})| \leq \xi a_0(\boldsymbol{Z}), \quad j = 1, \ldots, M, \tag{4.9}$$

where the error parameter $\xi \in (0,1)$, would be suitable. Of course, the change in propensity value is a stochastic quantity (that depends on the random reactions that take place within the time-step), and so attempting to enforce condition (4.9) directly will be somewhat awkward. Moreover, the condition (4.9) is inadequate, as it does not treat all reaction channels equally. This difficulty is best demonstrated with an example.

**Example 4.1.** Suppose a system comprises two reaction channels, $R_1$ and $R_2$. Further, suppose that at the start of a time-step, the propensity of $R_1$ is $a_1 = 1$ and that of $R_2$ is $a_2 = 99$. Then, the total propensity is $a_0 = 100$, and if $\xi = 0.01$, then $\tau$ should be chosen so that the propensities of $R_1$ and $R_2$ do not change by more than 1. This means that, during a single time-step, the propensity of $R_1$ is allowed to change by up to 100% of its value, whilst the propensity of $R_2$ can change by a little over 1% of its value during that step. ∎

### 4.3.2 The CGP method

The CGP method instead suggests that $\tau$ should be chosen so that we attain the following bound for all reaction channels:

$$|\Delta_\tau a_j(\boldsymbol{Z})| \leq \xi a_j(\boldsymbol{Z}), \quad j = 1, \ldots, M. \tag{4.10}$$

As the left hand side is a stochastic quantity, condition (4.10) is enforced with what Cao et al. [64] term 'high probability'. As the propensities are continuous functions of the underlying population values, condition (4.10) can be enforced by controlling the change in population values.

In order to enforce the leap condition 'with high probability', the CGP approach is to insist that $\tau$ be chosen such that

$$\tau \leq \frac{\max\{\xi_\ell/g_i \cdot Z_i, c_i\}}{|\sum_j \nu_{ij} a_j(\boldsymbol{Z})|} \quad \text{and} \quad \tau \leq \frac{\max\{\xi_\ell/g_i \cdot Z_i, c_i\}^2}{\sum_j \nu_{ij}^2 a_j(\boldsymbol{Z})}, \tag{4.11}$$

for all *reactant* species $S_i$ (i.e. where $Z_i$, the population of $S_i$, is an argument of some propensity function). Here $\xi_\ell$ is the control parameter described previously, $g_i$ is a weight function that indicates the relative effect of changes in $Z_i$ on the leap condition, and $c_i$ is the minimum expected change allowed. This approach bounds the expected change, and expected variance of the change, in the propensities of each reaction channel. For a more rigorous, probabilistic approach for selecting $\tau$, see

Moraes et al. [66].

In the original CGP paper, the value of $c_i$ was set to unity to avoid very small timesteps being returned. The authors claim that this is reasonable [64]. It is, however, important to be aware that, if $c_i$ is not set to zero, then as $\xi \downarrow 0$, the dynamics of the DM are *not* recovered. We illustrate why with an example:

**Example 4.2.** Consider the reaction network that comprises the single reaction:

$$R_1 : A \xrightarrow{1} \emptyset.$$

Then, with $c_1 = 1$, Formula (4.11) chooses $\tau$ to be

$$\tau = \frac{\max\{\xi \cdot Z, 1\}}{Z}.$$

As $\xi \downarrow 0$, then $\tau \to 1/Z$. Therefore, $\tau \not\downarrow 0$, and for all times $t \in [0, T]$, we have that

$$\mathbb{E}[Z(t)] < \mathbb{E}[X(t)].$$

Thus, in the limit $\xi \downarrow 0$, the summary statistics remain biased. The knock-on effect is that if the CGP method is used within the adaptive multi-level method, the minimum bias of a level will be limited by the CGP method. ∎

If a biased multi-level estimate is calculated, then an arbitrarily low overall bias cannot necessarily be attained. If the unbiased multi-level method is used, Equation (4.4) confirms that a suitable, unbiased estimate can still be calculated.

Returning to the original CGP method, the authors stated that special care should be taken in its implementation to mitigate the risk of a negative population being realised. Therefore, certain reaction channels are labelled as 'critical' and afforded special treatment. We overlook this complication and instead take

$$\tau = \min_{i \in I_r} \left\{ \frac{\max\{\xi_\ell/g_i \cdot Z_i, c_i\}}{|\sum_j \nu_{ij} a_j(\bm{Z})|}, \frac{\max\{\xi_\ell/g_i \cdot Z_i, c_i\}^2}{\sum_j \nu_{ij}^2 a_j(\bm{Z})} \right\}, \qquad (4.12)$$

with the parameters as previously defined (typically, $c_i = 1$), and $I_r$ as the set of reactant species. Since we do not mark any reaction channels as 'critical', an alternative method for managing negative populations is set out in Section 4.6.

Having described a method for choosing $\tau$, we are now in a position to implement the adaptive multi-level method.

## 4.4 Numerical examples

We again return to the motivating examples of Section 4.1, and implement the adaptive multi-level method to demonstrate the efficiency of our approach. In each case, results generated using the Gillespie DM and a fixed time-step multi-level implementation are compared against results determined with the adaptive multi-level method.

**Case Study 5.** We consider reaction network (4.1) of Case Study 5. We seek an unbiased estimate of $\mathcal{Q} = \mathbb{E}[X_3(30)]$ with the adaptive multi-level method. For the purposes of illustration, we choose a variety of ensembles, $\bm{\xi}$, by hand; these parameters are used to implement our chosen method of adaptively selecting $\tau$ [64]. The multi-level method is run, in full, a total of 100 times for each choice of $\bm{\xi}$. For each choice of $\bm{\xi}$, the left side of Figure 4.6 indicates the mean total simulation time required by the algorithm; the black lines indicate the range occupied by the 10-th to 90-th percentiles.

Most impressively, for control parameter ensemble $\bm{\xi} = [0.18, 0.06, 0.02]$, we achieve an estimate of $\mathbb{E}[X_3(30)]$ within an average of 72.0 seconds. This means the adaptive multi-level method can be nearly 27 times faster than the DM method. In addition, it is 9.4 times faster than the most efficient results presented in Figure 4.2 where the fixed time-step multi-level method was employed.

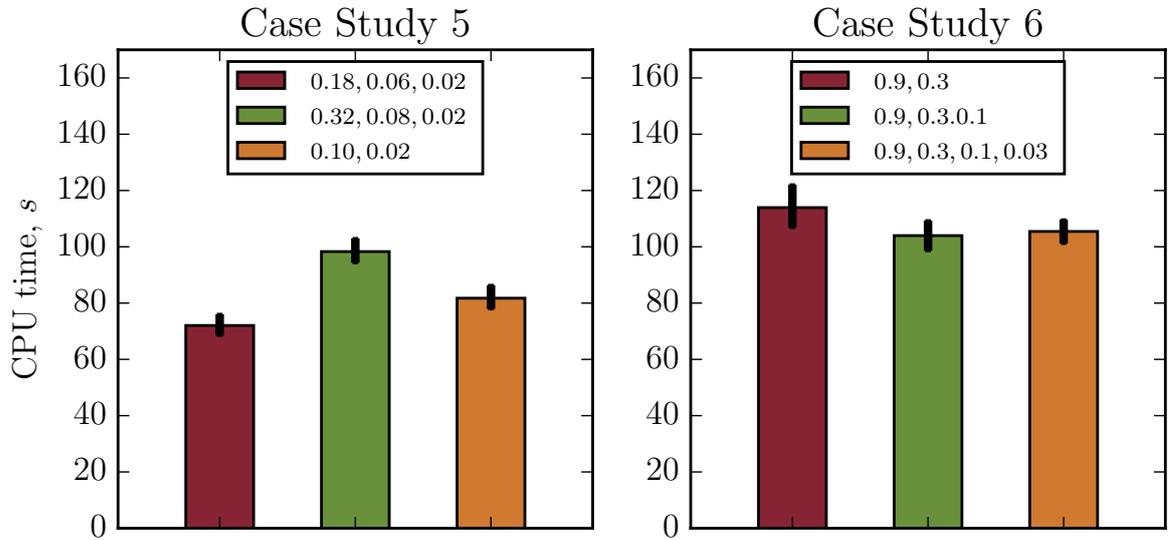

Figure 4.6: The simulation time for a range of system configurations for the adaptive time-step multi-level method used to estimate $\mathbb{E}[X_3(30)]$ for System (4.1) (on the left), and to estimate $\mathbb{E}[X_3(100)]$ for System (4.2) (on the right). The black bars indicate the ranges occupied by the 10-th to 90-th percentiles of the data.

**Case Study 6.** The second motivating example of interest is reaction network (4.2). In order to demonstrate the efficacy of our adaptive multi-level method, we seek an unbiased estimate of $\mathcal{Q} = \mathbb{E}[X_3(100)]$. A variety of suitable choices of $\boldsymbol{\xi}$ are chosen by hand, and the adaptive multi-level method is then run, in full, a total of 100 times for each choice of $\boldsymbol{\xi}$. In the right side of Figure 4.6, we show the mean simulation times required by the algorithm; the black lines indicate the range occupied by the 10-th to 90-th percentiles.

In particular, for the ensemble $\boldsymbol{\xi} = [1, 0.2, 0.04]$, the estimation of $\mathbb{E}[X_3(100)]$ is completed within 104.0 seconds, giving an estimated value of $1535.3 \pm 2.5$. Compared with the adaptive multi-level method, the Gillespie DM took 137 times longer to estimate this quantity. In addition, our calculation was completed 2.7 times quicker that the most efficient configuration of the fixed time-step multi-level approach we found.

## 4.5 Automatically choosing a parameter ensemble

From the previous section, it is clear that the adaptive multi-level method can speed up the Monte Carlo estimation of summary statistics. The adaptive method relies on an ensemble of control parameters, $\boldsymbol{\xi}$, to achieve its speed-ups. In this section, we describe a structured procedure for selecting a suitable ensemble of such parameters. Our aim is to provide immediate insights into the optimal choice of $\xi_0 > \xi_1 > \xi_2 > \ldots$, so that an *unbiased* summary statistic, $\mathcal{Q}$, can be estimated according to Equation (4.4). Our choice of $\xi_0 > \xi_1 > \xi_2 > \ldots$ should minimise the computational resources required. The method we develop is almost entirely automated, and is set out in steps below.

**Step one.** Suppose the adaptive scheme chooses $\tau$ according to a rule $h(\xi, \cdot)$, where $\xi$ is a control parameter. Let $\Xi$ be the range of values which $\xi$ might take on. Following on from Section 4.2, for the CGP method $\Xi$ is taken as $(0, 1)$ [64]. We simplify our problem by discretising $\Xi$ with a suitable mesh, which gives a finite list of possibilities for $\xi$. The mesh is labelled as $\widehat{\Xi}$. For example, with the CGP method outlined in Section 4.3, we might take $\widehat{\Xi} = \{0.01, 0.02, 0.03, \ldots, 0.99\}$. We will choose our control parameter ensemble out of the possibilities contained within $\widehat{\Xi}$.

**Step two.** A graphical representation of the chosen multi-level algorithm is now set out. Let $G(V, E)$ represent the graph of interest. The set of vertices, $V$, contains accuracy levels (i.e. choices of $\xi$) attainable with a multi-level configuration, as well as vertices that represent the start and the end of the algorithm. The set of edges, $E$, represents (sub-) estimates, $\mathcal{Q}_\ell$.

For example, suppose the adaptive multi-level method is implemented with $\boldsymbol{\xi}$ chosen to be $[0.1, 0.05]$. In this case, the multi-level method comprises three (sub)-estimators, $\mathcal{Q}_0$, $\mathcal{Q}_1$ and $\mathcal{Q}_2$: the base estimate ($\mathcal{Q}_0$), the tau-leap correction estimator ($\mathcal{Q}_1$), and the final correction estimator ($\mathcal{Q}_2$). The following graph is used to represent this implementation of the adaptive multi-level method:

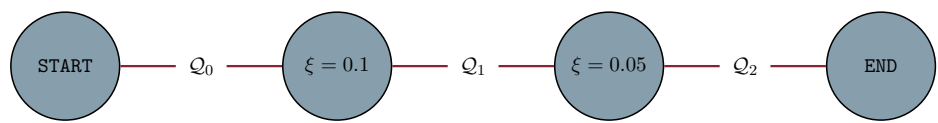

Note that, in accordance with Section 4.2, START formally corresponds with $\xi = \infty$ and END formally corresponds with $\xi = 0$. At the START, nothing is known about the summary statistic of interest, but once the END is reached, by estimating $\mathcal{Q}_0$, $\mathcal{Q}_1$ and $\mathcal{Q}_2$, we have an unbiased estimate of the summary statistic.

Of course, it is also possible to construct a multi-level implementation that skips a level out, thereby taking $\boldsymbol{\xi} = [0.1]$ or $\boldsymbol{\xi} = [0.05]$. As always, the DM can also be used to estimate $\mathcal{Q}$. These three further scenarios can be represented on a graph by including additional edges:

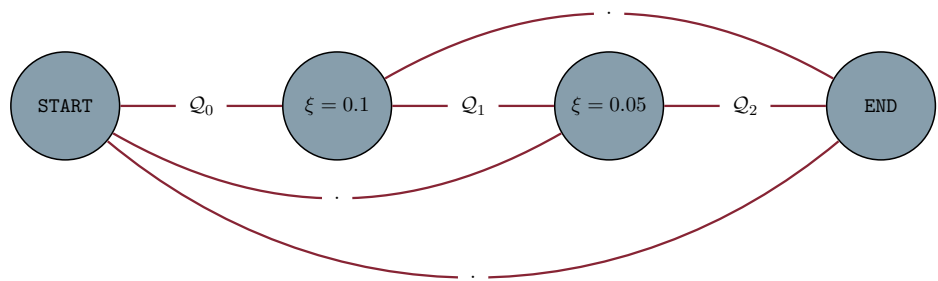

Therefore, every implementation of the multi-level method corresponds with a set of edges that trace a path from START to END.

**Step three.** The purpose of representing multi-level configurations as a graph is to help us choose suitable values of $\xi_0 > \xi_1 > \ldots$. Therefore, each edge, $e \in E$, will be assigned a length (or, alternatively, weight). The length is calculated with the following procedure. First, note that the computational cost of the multi-level method, $\mathcal{C}$, is given by Equation (3.13), which can be stated in the convenient form of

$$\mathcal{C} \propto \left( \sum_{\ell=0}^{\mathcal{L}+1} \sqrt{k_\ell} \right)^2, \tag{4.13}$$

where the relative cost of each level[3] is given by $k_\ell := \sigma_\ell^2 \cdot \mathcal{C}_\ell$. To minimise the total computational cost given by Equation (3.13), it is sufficient to minimise the square-root of the total computational cost,

$$\sqrt{\mathcal{C}} \propto \sum_{\ell=0}^{\mathcal{L}+1} \sqrt{k_\ell}.$$

We will assign values for $\xi_0, \xi_1, \ldots$ so that $\sqrt{\mathcal{C}}$ is minimised.

Define the length (or weight) associated with a given edge to equal $\sqrt{k} = \sigma_\ell \cdot \sqrt{\mathcal{C}_\ell}$, where $\sigma_\ell$ and $\mathcal{C}_\ell$ are determined through a number of trial simulations[4]. We call $\sqrt{k}$ the rooted relative cost (RRC) of the (sub)-estimator; this quantity represents the contribution of the estimator to $\sqrt{\mathcal{C}}$. For Case Study 5, $G$ can be drawn as[5]

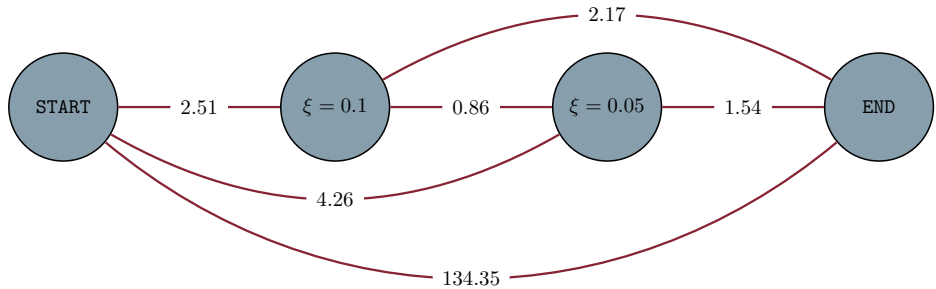

**Step four.** We now return to the complete problem. Our graph now contains a vertex for each $\xi \in \widehat{\Xi}$, where $\widehat{\Xi} = \{0.01, 0.02, 0.03, \ldots, 0.99\}$, and edges between each pair of vertices[6]. We search over this larger graph. We draw a path along the edges from START to END. As explained in step three, we want to draw the path with the minimal value of $\sum_{\ell=0}^{\mathcal{L}+1} \sqrt{k_\ell}$ (i.e. $\sqrt{\mathcal{C}}$). Note that the chosen path will comprise a total of $\mathcal{L} + 2$ edges.

---

[3] As mentioned in Chapter 3, the relative cost of a level indicates the contribution of the level to the total computational time given by Equation (3.13). Recall that $\mathcal{C}_\ell$ is the CPU time required to generate a single sample path on level $\ell$, whilst $\sigma_\ell^2$ indicates the sample variance.

[4] The number of simulations required will depend on the problem of interest.

[5] Therefore, in this diagram, we work with the mesh $\hat{\Xi} = \{0.1, 0.05\}$.

This problem is effectively a minimum length problem: we seek the minimum length path from START to END. A range of suitable algorithms can be implemented to find such a path. We use Dijkstra's Algorithm as it is relatively straightforward to implement [98], and note that performance can be improved through the use of methods such as the $A^*$ algorithm [99]. The Dijkstra algorithm is stated as Algorithm 4.3.

---

Algorithm 4.3: The Dijkstra algorithm finds the shortest path from a source to a target.

---

**Require:** graph, $G$, and source, $s$, and target, $t$
 1: **for** each vertex $v$ in $G \setminus s$ **do**
 2:     set dist$[v] \leftarrow \infty$ and previous$[v] \leftarrow$ `undefined`
 3: **end for**
 4: set dist$[s] \leftarrow 0$ and let $Q \leftarrow \{v \in G\}$, the set of vertices
 5: **while** $Q \neq \emptyset$ **do**
 6:     set $u \leftarrow \mathrm{argmin}_{w \in Q}\mathrm{dist}[w]$, and delete $u$ from $Q$
 7:     **for** each neighbour $v$ of $u$ **do**
 8:         $\alpha \leftarrow \mathrm{dist}[u] + \mathrm{length}[u, v]$
 9:         **if** $\alpha < \mathrm{dist}[v]$ **then**
10:             set dist$[v] \leftarrow \alpha$ and previous$[v] \leftarrow u$
11:         **end if**
12:     **end for**
13: **end while**
14:
15: let $S \leftarrow$ `empty sequence` and set $u \leftarrow t$
16: **while** prev$[u]$ is `defined` **do**
17:     insert $u$ at the start of $S$ and $u \leftarrow$ previous$[u]$
18: **end while**
19: insert $u$ at the start of $S$
20: **return** $S$

---

[6]In other words, the graph is complete.

We are now in a position to illustrate our method with an example.

### 4.5.1 Illustrating automatic selection

**Case Study 5.** We seek to choose $\xi_0 > \xi_1 > \cdots > \xi_\mathcal{L}$ to minimise the total CPU time required to estimate $\mathcal{Q} = \mathbb{E}[X_3(30)]$ (to a suitable level of statistical accuracy). The procedure that we follow with Case Study 5 is as follows:

- we use the CGP method (Section 4.3.2) to choose $\tau$, so we take $\Xi = (0, 1)$;

- the range $\Xi$ is discretised as $\widehat{\Xi} = \{0.01, 0.02, \ldots, 0.99\}$.

- the vertices START, $\xi = 0.99$, ..., $\xi = 0.01$, and END are placed in graph $G$.

- a number of simulations (we took $\mathcal{N} = 1000$ as this allows detailed investigations to be undertaken) are used to estimate the length of each edge in graph $G$.

- the Dijkstra algorithm finds the shortest path from START to END.

Our method suggests that $\boldsymbol{\xi} = [0.18, 0.03]$ is a suitable *control parameter ensemble*. In total, the Dijkstra algorithm chose the best option out of $2^{99}$ possible choices of $\boldsymbol{\xi}$.

We now conduct multi-level simulation with $\boldsymbol{\xi} = [0.18, 0.03]$ and compare our results with the trial-and-error approach adopted in Figure 4.6. The complete multi-level algorithm is run a total of 1000 times to study its computational performance. On average, the multi-level method takes 68.8 seconds of CPU time to run. This represents a factor 27.6 speed-up over the SSA. In Figure 4.7, a histogram shows the simulation times taken by different iterations of the multi-level method with our fixed choice of $\boldsymbol{\xi} = [0.18, 0.03]$. The 10-th and 90-th percentile CPU times correspond to 65.4 and 72.3 seconds, respectively.

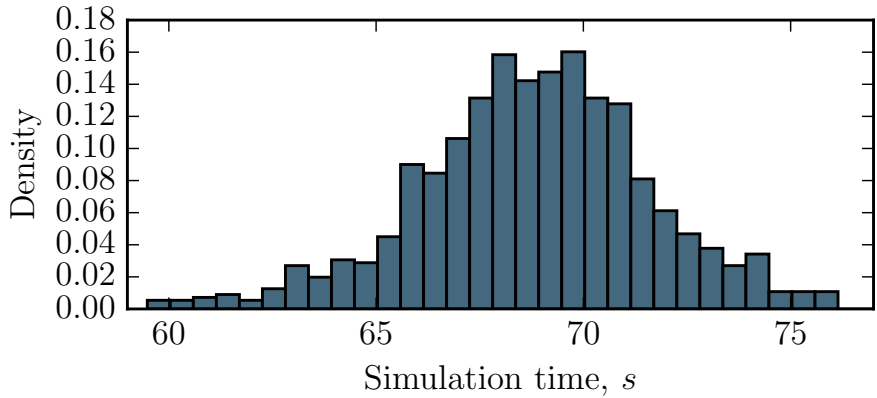

Figure 4.7: The simulation time required to estimate $\mathbb{E}[X_3(30)]$ of System (4.1) using an adaptive multi-level method with $\boldsymbol{\xi} = [0.18, 0.03]$ is shown as a histogram.

### 4.5.2 Towards a lightweight algorithm

Ultimately, the automatic selection procedure outlined in Section 4.5 has a computational cost. The computational burden of an in-depth and accurate search procedure could be significant, and this cost needs to be weighed against the benefits of using a particularly efficient choice of $\boldsymbol{\xi}$. In particular, we show that a less-exhaustive procedure for choosing $\boldsymbol{\xi}$ can be carried out quickly, and can deliver acceptable computational performance.

If the discretisation, $\widehat{\widehat{\Xi}}$, contains $\mathcal{D}$ points, then there are $\mathcal{D}$ potential base estimates (see Section 4.2.1), as well as $\binom{\mathcal{D}}{2}$ potential correction estimates (see Section 4.2.2), and $\mathcal{D}$ potential final estimates (see Section 4.2.3) to consider. For each of the aforementioned estimators the RRC, $\sqrt{k}$, is estimated with simulated data. This step dominates the computational cost of the optimisation procedure. For completeness, we note that the Dijkstra algorithm has a computational cost of $\mathcal{O}(|V|^2)$ (where $|V|$ counts the number of vertices; if a binary heap or similar is implemented, then the order of the method reduces): in our case, the Dijkstra algorithm will have order $\mathcal{O}(\mathcal{D}^2)$. For the case studies that we consider, the cost of running the actual Dijkstra algorithm itself is negligible.

Therefore, it is reasonable to choose a suitable, but coarsened discretisation of $\Xi$ to work with. We start by considering the following choice of $\widehat{\Xi}$ for Case Study 5:

$$\widehat{\Xi} = \{0.02, 0.06, 0.10, 0.14, \ldots, 0.98\}.$$

Then, we will start preparing the information needed to construct the graph defined in Section 4.5.1. We start by estimating the values of $\sqrt{k_0}$, the RRC of the base-level. Different values of $\xi \in \widehat{\Xi}$ are tested in increasing order, and values for $\sqrt{k_0}$ are calculated.

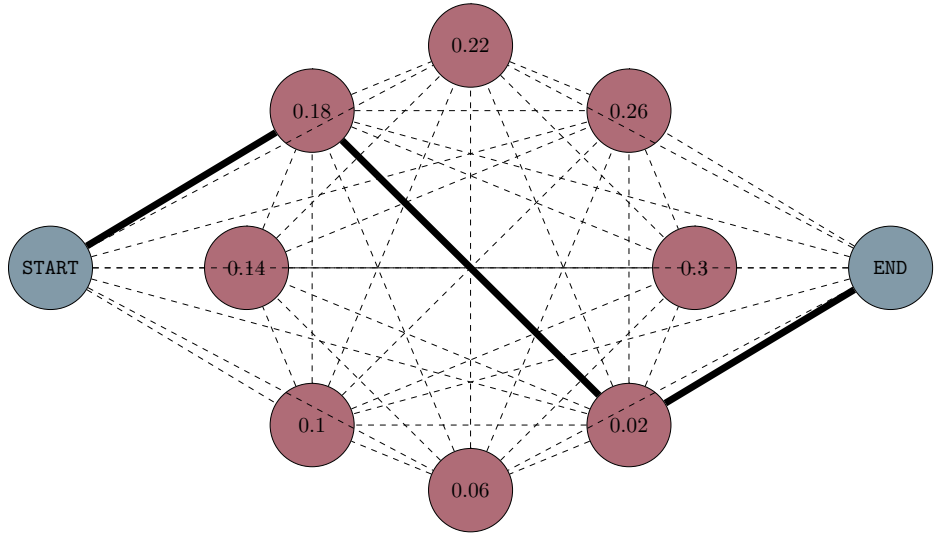

Figure 4.8: A graph that represents the adaptive multi-level method is shown. The dashed edges represent possible estimators, $\mathcal{Q}_{\cdot}$, that might be included in a multi-level method. Based on a number of preliminary simulations, a minimal cost multi-level algorithm is shown with solid lines.

In this case, $\sqrt{k_0}$ starts increasing when larger values of $\xi$ are used. Increases in $\xi$ are expected to reduce the simulation time, but as $\sqrt{k_0}$ is increasing, this means that the variance, $\sigma^2$, is increasing. An increase in $\sigma^2$ would, for example, be consistent with the tau-leap method becoming numerically unstable with such choices of $\xi$. For this case study, $\widehat{\Xi}$ can safely be restricted (details not shown) to the following eight choices:

$$\widehat{\Xi} = \{0.02, 0.06, 0.10, 0.14, 0.18, 0.22, 0.26, 0.30\}.$$

The values of $\sqrt{k_\ell}$, the RRCs of the tau-leap correction terms, are next to be estimated. There are a total of 28 quantities to be estimated at this step. Finally, the values of $\sqrt{k_{\mathcal{L}+1}}$, the RRCs of the final terms, are estimated.

**Lightweight searching.** The estimates for $\sqrt{k_0}$ are completed within 1.1 seconds. In each case, $\mathcal{N} = 100$ samples were used. The estimates for $\sqrt{k_\ell}$ took 10.1 seconds; again, $\mathcal{N} = 100$ samples were used. The final terms (i.e. $\sqrt{k_{\ell+1}}$) were estimated using $\mathcal{N} = 50$ samples, and this step required 27.8 seconds. This makes for a total computational cost of 39.0 seconds.

The Dijkstra algorithm is run to optimally choose $\boldsymbol{\xi}$ from the range of possibilities, $\Xi$. The vertices are given by $\widehat{\Xi}$, and the length between vertices, $\sqrt{k}$, is as calculated above. The CPU time taken by the Dijkstra algorithm is trivial (0.17 seconds). An example of such a graph is shown in Figure 4.8. Based on the estimated values of the RRCs, in this case, the optimal control parameter ensemble is given by $\boldsymbol{\xi} = [0.18, 0.02]$.

The optimal choice of $\boldsymbol{\xi}$ will obviously depend on the estimated values of the RRCs, $\sqrt{k_\ell}$. Relatively few samples have been used to estimate each RRC, and so different optimisation profiles are to be expected every time the algorithm is run with a different random seed. In Figure 4.9, we show what happens when the lightweight searching method is run a total of 1000 times. From these 1000 optimisations, a total of 17 different choices of $\boldsymbol{\xi}$ are made (this quantity is out of a total of $2^8 = 256$ theoretical possibilities for $\boldsymbol{\xi}$), but the two most prevalent choices, $\boldsymbol{\xi} = [0.18, 0.02]$ and $\boldsymbol{\xi} = [0.14, 0.02]$, together account for over half the runs of the search algorithm.

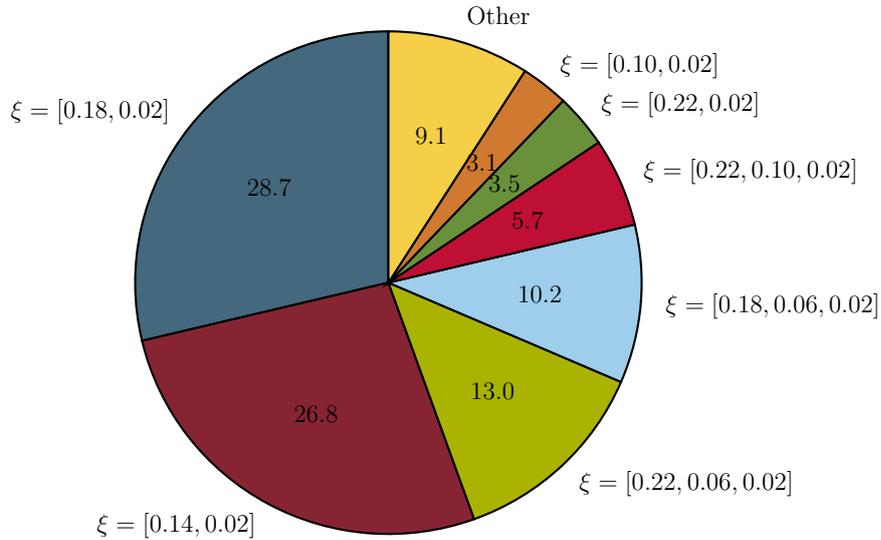

Figure 4.9: The choices of $\boldsymbol{\xi}$ proposed by the lightweight search method for use with Case Study 5 are shown. The search procedure was repeated 1000 times.

**Improving algorithmic efficiency.** As highlighted at the start of this section, we have sought to deliver immediate insights into the effects of different choices of $\boldsymbol{\xi}$. We have already mentioned that the $A^*$ algorithm can be used to improve computational performance; we now note that this is a consequence of the $A^*$ algorithm making use of a heuristic to efficiently search through the graph structure. It might also be possible to emulate the algorithmic performance with a suitable Gaussian Process, and then make use of Bayesian optimisation to ensure that the algorithm runs quickly [100]. A geometrically-spaced mesh over the set of possible control parameters could be tested for efficacy. Further directions for future improvement are detailed in Santner et al. [101].

We now demonstrate the computational performance of the multi-level method, when parametrised by the lightweight search method. The estimation of the RRC values and the running of the Dijkstra algorithm take an average of 39.2 seconds to complete. Then, having run the search method 1000 times, we proceed to use each search result to run the multi-level method. On average, 72.6 seconds are spent performing the multi-level simulation. This means that, from start to finish, the algorithm takes, on average, a total of 111.8 seconds to run. If we take into account the CPU time taken to choose $\boldsymbol{\xi}$, the multi-level method is 17 times faster than the DM.

The CPU time spent on the core simulation component of the multi-level method is detailed in Figure 4.10. Figure 4.10 also shows how the choice of $\boldsymbol{\xi}$ affects the multi-level simulation time. The most prevalent choice of $\boldsymbol{\xi}$, $\boldsymbol{\xi} = [0.18, 0.02]$ also leads to the best average (simulation-only) CPU time, of 70.0 seconds. The second most prevalent choice of $\boldsymbol{\xi}$, $\boldsymbol{\xi} = [0.14, 0.02]$, leads to the second-best (simulation-only) average time, that of 71.4 seconds, and the third most prevalent choice, $\boldsymbol{\xi} = [0.22, 0.06, 0.02]$, gives the fourth best (simulation-only) time at 72.9 seconds.

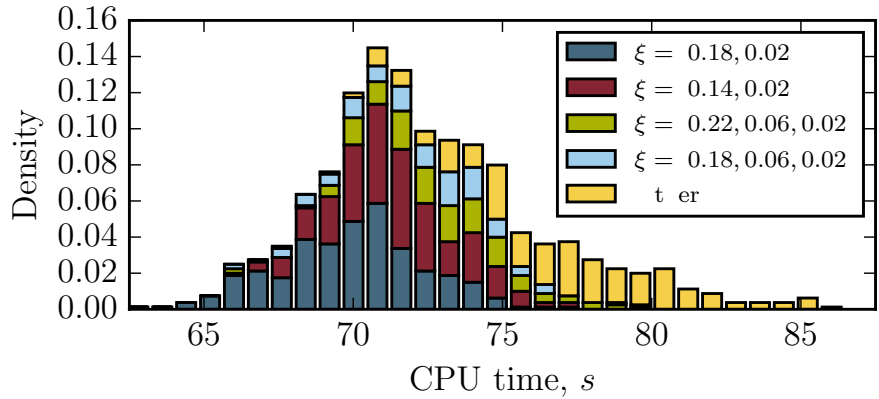

Figure 4.10: The simulation time required to estimate $\mathcal{Q}$ of Case Study 5 using an adaptive multi-level method is shown as a histogram. The choice of $\boldsymbol{\xi}$ has been made by the lightweight method; different colours show the effect of different choices of $\boldsymbol{\xi}$. The prevalence of each choice of $\boldsymbol{\xi}$ is indicated in Figure 4.9.

### 4.5.3 Further numerical results

The lightweight searching method is now applied to Case Study 6. We take $\widehat{\Xi}$ to be

$$\widehat{\Xi} = \{0.02, 0.10, 0.18, 0.26, 0.34, 0.42, 0.50, 0.58, 0.66, 0.74, 0.82, 0.90, 0.98\}.$$

We estimate $\mathbb{E}[X_3(100)]$, to within a 95% confidence interval of semi-length 2.5, a total of 1000 times. In each case, an ensemble, $\boldsymbol{\xi}$, is first chosen from the set $\widehat{\Xi}$, and the relevant multi-level algorithm is then run to completion.

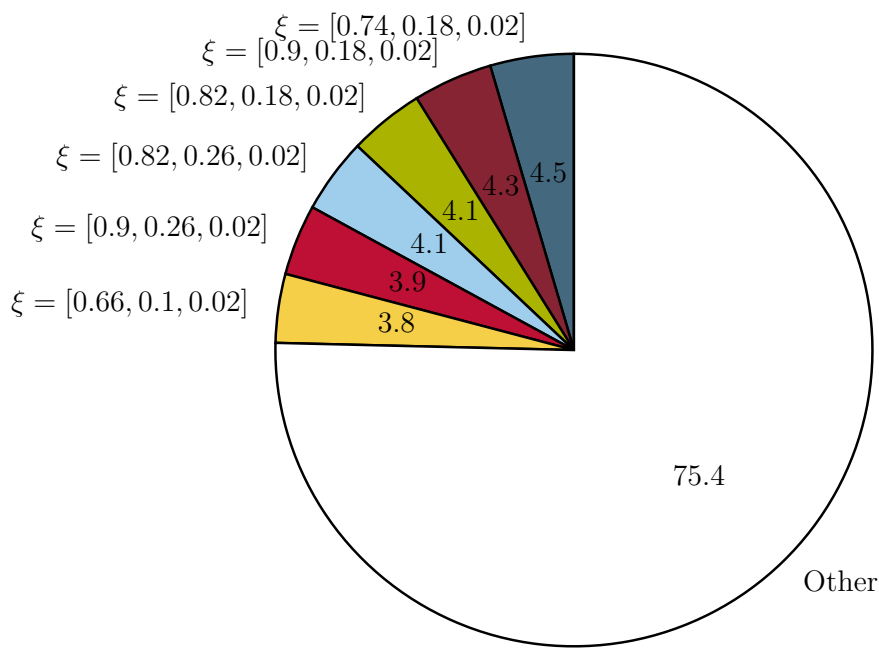

Figure 4.11: The choices of $\boldsymbol{\xi}$ proposed by the lightweight search method for use with Case Study 6 are shown. The search procedure was repeated 1000 times.

After running the lightweight searching algorithm a total of 1000 times, a total of 108 different choices of $\boldsymbol{\xi}$ are proposed. Therefore, 98.7% of the $2^{13} = 8192$ theoretically possible choices of $\boldsymbol{\xi}$ have been eliminated. In Figure 4.11, we show the distribution of the chosen ensemble, $\boldsymbol{\xi}$. For this case study, no single ensemble accounts for more than 5% of the total.

The estimation of the RRC values and the running of the Dijkstra algorithm take an average of 95.6 seconds to complete. Then, for each of the 1000 values of $\boldsymbol{\xi}$ chosen by the search method, we run the multi-level method. On average, 109.3 seconds of CPU time are spent on the core multi-level simulation algorithm. The CPU time spent on the core simulation component of the multi-level method is detailed in Figure 4.12.

Thus, by also taking into account the CPU time required to choose $\boldsymbol{\xi}$, we arrive at an overall average CPU time of 204.9 seconds. Therefore, on an all-inclusive basis, the multi-level method is 69 times faster than then DM.

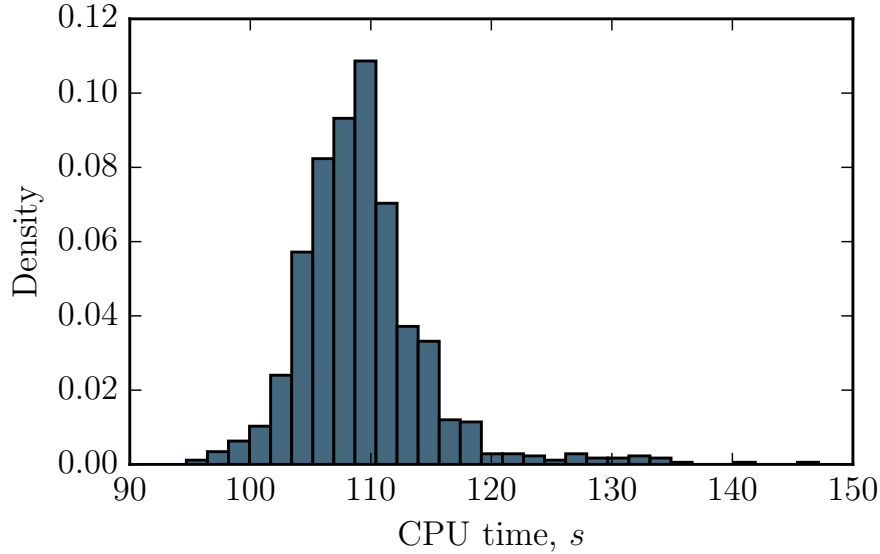

Figure 4.12: The simulation time required to estimate $\mathcal{Q}$ using an adaptive multi-level method is shown as a histogram. The choice of $\boldsymbol{\xi}$ has been made by the lightweight method.

## 4.6 Discussion

In this section, the results described in this chapter are examined in further detail, and a number of areas of particular interest are identified. This chapter is then drawn to a close with a number of conclusions.

### 4.6.1 Negative population values

In Section 2.4 it was mentioned that, occasionally, the tau-leap method generates non-physical, negative population values. Some authors have seen adaptive time-step methods as a means to avoid this problem. For example, algorithms have been described to minimise the probability of a population being driven negative [66, 67], and these tau-choosing methods can be implemented with the adaptive multi-level method.

A population is driven negative when more reactants than there are available are consumed during a time-step of the tau-leap method. Even with the most careful tau-choosing method, it is impossible to exclude completely the risk of a negative population being reached.

When a population reaches a negative value at one time-step, then one or more negative propensities are typically observed at the following time-step. Poisson variates cannot accept negative propensities as parameters: this simply does not accord with our understanding of the Poisson process[7]. The following time-step can therefore only be completed if the propensity is adjusted in some way. For the multi-level method, we can introduce boundary conditions to ensure that the tau-leap method sample paths never reach negative population values. We now explain how to implement these boundary conditions, so that the telescoping sums (4.3) and (4.4) can be used to estimate the required summary statistic, $\mathcal{Q}$.

As the following, commonly-overlooked example shows, boundary conditions need to be carefully considered before they are introduced.

**Example 4.3.** Consider the reaction network given by the following two reaction channels:

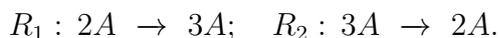

$$R_1 : 2A \rightarrow 3A; \quad R_2 : 3A \rightarrow 2A.$$

Let $X$ represent the population of $A$, and the initial condition be $X(0) = 10$. Clearly, $X = 1$ and $X = 0$ represent non-negative population states. However, the propensity of $R_2$, $a_2(X)$ is proportional to $X(X-1)(X-2)$, and therefore zero for $0 \leq X \leq 2$. Thus, taking into account the initial condition, the population states $X = 0$ and $X = 1$ should be impossible to obtain. The minimum achievable population is $X_{min} = 2$.

---

[7]Recall from Chapter 2 that a Poisson process counts the number of 'arrivals' that take place at a given, non-negative rate.

■

Therefore, for each tau-leap method sample path, $\boldsymbol{Z}_{h(\xi,\cdot)}$, just before the propensities are updated we enforce the boundary condition: `if` $X < X_{min}$, then `set` $X \leftarrow X_{min}$. This procedure must be consistently followed in Algorithms 2.3, 3.1, 3.2, 4.1, and 4.2 to ensure that the telescoping sums given by Equations (4.3) and (4.4) reduce to $\mathcal{Q}$.

Our example hints at the deeper problem of 'reachability'. Given two state vectors, $\boldsymbol{X}$ and $\boldsymbol{X}'$, it often is not clear whether, by starting from $\boldsymbol{X}$ and firing a number of reactions, that state $\boldsymbol{X}'$ can subsequently be reached [9]. This non-trivial problem can be studied in greater detail by representing the reaction network as a Petri net [102], and making use of existing algorithms for determining the reachability of different system states [103]. The concern is that the tau-leap method might not respect the 'reachability' rules of a given reaction network; however, this causes us no practical difficulties, and the telescoping sums given by Equations (4.3) and (4.4) indeed reduce to $\mathcal{Q}$.

### 4.6.2 Computational matters

We now compare the computational cost of the adaptive tau-leap coupling method with the fixed time-step coupling method. To draw such a comparison, we will need to consider two pairs of sample paths – one pair generated with an adaptive step method, and the second pair being generated with the fixed time-step method. Each pair comprises a coarse and a fine sample path, and, to compare the two competing coupling methods, we assume that both coarse sample paths (one from each pair) were generated with a similar number of time-steps. Furthermore, we assume that both fine sample paths (again, one from each pair) comprise a similar number of time-steps. With these assumptions, we see that the CPU time taken by the adaptive tau-leap coupling method to generate a pair of sample paths (Algorithm 4.1) is higher than the time taken by the comparable fixed time-step tau-leap method (Algorithm 3.1) to generate a comparable pair of sample paths.

The reason is as follows: where the fixed time-step method is used, the time-steps of the fine resolution paths can be neatly nested inside the coarse paths' time-steps. The total number of steps taken by the algorithm is equal to the number of fine resolution steps, which is the *maximum* number of time-steps taken within the pair of sample paths. However, with the adaptive time-stepping method described as Algorithm 4.1, there is no neat nesting, and time-steps almost surely do not co-incide. The overall number of steps is therefore equal to the *sum* of the number of coarse and fine time-steps.

However, for the reasons outlined in Section 4.1.3, the adaptive algorithm allows us to generate more *accurate* paths with fewer time-steps, and it is for this reason that the adaptive multi-level scheme is able to estimate summary statistics of stiff reaction networks effectively.

### 4.6.3 Conclusion

The multi-level method provides impressive time savings by combining a number of SSAs in an efficient manner to generate system statistics of interest. However, the original formulation of the algorithm required each sample path to be generated using a fixed value of $\tau$, and each of the levels to be nested, in the sense that $\tau_\ell = \tau_{\ell-1}/\mathcal{M}$ where $\mathcal{M} \in \{2, 3, \dots\}$.

In this chapter we have shown how to extend the multi-level method to remove these restrictions, and hence make it applicable to the study of systems where reaction activity varies substantially on the time-scale of interest. We have demonstrated the efficiency of our method using two example systems, and in each case used the CGP method to define the adaptive choice of $\tau$. However, our algorithm is general in the sense that it can accommodate almost any method for choosing $\tau$ adaptively.

Further, we have described an automated technique for implementing the adaptive multi-level method, so that user input is minimised and computational efficiency is improved.

In the following chapter, a new method for generating the sample paths on the *correction levels*, $\ell = 1, \ldots, \mathcal{L}$, is set out. This new approach leads to lower CPU times, and a more robust multi-level method. We also investigate the possibility of constructing an R-leap variant of the multi-level method.

# Chapter 5

# Robust Multi-level Methods

In this chapter, we revamp the fixed time-step multi-level method. Compared with the multi-level implementation presented in Chapter 3, our re-designed method will be more robust, reliable and computationally efficient. Broadly speaking, the aims of this chapter are realised under two headings. Firstly, we use a new variance reduction technique for generating tau-leap sample paths on the correction levels. Secondly, we investigate and explore the possibility of using R-leaping with the multi-level method.

## 5.1 Introduction

In Chapter 3, we described the multi-level method due to Giles [46], and Anderson and Higham [47]. We explained that a summary statistic of interest, $\mathcal{Q}$, is re-written as a sum of (sub)-estimators, $\mathcal{Q}_0, \ldots, \mathcal{Q}_{\mathcal{L} \ (+1)}$. Each sub-estimator, $\mathcal{Q}_\ell$ (for $\ell = 0, \ldots, \mathcal{L} \ (+1)^1$), is independently estimated with an SSA, and the estimates are summed up to arrive at an estimate of $\mathcal{Q}$. For this chapter, we focus our efforts in two directions: firstly, we will implement the Common process method (the 'CPM'), which allows us to estimate $\mathcal{Q}_1, \ldots, \mathcal{Q}_\mathcal{L}$ with a lower variance than that seen in Chapter 3;

---
[1]As before, '(+1)' indicates whether or not an unbiased estimator is calculated.

and, secondly, we show that we can also derive an R-leap variant of the multi-level method. Each aim is now introduced.

### 5.1.1 Introducing the Common process method

In Chapter 3, we explained that there are three types of level in the unbiased multi-level method: $\ell = 0$ refers to the *base level*, $\ell = 1, \ldots, \mathcal{L}$ refers to the *correction levels*, and $\ell = \mathcal{L} + 1$ refers to the *final correction level*. The multi-level method is efficient, because, for levels, $\ell = 1, \ldots, \mathcal{L} + 1$, we can use variance reduction techniques to reduce the CPU time required to estimate the required quantities. From Equation (3.13), we recall that the total simulation time is given by

$$\frac{1}{\varepsilon^2} \left\{ \sum_{\ell=0}^{\mathcal{L}(+1)} \sqrt{\mathcal{C}_\ell \cdot \sigma_\ell^2} \right\}^2.$$

For each level $\ell$, $\sigma_\ell^2$ is the sample variance of level $\ell$, and $\mathcal{C}_\ell$ is the CPU time required to generate each sample value. By reducing $\sigma_\ell^2$ (for $\ell = 1, \ldots, \mathcal{L}$ (+1)), the overall simulation time can potentially be reduced.

In Chapter 3, we used the variance reduction method described by Anderson and Higham [47] to implement the multi-level method. We will call that method the 'Split propensity method' (the 'SPM'). In Section 5.2, we will outline the 'Common process method' (the 'CPM'). We will argue that the CPM can, under appropriate conditions, provide better performance than the SPM. The CPM has, to our knowledge, yet to be used in a multi-level setting, but the CPM framework has previously been used to provide low-variance estimators for parameter sensitivity analysis [104].

### 5.1.2 Introducing the R-leap multi-level method

Thus far, we have only attempted to build a multi-level estimator by using the tau-leap method. The R-leap method is an alternative method for generating approximate sample paths of a chemical reaction network. As outlined in Chapter 2, at each step of the R-leap method, a fixed number of reactions are fired, with the precise combination of reactions and the time-step determined by random variates.

In this chapter, we develop an R-leap formulation of the multi-level algorithm, and we compare its numerical performance with a tau-leap formulation of the multi-level method.

### 5.1.3 Outline

This chapter is arranged as follows: in Section 5.2 we present the CPM variance reduction method for tau-leaping processes (the first aim of this chapter). The CPM is compared numerically with the SPM method in Section 5.3. A new multi-level framework that is built on the R-leap mechanism is set out and tested numerically in Section 5.4 (the second aim of this chapter). Our methods are discussed in detail in Section 5.5.

## 5.2 Common process method

In this section, we present the CPM variance reduction method. As mentioned, the CPM has been previously used to produce low variance estimates for parametric sensitivities of well-mixed, CME models [104].

As with the SPM, the CPM involves the *coupling* of two tau-leap processes that are required to provide sample paths for $\boldsymbol{Z}_{\tau_\ell}$ and $\boldsymbol{Z}_{\tau_{\ell-1}}$, so that we can efficiently estimate $\mathcal{Q}_\ell$, where $\ell = 1, \ldots, \mathcal{L}$. As before, we will refer to a pair of sample paths as comprising a 'coarse' (referring to $\boldsymbol{Z}_{\tau_{\ell-1}}$) and a 'fine' (referring to $\boldsymbol{Z}_{\tau_\ell}$) sample path.

Our variance reduction method will use the same random input, as far as possible, for both $\boldsymbol{Z}_{\tau_\ell}$ and $\boldsymbol{Z}_{\tau_{\ell-1}}$, in the manner we now set out.

### 5.2.1 Estimating $\mathcal{Q}_\ell$, where $\ell = 1, \ldots, \mathcal{L} + 1$

We use the RTCR provided by Equation (2.17) to represent the processes $\boldsymbol{Z}_{\tau_\ell}$ (a tau-leap sample path with time-step $\tau_\ell$) and $\boldsymbol{Z}_{\tau_{\ell-1}}$ (a tau-leap sample path with time-step $\tau_{\ell-1}$):

$$\boldsymbol{Z}_{\tau_\ell}(T) = \boldsymbol{Z}_{\tau_\ell}(0) + \sum_{k=0}^{K}\sum_{j=1}^{M} \mathcal{Y}_j\left(P_{\ell,k-1,j}, P_{\ell,k,j}\right) \cdot \boldsymbol{\nu}_j; \tag{5.1}$$

$$\boldsymbol{Z}_{\tau_{\ell-1}}(T) = \boldsymbol{Z}_{\tau_{\ell-1}}(0) + \sum_{k=0}^{K'}\sum_{j=1}^{M} \mathcal{Y}_j\left(P_{\ell-1,k-1,j}, P_{\ell-1,k,j}\right) \cdot \boldsymbol{\nu}_j, \tag{5.2}$$

where

$$P_{\ell,k,j} = \sum_{k'=0}^{k} a_j(\boldsymbol{Z}_{\tau_\ell}(\tau_\ell \cdot k')) \cdot \tau_\ell, \tag{5.3}$$

and the $\mathcal{Y}_j$ (for $j = 1, \ldots, M$) are unit-rate Poisson processes. The CPM method can produce a low variance estimate for $\mathcal{Q}_\ell$ by using the same set of $M$ Poisson processes, i.e. $\mathcal{Y}_1, \ldots, \mathcal{Y}_M$, for both sample paths $\boldsymbol{Z}_{\tau_\ell}$ and $\boldsymbol{Z}_{\tau_{\ell-1}}$.

This CPM scheme can be implemented by essentially running the tau-leap algorithm twice. We explain the procedure in detail in the following paragraphs, but the method can be summarised as follows:

- during the first phase, we simulate a sample path $\boldsymbol{Z}_{\tau_\ell}$ with time-step $\tau_\ell$. In doing so, the total number of times each Poisson process, $\mathcal{Y}_j$ (for $j = 1, \ldots, M$), has fired over each time-step is stored into memory (see Section 5.2.1.1);

- during the second phase, we simulate a sample path $\boldsymbol{Z}_{\tau_{\ell-1}}$, making use of the Poisson processes stored in memory and using interpolation (as phases one and two will use distinct time-steps) as required (see Section 5.2.1.2).

A sample path for process $\boldsymbol{Z}_\ell$ will be simulated according to the pseudo-code provided in Algorithm 5.1; a sample path for process $\boldsymbol{Z}_{\tau_{\ell-1}}$ will be simulated according to Algorithm 5.2.

Our algorithms differ substantially from the CPM methods developed to estimate parameter sensitivities. The previous parameter sensitivity analysis usage [104] of the CPM results in each reaction being individually fired. In this chapter, we devise a CPM method that works with the tau-leap method, which means that multiple reactions must be concurrently fired. We now set out the details of the simulation procedures.

#### 5.2.1.1 Phase one: simulating $\boldsymbol{Z}_{\tau_\ell}$

We first explain how to simulate a sample path of the tau-leap process $\boldsymbol{Z}_{\tau_\ell}$. We use the RTCR as a starting point. At time $t = 0$, the population is equal to the initial condition, $\boldsymbol{Z}(0)$. The populations at later times, $t = k \cdot \tau_\ell$ (for $k = 1, 2, \ldots$), are given by $\boldsymbol{Z}(k \cdot \tau_\ell)$. By setting $k = 1, 2, \ldots$, the value of $\boldsymbol{Z}(k \cdot \tau_\ell)$ can be recursively calculated according to Equation (2.18), that is

$$\boldsymbol{Z}_{\tau_\ell}(k \cdot \tau_\ell) = \boldsymbol{Z}_{\tau_\ell}((k-1) \cdot \tau_\ell) + \sum_{j=1}^{M} \mathcal{Y}_j\left(P_{\ell,k-1,j}, P_{\ell,k,j}\right) \cdot \boldsymbol{\nu}_j.$$

Recall that the quantity $\mathcal{Y}_j\left(P_{\ell,k-1,j}, P_{\ell,k,j}\right)$ represents the number of arrivals of the unit-rate Poisson process $\mathcal{Y}_j$ over the interval $(P_{\ell,k-1,j}, P_{\ell,k,j}]$. The value of $\mathcal{Y}_j\left(P_{\ell,k-1,j}, P_{\ell,k,j}\right)$ is Poisson distributed with parameter $P_{\ell,k,j} - P_{\ell,k-1,j}$. Note that, by Equation (2.14),

$$P_{\ell,k,j} - P_{\ell,k-1,j} = a_j(\boldsymbol{Z}_{\tau_\ell}(k \cdot \tau_\ell)) \cdot \tau_\ell.$$

Thus, the number of times reaction $R_j$ fires over the time span $((k-1) \cdot \tau_\ell, k \cdot \tau_\ell]$ is, as expected, given by the Poisson random variate, $\mathcal{P}(a_j(\boldsymbol{Z}_{\tau_\ell}(k \cdot \tau_\ell)) \cdot \tau_\ell)$.

The tau-leap method provided in Algorithm 5.1 is implemented. At each step of the

---

Algorithm 5.1: Phase one of the CPM-coupled tau-leap method. This simulates a single fine sample path, and records partial details of the Poisson processes associated with each reaction channel.

---

**Require:** initial conditions, $\mathbf{Z}(0)$, time-step, $\tau$, and terminal time, $T$.
1: set $\mathbf{Z} \leftarrow \mathbf{Z}(0)$ and set $t \leftarrow 0$
2: **while** $t < T$ **do**
3:     **for** each $R_j$ **do**
4:         calculate propensity value $a_j(\mathbf{Z})$
5:         generate $K_j \sim \mathcal{P}(a_j(\mathbf{Z}) \cdot \tau)$
6:         add the tuple $\langle a_j(\mathbf{Z}) \cdot \tau, K_j \rangle$ to the end of ordered list $\mathcal{F}_j$
7:     **end for**
8:     set $\mathbf{Z} \leftarrow \mathbf{Z} + \sum_{j=1}^{M} K_j \cdot \boldsymbol{\nu}_j$ and set $t \leftarrow t + \tau$
9: **end while**

---

algorithm ($k = 1, 2, \ldots$), we will store the tuple $\langle a_j(\mathbf{Z}_{\tau_\ell}(k \cdot \tau_\ell)) \cdot \tau, K_j \rangle$, where $K_j$ describes the number of times reaction $R_j$ occurs over the time-step. The tuple is stored in an ordered list $\mathcal{F}_j$. The ordered list $\mathcal{F}_j$ therefore stores details of the arrivals of the unit-rate Poisson process $\mathcal{Y}_j$. We give an example of a list, $\mathcal{F}_j$:

**Example 5.1.** Suppose that $\mathcal{F}_j = \{\langle 2.1, 3 \rangle, \langle 4.0, 7 \rangle, \langle 1.7, 3 \rangle, \ldots\}$. This means that we know that over the interval (of the unit-rate process) $(0.0, 2.1]$, there were three arrivals in the unit-rate Poisson process $\mathcal{Y}_j$. We also know that there were seven arrivals over the interval $(2.1, 6.1]$. Thus, over the interval $(0.0, 6.2]$, a total of ten arrivals were observed. ∎

Therefore, the ordered lists $\mathcal{F}_j$, for $j = 1, \ldots, M$, provide only *partial information* about the arrivals of the unit-rate Poisson process $\mathcal{Y}_j$. Each ordered list, $\mathcal{F}_j$, contains information about the total number of arrivals between specific positions, but does *not* contain information about the precise time at which each arrival is observed. Additional details will therefore be generated when they are needed for the coarse path. In the next section, we show how to use these ordered lists to generate a sample path of the process $\mathbf{Z}_{\tau_{\ell-1}}$ according to the CPM.

### 5.2.1.2 Phase two: simulating $Z_{\tau_{\ell-1}}$

The CPM method uses the same unit-rate Poisson process, $\mathcal{Y}_j$, to fire the $R_j$ reactions in each of $Z_\ell$ and $Z_{\tau_{\ell-1}}$. Therefore, in this section we describe how to generate a sample path of process $Z_{\tau_{\ell-1}}$ using the information stored in the ordered lists, $\mathcal{F}_j$ (for $j = 1, \ldots, M$). As summarised above, the ordered list, $\mathcal{F}_j$, contains only an *outline* of the Poisson process, $\mathcal{Y}_j$ (for $j = 1, \ldots, M$), and further details of these Poisson processes need to be filled in as required. The population of $Z_{\tau_{\ell-1}}$ is determined by setting $k = 1, 2, \ldots$, and recursively calculating

$$Z_{\tau_{\ell-1}}(k \cdot \tau_{\ell-1}) = Z_{\tau_{\ell-1}}((k-1) \cdot \tau_{\ell-1}) + \sum_{j=1}^{M} \mathcal{Y}_j \left(P_{\ell-1,k-1,j}, P_{\ell-1,k,j}\right) \cdot \boldsymbol{\nu}_j.$$

From the ordered list $\mathcal{F}_j$ we can directly read off the number of arrivals of the Poisson process $\mathcal{Y}_j$ at positions $P_{\ell,0,j}, P_{\ell,1,j}, \ldots, P_{\ell,k-1,j}$ (for $j = 1, \ldots, M$). At other positions, interpolation will be required. We use the following two facts to interpolate the Poisson process:

**Fact 3.** If there are $\mathcal{K}$ arrivals over the interval $(\alpha, \gamma)$ then, for $\beta$ between $\alpha$ and $\gamma$, the number of arrivals over the interval $(\alpha, \beta)$ is binomially distributed[2]

$$\mathcal{B}\left(\mathcal{K}, \frac{\beta - \alpha}{\gamma - \alpha}\right). \tag{5.4}$$

∎

**Fact 4.** If there are $\mathcal{K}_1$ arrivals over the interval $(\alpha, \gamma)$, and $\mathcal{K}_2$ arrivals over the interval $(\gamma, \beta)$, then interpolation using Fact 3 must be individually performed on the intervals $(\alpha, \gamma)$ and $(\gamma, \beta)$. Fact 3 cannot be directly applied to the entire interval $(\alpha, \beta)$ with $\mathcal{K}_1 + \mathcal{K}_2$ arrivals. ∎

---

[2] This follows as the $\mathcal{K}$ arrivals are uniformly distributed over $(\alpha, \gamma)$.

Facts 3 and 4 provide all the tools needed for Monte Carlo simulation. We first illustrate interpolation of a Poisson process with an example; pseudo-code is then provided in Algorithm 5.2.

**Example 5.2.** We return to our earlier example of an ordered list, $\mathcal{F}_j = \{\langle 2.1, 3\rangle, \langle 4.0, 7\rangle, \langle 1.7, 3\rangle, \ldots \}$. Suppose we wish to determine $\mathcal{Y}_j(0.0, 5.0]$, the number of arrivals in the unit-rate Poisson process by (normalised) time 5.0. We first apply Fact 4: there are three arrivals over the time span $(0.0, 2.1]$, and we need to determine how many further arrivals are observed over the time span $(2.1, 5.0]$. We apply Fact 3 as follows: over the interval $(2.1, 6.1]$ there are seven arrivals, and we want to know how many of these arrivals occur inside the sub-interval $(2.1, 5.0]$. A binomial variate, $\mathcal{B}(7, 2.9/4)$, is generated to interpolate the Poisson process. ∎

### 5.2.2 Estimating $\mathcal{Q}_{\mathcal{L}+1}$

The final estimator, $\mathcal{Q}_{\mathcal{L}+1}$, couples a tau-leap process $\boldsymbol{Z}_{\mathcal{L}}$ (with time-step $\tau_{\mathcal{L}}$), with an exact process, $\boldsymbol{X}$. The CPM implementation is outlined in the following paragraph:

To simulate a sample path of the exact process, $\boldsymbol{X}$, every reaction must be individually simulated. Therefore, we will need to determine every required arrival time of each Poisson process, $\mathcal{Y}_j$. The arrival times of each Poisson process are then saved into memory; the same set of Poisson processes is then used to generate a sample path for $\boldsymbol{Z}_{\tau_{\mathcal{L}}}$, the tau-leap process with time-step $\tau_{\mathcal{L}}$. A range of algorithms can be satisfactorily implemented to simulate the sample paths. We will adapt the Modified next reaction method (MNRM), as described by Anderson [63]. The CPM proceeds in two phases that can be outlined as follows:

- during the first phase, a sample path for process $\boldsymbol{X}$ is generated using the MNRM. The waiting times of each Poisson process $\mathcal{Y}_j$ (for $j = 1, \ldots, M$) are

---

Algorithm 5.2: Phase two of the CPM-coupled tau-leap method. This simulates a single coarse sample path from the Poisson processes stored during phase one.

---

**Require:** initial conditions, $\mathbf{Z}(0)$, time-step, $\tau$, ordered lists $\mathcal{F}_j$ (with $j = 1, \ldots, M$) and terminal time, $T$.
1: set $\mathbf{Z} \leftarrow \mathbf{Z}(0)$ and set $t \leftarrow 0$
2: **while** $t < T$ **do**
3:    **for** each $R_j$ **do**
4:       calculate propensity value $a_j(\mathbf{Z})$
5:       set $P \leftarrow 0$, $K \leftarrow 0$
6:       **while** $P < a_j(\mathbf{Z}) \cdot \tau$ **do**
7:          **if** $\mathcal{F}_j = \emptyset$ **then**
8:             break
9:          **else**
10:            read and then delete $(P', K')$ from the front of $\mathcal{F}_j$
11:            set $P \leftarrow P + P'$ and $K \leftarrow K + K'$
12:          **end if**
13:       **end while**
14:       **if** $P > a_j(\mathbf{Z}) \cdot \tau$ **then**
15:          generate $K' \sim \mathcal{B}(K', (P - a_j(\mathbf{Z}) \cdot \tau)/P')$
16:          set $K_j \leftarrow K - K'$
17:          add the tuple $\langle P - a_j(\mathbf{Z}) \cdot \tau, K' \rangle$ to the front of ordered list $\mathcal{F}_j$
18:       **else if** $P < a_j(\mathbf{Z}) \cdot \tau$ **then**
19:          generate $K' \sim \mathcal{P}(a_j(\mathbf{Z}) \cdot \tau - P)$
20:          set $K_j \leftarrow K + K'$
21:       **else**
22:          set $K_j \leftarrow K$
23:       **end if**
24:    **end for**
25:    set $\mathbf{Z} \leftarrow \mathbf{Z} + \sum_{j=1}^{M} K_j \cdot \boldsymbol{\nu}_j$ and $t \leftarrow t + \tau$
26: **end while**

---

recorded in an ordered list[3], $\mathcal{F}_j$ (Algorithm 5.3);

- during the second phase, a sample path for process $\mathbf{Z}$ is generated using the MNRM. The waiting times for Poisson process $\mathcal{Y}_j$ are determined using ordered list $\mathcal{F}_j$ (see Algorithm 5.4).

---

[3] For $\ell = \mathcal{L} + 1$, the ordered list $\mathcal{F}_j$ contains only numbers that correspond to the individual arrival times; but where $\ell = 1, \ldots, \mathcal{L}$, the list contains tuples.

## 5.3 Numerical experiments

In this section, we consider three representative case studies. In each case, we will test our two tau-leap based multi-level coupling methods: the SPM and the CPM. We will also compare the SPM and CPM implementations of the multi-level method with the regular DM.

Algorithmic performance will depend on the chosen algorithm parameters: $\mathcal{L}$, $\mathcal{M}$ and $\tau_0$. A simple search procedure can easily discard clearly inefficient algorithm parameter choices, leaving us with a range of possible values for $\mathcal{L}$, $\mathcal{M}$ and $\tau_0$ that should be investigated in greater detail. By using the algorithm parameters earmarked for further investigation, we compare the computational performance of the SPM and CPM variance reduction methods on the multi-level method. In each case study, we

---

Algorithm 5.3: Phase one of CPM-coupled DM and tau-leap method. This simulates a single, exact sample path, and records the Poisson processes associated with each reaction channel.

---

**Require:** initial conditions, $\boldsymbol{X}(0)$ and terminal time, $T$.
1: set $\boldsymbol{X} \leftarrow \boldsymbol{X}(0)$, and set $t \leftarrow 0$
2: for each $R_j$, set $A_j \leftarrow 0$, generate $T_j \leftarrow \text{Exp}(1)$, and store $T_j$ as the first element of $\mathcal{F}_j$
3: **loop**
4:     for each $R_j$, calculate propensity values $a_j(\boldsymbol{X})$ and calculate $\Delta_j$ as

$$\Delta_j = \frac{T_j - A_j}{a_j}$$

5:     set $\Delta \leftarrow \min_j \Delta_j$, and $k \leftarrow \text{argmin}_j \Delta_j$
6:     **if** $t + \Delta > T$ **then**
7:         **break**
8:     **end if**
9:     set $\boldsymbol{X}(t+\Delta) \leftarrow \boldsymbol{X}(t) + \boldsymbol{\nu}_j$, set $t \leftarrow t + \Delta$, and for each $R_j$, set $A_j \leftarrow A_j + a_j \cdot \Delta$
10:    generate $u \sim \text{Exp}(1)$, then set $T_k \leftarrow T_k + u$ and append $u$ to end of $\mathcal{F}_k$
11: **end loop**

---

Algorithm 5.4: Phase two of the CPM-coupled DM and tau-leap method. This simulates an approximate sample path from the Poisson processes stored during phase one.

---

**Require:** initial conditions, $\boldsymbol{Z}(0)$, terminal time, $T$, and ordered lists, $\mathcal{F}_j$ ($j = 1, \ldots, M$)

1: set $\boldsymbol{Z} \leftarrow \boldsymbol{Z}(0)$, $t \leftarrow 0$, $t^* \leftarrow \tau$
2: for each $R_j$, set $A_j \leftarrow 0$
3: for each $R_j$, set $T_j$ to be the first element of list $\mathcal{F}_j$, then delete the first element of $\mathcal{F}_j$
4: for each $R_j$, calculate propensity values $a_j(\boldsymbol{Z})$
5: **loop**
6:    for each $R_j$, calculate $\Delta_j$ as

$$\Delta_j = \frac{T_j - A_j}{a_j} \quad (5.5)$$

7:    set $\Delta \leftarrow \min_j \Delta_j$, and $k \leftarrow \operatorname{argmin}_j \Delta_j$
8:    **if** $t + \Delta > T$ **then**
9:       **break**
10:   **else if** $t + \Delta > t^*$ **then**
11:       set $t \leftarrow t^*$, and $t^* \leftarrow t^* + \tau$
12:       for each $R_j$, set $A_j \leftarrow A_j + a_j \cdot (t^* - t)$, then recalculate propensity $a_j$
13:   **else**
14:       set $\boldsymbol{Z}(t+\Delta) \leftarrow \boldsymbol{Z}(t)+\boldsymbol{\nu}_k$, set $t \leftarrow t+\Delta$, and for each $R_j$, set $A_j \leftarrow A_j+a_j \cdot \Delta$
15:       **if** $\mathcal{F}_k \neq \emptyset$ **then**
16:          let $u$ be the first element of $\mathcal{F}_k$: set $T_k \leftarrow T_k + u$, and then delete the first element of $\mathcal{F}_k$
17:       **else**
18:          generate $u \sim \operatorname{Exp}(1)$, then set $T_k \leftarrow T_k + u$
19:       **end if**
20:   **end if**
21: **end loop**

---

use the unbiased multi-level method to estimate an important summary statistic of the system.

### 5.3.1 A gene regulatory network

**Case Study 2.** We return once again to Case Study 2, a model of gene expression that has been used previously to test the multi-level method by Anderson and Higham

[47]. The reaction network is stated as:

$$R_1 : \emptyset \xrightarrow{25} M; \quad R_2 : M \xrightarrow{1000} M + P; \quad R_3 : P + P \xrightarrow{0.001} D;$$

$$R_4 : M \xrightarrow{0.1} \emptyset; \quad R_5 : P \xrightarrow{1} \emptyset.$$

We use the multi-level method to estimate $\mathbb{E}[X_3(1)]$, testing both the SPM and CPM implementations of the method. In Figure 5.1, we show the average CPU time taken by the respective multi-level method implementations. The values of $\mathcal{L}$, $\mathcal{M}$ and $\tau_0$ are varied; each average is computed over 100 test runs of the complete multi-level algorithm. Our fastest CPM configuration takes, on average, 120.9 seconds to estimate the dimer population (using $\mathcal{M} = 2$, $\tau_0 = 1/8$ and $\mathcal{L} = 7$), and our most efficient SPM implementation takes, on average, 156.5 seconds to estimate the same quantity (using $\mathcal{M} = 3$, $\tau_0 = 1/9$ and $\mathcal{L} = 5$). For System (2.10), our most efficient SPM-implementation of the multi-level method therefore takes approximately 29% longer to run than the comparable CPM implementation. In fact, for each test case that we considered, when compared with the CPM, the SPM implementation requires more CPU time to estimate the required summary statistic.

Whilst the CPM implementation can reduce the average duration of multi-level Monte Carlo simulation, another key benefit is that its run-time is far more predictable than the run-time of the SPM implementation. The black lines in Figure 5.1 indicate the range occupied by the 10-th to 90-th percentiles of the total CPU time. It is clear that the CPU times of the CPM implementation are very tightly clustered around the mean CPU time, whilst, for the SPM, the CPU times are of a higher variance. Moreover, the lower run-time of the CPM means that, even if a sub-optimal set of algorithm parameters is used, then the effect on the CPU time required is somewhat limited.

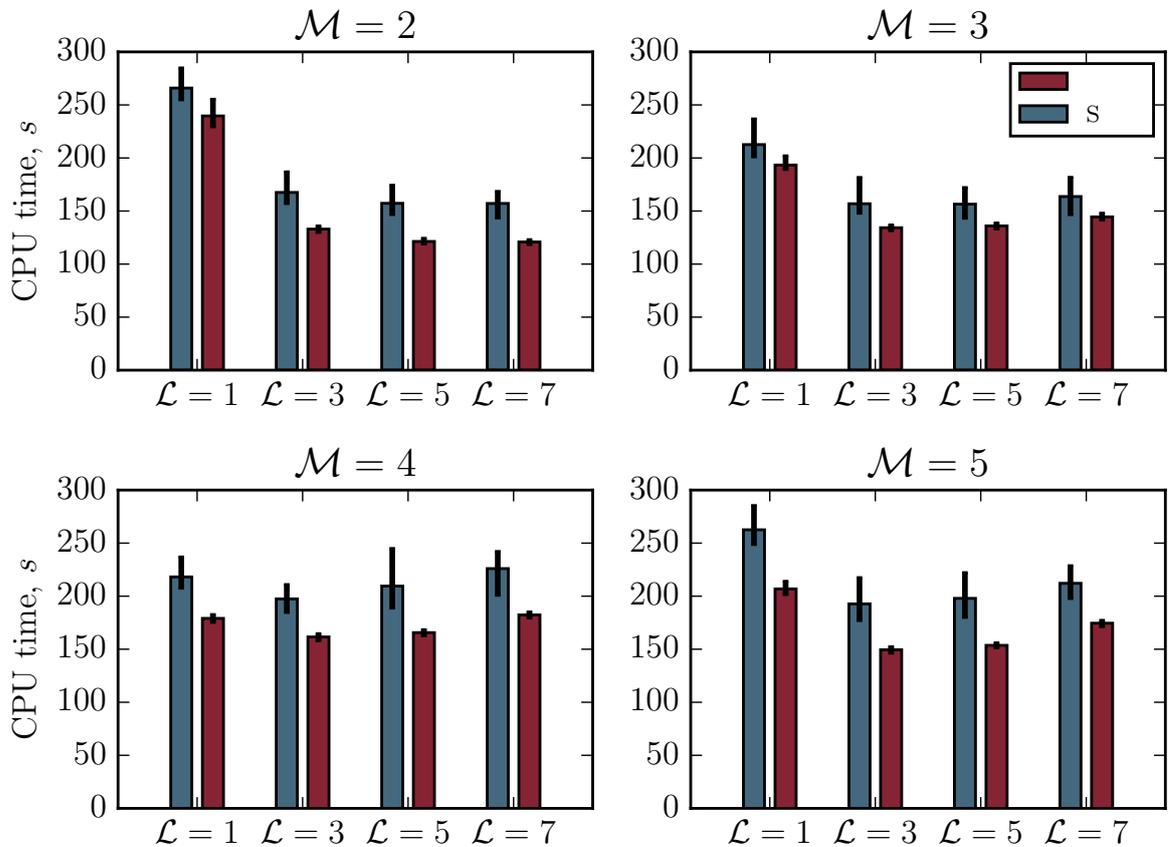

Figure 5.1: The average CPU time required by the multi-level method to estimate $\mathbb{E}[X_3(1)]$ for System (2.10). We vary $\mathcal{M}$ and $\mathcal{L}$; the estimator is unbiased. The black bars indicate the range occupied by the 10-th to 90-th percentiles of the data. The values of $\tau_0$ are: $\mathcal{M} = 2 \Rightarrow \tau_0 = 1/8$; $\mathcal{M} = 3 \Rightarrow \tau_0 = 1/9$; $\mathcal{M} = 4 \Rightarrow \tau_0 = 1/16$ and $\mathcal{M} = 5 \Rightarrow \tau_0 = 1/5$.

To compare the SPM with the CPM in more detail, we will need to concentrate on specific values of $\tau_0$, $\mathcal{M}$ and $\mathcal{L}$. We will not hand-pick algorithm parameters that give the CPM any potential advantage over the SPM. We take $\tau_0 = 1/9$, $\mathcal{M} = 3$ and $\mathcal{L} = 5$. With the aforementioned algorithm parameters, the CPM multi-level algorithm takes an average of 136.0 seconds to run (15.1 seconds slower than the fastest CPM configuration), whilst the SPM method runs in an average of 156.5 seconds (the fastest SPM configuration). In this case, the SPM takes approximately 15% longer to run than the CPM method.

In Figure 5.2, we show the empirical mean, sample variance, kurtosis and CPU time of each level estimator, $\mathcal{Q}_\ell$, when each of the SPM and CPM are used. The same base level ($\ell = 0$) is used for the SPM and CPM. The variances, $\sigma_\ell^2$, of the CPM are substantially lower than the SPM variances, but the CPU time taken to generate each sample, $\mathcal{C}_\ell$, is typically higher. For the higher levels, the use of the SPM results in a very high kurtosis (this effect has previously been described as representing a 'catastrophic decoupling' – see Section 3.9). A high kurtosis means that more of the sample variance can be attributed to infrequent but substantial deviations from the mean (as compared with frequent but modestly sized deviations from the mean).

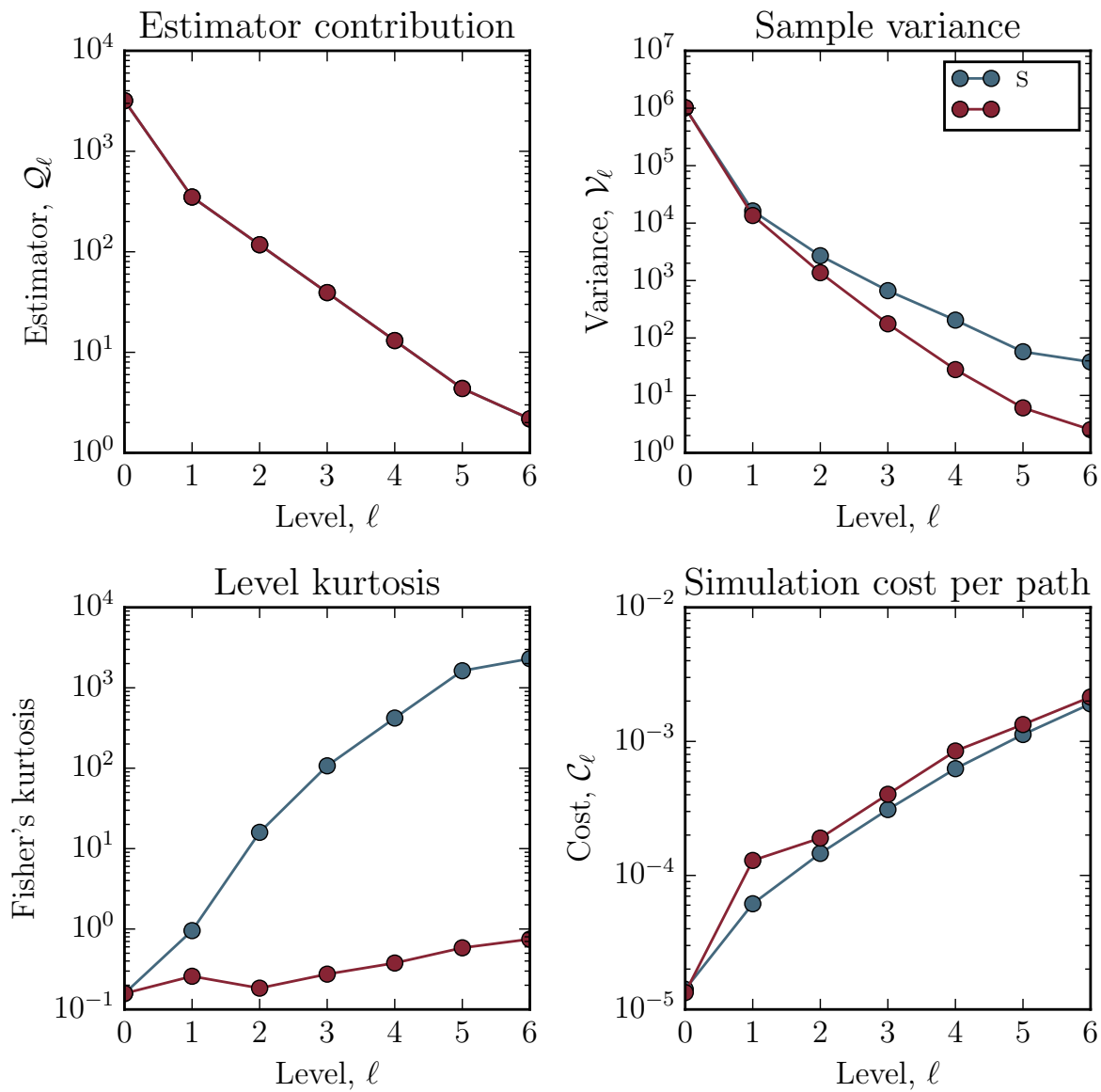

Figure 5.2: The expected mean, sample variance, normalised kurtosis and CPU time for each level $\ell$ in our multi-level simulator for System (2.10). We have taken $\tau_0 = 1/9$, $\mathcal{M} = 3$ and $\mathcal{L} = 5$, and used $10^5$ sample paths on each level. Note that the mean values for the SPM and CPM shown in the upper-left diagram overlap.

The effect of using the CPM implementation on the overall CPU time is outlined as follows. Subject to choosing $\mathcal{N}_\ell$ according to Equation (3.12), the total CPU time is given by Equation (3.13), i.e.

$$\frac{1}{\varepsilon^2} \left\{ \sum_{\ell=0}^{\mathcal{L}(+1)} \sqrt{\mathcal{C}_\ell \cdot \sigma_\ell^2} \right\}^2.$$

When the CPM is implemented, different values of $\sigma_\ell^2$ and $\mathcal{C}_\ell$ are inserted into Equation (3.13), and the result is that, with this case study, the total CPU time is reduced.

We now explore the variation in CPU times between successive iterations of the multi-level method. Estimated values of $\sigma_\ell^2$ and $\mathcal{C}_\ell$ are used to populate Equation (3.12), and variations in these estimates cause variations in the CPU time expended by the multi-level method. As highlighted in Chapter 3, there are two competing approaches for estimating $\sigma_\ell^2$ and $\mathcal{C}_\ell$:

- the 'one-step calibration' approach can be used. A small number of initial sample paths is generated for each level, and then Equation (3.12) is evaluated. The requisite number of sample paths are generated (see Section 3.5.1);

- the 'dynamic calibration' approach can also be used. In this case, the estimated values of $\sigma_\ell^2$ are repeatedly updated as the algorithm progresses, and increased numbers of sample paths are generated (see Section 3.5.2).

The CPU times shown in Figure 5.1 were generated with the 'dynamic calibration'

procedure. In Chapter 3, the 'one-step calibration' and 'dynamic calibration' procedures were compared, but the analysis presented in Chapter 3 focused on only the SPM. We now evaluate the effect of using the CPM on the one-step and dynamic calibration procedures.

**One-step calibration.** We first generate $\mathcal{N} = 10^2$ sample paths from which to estimate $\sigma_\ell^2$ and $\mathcal{C}_\ell$. In Figure 5.3 we compare the effect of using the one-step calibration procedure with the SPM and the CPM. For each of the SPM and CPM, the entire multi-level method is run, from start until finish, 1000 times. We show the absolute CPU time, and plot this against the resultant confidence interval semi-length (as before, we aim for a confidence interval of semi-length 1.0).

Our results indicate that the CPM implementation is far more likely to achieve the required estimator variance than the SPM implementation, and to do so with a broadly comparable CPU time. The mean CPU times for the SPM and CPM are 136.7 and 134.4 seconds, respectively. As highlighted in Chapter 3, the required confidence interval semi-length is not necessarily achieved; and it transpires that the estimated sample variances (based on $\mathcal{N} = 10^2$ initial paths) are not sufficiently accurate, and therefore too few sample paths are generated. The issue is far more pronounced for the SPM implementation as the required confidence interval is attained only 8% of the time. With the CPM, the proportion of runs that achieve the required confidence intervals rises substantially to 58%.

**Dynamic calibration.** The difficulties encountered with estimated sample variances, $\sigma_\ell^2$, can be mitigated by using the 'dynamic calibration' procedure: after the sample paths for each level are completed, the variance estimates can be updated as appropriate, and Equation (3.12) used to recalculate the number of sample paths required for each level estimator. Following the procedure of Section 3.5 a total of 1000 times (we used $\mathcal{N} = 10^2$ initial paths to start the dynamic algorithm), we plot our results in Figure 5.4. The average CPU times of the SPM and CPM are now 158.0 seconds and 134.7 seconds, respectively[4]. The average CPU time required by the CPM is broadly similar to the CPU time required without variance reallocation (see Figure 5.3), but the CPU time for the SPM method increases significantly.

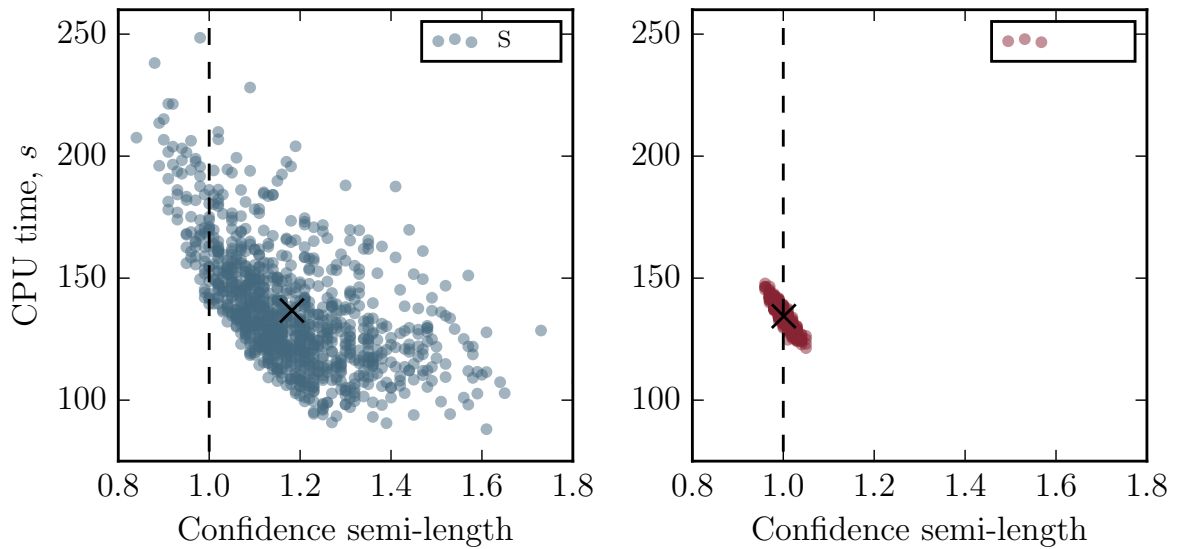

Figure 5.3: The full multi-level method is run 1000 times to estimate $\mathbb{E}[X_3(1)]$ for System (2.10) by following the 'one-step calibration' approach, and using both the SPM and CPM. The CPU times are plotted against the semi-length of the confidence intervals attained. The black crosses represent the mean values of the data. The target confidence interval size is shown with a dashed line. The CPM is clearly superior.

Therefore, in this case, the CPM remains superior to the SPM. A CPM implementation of the multi-level method results in a decreased CPU time, when compared with the SPM. The CPM is reliable, and can be used with sample variances estimated using a small number of preliminary sample paths.

---

[4]The averages differ slightly from those presented in Figure 5.1: the sample size is different.

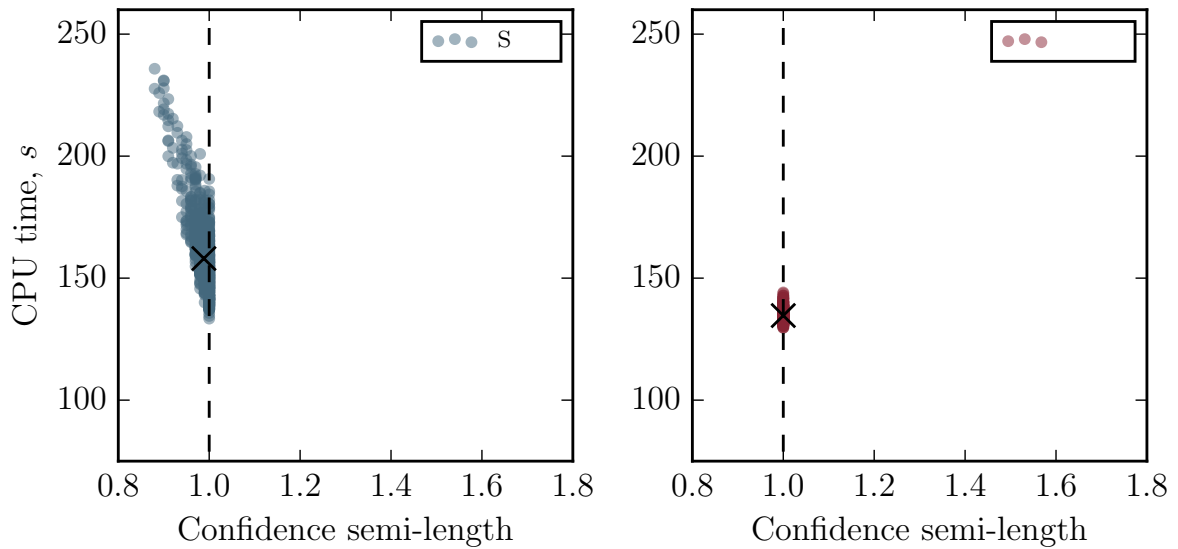

Figure 5.4: The full multi-level method is run 1000 times to estimate $\mathbb{E}[X_3(1)]$ for System (2.10) by following the 'dynamic calibration' approach, and using both the SPM and CPM. The CPU times are plotted against the semi-length of the confidence intervals attained. The black crosses represent the mean values of the data. The target confidence interval size is shown with a dashed line. This approach is clearly preferable to that shown in Figure 5.3. The CPM still outperforms the SPM.

### 5.3.2 A stochastic Lotka-Volterra model

**Case Study 7.** In this case study, we evaluate the performance of the multi-level method with a stochastic analogue of the Lotka-Volterra system [19]. The population dynamics of a predator, $A$, and its prey, $B$, are considered. The following reaction channels are defined:

$$R_1 : A \xrightarrow{10} \emptyset; \quad R_2 : A + B \xrightarrow{0.01} 2A; \quad R_3 : B \xrightarrow{10} 2B. \tag{5.6}$$

Initially, the population of $A$, $X_1$, and the population of $B$, $X_2$, are both set to equal 1200. We estimate the population levels of System (5.6) at time $T = 3$.

System (5.6) clearly exhibits oscillatory dynamics, with the amplitude of the oscillations being highly unstable [59]. To our knowledge, the multi-level method has yet to be successfully applied to a system that exhibits such dynamics, even over a short time-interval. The average predator population, given by $\mathbb{E}[X_1(T)]$, can be estimated with the DM. This calculation takes a little over 25 minutes (1523 seconds), and, with $1.94 \times 10^5$ sample paths, we estimate $\mathbb{E}[X_1(T)] = 783.4 \pm 1.0$.

We now use both the SPM- and CPM-controlled multi-level method to estimate $\mathbb{E}[X_1(T)]$. In Figure 5.5 we show the average CPU times for a range of choices of the refinement factor, $\mathcal{M}$. In each case, we show the average CPU time achieved with the most efficient choice of $\mathcal{L}$ and $\tau_0$ that we have found. The most efficient CPM algorithm we found requires 600.2 seconds of CPU time; this is 37% faster than the most efficient SPM algorithm, which requires 957.1 seconds. For each choice of $\mathcal{M}$, the most efficient SPM algorithm took substantially longer to run than the most efficient CPM method. Moreover, if we compare the most efficient sets of algorithm parameters for the CPM and SPM, with the DM, then the CPM method is 2.5 times faster than the DM, whereas the SPM by itself is 37% faster than the DM.

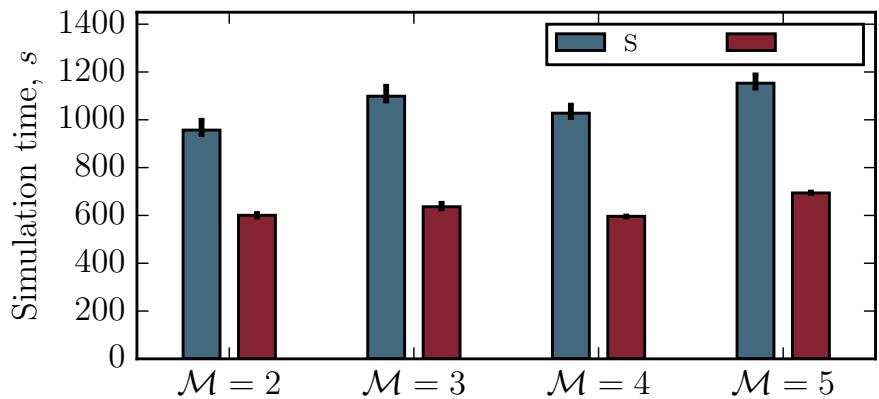

Figure 5.5: The average CPU time required by the multi-level method to estimate $\mathbb{E}[X_1(30)]$ for System (5.6). We vary $\mathcal{M}$ and $\mathcal{L}$; the estimator is unbiased. The black bars indicate the range occupied by the 10-th to 90-th percentiles of the data. The values of $\tau_0$ are: $\mathcal{M} = 2 \Rightarrow \tau_0 = T \cdot 2^{-12}$; $\mathcal{M} = 3 \Rightarrow \tau_0 = T \cdot 3^{-8}$; $\mathcal{M} = 4 \Rightarrow \tau_0 = T \cdot 2^{-12}$ and $\mathcal{M} = 5 \Rightarrow \tau_0 = T \cdot 2^{-5}$. In each case, taking $\mathcal{L} = 1$ is optimal.

### 5.3.3 A logistic growth model

**Case Study 1.** This example involves a return to Case Study 1, that is, a stochastic logistic growth model that comprises one species, and the following two reaction channels:

$$R_1 : A \xrightarrow{10} 2A; \quad R_2 : 2A \xrightarrow{0.01} A. \tag{5.7}$$

Initially, the population of $A$ is given as $X(0) = 50$. We will simulate System (3.8.1) until a terminal time $T = 3$, and estimate the mean population at that time, $\mathbb{E}[X(T)]$.

The DM estimates $\mathbb{E}[X(T)] = 999.6 \pm 1.0$ using 3800 sample paths in 16.2 seconds. We have run the multi-level algorithm with a wide variety of algorithm parameters (i.e. $\mathcal{L}$, $\mathcal{M}$, and $\tau_0$) on this system. Our efforts are briefly summarised in Figure 5.6, which is constructed as follows: for each choice of $\mathcal{M}$, we find the optimal $\mathcal{L}$ and $\tau_0$ for each of the SPM and CPM, and present our results.

We find that the multi-level method can be effectively implemented with the SPM, but the CPM approach is less efficient for this example. The SPM can estimate $\mathbb{E}[X(3)]$ to a 95% confidence interval of semi-length 1.0 within 2.1 seconds, which means that the DM requires 7.7 times as long to perform the calculation to the same level of accuracy. Our best result for the CPM is less efficient: our implementation requires 9.8 seconds of CPU time to estimate $\mathbb{E}[X(3)]$ to within the required statistical accuracy. Whilst the CPM substantially is slower than the SPM, the CPM is still faster than the DM. We provide a brief discussion of these results in Section 5.5.

## 5.4 R-leap-based multi-level Monte Carlo

In this section, we present a new implementation of the multi-level method. With a view to improving computational performance, we will use the R-leap method to construct an efficient multi-level algorithm. Once we have described our method, we demonstrate the performance of our algorithm with an example. We start by reviewing the R-leap method.

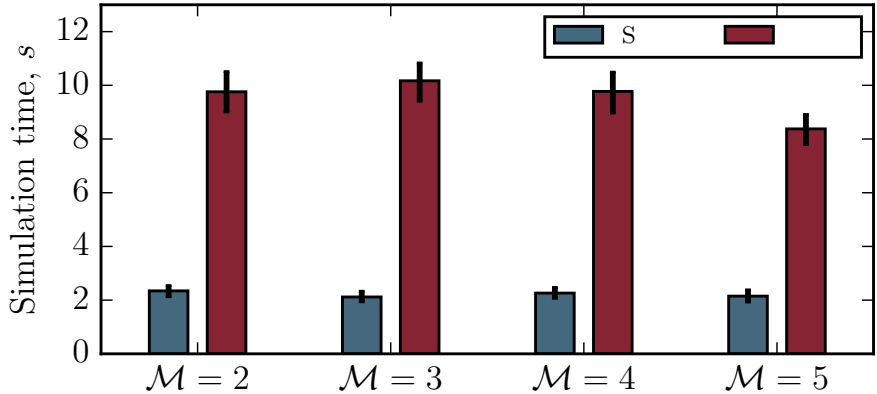

Figure 5.6: The average CPU time required by the multi-level method to estimate $\mathbb{E}[X(3)]$ for System (3.8.1). We vary $\mathcal{M}$ and $\mathcal{L}$; the estimator is unbiased. The black bars indicate the range occupied by the 10-th to 90-th percentiles of the data. The values of $\tau_0$ and $\mathcal{L}$ are as follows: for the SPM, $\mathcal{M} = 2 \Rightarrow \tau_0 = T \cdot 2^{-6}$ and $\mathcal{L} = 3$; $\mathcal{M} = 3 \Rightarrow \tau_0 = T \cdot 3^{-4}$ and $\mathcal{L} = 1$; $\mathcal{M} = 4 \Rightarrow \tau_0 = T \cdot 2^{-4}$ and $\mathcal{L} = 1$; and $\mathcal{M} = 5 \Rightarrow \tau_0 = T \cdot 5^{-3}$ and $\mathcal{L} = 1$. For the CPM, $\mathcal{M} = 2 \Rightarrow \tau_0 = T \cdot 2^{-9}$ and $\mathcal{L} = 1$; $\mathcal{M} = 3 \Rightarrow \tau_0 = T \cdot 3^{-6}$ and $\mathcal{L} = 1$; $\mathcal{M} = 4 \Rightarrow \tau_0 = T \cdot 2^{-9}$ and $\mathcal{L} = 1$; and $\mathcal{M} = 5 \Rightarrow \tau_0 = T \cdot 5^{-4}$ and $\mathcal{L} = 1$.

### 5.4.1 The R-leap method in detail

From Chapter 2, we recall that the R-leap method generates sample paths by simulating a user-specified number of reactions at each step of the algorithm. At each step of the R-leap method, $\Delta$, the time for the user-specified number of reactions (which we label as $\mathcal{K}$) to take place, is generated. Then, the precise combination of reactions that take place (e.g. three $R_1$ reactions, four $R_2$ reactions, and so on) is chosen.

The R-leap method is implemented as Algorithm 2.3. We now elaborate on the R-leap algorithm in more detail.

In the Gillespie DM (see Algorithm 2.1), the waiting time until the next reaction is given by an exponential variate with rate $a_0 = \sum_{j=1}^{M} a_j$. Consequently, given the fixed propensity values, the total waiting time for $\mathcal{K}$ reactions to fire in the R-leap method algorithm, $\Delta$, is given by the sum of $\mathcal{K}$ exponential variates[5]. This sum is Gamma distributed, i.e. $\Delta \sim \Gamma(\mathcal{K}, 1/a_0)$, with shape parameter $\mathcal{K}$ and scale parameter $1/a_0$.

Having calculated the distribution of the time period over which the next $\mathcal{K}$ reaction events occur, we must decide on the specific combination of reaction types that take place. We will be sampling *vectors* of random numbers from a multinomial distribution, the details of which we now set out.

**Multinomial distribution.** For $\mathcal{K}$ independent trials, each trial leading to one of $M$ possible outcomes, the multinomial distribution indicates how many trials result in each of the $M$ possible outcomes. We parametrise the distribution with fixed probabilities $p_1, \ldots, p_M$, as well as the previously-mentioned $\mathcal{K}$, the number of trials. A sample from a multinomial distribution is a $\mathcal{K}$-dimensional vector of random numbers, $(K_1, \ldots, K_M)$. Note that $\sum_{m=1}^{M} K_m = \mathcal{K}$.

One possible device for sampling from a multinomial distribution is the conditional binomial method [105] that proceeds as follows. For each of the $\mathcal{K}$ reaction events that we will simulate, the probability that it is a reaction of type $R_j$ is given by $a_j/a_0$. We start by considering $R_1$. A binomial random number, $\mathcal{B}(\mathcal{K}, a_1/a_0)$ is simulated, using $\mathcal{K}$ trials, each with probability $a_1/a_0$ of success. This provides the number of reactions of type $R_1$; we label this quantity as $K_1$. There are therefore $\mathcal{K} - K_1$ reactions that still need to be assigned to a type, and we know that these reactions are *not* $R_1$ reactions. Thus, we generate a binomial random number, $K_2 \sim \mathcal{B}(\mathcal{K} - K_1, a_2/(a_0 - a_1))$ to decide how many reactions of type $R_2$ have fired. In this case, there are $\mathcal{K} - K_1$ trials, each with conditional probability $a_2/(a_0 - a_1)$ of success. The number of reactions of type $R_3$ is given by $K_3 \sim \mathcal{B}(\mathcal{K} - K_1 - K_2, a_3/(a_0 - a_1 - a_2))$. This process repeated until all $\mathcal{K}$ reactions have been assigned a type.

---

[5]As the propensities are only updated every $\mathcal{K}$ reactions, $a_0$ is fixed for this many reactions.

Having expanded on the description of the R-leap method presented in Section 2.4, we are now in a position to outline the R-leap-based multi-level method.

## 5.4.2 Variance reduction with the R-leap method

Our description of the R-leap multi-level method will mimic the description of the tau-leap multi-level method described in Chapter 3. We write the summary statistic of interest, $\mathcal{Q}$, as the following telescoping sum[6]:

$$\mathcal{Q} = \sum_{\ell=0}^{\mathcal{L}+1} \mathcal{Q}_\ell.$$

The values of $\mathcal{Q}_\ell$, for $\ell = 0, \ldots, \mathcal{L}+1$, are now determined as follows:

- on the *base level*, where $\ell = 0$, we generate sample paths using the R-leap method, with $\mathcal{K}_0$ reactions simulated at each step. We use a large value of $\mathcal{K}_0$, so that we can quickly produce the sample paths required to estimate $\mathcal{Q}_0$. The $r$-th such sample paths is labelled as $\boldsymbol{Z}_0^{(r)}$, and we use the scalar $Z_0^{(r)}$ to represent the point statistic, $Z_0^{(r)} \coloneqq f\big(\boldsymbol{Z}_0^{(r)}\big)$. Accordingly, the estimator for level 0 is given by

$$\mathcal{Q}_0 \coloneqq \mathbb{E}\Big[Z_0^{(r)}\Big] \approx \frac{1}{\mathcal{N}_0} \sum_{r=1}^{\mathcal{N}_0} Z_0^{(r)};$$

- the *correction levels* ($\ell = 1, \ldots, \mathcal{L}$) involve the generation of pairs of sample paths, $(\boldsymbol{Z}_{\ell-1}, \boldsymbol{Z}_\ell)$. The 'fine' sample path $\boldsymbol{Z}_\ell$ is generated using the R-leap method, where $\mathcal{K}_\ell$ reactions are fired during each step. The 'coarse' sample path $\boldsymbol{Z}_{\ell-1}$ is also generated with the R-leap method, but with $\mathcal{K}_{\ell-1}$ reactions at each step. As with the tau-leap multi-level method, we choose a refinement factor, $\mathcal{M}$, so that $\mathcal{K}_\ell = \mathcal{K}_{\ell-1}/\mathcal{M}$. The estimator for level $\ell$ is

$$\mathcal{Q}_\ell := \mathbb{E}\left[Z_\ell^{(r)} - Z_{\ell-1}^{(r)}\right] \approx \frac{1}{\mathcal{N}_\ell} \sum_{r=1}^{\mathcal{N}_\ell} \left[Z_\ell^{(r)} - Z_{\ell-1}^{(r)}\right];$$

---

[6]Throughout the remainder of this chapter, we seek an unbiased estimate, $\widehat{\mathcal{Q}}$.

- finally, and optionally, the *final correction level*, $\mathcal{L}+1$, removes all remaining bias:

$$\mathcal{Q}_{\mathcal{L}+1} := \mathbb{E}\left[Z_{\mathcal{L}+1}^{(r)} - Z_{\mathcal{L}}^{(r)}\right] \approx \frac{1}{\mathcal{N}_{\mathcal{L}+1}} \sum_{r=1}^{\mathcal{N}_{\mathcal{L}+1}} \left[Z_{\mathcal{L}+1}^{(r)} - Z_{\mathcal{L}}^{(r)}\right].$$

The sample paths for $\mathcal{Q}_0$ are performed with the regular R-leap method; pseudo-code is provided in Algorithm 2.4. As before, we will use an algorithm that *couples* sample paths, so that $\mathcal{Q}_1, \ldots, \mathcal{Q}_{\mathcal{L}+1}$ can be estimated with a low variance. Note that, unlike the tau-leap method implementation, a special algorithm for the 'final estimator' (i.e. $\mathcal{Q}_{\mathcal{L}+1}$) is not required. If we set $\mathcal{K}_{\mathcal{L}+1} = 1$, then $\mathcal{Q}_{\mathcal{L}+1}$ can be estimated with exactly the same method as $\mathcal{Q}_1, \mathcal{Q}_2, \ldots, \mathcal{Q}_{\mathcal{L}}$. As before, the multi-level method will only reduce computational costs if we can efficiently estimate $\mathcal{Q}_1, \ldots, \mathcal{Q}_{\mathcal{L}+1}$: we now discuss techniques for doing so.

When we implement the R-leap method within the multi-level scheme, we use a coupling method that combines elements of the CPM and the SPM. We start by referring to Section 2.4, which states that at each step of the R-leap algorithm, two quantities are stochastically generated:

1. the time-period covered by that step;

2. the precise combination of reactions that fire during that time-period.

We will let sample paths $\mathbf{Z}_\ell$ (which we will call the fine path) and $\mathbf{Z}_{\ell-1}$ (which we will call the coarse path) advance by different time-periods at each step (point 1 above). A variance reduction technique is used to choose the time-periods that each sample path traverses. Then, to achieve maximal variance reduction, we will also ensure that, as far as possible, the same reactions fire in each sample path (point 2 above).

In order to simultaneously generate a pair of sample paths, $[\boldsymbol{Z}_\ell, \boldsymbol{Z}_{\ell-1}]$, at each step of the coupled simulation algorithm a total of $\mathcal{K}_\ell$ ($= \min\{\mathcal{K}_\ell, \mathcal{K}_{\ell-1}\}$) reaction events will take place in *each* sample path. In particular:

- a Gamma variate, $\Delta = \Gamma(\mathcal{K}_\ell, 1)$, is generated. The time-period spanned by this step in each of the coarse and fine sample paths is then determined by a rescaling argument. In distribution, $\Gamma(\mathcal{K}, \theta) \sim \theta \cdot \Gamma(\mathcal{K}, 1)$. Therefore, the time-period for fine path is given by $\Delta / \sum_{j=1}^{M} a_j^F$, and for the coarse system by $\Delta / \sum_{j=1}^{M} a_j^C$;

- the precise combination of reactions is chosen as follows. For each of the $\mathcal{K}_\ell$ reactions that take place in *each* sample path, the probability that it is a $R_j$ reaction is given by $a_j^C / \sum_{j'=1}^{M} a_{j'}^C$ (for the coarse path), and $a_j^F / \sum_{j'=1}^{M} a_{j'}^F$ (for the fine path). Our coupling method must therefore fire reactions with these probabilities. For each $R_j$ (for $j = 1, \ldots, M$), we define the probabilities $b_j^m$, where $m \in \{1, 2, 3\}$, as:

$$b_j^1 = \min\left\{ \frac{a_j^C}{\sum_{j'=1}^{M} a_{j'}^C}, \frac{a_j^F}{\sum_{j'=1}^{M} a_{j'}^F} \right\}; \tag{5.8}$$

$$b_j^2 = \frac{a_j^C}{\sum_{j'=1}^{M} a_{j'}^C} - b_j^1; \qquad b_j^3 = \frac{a_j^F}{\sum_{j'=1}^{M} a_{j'}^F} - b_j^1.$$

We interpret each probability, $b_j^m$, where $m \in \{1, 2, 3\}$ by noting[7]:

- $b_j^1$ represents the probability reaction $R_j$ takes place in both the coarse and the fine paths;

---
[7] Note that, for each $j$, at least one of $b_j^2$ and $b_j^3$ will be zero.

- $b_j^2$ represents the probability reaction $R_j$ takes place in only the coarse path;

- $b_j^3$ represents the probability reaction $R_j$ takes place in only the fine path;

As with the regular R-leap method, a conditional binomial method is used to choose the precise combination of reactions that take place. A total of $\mathcal{K}_\ell$ reaction events must take place in each of the coarse and the fine paths. We would, if possible, like the same reactions to fire in each sample path. Thus, we first consider the 'common' reactions – i.e. those that occur with probability $b_j^1$ (for $j = 1, \ldots, M$). Once the 'common' reactions have been determined, then we will determine the reactions specific to either the coarse or the fine path. Note that, as we are coupling sample paths, we expect that the propensities of the fine and coarse paths are similar for each $R_j$, i.e. $a_j^F \sim a_j^C$, so that $a_j^F / \sum_{j'=1}^M a_{j'}^F \sim a_j^C / \sum_{j'=1}^M a_{j'}^C$. Following Equation (5.8), we conclude that $b_j^1 \gg b_j^2, b_j^3$. Therefore, most events will be in the form of a 'common' reaction, and so will take place in both the coarse and the fine sample paths.

- the propensity values are updated as appropriate. The propensities associated with the fine path are updated at every step of the algorithm (i.e. after the required $\mathcal{K}_\ell$ reaction events have taken place in the fine path), whilst the propensities of the coarse path are updated every $\mathcal{M}$ steps of the algorithm (i.e. after $\mathcal{M} \cdot \mathcal{K}_\ell = \mathcal{K}_{\ell-1}$ reaction events have taken place in the coarse path).

As mentioned, once the common reactions have been completed, the remaining reactions events must be performed. The R-leap coupling method is presented as pseudo-code in Algorithm 5.5.

We now proceed to present numerical results in Section 5.4.3. The computational performance of our new method is compared with the tau-leap multi-level method and traditional simulation methods.

---

Algorithm 5.5: The coupled R-leap method. This simulates a pair of sample paths.

---

**Require:** initial conditions, $\mathbf{Z}(0)$, $\mathcal{K}$ $(=\mathcal{K}_\ell)$, $\mathcal{M}$ $(=\mathcal{K}_{\ell-1}/\mathcal{K}_\ell)$, and terminal time, $T$.
1: set $\mathbf{Z}^C \leftarrow \mathbf{Z}(0)$, $\mathbf{Z}^F \leftarrow \mathbf{Z}(0)$, $t^C \leftarrow 0$ and $t^F \leftarrow 0$
2: for each $R_j$, calculate propensity values $a_j^C(\mathbf{Z}^C)$ and $a_j^F(\mathbf{Z}^F)$
3: set flag $\leftarrow$ `false`, iter $\leftarrow 0$
4: **while** flag **is** `false` **do**
5:     set iter $\leftarrow$ iter $+ 1$
6:     generate $\Delta \sim \Gamma(\mathcal{K}, 1)$
7:     set $t^C \leftarrow t^C + \Delta/\sum_{j=1}^M a_j^C$ and $t^F \leftarrow t^F + \Delta/\sum_{j=1}^M a_j^F$
8:     **if** $\max\{t^C, t^F\} > T$ **then**
9:         set $\mathcal{K} \sim \mathcal{B}\big(\mathcal{K}-1, \min\big\{(T-t^C)/(\Delta/\sum_{j=1}^M a_j^C), (T-t^F)/(\Delta/\sum_{j=1}^M a_j^F) - 1\big)\big\}$
10:        set $\max\{t^C, t^F\} \leftarrow T$
11:        set flag $\leftarrow$ `true`
12:     **end if**
13:     set $K^* \leftarrow \mathcal{K}$
14:     for each $R_j$, calculate probabilities $b_j^1$, $b_j^2$ and $b_j^3$ according to Equations (5.8)
15:     **for** $j = 1, \ldots, M$ **do**
16:        generate $K_{j1} \sim \mathcal{B}\big(K^*, b_j^1/(1-\sum_{j'=0}^{j-1} b_{j'}^1)\big)$ and set $K^* \leftarrow K^* - K_j$
17:     **end for**
18:     set $K^{*2} \leftarrow K^*$ and $K^{*3} \leftarrow K^*$
19:     **for** $j = 1, \ldots, M$ **do**
20:        generate $K_{j2} \sim \mathcal{B}\big(K^{*2}, b_j^2/\sum_{j'=j}^M b_{j'}^2\big)$ and set $K^{*2} \leftarrow K^{*2} - K_{j2}$
21:        generate $K_{j3} \sim \mathcal{B}\big(K^{*3}, b_j^3/\sum_{j'=j}^M b_{j'}^3\big)$ and set $K^{*3} \leftarrow K^{*3} - K_{j3}$
22:     **end for**
23:     set $\mathbf{Z}^C \leftarrow \mathbf{Z}^C + \sum_{j=1}^M (K_{j1} + K_{j2}) \cdot \boldsymbol{\nu}_j$
24:     set $\mathbf{Z}^F \leftarrow \mathbf{Z}^F + \sum_{j=1}^M (K_{j1} + K_{j3}) \cdot \boldsymbol{\nu}_j$
25:     for each $R_j$, calculate propensity values $a_j^F(\mathbf{Z}^F)$
26:     if $\mathcal{M}$ divides iter, for each $R_j$, calculate propensity values $a_j^C(\mathbf{Z}^C)$
27: **end while**
28: **if** $t^C < T$ **then** use Algorithm 2.4 to simulate $\mathbf{Z}^C$ until time $T$. **end if**.
29: **if** $t^F < T$ **then** use Algorithm 2.4 to simulate $\mathbf{Z}^F$ until time $T$. **end if**.

---

### 5.4.3 Numerical experimentation

**Case Study 1.** We return to Case Study 1, previously discussed in Section 5.3.3. The reaction channels are

$$R_1 : A \xrightarrow{10} 2A, \quad R_2 : 2A \xrightarrow{0.01} A,$$

with initial conditions as described in Section 5.3.3. As before, we estimate $\mathcal{Q} = \mathbb{E}[X(3)]$; for reference, we note that the DM estimates $\mathcal{Q}$ to be $999.6 \pm 1.0$ within 16.2 seconds. The R-leap multi-level method has been implemented with a range of choices of $\mathcal{M}$ (the refinement factor) and $\mathcal{L}$ (that controls the number of levels). We seek an unbiased estimate, which means that $\mathcal{K}_0 = \mathcal{M}^{\mathcal{L}+1}$, $\mathcal{K}_1 = \mathcal{M}^{\mathcal{L}}$, ..., $\mathcal{K}_{\mathcal{L}} = \mathcal{M}$, and $\mathcal{K}_{\mathcal{L}+1} = 1$. Our results are collated in Figure 5.7. The most efficient R-leap multi-level implementation estimates $\mathcal{Q}$ using 3.6 seconds of CPU time. The DM method therefore takes approximately 4.5 times longer than our R-leap multi-level method.

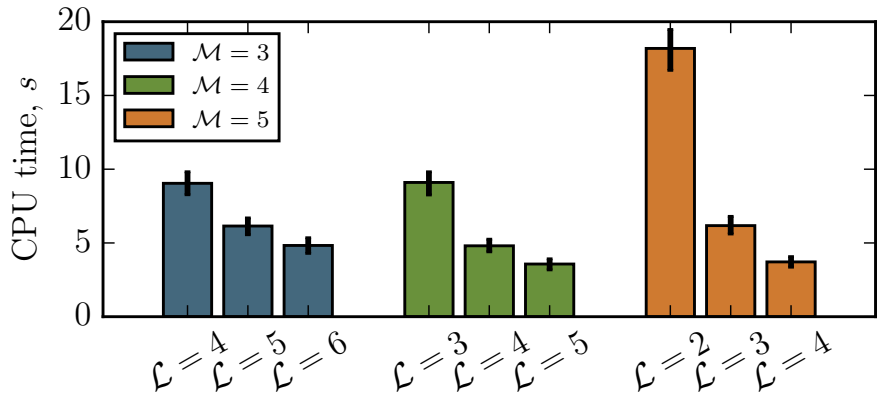

Figure 5.7: The average CPU time required by the R-leap multi-level method to estimate $\mathbb{E}[X(3)]$ for System (3.8.1). We vary $\mathcal{M}$ and $\mathcal{L}$; the estimator is unbiased. The black bars indicate the range occupied by the 10-th to 90-th percentiles of the data.

Our results demonstrate that the R-leap multi-level method is a feasible alternative to the tau-leap multi-level method. If we consider the numerical performance of Case Study 1 when the tau-leap multi-level method is used, it is clear that for this example, R-leap is superior to a CPM tau-leap method (which takes 9.8 seconds to estimate $\mathcal{Q}$), but inferior to a SPM implementation (which we recall takes 2.1 seconds to estimate $\mathcal{Q}$). The optimal method will depend on the particular reaction network and summary statistics of interest.

## 5.5 Discussion

At the outset of this chapter, we set out to develop the multi-level method into a robust, reliable and computationally efficient tool. Through the use of the SPM and the CPM approaches, the multi-level method has the potential to reduce the CPU time required to carry out Monte Carlo simulation dramatically. Whilst the CPM method does not always outperform the SPM method (although it often does), we have demonstrated its effectiveness and reliability. The R-leap multi-level method has also been demonstrated as a potential simulation technique.

The computational performance of the multi-level method may still depend on the particular reaction network and the chosen summary statistic, but our new methods provide additional tools for accelerating stochastic simulation.

### 5.5.1 Tau-leap multi-level simulation

In Chapter 3, the SPM method was used to implement the tau-leap multi-level method. Whilst the SPM approach has produced very impressive speed-ups and has been shown to be substantially more efficient than alternative simulation methods such as the DM, the CPM method can provide even better performance.

In Section 5.3.1, we demonstrated that, for the gene regulatory reaction network, the CPM is nearly 30% more efficient than the SPM. Furthermore, we noted that the kurtoses of levels $\ell = 1, \ldots, \mathcal{L} + 1$ are substantially higher when the SPM is used, when compared with the CPM. As explained in Section 5.3, the high kurtoses make it difficult to estimate the sample variances of the SPM method. There are two possible consequences:

- the sample variance, $\sigma_\ell^2$, is an under-estimate. The effect is that the required confidence interval semi-length is not faithfully attained;
- the sample variance $\sigma_\ell^2$, is an over-estimate. In this case, too many sample paths are generated, and the algorithm takes substantially longer to run.

A more robust approach is therefore provided by the CPM. Over a wide range of algorithm parameters, the CPM is able to outperform the SPM.

In Section 5.3.2, the CPM clearly outperformed the SPM. The Lotka-Volterra model, System (5.6), was a particularly challenging test case, and we follow Gillespie [59] in explaining why. Gillespie [59] argues that if an ODE modelling approach is followed, then, in the $X_1$-$X_2$ plane, the solution trajectories of System (5.6) are closed orbits. Gillespie [59] then says that the addition of microscopic fluctuations (due to using a stochastic, CME model) induces a 'drunkard's walk' over the continuum of deterministic orbits, thereby resulting in unstable behaviour. This feature makes it difficult to ensure that pairs of sample paths are tightly coupled. In particular, if a pair of sample paths differs slightly in their state vectors, then the difference in state vectors continues to increase with the SPM method. The CPM does not suffer the same defect, and, as such, it is more efficient than the DM.

In Section 5.3.3, a case study of a logistic growth model was presented. In this case, the SPM outperformed the CPM. This is due to the choice of estimator, $\mathcal{Q} = \mathbb{E}[X(T)]$. By the terminal time, $T = 3$, $X$ is fluctuating rapidly about a steady state. As such, the state vectors at earlier times, $X(t)$, tend to have a very limited effect on $\mathcal{Q}$, and so the CPM method is of limited benefit. If, however, one is concerned with the time-averaged population of System (2.10), given by

$$\mathcal{Q} = \mathbb{E}\left[\frac{1}{T}\int_0^T X(T)\,\mathrm{d}t\right],$$

then the CPM outperforms the SPM (details not shown).

## 5.5.2 Comparing tau-leap and R-leap multi-level methods

By treating the time traversed by each algorithm step of the R-leap method, and the particular combination of reactions that take place during that step, as two distinct and unrelated quantities that must be determined by Monte Carlo simulation, we have described a new variance reduction technique for multi-level simulation. The R-leap multi-level method was assessed by considering Case Study 1. We demonstrated that the R-leap approach performed slightly better than the CPM-driven, tau-leap multi-level method, but that it did not perform as well as the SPM-driven method. In summary, the R-leap multi-level method has the potential to provide good computational performance.

The crucial difference between the tau-leap and R-leap multi-level methods is seen by comparing Equations (3.16) and (5.8). The tau-leap method couples two distinct sample paths by considering the difference in *absolute propensity values* of corresponding reaction channels in the sample paths, whilst the R-leap method considers the difference in the corresponding *proportion of the total propensity* (or probability) attributable to each reaction channel. Variance reduction in the tau-leap and R-leap multi-level techniques arises in a different format, and the computational performance is therefore different.

Just as the SPM-based multi-level method can experience a catastrophic decoupling,

so can the R-leap method. As the following example shows, a decoupling might arise where the propensity functions of different reaction channels differ substantially.

**Example 5.3.**  We consider the following reaction network that comprises two reaction channels:

$$R_1 : \emptyset \xrightarrow{1} A; \quad R_2 : A \xrightarrow{100} A + B.$$

Initially, suppose that the populations of $A$ and $B$ particles are both zero. We attempt to use the R-leap coupling to generate a pair of sample paths, $[\boldsymbol{Z}_\ell, \boldsymbol{Z}_{\ell-1}]$. At the first step of Algorithm 5.5, reaction $R_1$ fires $\mathcal{K}_\ell = \min\{\mathcal{K}_\ell, \mathcal{K}_{\ell-1}\}$ times. The propensities are immediately updated for the sample path of $\boldsymbol{Z}_\ell$, but the propensities of sample path $\boldsymbol{Z}_{\ell-1}$ will be updated at a later step. Then, we note that

$$\frac{a_1^F}{a_1^F + a_2^F} = \frac{1}{1 + 100 \cdot \mathcal{K}_\ell} \quad \text{but} \quad \frac{a_1^C}{a_1^C + a_2^C} = 1.$$

At the second step of Algorithm 5.5, a further $\mathcal{K}_\ell$ reactions are simulated. At this step, in the coarse path, *all* the reactions are of type $R_1$, but in the fine path, fewer than one percent of reactions will be of type $R_1$. At the end of this step, it is very likely that the $A$ populations in the coarse and fine sample paths will have diverged. This will not be corrected; the $A$ populations will remain different for all time. Consequently, the $B$ populations will diverge. ∎

### 5.5.3 Outlook

The CPM provides a natural framework for implementing the multi-level method. It is able to mitigate some of the difficulties associated with the previously-used SPM implementation. Future work will include categorising reaction networks and summary statistics in order to derive criteria to decide whether the CPM or SPM should be used for that particular problem. Hybrid approaches that combine the SPM and CPM can also be implemented. The R-leap method has been successfully imple-mented, and future work will determine the problems it is most suited to handling. Ultimately, a refined multi-level method will dramatically reduce the computational burden of Monte Carlo simulation.

In the next chapter, we re-purpose the CPM and SPM methods. Through a bespoke variance reduction scheme, we provide a computationally-efficient method for calculating parametric sensitivities of spatially-inhomogeneous reaction networks.

# Chapter 6

# Parameter sensitivity analysis for spatially-extended models

The behaviour of a spatially-dependent biological model will necessarily depend on experimentally-derived input parameters. Therefore, to understand how the dynamics of a reaction-diffusion model are affected by changes in its input parameters, efficient methods for computing parametric sensitivities are required. In this chapter, we use efficient variance reduction techniques to develop Monte Carlo methods for estimating parametric sensitivities of spatially-extended reaction-diffusion models.

## 6.1 Introduction

In this chapter, we will focus on voxel-based or lattice-based stochastic models [80]. By exploiting the characteristic dynamics of spatially-extended reaction networks, we are able to adapt existing variance reduce schemes to robustly and efficiently estimate parametric sensitivities in a spatially-extended network.

As outlined in Section 2.5, we are interested in a volume, $\Omega$, that is discretised into a finite number of voxels. As before, each particle of the system is located within a voxel, and moves (diffuses) by transferring into a neighbouring voxel. Within each voxel, the particles are 'well-stirred' and can react with one another. As specified in Section 2.5, this framework is described by the RDME.

A variety of analytical tools have been developed for performing parameter sensitivity analysis on spatially-homogeneous systems. In this case, the particles can be thought of as being contained within a large, single voxel. The dynamics of such well-mixed systems are, as mentioned in Chapter 2, described by the CME. Finite difference methods for parameter sensitivity analysis have been developed by Rathinam et al. [104] and Anderson [106]. Exact, likelihood ratio or path-wise methods have been described by Plyasunov and Arkin [107], and Sheppard et al. [108]. Meanwhile, Liao et al. [40] have described a tensor-based method for calculating sensitivities for CME systems. Further advances in the methodology have been described by Morshed et al. [109], Thanh et al. [110], and Gupta et al. [111].

Methods of efficient exploration of parametric sensitivities in stochastic, spatially-extended networks are less well developed. Mathematically, CME and RDME models are both continuous-time, discrete-state Markov chains. This means that, in principle, any parameter sensitivity analysis method developed for well-mixed systems can be used for spatially-extended systems. However, this does not necessarily guarantee that the parameter sensitivity analysis method will be efficient. In this regard, this chapter makes three contributions. Firstly, we demonstrate that a finite difference scheme can be used to efficiently estimate parametric sensitivities for a spatially-extended model, under a range of circumstances. Secondly, we exploit particular features of the model of interest to describe a novel *Grouped sampling* method. Finally, we implement what we call the *Multi-choice* technique that dynamically combines different finite difference simulation schemes to efficiently estimate parametric sensitivities.

### 6.1.1 Outline

In Section 6.2 we describe finite difference methods for the parameter sensitivity analysis of spatially-inhomogeneous models. We compare and contrast simulation methods with a number of case studies in Section 6.3. Grouped sampling is implemented in Section 6.4, and the multi-choice method is described in Section 6.5. We conclude this chapter with Section 6.6.

## 6.2 Finite difference methods

In this section we describe how to carry out an efficient parameter sensitivity analysis on a RDME model. Throughout this section, we draw on the information contained within Section 2.5 of Chapter 2, where we described the RDME framework and discussed how one might generate sample paths of such models.

We start by considering the effect of a change in the value of an individual input parameter on suitable model summary statistics. As before, the summary statistic of interest is written as

$$\mathbb{E}\left[f\left(\boldsymbol{X}\right)\right], \tag{6.1}$$

where $\boldsymbol{X}$ represents a sample path of our RDME model and $f(\boldsymbol{X})$ is a suitable function of interest. We will estimate a partial derivative with respect to a change in the input parameter $A = \alpha$ as

$$\left.\frac{\partial \mathbb{E}\left[f\left(\boldsymbol{X}\right)\right]}{\partial A}\right|_{A=\alpha}. \tag{6.2}$$

This partial derivative can be estimated for different values of $\alpha$, and choices of parameter $A$. Unsurprisingly, analytic expressions for (6.2) will not, in general, be obtainable, and stochastic methods must be used to estimate the partial derivative.

We use a finite difference method to estimate the quantity described by (6.2). Suppose that Systems $X$ and $Y$ are identical, except that the value of parameter $A$ is perturbed. Pick $\varepsilon \ll 1$, and take $A := \alpha - \varepsilon/2$ in System $X$. For System $Y$, take $A := \alpha + \varepsilon/2$. Then [106],

$$\frac{\partial \mathbb{E}\left[f\left(\boldsymbol{X}\right)\right]}{\partial A} = \frac{\mathbb{E}\left[f\left(\boldsymbol{Y}\right)\right] - \mathbb{E}\left[f\left(\boldsymbol{X}\right)\right]}{\varepsilon} + \mathcal{O}(\varepsilon^2). \tag{6.3}$$

The centred finite difference approximation[1] is given by $\mathbb{E}\left[f\left(\boldsymbol{Y}\right) - f\left(\boldsymbol{X}\right)\right]/\varepsilon$. It might be tempting to generate sample paths to estimate $\mathbb{E}\left[f\left(\boldsymbol{Y}\right)\right]$ and $\mathbb{E}\left[f\left(\boldsymbol{X}\right)\right]$ independently: but this is usually very inefficient. As such, we explain how to estimate $\mathbb{E}\left[f\left(\boldsymbol{Y}\right) - f\left(\boldsymbol{X}\right)\right]$ with a minimal level of computational resources.

We simulate $\mathcal{N}$ pairs of sample paths, which we enumerate as

$$\left\{[\boldsymbol{X}, \boldsymbol{Y}]^{(r)}, r = 1, \ldots, \mathcal{N}\right\},$$

and then take $\widehat{\mathcal{Q}}$ as our estimate for $\mathbb{E}\left[f\left(\boldsymbol{Y}\right) - f\left(\boldsymbol{X}\right)\right]$, where $\widehat{\mathcal{Q}}$ is defined as

$$\widehat{\mathcal{Q}} := \frac{1}{\mathcal{N}} \sum_{r=1}^{\mathcal{N}} \left[f\left(\boldsymbol{Y}^{(r)}\right) - f\left(\boldsymbol{X}^{(r)}\right)\right]. \tag{6.4}$$

We will use variance reduction to ensure that $\widehat{\mathcal{Q}}$ is accurately estimated, even if $\mathcal{N}$ is small.

We now discuss two methods for producing low variance estimators: the Coupled finite difference method, described by Anderson [106], and the Common reaction path method, described by Rathinam et al. [104]. As with Chapter 5, we will distinguish clearly between different finite difference methods that make use of different variance-reducing, coupling techniques. As such, and to keep terminology consistent, we refer to the Coupled finite difference method as the SPM, and the Common reaction path method as the CPM.

---

[1] If $A = \alpha$ in System $X$, and $A = \alpha + \varepsilon$ in System $Y$, then the *forward difference* gives a bias of $\mathcal{O}(\varepsilon)$.

The SPM and CPM techniques described in this chapter are derived from the same fundamental hypotheses as the SPM and CPM methods presented in Chapter 5. There are, however, a number of important differences.

**The SPM and CPM with the tau-leap method.** In Chapter 5, the CPM and SPM were used to generate correlated sample paths, $\boldsymbol{Z}_{\tau_\ell}$ and $\boldsymbol{Z}_{\tau_{\ell-1}}$. The sample paths are generated by using the tau-leap approximation. Multiple reactions are fired at once. The parameters used to generate sample paths $\boldsymbol{Z}_{\tau_\ell}$ and $\boldsymbol{Z}_{\tau_{\ell-1}}$ are the same, but the tau-leap approximations use different time-steps.

**The SPM and CPM with an exact SSA.** In Sections 6.2.1 and 6.2.2 of this chapter, the SPM and CPM are used to generate correlated sample paths, $\boldsymbol{X}$ and $\boldsymbol{Y}$. The sample paths are generated with an exact SSA, which means that every change to $\boldsymbol{X}$ and $\boldsymbol{Y}$ is individually simulated. The parameters used to generate sample paths $\boldsymbol{X}$ and $\boldsymbol{Y}$ are, in line with Equation (6.3), different.

Therefore, the re-purposing of the SPM and CPM hypotheses will result in very different implementations, which we now set out.

### 6.2.1 Split propensity method

In this section, we discuss the SPM proposed by Anderson [106] for well-mixed systems. As described in Section 6.2, we consider Systems $X$ and $Y$ that differ only in their value for parameter $A$. Thus, for each event[2] $\zeta_j$ in the set of possible events $(\zeta_j)_{j \in 1,\ldots,J}$, let $\zeta_j^X$ refer to the event taking place in System $X$, and $\zeta_j^Y$ the same for System $Y$. Furthermore, suppose each $\zeta_j^X$ has propensity $a_j^X$, and, likewise, each $\zeta_j^Y$ has propensity $a_j^Y$. We will simultaneously simulate pairs of sample paths for Systems $X$ and $Y$, and will enumerate our samples as $[\boldsymbol{X}, \boldsymbol{Y}]^{(r)}$, where $r$ refers to the specific repeat. If $f(\cdot)$ represents the summary statistic of interest, we will use Equation (6.3) to produce partial derivative estimators for Equation (6.2).

---

[2]Recall that, from Section 2.5, the set $\left(\zeta_j\right)_{j\in\{1,\dots,J\}}$ includes the diffusion of particles, and the various reactions that can take place.

The SPM will be implemented with a suitable simulation algorithm, such as the DM or MNRM. We will simultaneously generate sample paths for Systems $X$ and $Y$, using either Algorithm 2.1 (the DM) or the Algorithm 2.2 (the MNRM).

At each time $t$ in the interval $[0, T]$, we consider four possibilities:

- no event takes place in either of Systems $X$ or $Y$;
- the event $\zeta_k$ takes place in both Systems $X$ and $Y$;
- the event $\zeta_k$ takes place in System $X$ only;
- the event $\zeta_k$ takes place in System $Y$ only.

As there are $J$ different events in total, there are $3 \cdot J$ possible ways in which the populations can change.

Stochastic simulation is carried out as follows. For each $\zeta_j$, where $j \in \{1, 2, \ldots, J\}$, define the following propensities:

$$b_j^C = \min\{a_j^X, a_j^Y\}; \quad b_j^X = a_j^X - b_j^C; \quad b_j^Y = a_j^Y - b_j^C. \quad (6.5)$$

The set $(b_j^Z)$, where $j \in \{1, 2, \ldots, J\}$ and $Z \in \{C, X, Y\}$, provides the propensities that we will simulate using the DM or MNRM. If the channel with propensity value $b_k^C$ fires, then event $\zeta_k$ takes place in both Systems $X$ and $Y$ (i.e. events $\zeta_k^X$ and $\zeta_k^Y$ take place). If the channel associated with propensity $b_k^X$ fires, then event $\zeta_k$ takes place in System $X$ only (i.e. event $\zeta_k^X$ takes place). Similarly, if the channel associated with propensity $a_k^Y$ fires, then System $Y$ is updated with event $\zeta_k$ (i.e. event $\zeta_k^Y$ takes place). An implementation of this simulation algorithm is provided as Algorithm 6.1.

Note that for each $j$, one of $b_j^X$ and $b_j^Y$ will be zero [106], and the other term should be much smaller than $b_j^C$. When the SPM is run, the channels with propensity of the form $b_j^C$ will typically fire far more often that the channels with propensities $b_j^X$ and $b_j^Y$. This means that, most of the time, events occur simultaneously in Systems $X$ and $Y$, and the sample paths should remain close together. In turn, this can lead to a low variance estimator. Note that, even though Algorithms 3.1 (the SPM for the multi-level method) and 6.1 are quite different, the arguments justifying their use as variance reduction techniques are essentially identical.

---

Algorithm 6.1: The SPM produces a pair of correlated sample paths: one for System $X$ and another for System $Y$.

---

**Require:** initial conditions, $\boldsymbol{X}(0) = \boldsymbol{Y}(0)$, parameters and terminal time, $T$.
 1: set $\boldsymbol{X} \leftarrow \boldsymbol{X}(0)$, and $\boldsymbol{Y} \leftarrow \boldsymbol{Y}(0)$
 2: **loop**
 3:     calculate propensity values $a_j^X$, $a_j^Y$ and thus $b_j^C$, $b_j^X$ and $b_j^Y$ (per (6.5))
 4:     choose the time to next event, $\Delta$, as $\Delta \sim \text{Exp}\left(\sum_j b_j^C + b_j^X + b_j^Y\right)$
 5:     **if** $t + \Delta > T$ **then**
 6:         break
 7:     **else** set $t \leftarrow t + \Delta$
 8:     **end if**
 9:     choose $a_k^Z$ in proportion to its value, with $k \in \{1, \ldots, J\}$ and $Z \in \{C, X, Y\}$
10:     **if** $Z \in \{C, X\}$ **then**
11:         set $\boldsymbol{X} \leftarrow \boldsymbol{X} + \boldsymbol{\nu}_k$
12:     **end if**
13:     **if** $Z \in \{C, Y\}$ **then**
14:         set $\boldsymbol{Y} \leftarrow \boldsymbol{Y} + \boldsymbol{\nu}_k$
15:     **end if**
16: **end loop**

---

## 6.2.2 Common process method

In this section, we describe the CPM described by Rathinam et al. [104]. Consider Systems $X$ and $Y$, where a chosen parameter $A = \alpha$ has been perturbed, and we are estimating the parametric sensitivity given by (6.3). We can use the Kurtz RTCR (see Equation (2.6)) to describe the time evolution of a pair of sample paths for Systems $X$ and $Y$. We write

$$\boldsymbol{X}(T) = \boldsymbol{X}(0) + \sum_{j=1}^{J} \mathcal{Y}_j \left(0, \int_0^T a_j^X(\boldsymbol{X}(t))\mathrm{d}t\right) \cdot \boldsymbol{\nu}_j,$$

$$\boldsymbol{Y}(T) = \boldsymbol{Y}(0) + \sum_{j=1}^{J} \mathcal{Y}_j \left(0, \int_0^T a_j^Y(\boldsymbol{Y}(t))\mathrm{d}t\right) \cdot \boldsymbol{\nu}_j,$$

where the $\mathcal{Y}_j$ are unit-rate Poisson processes[3]. The CPM produces a low variance estimate for Equation (6.3) by using the same set of Poisson processes[4], $\left(\mathcal{Y}_j\right)_{\{j \in 1,\ldots,J\}}$, for sample paths $\boldsymbol{X}$ and $\boldsymbol{Y}$.

This CPM scheme can be implemented by essentially running the MNRM algorithm (see Algorithm 2.2) twice. The procedure is as follows:

- firstly, simulate a sample path for System $X$ with the MNRM. The waiting times for each Poisson process, $\mathcal{Y}_j$, are recorded in an ordered list, $\mathcal{F}_j$ (see Algorithm 6.2);

- secondly, simulate System $Y$ with the MNRM method, but making use of the recorded waiting times: for Poisson process $\mathcal{Y}_j$, the waiting times are read from the list $\mathcal{F}_j$ (see Algorithm 6.3).

---

[3] Recall that a Poisson process $\mathcal{Y}(t)$ counts the number of arrivals that occur over the time-interval $(0, t]$, where the time between arrivals is exponentially distributed, with parameter 1. We can therefore identify a Poisson process with an ordered list of Exp(1) random variables.

[4] Equivalently, the same set of Exp(1) waiting times.

---

Algorithm 6.2: This simulates a path $\boldsymbol{X}$, and preserves the firing times of the underlying Poisson processes. This algorithm has been adapted from Algorithm 2.2.

---

**Require:** initial conditions, $\boldsymbol{X}(0)(= \boldsymbol{Y}(0))$ and terminal time, $T$.
1: set $\boldsymbol{X} \leftarrow \boldsymbol{X}(0)$, and set $t \leftarrow 0$
2: for each $\zeta_j$, set $A_j \leftarrow 0$, generate $T_j \leftarrow \text{Exp}(1)$, and store $T_j$ as the first element of list $\mathcal{F}_j$
3: **loop**
4:     for each $\zeta_j$, calculate propensity values $a_j(\boldsymbol{X}(t))$ and calculate $\Delta_j$ as

$$\Delta_j = \frac{T_j - A_j}{a_j}$$

5:     set $\Delta \leftarrow \min_j \Delta_j$, and $k \leftarrow \text{argmin}_j \Delta_j$
6:     **if** $t + \Delta > T$ **then**
7:         **break**
8:     **end if**
9:     set $\boldsymbol{X}(t+\Delta) \leftarrow \boldsymbol{X}(t) + \boldsymbol{\nu}_k$, set $t \leftarrow t + \Delta$, and for each $\zeta_j$, set $A_j \leftarrow A_j + a_j \cdot \Delta$
10:    generate $u \sim \text{Exp}(1)$, set $T_k \leftarrow T_k + u$ and append $u$ to the end of list $\mathcal{F}_k$
11: **end loop**

---

System $X$ is therefore simulated according to the pseudo-code provided in Algorithm 6.2; System $Y$ is then simulated according to Algorithm 6.3. The extension to RDME networks is straightforward [88].

Note that, in Chapter 5, we did not necessarily determine every inter-arrival time of the Poisson process, $\mathcal{Y}_j$. In this chapter, each reaction of Systems $X$ and $Y$ is fired individually, and, therefore, until the terminal time, $T$, is reached, each inter-arrival time of the Poisson process $\mathcal{Y}_j$ will need to be generated. Compared with the approaches demonstrated in Chapter 5, the algorithms of this chapter are simpler to state and implement.

### 6.2.3 Mathematically representing coupling methods

A useful relationship between the SPM and CPM framework has been rigorously derived by Anderson and Koyama [112]. A time mesh, $\pi = \{0 = s_0 < s_1, \ldots s_L = T\}$,

---

Algorithm 6.3: This simulates a path $\boldsymbol{Y}$, using the waiting times previously generated when simulating a path $\boldsymbol{X}$.

---

**Require:** initial conditions, $\boldsymbol{Y}(0) = \boldsymbol{X}(0)$, terminal time, $T$, and lists $\mathcal{F}_j$
1: set $\boldsymbol{Y} \leftarrow \boldsymbol{Y}(0)$, and set $t \leftarrow 0$
2: for each $\zeta_j$, set $A_j \leftarrow 0$
3: for each $\zeta_j$, set $T_j$ to be the first element of list $\mathcal{F}_j$, then delete the first element of $\mathcal{F}_j$
4: **loop**
5:     for each $\zeta_j$, calculate propensity values $a_j(\boldsymbol{Y}(t))$ and calculate $\Delta_j$ as

$$\Delta_j = \frac{T_j - A_j}{a_j} \tag{6.6}$$

6:     set $\Delta \leftarrow \min_j \Delta_j$, and $k \leftarrow \operatorname{argmin}_j \Delta_j$
7:     **if** $t + \Delta > T$ **then**
8:         **break**
9:     **end if**
10:    set $\boldsymbol{Y}(t+\Delta) \leftarrow \boldsymbol{Y}(t) + \boldsymbol{\nu}_k$, set $t \leftarrow t+\Delta$, and for each $\zeta_j$, set $A_j \leftarrow A_j + a_j \cdot \Delta$
11:    **if** $\mathcal{F}_k \neq \emptyset$ **then**
12:        let $u$ be the first element of $\mathcal{F}_j$: set $T_k \leftarrow T_k + u$, and then delete the first element of $\mathcal{F}_k$
13:    **else**
14:        generate $u \sim \text{Exp}(1)$, then set $T_k \leftarrow T_k + u$
15:    **end if**
16: **end loop**

---

is chosen and Equation (2.6) is then restated to represent the evolution of $\boldsymbol{X}$ and $\boldsymbol{Y}$ as

$$\boldsymbol{X}(T) = \boldsymbol{X}(0) + \sum_{j=1}^{J} \sum_{\ell=0}^{L-1} \mathcal{Y}_{j\ell}\left(0, \int_{s_\ell}^{s_{\ell+1}} a_j^X(\boldsymbol{X}(t))\mathrm{d}t\right) \cdot \boldsymbol{\nu}_j, \tag{6.7}$$

$$\boldsymbol{Y}(T) = \boldsymbol{Y}(0) + \sum_{j=1}^{J} \sum_{\ell=0}^{L-1} \mathcal{Y}_{j\ell}\left(0, \int_{s_\ell}^{s_{\ell+1}} a_j^Y(\boldsymbol{Y}(t))\mathrm{d}t\right) \cdot \boldsymbol{\nu}_j, \tag{6.8}$$

where $\mathcal{Y}_{j\ell}$ are independent, unit-rate Poisson processes. We do not provide a derivation here, but Anderson and Koyama [112] demonstrate that if we take $L = 1$, then the CPM is recovered. Furthermore, if the mesh is uniformly spaced then, as $L \to \infty$, the SPM is recovered [112]. The multi-choice method that we describe in Section 6.5 can be viewed as a method for dynamically choosing a time mesh for each individual Poisson process of Equations (6.7) and (6.8).

### 6.2.4 Comparing simulation methods

To illustrate the SPM and CPM simulation methods, we begin with an elementary example. The volume $\Omega$ is partitioned into $K = 101$ equally-sized voxels. A single particle is placed in the central voxel, and is allowed to diffuse. Zero-flux boundary conditions are implemented. We will study the effect of perturbing the diffusion coefficient of this particle. We make two copies of the model, which we label as Systems $X$ and $Y$. The jump rate for System $X$ is given by $d_x = 0.9$, and for System $Y$ by $d_y = 1.0$. We will work in non-dimensional time.

In order to apply the SPM and CPM, we will estimate $\mathbb{E}[f(\cdot)]$, where $f(\cdot)$ provides the voxel that contains the particle[5], at times $t \in \{1, 2, \ldots, 10^3\}$. As $\mathbb{E}[f(\boldsymbol{X})] = \mathbb{E}[f(\boldsymbol{Y})]$, the true value of $\mathbb{E}[f(\boldsymbol{Y}) - f(\boldsymbol{X})]$ will be zero. We will now compare and contrast the SPM and CPM simulation methods.

**Example 6.1.** The SPM is implemented first. Initially, our model can be described as[6] $X^{51} = Y^{51} = 1$. Thus, the particle can either jump to voxel $V^{52}$ (on the right-hand side of $V^{51}$) or to $V^{50}$ (on the left-hand side). This means that only the events $S^{51} \to S^{52}$ and $S^{51} \to S^{50}$ have non-zero propensities, and therefore the propensities for the combined process envisaged by the SPM (see Equations (6.5)), are given by

$$b^C = 0.9, \quad b^X = 0.0, \quad b^Y = 0.1.$$

There are therefore two possibilities for the first simulation event:

---

[5]For completeness, we note that $f(\boldsymbol{X}) = \sum_{k=1}^{K} k \cdot X^k(t)$.
[6]As there is only one species, we will suppress the subscript.

1. diffusion occurs in Systems $X$ and $Y$ (as $b^C = 0.9$ for both $S^{51} \to S^{52}$ and $S^{51} \to S^{50}$);

2. diffusion occurs only in System $Y$ (as $b^Y = 0.1$ for both $S^{51} \to S^{52}$ and $S^{51} \to S^{50}$).

Note that it is impossible, with this construction, for a diffusion event to occur only in System $X$ (as $a^X = 0$ for both $S^{51} \to S^{52}$ and $S^{51} \to S^{50}$). The propensity values show that possibility 2 occurs with probability 0.1. This could happen if, for example, the particle diffuses to the right in System $Y$ only. Then, the dynamics of System $X$ and $Y$ are no longer coupled, as $b^C = 0$ for all possible events $\zeta_j \in \{1, \ldots, J\}$. This decoupling continues until, after a random period of time, the particles of System $X$ and $Y$ occupies the same voxel. ∎

**Example 6.2.** We now implement the CPM. Consider System $X$. At the first simulation step, the particle can either diffuse to voxel $V^{52}$ (with event $S^{51} \to S^{52}$ taking place) or to $V^{50}$ (when event $S^{51} \to S^{50}$ takes place). According to Algorithm 2.2, the time to the next event is calculated using Equation 6.6, which gives $\Delta^X = \min\{\Delta^X_{S^{51} \to S^{52}}, \Delta^X_{S^{51} \to S^{50}}\}$, as the propensity values of all other events are zero. The appropriate event, $j$, is implemented at time $\Delta^X$, and this process is repeated as per Algorithm 2.2 until the terminal time. Now consider System $Y$. At the first simulation step, the time to the next event is given by: $\Delta^Y = \min\{\Delta^Y_{S^{51} \to S^{52}}, \Delta^Y_{S^{51} \to S^{50}}\}$. The Equations that describe the time to the next reaction for Systems $X$ and $Y$ are linearly scaled versions of one another. Thus, at each step, $\Delta^Y$ will take on a different value to $\Delta^X$, but the choice of event, $j$, will be the same. The CPM therefore ensures that events occur in the same order in System $X$ as in System $Y$ (that is, the same order of the particle moving left, right, left, left, and so on), but at different times. ∎

**Comparing numerical performance.** We are now in a position to compare the performance of the SPM and CPM. In Figure 6.1 we show estimates of the value of $\mathbb{E}[f(\boldsymbol{Y}) - f(\boldsymbol{X})]$. In this case, the CPM ensures that the sample values of $[f(\boldsymbol{Y}) - f(\boldsymbol{X})]$ are typically much closer to zero, the expected value, and are therefore more tightly coupled than the simulations produced with the SPM. One might think of System $X$ as a slower version of System $Y$. As the CPM uses the same randomness to simulate the $n$-th event in both Systems $X$ and $Y$ (even if that event takes place at different times in Systems $X$ and $Y$), it performs better than the SPM.

In the discussion (Section 6.6.2) we explain how the CPM is better at coupling samples paths than the SPM due to the different natural time-scales of Systems $X$ and $Y$.

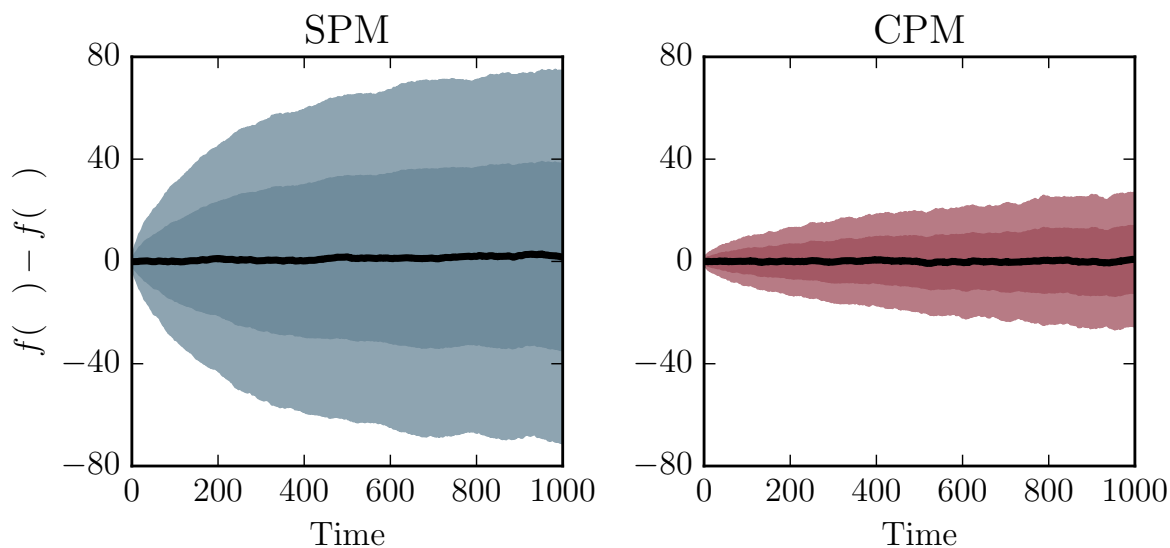

Figure 6.1: We compare the SPM (left) and CPM (right) schemes. Each diagram shows the mean value of $f(\boldsymbol{Y}) - f(\boldsymbol{X})$ at different times in black, one standard deviation from the mean in dark shading and two standard deviations in light shading. The estimator is as described in the text and $\mathcal{N} = 10^4$ simulations have been used to produce each plot. The CPM has a substantially lower variance at each time point.

## 6.3 Comparing the CPM and SPM techniques

In this section, we compare the CPM and SPM techniques across two case studies.

### 6.3.1 A simple reaction-diffusion model

**Case Study 8.** In this case study, we simulate a simple example of a reaction-diffusion network that comprises multiple particles. Suppose $\Omega$, of dimensions $L \times a \times a$, where $a \ll L$, is partitioned into $K = 101$ equally-sized voxels along the first dimension. The domain contains particles of species $S_1$ and $S_2$. Suppose particles of species $S_1$ each diffuse with rate $d = D \cdot K^2$, and react to produce particles of type $S_2$ in the following way:

$$S_1 + S_1 \xrightarrow{r} S_1 + S_1 + S_2. \tag{6.9}$$

We will take $d = 1$ and $r = 0.01$, and we work with non-dimensional time.

As an initial condition, let the central voxel have population $X_1^{51} = 250$, with all other voxels empty. In this example, the summary statistic we are looking to study is the expected total population of species $S_2$ at a terminal time $T$, which is given by summing the $S_2$ populations within each voxel. Thus take $f$ to be

$$f(\boldsymbol{X}) = \sum_{k=1}^{K} X_2^k(T), \tag{6.10}$$

so that $\mathcal{Q} = \mathbb{E}\left[f(\boldsymbol{X}(T))\right]$ and we choose $T = 50$. We will evaluate the parametric sensitivities

$$\left.\frac{\partial \mathcal{Q}}{\partial r}\right|_{r=0.01} \quad \text{and} \quad \left.\frac{\partial \mathcal{Q}}{\partial d}\right|_{d=1}, \tag{6.11}$$

with a suitable confidence interval length.

The finite difference scheme of Equation (6.3) is implemented to estimate the partial derivatives described by Equation (6.11). The numerical simulations can be simplified by not explicitly modelling the diffusion of the $S_2$ particles, because neither the summary statistic given in Equation (6.10), nor reaction (6.9) is affected by the diffusion of $S_2$. We start by estimating $\partial \mathcal{Q}/\partial r$. To do so, we need to choose the value of the simulation parameter, $\varepsilon$. Table 6.1 confirms that the SPM and CPM can both be used to accurately estimate $\partial \mathcal{Q}/\partial r$, and for each choice of $\varepsilon$ the SPM and CPM both produce simulations with roughly equal sample variances. The SPM and CPM techniques therefore require similar numbers of sample paths to produce estimates with the same confidence interval size.

|  | Parameter $\varepsilon$ | Sensitivity estimate | Mean of $[f(\boldsymbol{Y}) - f(\boldsymbol{X})]$ | Variance of $[f(\boldsymbol{Y}) - f(\boldsymbol{X})]$ | Simulations required |
|---|---|---|---|---|---|
| SPM | $2.50 \times 10^{-4}$ | $175882 \pm 502$ | 43.97 | 45.46 | 11098 |
| SPM | $5.00 \times 10^{-4}$ | $175709 \pm 496$ | 87.85 | 92.54 | 5774 |
| SPM | $7.50 \times 10^{-4}$ | $175844 \pm 486$ | 131.88 | 139.97 | 4053 |
| SPM | $10.00 \times 10^{-4}$ | $175509 \pm 501$ | 175.51 | 199.28 | 3055 |
| CPM | $2.50 \times 10^{-4}$ | $175754 \pm 500$ | 43.94 | 46.43 | 11422 |
| CPM | $5.00 \times 10^{-4}$ | $175586 \pm 498$ | 87.79 | 91.44 | 5675 |
| CPM | $7.50 \times 10^{-4}$ | $175052 \pm 498$ | 131.29 | 141.06 | 3888 |
| CPM | $10.00 \times 10^{-4}$ | $175321 \pm 498$ | 175.32 | 189.83 | 2944 |

Table 6.1: Estimated values for $\partial \mathcal{Q}/\partial r$ at $r = 0.01$, estimated using Equation (6.3). We have aimed to produce a confidence interval of semi-length 500. The sensitivities appear large: this is because we are working with dimensional quantities.

There are two further points to note. Firstly, as $\varepsilon \downarrow 0$ the sample variance of $[f(\boldsymbol{Y}) - f(\boldsymbol{X})]$, which we denote as $\sigma^2$, tends to zero. As we are estimating a partial derivative according to (6.3), the size of the confidence interval given by Equation (2.8) (with $\mathcal{N}$ fixed) scales as $\sigma/\varepsilon$. Table 6.1 shows that $\sigma$ decreases at a slower rate than $\varepsilon$, and consequently more simulations are required to produce estimates with a given confidence interval for smaller choices of $\varepsilon$. The second point we make is that the CPM and SPM are substantially more efficient than an uncoupled method. If $f(\boldsymbol{X})$ and $f(\boldsymbol{Y})$ were to be estimated independently, then we have $\sigma^2 \approx 8500$. If $\varepsilon = 2.50 \times 10^{-4}$, then the CPM and SPM each require approximately 185 times fewer sample paths than an uncoupled method would require for the same level of statistical accuracy.

We now consider the partial derivative $\partial \mathcal{Q}/\partial d$ evaluated at $d = 1$. Again, the finite difference scheme of Equation (6.3) is implemented, and Table 6.2 shows a range of estimates for $\partial \mathcal{Q}/\partial d$. As before, a range of choices for $\varepsilon$ are tested. In this case, the CPM produces a substantially lower estimator variance than the SPM, and should therefore be used in preference. The relative benefits of the CPM over the SPM are most noticeable when low bias estimates are required (equivalently, when $\varepsilon$ is small). With $\varepsilon = 2.50 \times 10^{-2}$ the CPM is 6.2 times more efficient (in terms of estimator variance) than the SPM, but when $\varepsilon = 10.00 \times 10^{-2}$, the CPM is only 2.6 times as efficient as the SPM method. The CPM is therefore particularly useful when low bias estimates for $\partial \mathcal{Q}/\partial d$ are required. Again, both the SPM and CPM are significantly more efficient than an uncoupled method.

|  | Parameter $\varepsilon$ | Sensitivity estimate | Mean of $[f(\boldsymbol{Y}) - f(\boldsymbol{X})]$ | Variance of $[f(\boldsymbol{Y}) - f(\boldsymbol{X})]$ | Sim's required |
|---|---|---|---|---|---|
| SPM | $2.50 \times 10^{-2}$ | $-874.67 \pm 10.00$ | -21.87 | 265.95 | 16342 |
| SPM | $5.00 \times 10^{-2}$ | $-871.86 \pm 9.90$ | -43.59 | 369.09 | 5783 |
| SPM | $7.50 \times 10^{-2}$ | $-877.42 \pm 10.10$ | -65.81 | 492.56 | 3296 |
| SPM | $10.00 \times 10^{-2}$ | $-888.00 \pm 10.06$ | -88.80 | 605.77 | 2298 |
| CPM | $2.50 \times 10^{-2}$ | $-874.97 \pm 10.14$ | -21.87 | 43.76 | 2618 |
| CPM | $5.00 \times 10^{-2}$ | $-885.38 \pm 10.05$ | -44.27 | 97.20 | 1479 |
| CPM | $7.50 \times 10^{-2}$ | $-877.60 \pm 9.78$ | -65.82 | 150.38 | 1073 |
| CPM | $10.00 \times 10^{-2}$ | $-881.90 \pm 9.57$ | -88.19 | 210.37 | 882 |

Table 6.2: Estimated values for $\partial \mathcal{Q}/\partial d$ at $d = 1$, estimated using Equation (6.3). We have aimed to produce a confidence interval of semi-length 10.00. The sensitivities appear large: this is because we are working with dimensional quantities.

In Section 6.6 we discuss the differences between the CPM and SPM, and consider intuitive reasons as to why the CPM provides better performance. We also compare the total CPU times required by the CPM and SPM.

## 6.3.2 A travelling wave-front

**Case Study 3.** We return to Case Study 3, which is a stochastic model of the Fisher-KPP wave, which has been used to model the spread of various biological populations [89]. We divide a volume $L \times a \times a$ into $K = 101$ equally-sized voxels along the first dimension. The particles, which are all of the same species, diffuse at a rate $d$ throughout the domain. Within a voxel, the particles interact through the following two reaction channels:

$$R_1 : S \xrightarrow{r_1} S + S; \quad R_2 : S + S \xrightarrow{r_{-1}} S.$$

In order to study this system, we place $10^4$ particles in the left-most voxel (formally, $X^1 = 10^4$), with the remaining voxels left empty. We take $d = 0.1$, $r_1 = 1$ and $r_{-1} = 0.01$; and generate paths until time $T = 25$. As the diffusion term, $d$, is non-zero, the particles will eventually be able to colonise the whole domain. We focus on two summary statistics of interest:

1. the expected total number of particles in the system at time $T = 25$. This is given by
$$\mathcal{Q}_1 = \mathbb{E}\left[\sum_{k=1}^{K} X^k\right]; \qquad (6.12)$$

2. the expected total number of voxels colonised by the population at time $T = 25$. This is evaluated as
$$\mathcal{Q}_2 = \mathbb{E}\left[\sum_{k=1}^{K} \mathbb{I}_{\{X^k > 0\}}\right]. \qquad (6.13)$$

Suppose that the diffusion term, $d$, is perturbed. As before, the sensitivity of summary statistics given by Equations (6.12) and (6.13), with respect to a changing diffusion constant, is to be estimated by Equation (6.3). In Table 6.3 we show estimated values for $\partial \mathcal{Q}_1 / \partial d$ (as per Equation (6.12)) and in Table 6.4 we show estimated values for $\partial \mathcal{Q}_2 / \partial d$ (as per Equation (6.13)). In both cases, the CPM outperforms the SPM.

|     | Parameter $\varepsilon$ | Sensitivity estimate | Mean of $[f(\boldsymbol{Y}) - f(\boldsymbol{X})]$ | Variance of $[f(\boldsymbol{Y}) - f(\boldsymbol{X})]$ | Simulations required |
|-----|---|---|---|---|---|
| SPM | $2.50 \times 10^{-3}$ | $6230 \pm 251$ | 15.57 | 2512.10 | 24464 |
|     | $5.00 \times 10^{-3}$ | $6406 \pm 251$ | 32.03 | 4809.75 | 11715 |
|     | $7.50 \times 10^{-3}$ | $6348 \pm 245$ | 47.61 | 6145.21 | 6991 |
|     | $10.00 \times 10^{-3}$ | $6220 \pm 250$ | 62.20 | 8478.50 | 5195 |
| CPM | $2.50 \times 10^{-3}$ | $6303 \pm 249$ | 15.76 | 591.91 | 5878 |
|     | $5.00 \times 10^{-3}$ | $6509 \pm 252$ | 32.54 | 1156.78 | 2800 |
|     | $7.50 \times 10^{-3}$ | $6317 \pm 250$ | 47.38 | 1571.85 | 1721 |
|     | $10.00 \times 10^{-3}$ | $6266 \pm 247$ | 62.66 | 1868.66 | 1176 |

Table 6.3: Estimated values for $\partial \mathcal{Q}_1 / \partial d$ at $d = 1$, estimated using (6.3). We have aimed to produce a confidence interval of semi-length 250. The sensitivities appear large: note this is because we are working with dimensional quantities.

In this section, we have shown that the optimal finite difference method depends on the model of interest. This is a departure from previous experience, which suggested that the SPM should be preferred [106]. We now discuss two novel simulation

|     | Parameter $\varepsilon$ | Sensitivity estimate | Mean of $[f(\boldsymbol{Y}) - f(\boldsymbol{X})]$ | Variance of $[f(\boldsymbol{Y}) - f(\boldsymbol{X})]$ | Simulations required |
|-----|---|---|---|---|---|
| SPM | $2.50 \times 10^{-3}$ | $71.83 \pm 2.51$ | $17.96 \times 10^{-2}$ | $40.17 \times 10^{-2}$ | 39140 |
|     | $5.00 \times 10^{-3}$ | $72.40 \pm 2.51$ | $36.20 \times 10^{-2}$ | $75.05 \times 10^{-2}$ | 18378 |
|     | $7.50 \times 10^{-3}$ | $72.63 \pm 2.48$ | $54.47 \times 10^{-2}$ | $98.12 \times 10^{-2}$ | 10936 |
|     | $10.00 \times 10^{-3}$ | $72.67 \pm 2.50$ | $72.67 \times 10^{-2}$ | $130.20 \times 10^{-2}$ | 7977 |
| CPM | $2.50 \times 10^{-3}$ | $74.34 \pm 2.51$ | $18.58 \times 10^{-2}$ | $16.44 \times 10^{-2}$ | 16051 |
|     | $5.00 \times 10^{-3}$ | $71.67 \pm 2.50$ | $35.84 \times 10^{-2}$ | $26.82 \times 10^{-2}$ | 6594 |
|     | $7.50 \times 10^{-3}$ | $71.25 \pm 2.49$ | $53.44 \times 10^{-2}$ | $32.86 \times 10^{-2}$ | 3634 |
|     | $10.00 \times 10^{-3}$ | $74.22 \pm 2.51$ | $74.22 \times 10^{-2}$ | $38.10 \times 10^{-2}$ | 2320 |

Table 6.4: Estimated values for $\partial \mathcal{Q}_2 / \partial d$ at $d = 1$, estimated using (6.3). We have aimed to produce a confidence interval of semi-length 250. The sensitivities appear large: this is because we are working with dimensional quantities.

strategies.

## 6.4 Grouped sampling method

This new implementation reduces the sample variance of Equation (6.3) for spatially-extended reaction networks. Consider again the SPM method, where we have enumerated the events that change the state matrix, Equation (2.24), as $\left(\zeta_j\right)_{j \in J}$. The propensity values of two systems, labelled as Systems $X$ and $Y$, are inserted into Equations (6.5), and a sample path for each system is then generated. We argued by example in Section 6.2.4 that the values given by Equation (6.5) are very sensitive to the exact location of particles, and so it can be difficult to generate tightly coupled sample paths. The Grouped sampling method (GSM) is designed to be less sensitive to the exact configuration of each of Systems $X$ and $Y$. By essentially re-ordering the key steps of the Next subvolume method [113], the GSM achieves a lower sample variance under a variety of circumstances.

We explain the GSM by first considering the simulation of a single system. We can partition the set of events $\left(\zeta_j\right)$ that change the state matrix into $(M+2)$ groups. The groups are as follows: $\Gamma_1$ contains all $R_1$ reaction events (so that there is an entry for each voxel, meaning $K$ events are contained in $\Gamma_1$), $\Gamma_2$ contains all $R_2$, ..., $\Gamma_M$ contains all $R_M$ reaction events, $\Gamma_{M+1}$ contains all diffusive jumps in which particles diffuse to the left, and, $\Gamma_{M+2}$ contains events where particles diffuse to the right. Furthermore, we order the events inside each group according to the voxel in which they take place (the importance of this will soon become clear). For example, an event that takes place in voxel $V^3$ will be 'next to' an event that takes place in voxel $V^4$. We can therefore simulate the events that take place in a single sample path by:

1. randomly selecting a group, $\Gamma_g$, where the probability of group $\Gamma_g$ being chosen is proportional to the sum of the propensities of the events inside that group;

2. randomly choosing an event $\zeta_k$ in group $\Gamma_g$, where the probability of $\zeta_k$ being chosen is proportional to its propensity value.

This approach clearly produces the same dynamics as Algorithms 2.1 or 2.2. We now show how to use this two-step method to simulate a correlated pair of sample paths. If we follow the two-step procedure, there are two opportunities to share random numbers between Systems $X$ and $Y$. At the first step outlined above, we effectively choose an event type: be it diffusion to the left, to the right, or a reaction of type $R_j$. This step will be accomplished by using the SPM described in Section 6.2.1. The SPM decides whether an event takes place in System $X$, $Y$ or both.

At the second step, we choose which voxel the event takes place in. This step will be performed with an inverse transform method[7]. This means that it is possible for the same event (e.g. diffusion to the left) to take place at the same time in both Systems $X$ and $Y$, even if that event takes place in a slightly different voxel (note that if we did not order the events inside each group, this would not be possible).

In order to use the SPM to perform step (1) of the simulation, for each group $\Gamma_g$, we define the following propensities

$$a_g^C = \min\left\{\sum_{\zeta_j \in \Gamma_g} a_j^X, \sum_{\zeta_j \in \Gamma_g} a_j^Y\right\}, \qquad a_g^X = \sum_{\zeta_j \in \Gamma_g} a_j^X - a_j^C, \qquad a_g^Y = \sum_{\zeta_j \in \Gamma_g} a_j^Y - a_j^C, \tag{6.14}$$

where $a_j^X$ and $a_j^Y$ refer to the propensity of event $\zeta_j$ in Systems $X$ and $Y$, respectively.

To perform step (2), we describe an inverse transform method, $\text{inv}(u, g)$, where $u$ is a uniformly generated $(0, 1)$ random variable, and $g$ refers to the index of the ordered group $\Gamma_g$. Pseudo-code is provided in Algorithm 6.4; we implement the algorithm in the next section.

---

[7]See Section 2.3 for further information.

---

Algorithm 6.4: The grouped sampling method is a two-step simulation method. We share randomness between Systems $X$ and $Y$ at each of the two steps. A group $\Gamma_g$ is chosen according to the SPM; the SPM also decides whether an event takes place in both Systems $X$ and $Y$, or only one. At the second step, the inverse transform method chooses the appropriate voxel.

---

**Require:** initial conditions, $\mathbf{X}(0) = \mathbf{Y}(0)$, terminal time, $T$, and event groups $(\Gamma_g)$.
1: set $\mathbf{X} \leftarrow \mathbf{X}(0)$, $\mathbf{Y} \leftarrow \mathbf{Y}(0)$ and $t \leftarrow 0$
2: **loop**
3:     for each $\Gamma_g$, calculate $a_g^C$, $a_g^X$ and $a_g^Y$ according to Equations (6.14)
4:     set $a_0 \leftarrow \sum_g \left( a_g^C + a_g^X + a_g^Y \right)$ and take $\Delta \leftarrow \mathrm{Exp}(a_0)$
5:     **if** $t + \Delta > T$ **then**
6:        break
7:     **end if**
8:     choose $(g, Z)$ with probability $a_g^Z / a_0$, where $Z \in \{C, X, Y\}$.
9:     **if** $Z = C$ **then**
10:        set $u \leftarrow U(0,1)$, and choose $k^X$, $k^Y$ with inverse transform method $\mathrm{inv}(u, g)$ (see main text)
11:        set $\mathbf{X} \leftarrow \mathbf{X} + \boldsymbol{\nu}_{k^X}$ and $\mathbf{Y} \leftarrow \mathbf{Y} + \boldsymbol{\nu}_{k^Y}$
12:     **else if** $Z = X$ **then**
13:        choose $k$ using $\mathrm{inv}(u, g)$, and set $\mathbf{X} \leftarrow \mathbf{X} + \boldsymbol{\nu}_k$
14:     **else if** $Z = Y$ **then**
15:        choose $k$ using $\mathrm{inv}(u, g)$, and set $\mathbf{Y} \leftarrow \mathbf{Y} + \boldsymbol{\nu}_k$
16:     **end if**
17:     set $t \leftarrow t + \Delta$
18: **end loop**

---

### 6.4.1 Illustrating grouped sampling

**Case Study 9.** The GSM is now tested and compared with the un-grouped SPM and CPM. We start by referring to Case Study 8, where species $S_1$ diffuses through a volume of $K = 101$ voxels, and reacts to form $S_2$ particles. In order to ensure that the simulation results presented in Table 6.1 and Table 6.2 are not simply a consequence of the symmetry of Case Study 8, we introduce stochastic drift into the model, so that the movement of the $S_1$ particles is biased to either the left or the right. Thus, for each voxel, $V^k$, the $S_1$ particles diffuse according to

$$S_1^k \xrightarrow{d_\ell} S_1^{k-1}; \qquad\qquad S_1^k \xrightarrow{d_r} S_1^{k+1}; \qquad (6.15)$$

where $d_\ell$ and $d_r$ are the appropriate biased transition rates. As before, we enforce zero-flux conditions. We take $d_\ell = 3.5$, $d_r = 1.0$ (meaning a net drift rate of 2.5), and $T = 50$. We re-use the remaining parameters and initial conditions from Case Study 8. We consider again the expected total number of $S_2$ particles in the system (given by Equation (6.10)), and we estimate the value of

$$\left.\frac{\partial \mathcal{Q}}{\partial d_\ell}\right|_{d_\ell=3.5}, \qquad (6.16)$$

where $\mathcal{Q} = \mathbb{E}[f(\boldsymbol{X})]$. The results of using the finite difference scheme given by Equation (6.11) are given in Table 6.5. The results show that the GSM is substantially more efficient than the SPM (requiring approximately three times fewer sample paths), and provides a modest improvement over the CPM (a saving of approximately 30%).

|  | Parameter $\varepsilon$ | Sensitivity estimate | Mean of $[f(\boldsymbol{Y}) - f(\boldsymbol{X})]$ | Variance of $[f(\boldsymbol{Y}) - f(\boldsymbol{X})]$ | Simulations required |
|---|---|---|---|---|---|
| SPM | 0.25 | $4589 \pm 10$ | 1147.27 | 6451.29 | 3962 |
|  | 0.50 | $4598 \pm 10$ | 2298.83 | 13269.47 | 2028 |
|  | 0.75 | $4616 \pm 10$ | 3462.18 | 18696.09 | 1271 |
| CPM | 0.25 | $4588 \pm 10$ | 1146.98 | 3424.79 | 2078 |
|  | 0.50 | $4593 \pm 10$ | 2296.26 | 6737.83 | 1040 |
|  | 0.75 | $4612 \pm 10$ | 3459.29 | 10126.12 | 685 |
| GSM | 0.25 | $4587 \pm 10$ | 1146.79 | 2237.95 | 1376 |
|  | 0.50 | $4599 \pm 10$ | 2299.58 | 4205.92 | 647 |
|  | 0.75 | $4612 \pm 10$ | 3458.76 | 7363.30 | 500 |

Table 6.5: Estimated values for $\partial \mathcal{Q}/\partial d_\ell$ at $d_\ell = 3.5$, estimated using Equation (6.3). We have aimed to produce a confidence interval of semi-length 10.

## 6.5 Multi-choice method

This is our second development. We have already demonstrated that the relative performance of the CPM and SPM methods depend on the problem to be investigated. In this section, we describe a hybrid switching scheme, which we call the Multi-choice (MC) method. The benefits of the multi-choice method are two-fold: firstly, it lets us choose the best coupling method (either the SPM or CPM) for each event; and, secondly, it allows us to change this decision dynamically. This flexibility enables an even greater reduction in sample variance. Our method has been designed without any specific problem in mind, but a heuristic justification for our approach is included within the discussion (see Section 6.6.3).

To improve on the SPM and CPM techniques, we will dynamically assign each event $\zeta_j$, $j \in \{1, \ldots, J\}$, to either the 'SPM part' or 'CPM part' of the algorithm, to determine how it should be simulated. This assignment may change, depending on the state matrix (2.24) of the system. To simplify matters, we will do this on a voxel-by-voxel basis, so that events with reactants in the same voxel will all be simulated with the SPM, or alternatively, all with the CPM. We first provide a broad overview of the method, and then explain how it might be implemented. Our method for switching between the SPM and CPM is intricate, because it is constructed in a way that ensures there is no additional bias introduced into the estimation of summary statistics. We return to this point in Section 6.5.2.

### 6.5.1 Switching between the SPM and CPM

We might think of using the voxel populations to decide whether the SPM or CPM should be used for each voxel, $V^k$, and then labelling our decision as $\Phi(X^k, Y^k)$. This suggests that every time the state matrix of System $X$ or $Y$ changes, we check if the new populations lead to a different choice. If necessary, we immediately change coupling methods. In Section 6.5.2, we show that this method will result in a further, uncontrolled bias in summary statistics.

We avoid biasing the statistics by implementing the following procedure. Consider only System $X$. For each value of $X^k$ (the population values in voxel $V^k$), we will determine which of the SPM or CPM is likely to be the better coupling method to implement (without explicitly considering $Y^k$). We label this method as $\Psi(X^k)$, and impose the method for voxel $V^k$ in System $X$. When the value of $X^k$ changes, we see if the coupling choice for voxel $V^k$ changes. The same procedure is followed for System $Y$ (with the label $\Psi(Y^k)$). When the populations $X^k$ and $Y^k$ both suggest the same method (either the SPM or CPM be used) for voxel $V^k$, then this method is implemented. However, there can be an interface period where the values of $X^k$ and $Y^k$ mean require different coupling methods to be implemented, and a bespoke simulation approach is needed for this *interface region*. This interface region is required to make sure that no bias is introduced. The upside is that in scenarios we have encountered, the size of the interface region is small relative to the time-scale of the sample path. We explain the details of the multi-choice method in two steps. We first describe the multi-choice method for just a single voxel. The second step is to implement the multi-choice method on a system with many voxels.

**Considering a single voxel**

Initially, suppose that $\Psi(X) = \Psi(Y) = \text{CPM}$. The CPM is implemented, and therefore:

- Algorithm 6.2 is used in System $X$ and Algorithm 6.3 in System $Y$.

Now suppose that $\Psi(X)$ changes to SPM, but $\Psi(Y)$ remains as the CPM. This is an interface region, which is characterised by System $X$ transitioning from the CPM to the SPM. Thus,

- Algorithm 2.2 is used in System $X$ and Algorithm 6.3 continues to be used in System $Y$.

Note that the lists of arrival times ($\mathcal{F}_j$ for $j \in \{1, \ldots, J\}$) generated by Algorithm 6.2 are still used at this step. Next, suppose that $\Psi(Y)$ changes to SPM. This means that we can couple the paths and

- Algorithm 6.1 is used for both Systems $X$ and $Y$.

The lists $(\mathcal{F}_j)$ for $j \in \{1, \ldots, J\}$ are deleted. Next, suppose $\Psi(X)$ changes to CPM, but $\Psi(Y)$ remains as the SPM. We are again in an interface region, and therefore

- Algorithm 6.2 is used in System $X$, and Algorithm 2.2 in System $Y$.

Finally, suppose that $\Psi(Y)$ changes to CPM. Full coupling of the paths can now be achieved, and:

- Algorithm 6.2 is used in System $X$; Algorithm 6.3 is restarted in System $Y$.

Note that Algorithm 6.3 uses new lists of arrival times $(\mathcal{F}_j)$. This scenario is graphically illustrated in Figure 6.2.

The above scenario described two kinds of interface scenarios: (1) a system has moved from the CPM to the SPM, with the other system to follow; and (2) a system has moved from the SPM to the CPM, with the other system to follow. This exhausts all possible ways in which an interface region can arise.

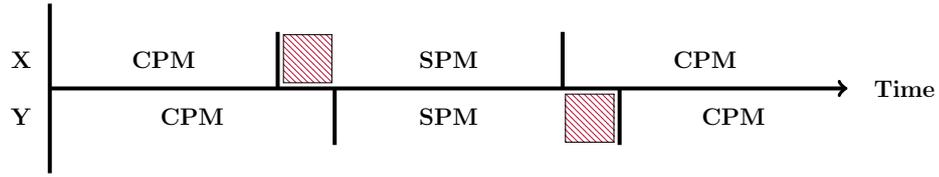

Figure 6.2: This diagram illustrates the multi-choice coupling method. The hatching illustrates an interface region: see Section 6.5.1 for further information.

**Considering multiple voxels**

The multi-choice method is now implemented across multiple voxels. There are multiple ways of achieving this, but we describe a method that is relatively easy to implement. Our procedure is to first simulate a sample path for System $X$ using Algorithm 6.2. We record the state matrix $\boldsymbol{X}$ at each time $t$, and save the firing times of the Poisson processes, $\mathcal{Y}_j$, used to generate the sample path for $X$. Now that sample path for System $X$ has been simulated, we must produce a sample path for System $Y$. For each voxel of System $Y$, there are three possible simulation algorithms:

1. $\Psi(X^k) = \Psi(Y^k) = $ CPM, so that Algorithm 6.3 is implemented;

2. $\Psi(X^k) \neq \Psi(Y^k)$, so either Algorithm 6.3 or Algorithm 2.2 is required;

3. $\Psi(X^k) = \Psi(Y^k) = $ SPM, so that the SPM must be used.

Recall that at this stage, we have fully simulated the sample path for System $X$. Cases (1) and (2) above are straightforward to deal with. To deal with the third possibility, we will need to reverse-engineer the SPM to deduce the sample path for System $Y$.

The SPM is reverse-engineered as follows: for each event that changes the population levels of System $X$, we need to decide whether it also takes place in System $Y$. At each simulation step, we compare propensities. If $a_j^X < a_j^Y$, then:

Algorithm 6.5: The SPM simulates path $\boldsymbol{Y}$ by sharing randomness with a previously generated path $\boldsymbol{X}$. This method produces the same dynamics as Algorithm 6.1.

**Require:** initial conditions, $\boldsymbol{Y}(0) = \boldsymbol{X}(0)$, terminal time, $T$, and lists $\mathcal{F}_j$
1: set $\boldsymbol{Y} \leftarrow \boldsymbol{Y}(0)$, and set $t \leftarrow 0$
2: for each $\zeta_j$, set $A_j^X \leftarrow 0$
3: for each $\zeta_j$, set $T_j^X$ to be the first element of list $\mathcal{F}_j$, then delete the first element of $\mathcal{F}_j$
4: for each $\zeta_j$, set $A_j^Y \leftarrow 0$, and generate $T_j^Y \leftarrow \text{Exp}(1)$
5: **loop**
6:     for each $\zeta_j$, calculate propensity values $a_j^X(\boldsymbol{X}(t))$, $a_j^Y(\boldsymbol{Y}(t))$
7:     for each $\zeta_j$ and $Z \in \{X, Y\}$, calculate $\Delta_j^Z$ as
$$\Delta_j^X = \frac{T_j^X - A_j^X}{a_j^X}, \qquad \Delta_j^Y = \frac{T_j^Y - A_j^Y}{\max\{0, a_j^Y - a_j^X\}}$$
8:     set $\Delta \leftarrow \min_{j,Z} \Delta_j^Z$ (where the minimum is over $Z$ and $j$), and let $k \leftarrow \text{argmin}_{j,Z} \Delta_j^Z$
9:     **if** $t + \Delta > T$ **then**
10:         break
11:     **end if**
12:     **if** $Z = X$ **then**
13:         **if** $a_j^X \leq a_j^Y$ **then**
14:             set $\boldsymbol{Y}(t+\Delta) \leftarrow \boldsymbol{Y}(t) + \boldsymbol{\nu}_k$
15:         **else if** $a_j^X > a_j^Y$ **then**
16:             with probability $a_j^y/a_j^X$, set $\boldsymbol{Y}(t+\Delta) \leftarrow \boldsymbol{Y}(t) + \boldsymbol{\nu}_k$
17:         **end if**
18:         set $t \leftarrow t + \Delta$
19:         **if** $\mathcal{F}_k \neq \emptyset$ **then**
20:             let $u$ be the first element of $\mathcal{F}_j$: set $T_k^X \leftarrow T_k^X + u$, and then delete $u$
21:         **else**
22:             generate $u \sim \text{Exp}(1)$, then set $T_k^X \leftarrow T_k + u$
23:         **end if**
24:     **else if** $Z = Y$ **then**
25:         set $\boldsymbol{Y}(t+\Delta) \leftarrow \boldsymbol{Y}(t) + \boldsymbol{\nu}_k$, and set $t \leftarrow t + \Delta$
26:         generate $u \sim \text{Exp}(1)$, then set $T_k \leftarrow T_k + u$
27:     **end if**
28:     for each $\zeta_j$, set $A_j^X \leftarrow A_j^X + a_j^X \cdot \Delta$ and set $A_j^Y \leftarrow A_j^Y + \max\{0, a_j^Y - a_j^X\} \cdot \Delta$
29: **end loop**

- any $\zeta_j$ that takes place in System $X$ necessarily also takes place in System $Y$ (see Equation (6.5), where $a_j^X = 0$); and

- it is possible for an event $\zeta_j$ to fire only in System $Y$ (in terms of Equation (6.5), $a_j^Y \geq 0$).

However, if $a_j^X > a_j^Y$:

- if an event $\zeta_j$ fires in System $X$, then it fires with probability $a_j^Y/a_j^X$ in System $Y$ (in terms of Equation (6.5), $a_j^C = a_j^Y$, $a_j^X \geq 0$ and $a_j^Y = 0$).

Please see Algorithm 6.5 for a pseudo-code implementation of the SPM.

Finally, we are in a position to describe the overall multi-choice method. The aforementioned algorithms are modifications of Algorithm 2.2, and so combining them into a single algorithm is natural. This technique is described in full in Algorithm 6.6.

### 6.5.2 A warning about model bias

In this section, we briefly explain why deciding on the coupling method based on the current populations $X^k$ and $Y^k$ together can lead to a model bias. Consider Systems $X$ and $Y$. For each voxel $V^k$, we might think of using the voxel populations to decide whether the SPM or CPM should be used for that voxel, and then labelling our decision as $\Phi(X^k, Y^k)$. Every time one of $X^k$ or $Y^k$ changes, $\Phi$ is re-evaluated. When our choice of $\Phi$ changes, we might then hope to immediately change the coupling method by relying on the memory-less property of exponential variates. Unfortunately, this implementation we have described leads to a model bias. Suppose that at time $t = 0$, $\Phi(X^k, Y^k) = \text{CPM}$, and so the events taking place in $V^k$ are simulated by explicitly considering the arrival times of Poisson processes (recall the System $X$ and $Y$ share Poisson processes). If an event fires at time $t = t^*$ that results in $\Phi(X^k, Y^k) = \text{SPM}$, we immediately switch to the SPM. Over the time-interval $(0, t^*]$ the Poisson processes associated with voxel $V^k$ might have fired a different number of times in Systems $X$

---

Algorithm 6.6: The multi-choice algorithm simulates path $Y$ by using randomness from path $\boldsymbol{X}$. Different simulation methods are used for different voxels.

---

**Require:** initial conditions, $\boldsymbol{Y}(0) = \boldsymbol{X}(0)$, terminal time, $T$, and complete details of $\boldsymbol{X}$
1: set $\boldsymbol{Y} \leftarrow \boldsymbol{Y}(0)$, and set $t \leftarrow 0$
2: for each $V^k$, let $M_k$ be simulation method implied by $\Psi(X^k)$ and $\Psi(Y^k)$;
3: **for** each voxel $V^k$ **do**
4:     configure $A_j$, $T_j$, etc. as appropriate per $M_k$
5: **end for**
6: **loop**
7:     **for** each voxel $V^k$ **do**
8:         calculate propensities, required internal values, set $\underline{\Delta}^k$ to be time to next event
9:     **end for**
10:     **if** $t + \min \underline{\Delta}^k > T$ **then**
11:         break
12:     **end if**
13:     set $t \leftarrow t + \min \underline{\Delta}^k$ and update $\boldsymbol{Y}$ per argmin $\underline{\Delta}^k$
14:     **for** each voxel $V^k$ **do**
15:         perform housekeeping as required by $M_k$
16:         recalculate $\Psi(X^k)$, $\Psi(Y^k)$ and so $M_k$ at time $t$
17:     **end for**
18: **end loop**

---

and $Y$. Let us suppose, without loss of generality, that the Poisson process $\mathcal{Y}_j$ fires more times in System $X$ than in System $Y$. By immediately switching to the SPM, we stop using the CPM, and so the firings of $\mathcal{Y}_j$ that have been ear-marked to occur in System $Y$, do not take place. The difficulty is that these ear-marked arrivals have already affected the value of $X^k$, thereby contributing to the choice $\Phi(X^k, Y^k) = \text{SPM}$. As the ear-marked values play a role in changing $\Phi$, when we observe the change in $\Phi$ we gain information as to the distribution of the ear-marked arrival times, and can no longer assume that they are exponentially distributed. We therefore cannot use the memory-less property on these arrival times without introducing a bias.

The multi-choice method will not bias model statistics for the following reason: when an individual system changes coupling methods (from the CPM to SPM, for example), this is done on the basis of the random numbers that have already been simulated and used in producing that sample path. The random numbers that will be simulated in future have no role in the coupling method changing, and we can therefore safely discard them.

### 6.5.3 A travelling wave

**Case Study 3.** We return again to a stochastic model of a Fisher-KPP wave. There are two distinct behaviours to consider. Between the wave-front and the left boundary, high molecular populations are maintained (and particle numbers exhibit quasi-steady-state dynamics). At the wave-front, diffusion drives the wave to the right. The colonisation of the domain is due to a small number of molecules jumping to the right. We postulate that the SPM will work better for simulating events in voxels that have been colonised (and are therefore characterised by high molecular populations). This follows as the SPM is a memory-less coupling method: see Section 6.6.2 for further information. The CPM should be preferred in the remaining voxels (that are characterised by low molecular populations). This is because the CPM considers the natural time-scale of both Systems $X$ and $Y$ as the CPM; again, see Section 6.6.2 for further information. Thus, we summarise our choice of coupling method, $\Psi$, as

$$\Psi(X^k) = \begin{cases} \text{CPM}, & \text{if } X^k \leq \alpha; \\ \text{SPM}, & \text{if } X^k > \alpha. \end{cases}$$

where $\alpha$ is a chosen threshold. We have worked with $\alpha = 67$, and will use this throughout the rest of this section. We have chosen $\alpha$ to be well away from the favourable state, but equally, not so low so that the benefits of the CPM cannot be realised. Further information as to the heuristics of choosing a coupling method are provided in the discussion. In our experience, the algorithm is not particularly sensitive to the precise choice of $\alpha$.

The multi-choice method is now implemented. The summary statistics of interest are the total number of particles (see (6.12)) and the number of non-empty voxels (see (6.13)). In Table 6.6 we set out the results of our investigation into the partial derivatives given by (6.12) and (6.13) with respect to a change in the diffusion term, $d$. We compare the simulation results in Table 6.3 and Table 6.4. In the case of the sensitivity of the total number of particles (see (6.12)), we see that the multi-choice method can be up to 5.6 times more efficient as the CPM, and 23 times more efficient as the SPM. When considering the sensitivity of the total number of voxels occupied (see (6.13)), we see that, as expected, the multi-choice method provides roughly equivalent performance. These speed-ups are shown in Figure 6.3.

## 6.6 Discussion

In this chapter, we have shown that the SPM and CPM techniques for estimating parametric sensitivities in well-mixed systems can be naturally extended to study spatially-inhomogeneous RDME models. Other researchers proceeded on the assumption that the SPM provides lower-variance estimates than the CPM, and should therefore be preferred [106]. We have shown that the relative performance of each method depends on the model of interest, as well as the summary statistics that are to be computed. In addition, we have presented two new simulation strategies: firstly, the GSM; and, secondly, the MC method that dynamically combines the SPM and CPM approaches. The efficiency of these novel methods have been demonstrated with numerical examples.

In the remainder of this chapter, we discuss a number of unresolved issues and challenges. We provide some intuition as to the circumstances under which the CPM outperforms the SPM, and when the grouped sampling or the multi-choice methods

|  | Parameter $\varepsilon$ | Sensitivity estimate | Mean of $[f(\boldsymbol{Y}) - f(\boldsymbol{X})]$ | Variance of $[f(\boldsymbol{Y}) - f(\boldsymbol{X})]$ | Simulations required |
|---|---|---|---|---|---|
| $\partial \mathcal{Q}_1/\partial d$ | $2.50 \times 10^{-3}$ | $6446 \pm 250$ | 16.12 | 107.22 | 1057 |
| | $5.00 \times 10^{-3}$ | $6657 \pm 252$ | 33.28 | 215.90 | 523 |
| | $7.50 \times 10^{-3}$ | $6572 \pm 260$ | 49.29 | 365.00 | 370 |
| | $10.00 \times 10^{-3}$ | $6066 \pm 241$ | 60.66 | 469.35 | 310 |
| $\partial \mathcal{Q}_2/\partial d$ | $2.50 \times 10^{-3}$ | $73.40 \pm 2.53$ | $18.35 \times 10^{-2}$ | $16.63 \times 10^{-2}$ | 16000 |
| | $5.00 \times 10^{-3}$ | $71.23 \pm 2.50$ | $35.62 \times 10^{-2}$ | $26.86 \times 10^{-2}$ | 6612 |
| | $7.50 \times 10^{-3}$ | $72.30 \pm 2.52$ | $54.22 \times 10^{-2}$ | $34.84 \times 10^{-2}$ | 3753 |
| | $10.00 \times 10^{-3}$ | $72.14 \pm 2.52$ | $72.14 \times 10^{-2}$ | $39.96 \times 10^{-2}$ | 2416 |

Table 6.6: Estimated values for $\partial \mathcal{Q}_1/\partial d$ and $\partial \mathcal{Q}_2/\partial d$ at $d = 1.0$, estimated using (6.3) and the multi-choice method. Appropriate confidence intervals have been constructed.

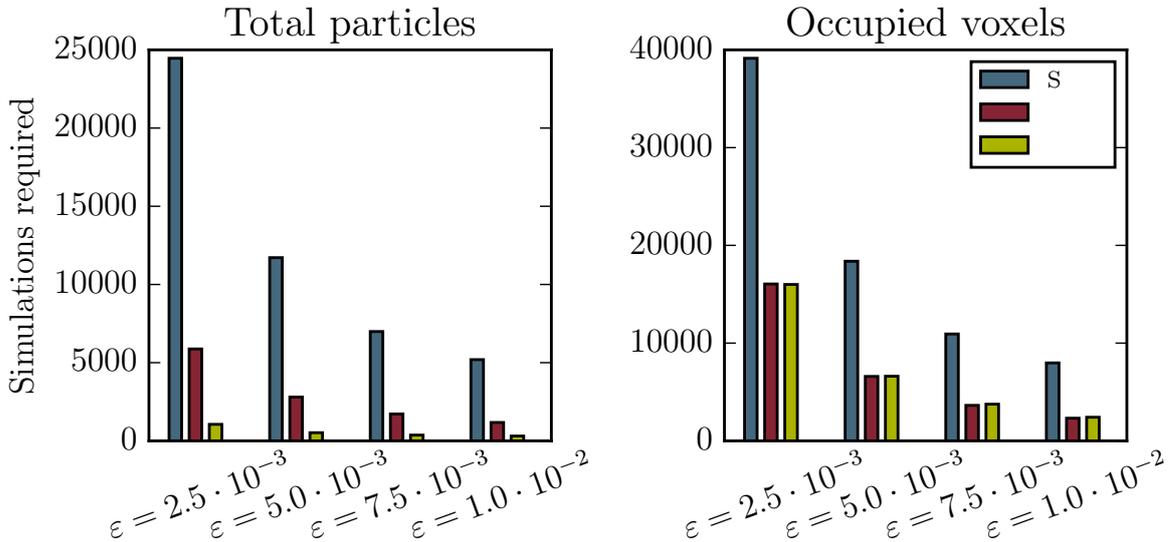

Figure 6.3: This shows the number of simulations required to estimate partial derivatives for the Fisher-KPP system, $\partial \mathcal{Q}_1/\partial d$ and $\partial \mathcal{Q}_2/\partial d$ at $d = 1.0$, for a constant level of statistical accuracy. For further information, see Table 6.3, Table 6.4, and Table 6.6.

are required. We then discuss a number of implementation issues.

## 6.6.1 Detailing the CPU times required by the SPM and CPM

Building on the material presented in Section 6.3.1, we return to Case Study 8 to illustrate the different CPU times taken by the SPM and CPM methods.

**Case Study 8.** We recall that Case Study 8 described a simple reaction-diffusion model. In Table 6.2, we presented estimated values for $\partial \mathcal{Q}/\partial d$, where $\mathcal{Q}$ counts a specific biological population (see Equation (6.10)), and $d = 1$. We argued that the CPM was more efficient, as it can achieve a given confidence interval with fewer simulations than the SPM method would require.

We have concerned ourselves with the number of simulations required to estimate the required sensitivities, and not the overall simulation time. The rationale for this approach is that the performance of the various simulation methods depends heavily on the particular implementation chosen; this is particularly evident where a spatially-extended model is considered. Therefore, considering the number of sample paths provides a safe means for benchmarking our results.

However, in Table 6.7, we show the CPU time required to generate a pair of sample paths, $[\boldsymbol{X}, \boldsymbol{Y}]^{(r)}$, using the SPM and the CPM method. For each choice of $\varepsilon$, we ran each of the SPM and CPM $\mathcal{N} = 1000$ times. We show the average CPU time required per pair of sample paths, together with the 10-th to 90-th percentiles. The CPM method requires approximately 16% more CPU time than the SPM method.

Table 6.8 shows the overall CPU time required to estimate $\partial \mathcal{Q}/\partial d$, as per Table 6.2. We show that, when comparing CPU times, the CPM outperforms the SPM, and that such effects are more noticeable when $\varepsilon$ (and therefore the bias) is small.

We now continue with a more general discussion about the SPM and CPM methods.

### 6.6.2 Intuitive differences between the SPM and CPM

Suppose that we wish to generate sample paths for Systems $X$ and $Y$ over a time-interval $[0, T]$. Informally, the SPM compares the propensities of each event, $\zeta_j$, in Systems $X$ and $Y$. If the propensities for event $\zeta_j$ are exactly the same, then we can insist that whenever $\zeta_j$ takes place in one of Systems $X$ or $Y$, it also takes place in the other system. If the propensities are different, then if $\zeta_j$ takes place in one system, it also takes place in the other system with some probability. If the propensities are similar, then the aforementioned probability will be high, and so we expect that the processes will be tightly coupled. This procedure is Markovian in the sense that it only depends on the current propensity values.

| Parameter, $\varepsilon$ | SPM – CPU time | |
|---|---|---|
| $2.50 \times 10^{-2}$ | $4.19 \times 10^{-2}$ s | $(4.14, 4.23) \times 10^{-2}$ s |
| $5.00 \times 10^{-2}$ | $4.25 \times 10^{-2}$ s | $(4.19, 4.31) \times 10^{-2}$ s |
| $7.50 \times 10^{-2}$ | $4.26 \times 10^{-2}$ s | $(4.22, 4.31) \times 10^{-2}$ s |
| $10.00 \times 10^{-2}$ | $4.27 \times 10^{-2}$ s | $(4.22, 4.32) \times 10^{-2}$ s |

| Parameter, $\varepsilon$ | CPM – CPU time | |
|---|---|---|
| $2.50 \times 10^{-2}$ | $4.99 \times 10^{-2}$ s | $(4.86, 5.13) \times 10^{-2}$ s |
| $5.00 \times 10^{-2}$ | $5.01 \times 10^{-2}$ s | $(4.88, 5.14) \times 10^{-2}$ s |
| $7.50 \times 10^{-2}$ | $5.06 \times 10^{-2}$ s | $(4.91, 5.21) \times 10^{-2}$ s |
| $10.00 \times 10^{-2}$ | $5.06 \times 10^{-2}$ s | $(4.93, 5.19) \times 10^{-2}$ s |

Table 6.7: The CPU time required to generate a pair of sample paths, $[\boldsymbol{X}, \boldsymbol{Y}]^{(r)}$, so that $\partial \mathcal{Q}/\partial d$ can be estimated. The CPM and SPM are tested; the SPM is slightly faster.

| Parameter, $\varepsilon$ | SPM | CPM | Extra time |
|---|---|---|---|
| $2.50 \times 10^{-2}$ | 684 s | 131 s | 423% |
| $5.00 \times 10^{-2}$ | 246 s | 74 s | 231% |
| $7.50 \times 10^{-2}$ | 140 s | 54 s | 159% |
| $10.00 \times 10^{-2}$ | 98 s | 45 s | 120% |

Table 6.8: The total CPU time required to estimate $\partial \mathcal{Q}/\partial d$ to within a confidence interval of semi-length 500.0 is shown. The overall CPU time expended by the CPM is lower than the CPU time required by the SPM.

The CPM is different. For each event, $\zeta_j$, a single, unit-rate Poisson process is used to simulate events for both Systems $X$ and $Y$. We determine the firing times of event $\zeta_j$ by keeping track of the internal times of each reaction channel (see Equation (2.6)), and comparing them with the firing times of each unit-rate Poisson process, $\mathcal{Y}_j$. The internal times depend on the entire history of the sample path, and not only the present value. The sequence of arrival times for $\mathcal{Y}_j$ is kept the same, and the CPM coupling method therefore uses the same arrival time for the $n$-th firing of the Poisson process. Unlike the SPM, this coupling is not explicitly time-based.

Sometimes the SPM and CPM techniques produce estimates for Equation (6.3) that have similar variances, as seen in Case Study 8 with a perturbed reaction rate, $r$. In cases where a perturbed parameter means one process has a different natural time-scale to the other, then the CPM provides better performance. In Case Study 8, when the diffusion term, $d$, was perturbed, System $X$ and $Y$ operated on different natural time-scales, with System $Y$ effectively a faster version of System $X$. A time-based coupling provides inferior performance. There are conditions under which the SPM is likely to outperform the CPM technique. In particular, where a steady state is expected, the SPM coupling is memory-less, which allows for mean-reversion effects. The summary statistic of interest will also have an effect on the choice between the SPM and CPM.

### 6.6.3 Justifying grouped sampling and multi-choice method

We illustrated the GSM with Case Study 9. In this model of biased diffusion, the GSM substantially reduced the sample variance compared with the SPM. The two-tier, GSM simulation procedure, meant that, as far as possible, the same ratio of left diffusion to right diffusion events could be maintained in both systems. By ensuring that the precise location of the diffusing particle is not as important as the direction in which the particle diffuses, a decreased variance was achieved.

The multi-choice method is useful for situations where there are substantial qualitative differences in stochastic behaviour in different voxels. With Case Study 3, the system dynamics in front of the wave-front are quite different to the dynamics behind the wave-front. The multi-choice method can choose between the SPM and CPM according to the stochastic behaviour of the particular sample path. The multi-choice method therefore explicitly accounts for the spatial variation inherent in problems modelled with the RDME by using different coupling methods for the events taking place in different voxels.

### 6.6.4 Outlook

The SPM and CPM can both provide accurate estimates of the parametric sensitivities of spatially-extended stochastic models. The grouped sampling and multi-choice extensions explicitly consider the characteristic dynamics of a spatially-extended network, thereby offering increased efficiency and flexibility. Future work should concentrate on rigorously evaluating which parameter sensitivity estimation method should be preferred for particular sets of circumstances.

Having completed this final investigative chapter, we turn to summarising the outcomes of this thesis.

# Chapter 7

# Discussion

Even a seemingly-innocuous chemical reaction network model might prove to be analytically intractable. Therefore, Monte Carlo methods often provide the only means by which model dynamics can be explored. Throughout this thesis, we have implemented a range of variance reduction methods to reduce the CPU time required to estimate summary statistics with a Monte Carlo approach. By building upon existing and well-understood simulation algorithms, we developed carefully-structured simulation methods of our own, making it easier to characterise and understand stochastic models of biological processes. We have focused our efforts on developing general tools that answer specific, testing questions about chemical reaction networks.

In this final chapter, we summarise the findings of this thesis, and suggest directions in which our work can be extended. In Section 7.1, the research findings of this thesis are discussed under two headings: first, we consider the multi-level method chapters (Chapters 3, 4 and 5), and then, we proceed to discussing our new parameter sensitivity methods (Chapter 6). In Section 7.2, we explain how our findings can be adapted and re-purposed to answer further, key questions in Systems Biology. Our final conclusions are stated in Section 7.3.

## 7.1 Review

This thesis started by describing the theoretical results that underpin the research chapters. In order to compare our new methods with traditional simulation techniques, in Chapter 2 we outlined a number of well-established and reliable Monte Carlo methods that generate sample paths of models described by the CME or RDME. The Kurtz RTCR was introduced; this representation of a chemical reaction network leads naturally to the CPM coupling methods implemented in Chapters 5 and 6. Crucially, we explained how the RTCR can describe a sample path generated with the tau-leap method, which is especially important for the algorithms described in Chapter 5.

In each of Chapters 3, 4, 5 and 6, we used variance reduction methods to efficiently estimate important summary statistics. In Chapters 3, 4 and 5, we focused on spatially-homogeneous multi-level methods. In Chapter 6, we shifted focus and considered a spatially-inhomogeneous model, for which it was necessary to estimate parametric sensitivities. We now discuss each of these two research focuses.

### 7.1.1 Multi-level methods

Chapter 3 started with an outline of the multi-level approach due to Giles [46]. The key research focus of this chapter was to establish confidence in the implementation of the multi-level method. We presented a number of sensible, practical and software orientated results that, together, render a more efficient and reliable algorithm. In particular, the dynamic calibration scheme presented as Algorithm 3.3 was to become especially important in subsequent chapters. We proceeded to compare algorithm performance across a range of test cases and algorithm parameters. In addition to studying the expected CPU time required by the multi-level method to estimate a summary statistic, we sought to understand how the CPU time required to generate the required sample paths might vary as the initial random number seed is changed. Finally, we tested the multi-level method with more complicated networks and summary statistics.

In Chapter 4, the adaptive multi-level method was described. As with the regular multi-level method, algorithm parameters need to be chosen before the adaptive multi-level method can be used. We highlight Section 4.5.2, which explains an automatic selection procedure for choosing the algorithm parameters, as one of the key contributions of this chapter. The automation procedure uses a topological, structured approach for optimising the CPU time by searching efficiently over a graph of possible algorithm parameters.

In Chapter 5, we developed the multi-level method in two directions: firstly, we implemented the CPM variance reduction method that couples two tau-leap sample paths more reliably; and, secondly, we compared our new coupling method with a multi-level approach that uses the R-leap method. The CPM we proposed has two steps: at the first step, a sample path is generated, and the Poisson processes used to generate the reactions are traced out and stored. At the second step, the second sample path is generated using the stored Poisson processes of the first path. We exploited the statistical properties of Poisson processes to avoid having to fire reactions individually, thereby ensuring the efficiency of our SSA. In the second strand of this chapter, we demonstrated that the multi-level approach is highly flexible and versatile, and that the R-leap method can be successfully incorporated into the multi-level framework.

### 7.1.2 Parameter sensitivity analysis

With Chapter 6, we continued with the theme of variance reduction methods, but our focus shifted onto spatially-inhomogeneous models. In this chapter, we explained how we can adapt existing finite difference schemes to robustly estimate parametric sensitivities in an RDME model. We showed that algorithmic performance depends on the dynamics of the given network and the choice of summary statistics. The key contribution of this chapter is the description of a hybrid technique that dynamically chooses the most appropriate simulation method for the network of interest.

## 7.2 Further directions and extensions

A number of further research directions and extensions are now discussed, and a range of important applications are highlighted.

### 7.2.1 Distribution construction

In this thesis, we have focused on estimating only point statistics of a chemical reaction network. For certain reaction networks of interest, point statistics provide limited information as to the system dynamics. In particular, for systems that comprise multiple favourable states, it might be necessary to adapt the multi-level method to generate an empirical probability distribution to describe the model.

**Example 7.1.** One particularly interesting challenge is to deal with systems that comprise multiple favourable states, such as the Schloegl system [114], for which the mean is a poor descriptor of the molecular populations. ∎

We mention two possible ways in which the multi-level method can be used to generate empirical probability distributions.

- the multi-level method can be used to estimate the $r$-th moment of a population of interest, $\mu_r = \mathbb{E}[X^r]$, for $r = 1, \ldots, C$. The moments can then be used to construct an empirical probability density function. For example, the Method of Moments can be used to generate such a distribution [115]. The choice of $C$ is non-trivial and must be carefully considered [116]. It is also possible to estimate the $r$-th central moment, $\mathbb{E}[(X - \mathbb{E}[X])^r]$;

- the multi-level method can be used to estimate a family of summary statistics, $\left(\mathcal{Q}_r\right)_{r \in 1, \ldots R}$, where
$$\mathcal{Q}_r := \mathbb{E}\left[\mathbb{I}_{\{\alpha_{r-1} \leq X < \alpha_r\}}\right].$$

An empirical histogram can then be constructed; the choices of $\alpha_0, \ldots, \alpha_R$ will determine the bin sizes of the histogram.

### 7.2.2 Multi-level model reduction methods

For certain chemical reaction networks, the network dynamics can be simplified, so that sample paths can be generated more efficiently. The network dynamics can be simplified in many ways. For example, instead of treating every reaction channel as a stochastic process, a subset of reaction channels can be modelled deterministically, meaning that it becomes unnecessary to generate random inputs for these channels. However, the deterministic description of a reaction channel will introduce a bias.

A multi-level model reduction method could proceed as follows: set $\ell = 0$, and decide that certain channels are labelled as 'deterministic' whilst others are 'stochastic'. As $\ell$ is increased, reaction channels are moved from the 'deterministic' set and into the 'stochastic' set. The process is repeated until no 'deterministic' channels remain. We provide an example that has been adapted from Chapter 3; an additional example can be found in Anderson and Higham [47].

**Example 7.2.** In the Michaelis-Menten system, the species are labelled as substrate ('S'), enzyme ('E'), complex ('ES') and product ('P'). The reaction channels are given as:

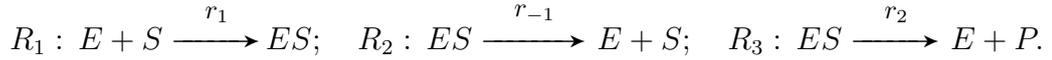

$$R_1 : E + S \xrightarrow{r_1} ES; \quad R_2 : ES \xrightarrow{r_{-1}} E + S; \quad R_3 : ES \xrightarrow{r_2} E + P.$$

The Michaelis-Menten assumption is to replace $R_1$, $R_2$ and $R_3$ with a *single* reaction channel, $R_*$, which is described in Equation (3.31). The single reaction is given as

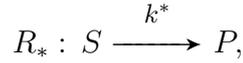

$$R_* : S \xrightarrow{k^*} P,$$

where $k^* = k_2 E_0 / (S + (r_{-1} + r_2)/r_1)$, and with $E_0$ representing the initial enzyme population. ∎

A numerical investigation shows that generating sample paths for the *complete* process with reactions $R_1, R_2, R_3$ takes a long time, whereas simulating the *reduced* process with reaction $R_*$ only is substantially faster. The following multi-level algorithm could therefore be implemented:

- on level $\ell = 0$, we estimate $\mathcal{Q}_0$ by simulating the reduced system (with $R_*$ only);

- on level $\ell = 1$, we estimate $\mathcal{Q}_1$ by comparing (a) the *reduced* system, with (b) the *complete* system (i.e. with $R_1, R_2, R_3$). We use the same unit-rate Poisson process for $R_3$ (in the complete, DM system) and $R_*$ (in the reduced system). Reactions $R_1$ and $R_2$ are left uncoupled.

### 7.2.3 Multi-level methods and Bayesian inference

In this thesis, we characterised the effects of changes in input parameters through the use of a parameter sensitivity analysis. That leaves the question of how one might derive model parameters from raw data as unanswered. Suppose we are to choose model parameters, $\theta$, based on observed data, $\mathcal{D}$. We therefore wish to know the probability of a parameter choice, $\theta$, given the data $\mathcal{D}$. This probability, $\mathbb{P}[\theta \mid \mathcal{D}]$, is known as the *posterior*. Bayes' Theorem expresses the posterior in terms of a *likelihood*, $\mathbb{P}[\mathcal{D} \mid \theta]$ and a *prior*, $\pi(\theta)$, as [58]

$$\mathbb{P}[\theta \mid \mathcal{D}] = \frac{\mathbb{P}[\mathcal{D} \mid \theta]}{\mathbb{P}[\mathcal{D}]} \cdot \pi(\theta). \tag{7.1}$$

The prior, $\pi(\theta)$, encodes any existing beliefs about the parameters, $\theta$. For a CME model, the likelihood, $\mathbb{P}[\mathcal{D} \mid \theta]$, is often analytically intractable, and alternative inference methods are required. We give one example of a likelihood-free method.

**Example 7.3.** Approximate Bayesian Computation (ABC) approximates the probability distribution $\mathbb{P}[\theta \mid \mathcal{D}]$ [117]. In its simplest form, the ABC algorithm proceeds by repeating the following steps:

- a parameter $\widehat{\theta}$ is sampled from the prior, $\pi(\cdot)$;

- synthetic data, $\overline{\mathcal{D}}$, is generated with an SSA by using parameter $\widehat{\theta}$;

- if $\mathcal{D}$ and $\overline{\mathcal{D}}$ are sufficiently similar, that is, for a comparison function $d(\cdot, \cdot)$ and tolerance, $\varepsilon$, $d(\mathcal{D}, \overline{\mathcal{D}}) < \varepsilon$, then $\widehat{\theta}$ is accepted, otherwise, it is rejected.

Thus, the accepted values of $\widehat{\theta}$ are sampled from the distribution $\mathbb{P}[\theta \mid d(\mathcal{D}, \overline{\mathcal{D}}) < \varepsilon]$. The comparison function, $d(\cdot, \cdot)$, needs to be carefully chosen; for example, $d(\cdot, \cdot)$, might compare a basket of summary statistics drawn from $\mathcal{D}$ and $\overline{\mathcal{D}}$ [118]. ∎

By approximating $\mathbb{P}[\theta \mid \mathcal{D}]$ with $\mathbb{P}[\theta \mid d(\mathcal{D},\overline{\mathcal{D}}) < \varepsilon]$, ABC methods have enjoyed particular success with CME and RDME models, but this has come at a high computational cost [119, 120].

The multi-level methodology can be used to accelerate ABC. Warne et al. [121] work with the CDF of the ABC posterior,

$$F_\varepsilon(\theta) = \int^\theta \mathbb{P}[\theta' \mid d(\mathcal{D},\overline{\mathcal{D}}) < \varepsilon]\,\mathrm{d}\theta'. \qquad (7.2)$$

As indicated, the CDF is parametrised by a choice of $\varepsilon$. Sufficiently many sample paths must be generated to estimate the integral contained in Equation (7.2). As $\varepsilon$ decreases, we expect the bias to decrease, but more sample paths, and consequently, more CPU time, is required to ensure statistical accuracy of our chosen CDF. A speed-up can be attained by using a multi-level style, hierarchy of choices of $\varepsilon$ for ABC computation [121].

## 7.3 Outlook

Computational resources are not limitless: only through the design and development of considered Monte Carlo simulation schemes can the best use of CPU time be made.

We have therefore sought to develop efficient, usable and modular simulation strategies to investigate stochastic models that arise in Systems Biology. Through our developments to the multi-level method, and our spatially-inhomogeneous parameter sensitivity method, we make it easier to design and interpret realistic biological models.

# Bibliography

[1] Brenner, S. Sequences and consequences. *Philosophical Transactions of the Royal Society of London B: Biological Sciences*, **365**(1537):207–212, 2010.

[2] Bruggeman, F. J. and Westerhoff, H. V. The nature of systems biology. *Trends in Microbiology*, **15**(1):45–50, 2007.

[3] Kell, D. B. Metabolomics and systems biology: making sense of the soup. *Current Opinion in Microbiology*, **7**(3):296–307, 2004.

[4] Kitano, H. Systems biology: a brief overview. *Science*, **295**(5560):1662–1664, 2002.

[5] Liepe, J., Filippi, S., Komorowski, M., and Stumpf, M. P. Maximizing the information content of experiments in systems biology. *PLoS Computational Biology*, **9**(1):e1002888, 2013.

[6] Westerhoff, H. V. and Palsson, B. O. The evolution of molecular biology into systems biology. *Nature Biotechnology*, **22**(10):1249–1252, 2004.

[7] Ideker, T., Galitski, T., and Hood, L. A new approach to decoding life: systems biology. *Annual Review of Genomics and Human Genetics*, **2**(1):343–372, 2001.

[8] Gardiner, C. W. *Handbook of Stochastic Methods*, volume 3. Springer, Berlin, 1985.

[9] Wilkinson, D. J. *Stochastic Modelling for Systems Biology*. CRC press, London, 2011.

[10] Elowitz, M. B., Levine, A. J., Siggia, E. D., and Swain, P. S. Stochastic gene expression in a single cell. *Science Signalling*, **297**(5584):1183, 2002.

[11] Hanna, J., Saha, K., Pando, B., Van Zon, J., Lengner, C. J., Creyghton, M. P., van Oudenaarden, A., and Jaenisch, R. Direct cell reprogramming is a stochastic process amenable to acceleration. *Nature*, **462**(7273):595–601, 2009.

[12] Thattai, M. and Van Oudenaarden, A. Intrinsic noise in gene regulatory networks. *Proceedings of the National Academy of Sciences*, **98**(15):8614–8619, 2001.

[13] Fedoroff, N. and Fontana, W. Small numbers of big molecules. *Science*, **297**(5584):1129–1131, 2002.

[14] Arkin, A., Ross, J., and McAdams, H. H. Stochastic kinetic analysis of developmental pathway bifurcation in phage $\lambda$-infected Escherichia coli cells. *Genetics*, **149**(4):1633–1648, 1998.

[15] Barrio, M., Burrage, K., Leier, A., and Tian, T. Oscillatory regulation of Hes1: discrete stochastic delay modelling and simulation. *PLoS Computational Biology*, **2**(9):e117, 2006.

[16] Lande, R., Engen, S., and Saether, B.-E. *Stochastic Population Dynamics in Ecology and Conservation*. Oxford University Press, Oxford, 2003.

[17] Székely, T., Burrage, K., Erban, R., and Zygalakis, K. C. A higher-order numerical framework for stochastic simulation of chemical reaction systems. *BMC Systems Biology*, **6**(1):85, 2012.

[18] Erban, R., Chapman, S. J., Kevrekidis, I., and Vejchodský, T. Analysis of a stochastic chemical system close to a SNIPER bifurcation of its mean-field model. *SIAM Journal on Applied Mathematics*, **70**(3):984–1016, 2009.

[19] Murray, J. D. *Mathematical Biology*. Springer, Berlin, 2002.

[20] Paulsson, J., Berg, O. G., and Ehrenberg, M. Stochastic focusing: fluctuation-enhanced sensitivity of intracellular regulation. *Proceedings of the National Academy of Sciences*, **97**(13):7148–7153, 2000.

[21] Hou, Z. and Xin, H. Internal noise stochastic resonance in a circadian clock system. *Journal of Chemical Physics*, **119**:11508, 2003.

[22] Erban, R. and Chapman, S. J. Stochastic modelling of reaction–diffusion processes: algorithms for bimolecular reactions. *Physical Biology*, **6**(4):046001, 2009.

[23] Higham, D. J. Modeling and simulating chemical reactions. *SIAM Review*, **50**(2):347–368, 2008.

[24] Gillespie, D. T., Hellander, A., and Petzold, L. R. Perspective: Stochastic algorithms for chemical kinetics. *Journal of Chemical Physics*, **138**(17):170901, 2013.

[25] Bernstein, D. Simulating mesoscopic reaction-diffusion systems using the Gillespie algorithm. *Physical Review E*, **71**(4):041103, 2005.

[26] van Zon, J. S. and Ten Wolde, P. R. Green's-function reaction dynamics: a particle-based approach for simulating biochemical networks in time and space. *Journal of Chemical Physics*, **123**(23):234910, 2005.

[27] Grimmett, G. and Stirzaker, D. *Probability and Random Processes.* Oxford University Press, Oxford, 2001.

[28] Jahnke, T. and Huisinga, W. Solving the chemical master equation for monomolecular reaction systems analytically. *Journal of Mathematical Biology*, **54**(1):1–26, 2007.

[29] Cotter, S. L., Vejchodský, T., and Erban, R. Adaptive finite element method assisted by stochastic simulation of chemical systems. *SIAM Journal on Scientific Computing*, **35**(1):8107–8131, 2013.

[30] Jahnke, T. and Huisinga, W. A dynamical low-rank approach to the chemical master equation. *Bulletin of Mathematical Biology*, **70**(8):2283–2302, 2008.

[31] Engblom, S. Spectral approximation of solutions to the chemical master equation. *Journal of Computational and Applied Mathematics*, **229**(1):208–221, 2009.

[32] Jahnke, T. and Udrescu, T. Solving chemical master equations by adaptive wavelet compression. *Journal of Computational Physics*, **229**(16):5724–5741, 2010.

[33] Gillespie, D. T. A general method for numerically simulating the stochastic time evolution of coupled chemical reactions. *Journal of Computational Physics*, **22**(4):403 – 434, 1976.

[34] Cao, Y., Gillespie, D. T., and Petzold, L. R. The slow-scale stochastic simulation algorithm. *Journal of Chemical Physics*, **122**(1):014116, 2005.

[35] Cotter, S. L. and Erban, R. Error analysis of diffusion approximation methods for multiscale systems in reaction kinetics. *SIAM Journal on Scientific Computing*, **38**(1):8144–8163, 2016.

[36] Cotter, S. L., Zygalakis, K. C., Kevrekidis, I. G., and Erban, R. A constrained approach to multiscale stochastic simulation of chemically reacting systems. *Journal of Chemical Physics*, **135**(9):094102, 2011.

[37] Duncan, A., Erban, R., and Zygalakis, K. Hybrid framework for the simulation of stochastic chemical kinetics. *Journal of Computational Physics*, **326**:398–419, 2016.

[38] Drawert, B., Engblom, S., and Hellander, A. URDME: a modular framework for stochastic simulation of reaction-transport processes in complex geometries. *BMC Systems Biology*, **6**(1):76, 2012.

[39] Fan, S., Geissmann, Q., Lakatos, E., Lukauskas, S., Ale, A., Babtie, A. C., Kirk, P. D., and Stumpf, M. P. MEANS: python package for Moment Expansion Approximation, iNference and Simulation. *Bioinformatics*, **32**(18):2863–2865, 2016.

[40] Liao, S., Vejchodský, T., and Erban, R. Tensor methods for parameter estimation and bifurcation analysis of stochastic reaction networks. *Journal of The Royal Society Interface*, **12**(108):20150233, 2015.

[41] Drawert, B., Hellander, A., Bales, B., Banerjee, D., Bellesia, G., Daigle Jr, B. J., Douglas, G., Gu, M., Gupta, A., Hellander, S., et al. Stochastic Simulation Service: Bridging the Gap between the Computational Expert and the Biologist. *PLoS Computational Biology*, **12**(12):e1005220, 2016.

[42] Klingbeil, G., Erban, R., Giles, M., and Maini, P. K. STOCHSIMGPU: parallel stochastic simulation for the systems biology toolbox 2 for Matlab. *Bioinformatics*, **27**(8):1170–1171, 2011.

[43] Klingbeil, G., Erban, R., Giles, M., and Maini, P. K. Fat versus thin threading approach on gpus: Application to stochastic simulation of chemical reactions. *IEEE Transactions on Parallel and Distributed Systems*, **23**(2):280–287, 2012.

[44] Glasserman, P. *Monte Carlo Methods in Financial Engineering*. Springer, New York, 2003.

[45] Gillespie, D. T. Approximate accelerated stochastic simulation of chemically reacting systems. *Journal of Chemical Physics*, **115**(4):1716–1733, 2001.

[46] Giles, M. B. Multilevel Monte Carlo path simulation. *Operations Research*, **56**(3):607–617, 2008.

[47] Anderson, D. F. and Higham, D. J. Multi-level Monte Carlo for continuous time Markov chains, with applications in biochemical kinetics. *SIAM Multiscale Modeling and Simulation*, **10**(1):146–179, 2012.

[48] Lester, C., Baker, R. E., Giles, M. B., and Yates, C. A. Extending the multi-level method for the simulation of stochastic biological systems. *Bulletin of Mathematical Biology*, **78**(8):1640–1677, 2016.

[49] Lester, C., Yates, C. A., Giles, M. B., and Baker, R. E. An adaptive multi-level simulation algorithm for stochastic biological systems. *Journal of Chemical Physics*, **142**(2):024113, 2015.

[50] Lester, C., Yates, C. A., and Baker, R. E. Efficient parameter sensitivity computation for spatially-extended reaction networks. *Journal of Chemical Physics*, **146**(4):044106, 2017.

[51] Lester, C., Yates, C. A., and Baker, R. E. Robustly simulating chemical reaction kinetics with multi-level Monte Carlo. *arXiv preprint arXiv:1707.09284*, 2017.

[52] Van Kampen, N. G. *Stochastic Processes in Physics and Chemistry*. Elsevier, Amsterdam, 1992.

[53] Fange, D., Berg, O. G., Sjöberg, P., and Elf, J. Stochastic reaction-diffusion kinetics in the microscopic limit. *Proceedings of the National Academy of Sciences*, **107**(46):19820–19825, 2010.

[54] Tian, T., Burrage, K., Burrage, P. M., and Carletti, M. Stochastic delay differential equations for genetic regulatory networks. *Journal of Computational and Applied Mathematics*, **205**(2):696–707, 2007.

[55] Gillespie, D. T. Stochastic Chemical Kinetics. In S. Yip, editor, *Handbook of Materials Modeling*, pages 1735–1752. Springer, Berlin, 2005.

[56] Kurtz, T. G. Representations of Markov processes as multiparameter time changes. *The Annals of Probability*, **8**(4):682–715, 1980.

[57] Norris, J. R. *Markov Chains*. Cambridge University Press, Cambridge, 1998.

[58] Garthwaite, P. H., Jolliffe, I. T., and Jones, B. *Statistical Inference*. Oxford University Press, Oxford, 2002.

[59] Gillespie, D. T. Exact stochastic simulation of coupled chemical reactions. *Journal of Physical Chemistry*, **81**(25):2340–2361, 1977.

[60] Cao, Y., Li, H., and Petzold, L. R. Efficient formulation of the stochastic simulation algorithm for chemically reacting systems. *Journal of Chemical Physics*, **121**(9):4059–4067, 2004.

[61] McCollum, J. M., Peterson, G. D., Cox, C. D., Simpson, M. L., and Samatova, N. F. The sorting direct method for stochastic simulation of biochemical systems with varying reaction execution behavior. *Computational Biology and Chemistry*, **30**(1):39–49, 2006.

[62] Li, H. and Petzold, L. R. Logarithmic direct method for discrete stochastic simulation of chemically reacting systems. Technical report, UCSB, 2006.

[63] Anderson, D. F. A modified next reaction method for simulating chemical systems with time-dependent propensities and delays. *Journal of Chemical Physics*, **127**(21):214107, 2007.

[64] Cao, Y., Gillespie, D. T., and Petzold, L. R. Efficient step size selection for the tau-leaping simulation method. *Journal of Chemical Physics*, **124**(4):044109, 2006.

[65] Gillespie, D. T. and Petzold, L. R. Improved leap-size selection for accelerated stochastic simulation. *Journal of Chemical Physics*, **119**(16):8229–8234, 2003.

[66] Moraes, A., Tempone, R., and Vilanova, P. Hybrid Chernoff tau-leap. *Multiscale Modeling and Simulation*, **12**(2):581–615, 2014.

[67] Yates, C. A. and Burrage, K. Look before you leap: a confidence-based method for selecting species criticality while avoiding negative populations in $\tau$-leaping. *Journal of Chemical Physics*, **134**(8):084109, 2011.

[68] Anderson, D. F. Incorporating postleap checks in tau-leaping. *Journal of Chemical Physics*, **128**(5):054103, 2008.

[69] Cao, Y., Gillespie, D. T., and Petzold, L. R. Adaptive explicit-implicit tau-leaping method with automatic tau selection. *Journal of Chemical Physics*, **126**(22):224101, 2007.

[70] Hammouda, C. B., Moraes, A., and Tempone, R. Multilevel hybrid split-step implicit tau-leap. *Numerical Algorithms*, **74**(2):527–560, 2017.

[71] Chatterjee, A., Vlachos, D. G., and Katsoulakis, M. A. Binomial distribution based $\tau$-leap accelerated stochastic simulation. *Journal of Chemical Physics*, **122**(2):024112, 2005.

[72] Tian, T. and Burrage, K. Binomial leap methods for simulating stochastic chemical kinetics. *Journal of Chemical Physics*, **121**:10356, 2004.

[73] Auger, A., Chatelain, P., and Koumoutsakos, P. R-leaping: accelerating the stochastic simulation algorithm by reaction leaps. *Journal of Chemical Physics*, **125**(8):084103, 2006.

[74] Kloeden, P. E. and Platen, E. Higher-order implicit strong numerical schemes for stochastic differential equations. *Journal of Statistical Physics*, **66**(1):283–314, 1992.

[75] Higham, D. J. An algorithmic introduction to numerical simulation of stochastic differential equations. *SIAM Review*, **43**(3):525–546, 2001.

[76] Li, T. Analysis of explicit tau-leaping schemes for simulating chemically reacting systems. *SIAM Multiscale Modeling and Simulation*, **6**(2):417–436, 2007.

[77] Williams, D. *Probability with Martingales*. Cambridge University Press, Cambridge, 1991.

[78] Thompson, R. N., Yates, C. A., and Baker, R. E. Modelling cell migration and adhesion during development. *Bulletin of Mathematical Niology*, **74**(12):2793–2809, 2012.

[79] Johnston, S. T., Baker, R. E., and Simpson, M. J. Filling the gaps: A robust description of adhesive birth-death-movement processes. *Physical Review E*, **93**(4):042413, 2016.

[80] Erban, R., Chapman, S. J., and Maini, P. K. A practical guide to stochastic simulations of reaction-diffusion processes. *arXiv preprint arXiv:0704.1908*, 2007.

[81] Hellander, S., Hellander, A., and Petzold, L. Reaction-diffusion master equation in the microscopic limit. *Physical Review E*, **85**(4):042901, 2012.

[82] Isaacson, S. A. A convergent reaction-diffusion master equation. *Journal of Chemical Physics*, **139**(5):054101, 2013.

[83] Cao, Y. and Erban, R. Stochastic Turing patterns: Analysis of compartment-based approaches. *Bulletin of Mathematical Biology*, **76**(12):3051–3069, 2014.

[84] Erban, R. and Chapman, S. J. Reactive boundary conditions for stochastic simulations of reaction–diffusion processes. *Physical Biology*, **4**(1):16, 2007.

[85] Erban, R. and Chapman, S. J. Time scale of random sequential adsorption. *Physical Review E*, **75**(4):041116, 2007.

[86] Berg, H. C. *Random Walks in Biology*. Princeton University Press, Princeton, 1993.

[87] Redner, S. *A guide to first-passage processes*. Cambridge University Press, Cambridge, 2001.

[88] Bauer, P. and Engblom, S. Sensitivity estimation and inverse problems in spatial stochastic models of chemical kinetics. In *Numerical Mathematics and Advanced Applications-ENUMATH 2013*, pages 519–527. Springer, 2015.

[89] Robinson, M., Flegg, M., and Erban, R. Adaptive two-regime method: application to front propagation. *Journal of Chemical Physics*, **140**(12):124109, 2014.

[90] Anderson, D. F., Higham, D. J., and Sun, Y. Complexity of multilevel Monte Carlo tau-leaping. *SIAM Journal on Numerical Analysis*, **52**(6):3106–3127, 2014.

[91] Moraes, A., Tempone, R., and Vilanova, P. Multilevel hybrid Chernoff tau-leap. *BIT Numerical Mathematics*, **56**(1):189–239, 2016.

[92] Moraes, A., Tempone, R., and Vilanova, P. A multilevel adaptive reaction-splitting simulation method for stochastic reaction networks. *SIAM Journal on Scientific Computing*, **38**(4):A2091–A2117, 2016.

[93] Bierig, C. and Chernov, A. Approximation of probability density functions by the Multilevel Monte Carlo Maximum Entropy method. *Journal of Computational Physics*, **314**:661–681, 2016.

[94] Gibson, M. A. and Bruck, J. Efficient exact stochastic simulation of chemical systems with many species and many channels. *Journal of Physical Chemistry A*, **104**(9):1876–1889, 2000.

[95] Tian, T. and Song, J. Mathematical modelling of the MAP kinase pathway using proteomic datasets. *PLoS One*, **7**(8):e42230, 2012.

[96] MacNamara, S., Bersani, A. M., Burrage, K., and SiDesmond Je, R. B. Stochastic chemical kinetics and the total quasi-steady-state assumption: application to the stochastic simulation algorithm and chemical master equation. *Journal of Chemical Physics*, **129**(9):095105, 2008.

[97] Huang, C.-Y. and Ferrell, J. E. Ultrasensitivity in the mitogen-activated protein kinase cascade. *Proceedings of the National Academy of Sciences*, **93**(19):10078–10083, 1996.

[98] Dijkstra, E. W. A note on two problems in connexion with graphs. *Numerische Mathematik*, **1**(1):269–271, 1959.

[99] Hart, P. E., Nilsson, N. J., and Raphael, B. A formal basis for the heuristic determination of minimum cost paths. *IEEE transactions on Systems Science and Cybernetics*, **4**(2):100–107, 1968.

[100] Snoek, J., Larochelle, H., and Adams, R. P. Practical Bayesian Optimization of Machine Learning Algorithms. In *Advances in Neural Information Processing Systems 25*, pages 2951–2959. Curran Associates, New York, 2012.

[101] Santner, T. J., Williams, B. J., and Notz, W. I. *The design and analysis of computer experiments*. Springer, 2013.

[102] Heiner, M., Gilbert, D., and Donaldson, R. Petri nets for Systems and Synthetic Biology. In *International School on Formal Methods for the Design of Computer, Communication and Software Systems*, pages 215–264. Springer, Berlin, 2008.

[103] Mayr, E. W. An algorithm for the general Petri net reachability problem. *SIAM Journal on Computing*, **13**(3):441–460, 1984.

[104] Rathinam, M., Sheppard, P. W., and Khammash, M. Efficient computation of parameter sensitivities of discrete stochastic chemical reaction networks. *Journal of Chemical Physics*, **132**(3):034103, 2010.

[105] Devroye, L. *Non-Uniform Random Variate Generation*. Springer, New York, 1986.

[106] Anderson, D. F. An efficient finite difference method for parameter sensitivities of continuous time Markov chains. *SIAM Journal on Numerical Analysis*, **50**(5):2237–2258, 2012.

[107] Plyasunov, S. and Arkin, A. P. Efficient stochastic sensitivity analysis of discrete event systems. *Journal of Computational Physics*, **221**(2):724–738, 2007.

[108] Sheppard, P. W., Rathinam, M., and Khammash, M. A pathwise derivative approach to the computation of parameter sensitivities in discrete stochastic chemical systems. *Journal of Chemical Physics*, **136**(3):034115, 2012.

[109] Morshed, M., Ingalls, B., and Ilie, S. An efficient finite-difference strategy for sensitivity analysis of stochastic models of biochemical systems. *Biosystems*, **151**:43–52, 2017.

[110] Thanh, V. H., Zunino, R., and Priami, C. Efficient Finite Difference Method for Computing Sensitivities of Biochemical Reactions. *arXiv preprint arXiv:1707.09193*, 2017.

[111] Gupta, A., Rathinam, M., and Khammash, M. Estimation of parameter sensitivities for stochastic reaction networks using tau-leap simulations. *arXiv preprint arXiv:1703.00947*, 2017.

[112] Anderson, D. F. and Koyama, M. An asymptotic relationship between coupling methods for stochastically modeled population processes. *IMA Journal of Numerical Analysis*, **35**(4):1757–1778, 2014.

[113] Elf, J. and Ehrenberg, M. Spontaneous separation of bi-stable biochemical systems into spatial domains of opposite phases. *IEE Proceedings-Systems Biology*, **1**(2):230–236, 2004.

[114] Vellela, M. and Qian, H. Stochastic dynamics and non-equilibrium thermodynamics of a bistable chemical system: the Schlögl model revisited. *Journal of The Royal Society Interface*, **6**(39):925–940, 2009.

[115] Andreychenko, A., Mikeev, L., and Wolf, V. Model reconstruction for moment-based stochastic chemical kinetics. *ACM Transactions on Modeling and Computer Simulation (TOMACS)*, **25**(2):12, 2015.

[116] Wilson, D. and Baker, R. E. Multi-level methods and approximating distribution functions. *AIP Advances*, **6**(7):075020, 2016.

[117] Toni, T., Welch, D., Strelkowa, N., Ipsen, A., and Stumpf, M. P. Approximate Bayesian computation scheme for parameter inference and model selection in dynamical systems. *Journal of the Royal Society Interface*, **6**(31):187–202, 2009.

[118] Harrison, J. U. and Baker, R. E. An automatic adaptive method to combine summary statistics in approximate Bayesian computation. *arXiv preprint arXiv:1703.02341*, 2017.

[119] Csilléry, K., Blum, M. G., Gaggiotti, O. E., and François, O. Approximate Bayesian computation (ABC) in practice. *Trends in Ecology & Evolution*, **25**(7):410–418, 2010.

[120] Del Moral, P., Doucet, A., and Jasra, A. An adaptive sequential Monte Carlo method for approximate Bayesian computation. *Statistics and Computing*, **22**(5):1009–1020, 2012.

[121] Warne, D. J., Baker, R. E., and Simpson, M. J. Accelerating computational Bayesian inference for stochastic biochemical reaction network models using multilevel Monte Carlo sampling. *bioRxiv: 064170*, 2016.